\documentclass[rmp,aps,twocolumn,nofootinbib,floatfix]{revtex4}
\usepackage{graphicx}
\usepackage{epsfig}

\begin{document}

\title{Magnetic dipole excitations in nuclei:
 elementary modes of nucleonic motion}

\author{Kris Heyde}\email{kris.heyde@ugent.be}
\affiliation{Department of Subatomic and Radiation Physics,
University of Gent, Proeftuinstraat 86, B-9000 Gent, Belgium}
\author{Peter von Neumann-Cosel} \email{vnc@ikp.tu-darmstadt.de}
\affiliation{Institut f\"ur Kernphysik, Technische Universit\"at
Darmstadt, Schlossgartenstrasse 9, D-64289 Darmstadt, Germany}
\author{Achim Richter}\email{richter@ikp.tu-darmstadt.de}
\affiliation{Institut f\"ur Kernphysik, Technische Universit\"at
Darmstadt, Schlossgartenstrasse 9, D-64289 Darmstadt, Germany}
\affiliation{ECT*, Villa Tambosi, I-38123 Villazzano (Trento), Italy}

\begin{abstract}
The nucleus is one of the most multi-faceted many-body systems in the
universe. It exhibits a multitude of responses depending on the way one
'probes' it. With increasing technical advancements of beams at the
various accelerators and of detection systems the nucleus has, over and
over again, surprised us by expressing always new ways of 'organized'
structures and layers of complexity. Nuclear magnetism is one of those
fascinating faces of the atomic nucleus we discuss in the present
review. We shall not just limit ourselves to presenting the by now very
large data set that has been obtained in the last two decades using
various probes, electromagnetic and hadronic alike and that presents
ample evidence for a low-lying orbital scissors mode around 3 MeV,
albeit fragmented over an energy interval of the order of $1.5$ MeV,
and higher-lying spin-flip strength in the energy region $5-9$ MeV in
deformed nuclei, nor to the presently discovered evidence for low-lying
proton-neutron isovector quadrupole excitations in spherical nuclei. To
the contrary, we put the experimental evidence in the perspectives of
understanding the atomic nucleus and its various structures of
well-organized modes of motion and thus enlarge our discussion to more
general fermion and bosonic many-body systems.
\end{abstract}

\date{\today}
\maketitle

\tableofcontents

\section{INTRODUCTION}
\label{sec:introduction}

\subsection{General remarks}
\label{generalremarks}

Nucleons moving inside the atomic nucleus naturally generate orbital
and spin magnetism. In certain mass regions - in particular for nuclei
between closed shells - the orbital magnetism can give rise to
cooperative effects between the many nucleons in the nucleus.
Collective modes might result from the out-of-phase motion of protons
and neutrons and of those, magnetic dipole modes at fairly low energies
($0\hbar\omega$ excitations) excited with electromagnetic probes are
one of the most pronounced ones. Besides these, at higher excitation
energies, cooperative effects may even lead to collective spin-flip
modes, as well as to even higher-lying genuine collective dipole modes
($2\hbar\omega$ excitations ) which so far have not even been seen
directly in experiments. In the present article, we start from a
succinct discussion of the early attempts in order to describe
nucleonic out-of-phase motion leading to magnetic collective
excitations (Sec.~\ref{sec:earlyth}) before entering into a discussion
of magnetic dipole excitations in heavy nuclei
(Sec.~\ref{sec:expheavy}). The experimental evidence that has been
accumulated over the years concerning the observation of a so-called
scissors mode in which the neutrons and protons in a deformed nucleus
perform small angle vibrations in a scissors-like motion with respect
to each other, using both electromagnetic and hadronic probes, is
listed. Theoretical concepts concerning the description of the
low-lying orbital scissors strength in even-even nuclei are presented.
Collective (geometric and algebraic) and microscopic (shell-model and
Quasi-Particle Random Phase Approximation (QRPA) studies) models are
discussed and they are related to one another in order to better
understand both the complementarity and the specific model effects. The
difficult problem of addressing the observed fragmentation of orbital
magnetic strength will be looked at also in the light of collective and
microscopic approaches.

The aspects related to the experimental evidence and the derivation of
a theoretical description of the concentration of spin-flip strength at
higher excitation energies is presented in Sec.~\ref{sec:expheavy} as
well, using mainly QRPA and shell-model calculations. In
Sec.~\ref{sec:magnetic-dipol-heavy} magnetic dipole excitations in
heavy odd-mass nuclei are discussed. There the problem of missing
strength in the measured spectra is looked at in some detail and the
need for a better theoretical description of the fragmented transition
strength is pointed out. Section \ref{sec:magnetic-dipol-light} deals
with experimental examples from magnetic dipole excitations in light
and medium-heavy nuclei and their theoretical treatment in terms of the
shell model and the QRPA. In Sec.~\ref{sec:isovector}, we discuss the
magnetic dipole isovector transitions in vibrational nuclei,
illuminating the intimate connection with the scissors mode typical to
rotational nuclei. In Sec.~\ref{sec:manybody}, we bring the former
discussion within a broader context of general many-body systems e.g.
deformed metallic clusters, quantum dots and scissors motion in trapped
Bose-Einstein condensates. In the final Sec.~\ref{sec:outlook},
conclusions and an outlook are given concerning the issue of magnetic
dipole excitations within the broader context of past and future
nuclear physics research. Relevant literature on the subject that
appeared until the end of 2009 has been considered in this review.

\subsection{Magnetic dipole response in atomic nuclei: a qualitative overview}
\label{intromagdipol}

Even on rather general grounds, one can make a fingerprint figure for
the magnetic dipole response in heavy rare-earth deformed nuclei on one
side and more spherical and light nuclei on the other side. We
illustrate the salient features of this response in two figures that
shall be referred to quite often in the present review.
\begin{figure}[tbh]
\includegraphics[angle=0,width=8.5cm]{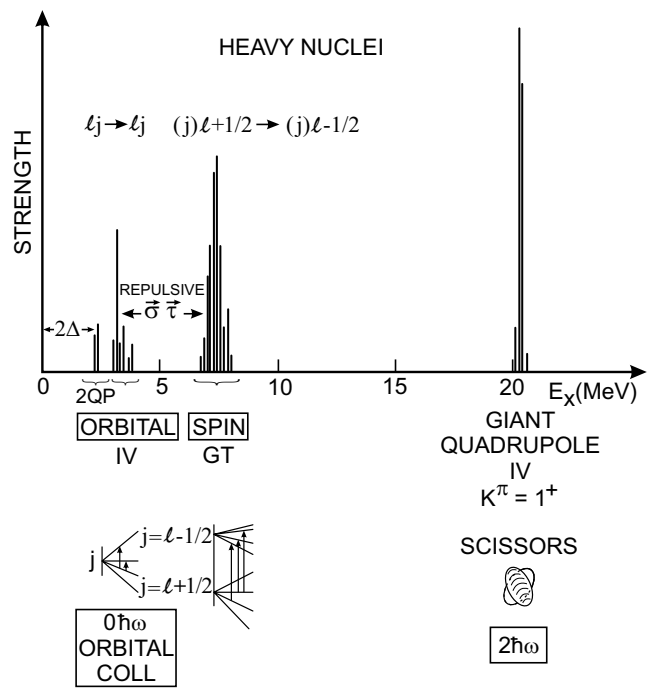}
\caption{Schematic illustration of the magnetic dipole strength
distribution in even-even heavy deformed nuclei and its model
character \cite{Richter:1990a}.}\label{Fig1}
\end{figure}

In even-even strongly-deformed (rare-earth and actinide) nuclei,
particularly due to the lifting of the spherical symmetry and the
associated degeneracy in the various $m$-components for a spherical
orbital with angular momentum $j$, one can separate four different
energies regions (see Fig.~\ref{Fig1}): \\
(i) At the excitation energy of about twice the pairing gap
($2\Delta\simeq2-2.5$ MeV),
two quasi-particle (2qp) $J^\pi = 1^{+}$ excitations show up with a
very specific shell-model structure, and thus, if these stay rather
pure, can be detected in electromagnetic decay and selective transfer
reactions. \\
(ii) At the excitation energy around 3 MeV, one observes a
concentration of orbital magnetic dipole strength, built up from
various 2qp configurations, into a weakly collective $0\hbar\omega$
mode, called the scissors mode. Here, a number of proton and neutron
2qp configurations $(lj)\rightarrow(lj)$ contribute in a more or less
coherent way, depending on external quantities like nuclear deformation
and the position of the Fermi level (number of protons $Z$ and the
number of neutrons $N$) in the Nilsson deformed single-particle
spectrum. \\
(iii) In the excitation energy interval of $6-8$ MeV, one
starts observing the spin Gamow-Teller giant resonance (with a bound
part, depending on the precise location in energy, and a resonance
part) resulting from particle-hole ($p-h$) excitations across the major
closed shells. In particular, shell-model transitions of the type
$(j=l+\frac{1}{2})\rightarrow(j=l-\frac{1}{2}$) play a major role. In
the rare-earth region, depending on the precise proton and neutron
number, the $1g_{9/2}\rightarrow1g_{7/2}$,
$1h_{11/2}\rightarrow1h_{9/2}$, and $1i_{13/2}\rightarrow1i_{11/2}$
transitions, respectively, contribute most to the $1^{+}$ spin mode.
Moreover, the residual
$\vec{\sigma}\cdot\vec{\sigma}\vec{\tau}\cdot\vec{\tau}$ repulsive part
of the effective nucleon-nucleon interaction concentrates the spin
strength from the lower-lying 2qp states around $2-4$ MeV into this
state. The final result is a large concentration of spin $M1$ strength.
(iv) Still higher in excitation energy, near 20 MeV, the $K^{\pi} =
1^{+}$ component of the isovector giant quadrupole resonance should
eventually show up. This particular mode, built in a microscopic way
from a coherent superposition of $2\hbar\omega$  configurations, has
originally been studied in macroscopic collective models and this state
would correspond to the 'real scissors mode' in strongly deformed,
rotational nuclei. Unfortunately, due to the very high excitation
energy and due to the fact that its transition strength vanishes at the
photon point because the $M1$ operator contains no radial dependence,
such a state in the continuum is very difficult to observe
unambiguously.

The excitation mechanisms described in (i)-(iv) should thus lead to the
typical fingerprint pattern of the magnetic dipole response in deformed
rare-earth nuclei, in the actinides and partly also in the medium-heavy
deformed nuclei in the $fp$-shell region as illustrated schematically
in Fig.~\ref{Fig1}. In the following discussion, we concentrate on
these various patterns, and give a thorough but succinct discussion of
the experimental facts validating this former, more 'idealized'
picture, emerging from the basic underlying shell structure in deformed
nuclei in combination with the essential multipole components of the
residual two-body interactions : the pairing component, the repulsive
$\vec{\sigma}\cdot\vec{\sigma}\vec{\tau}\cdot\vec{\tau}$ component, and
the long-range quadrupole force.
\begin{figure}[tbh]
\includegraphics[angle=0,width=8.5cm]{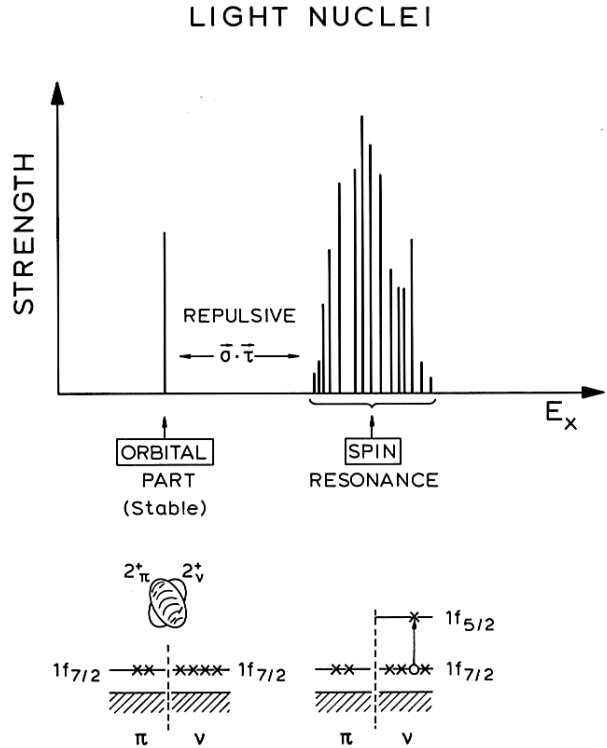}
\caption{Schematic representation of the magnetic dipole strength
distribution in even-even light nuclei and its interpretation in
terms of the shell model. The subscripts $\pi$ and $\nu$ denotes the
proton and neutron contribution, respectively
\cite{Richter:1990a}.}\label{Fig2}
\end{figure}

\begin{figure}[tbh]
\includegraphics[angle=0,width=8.5cm]{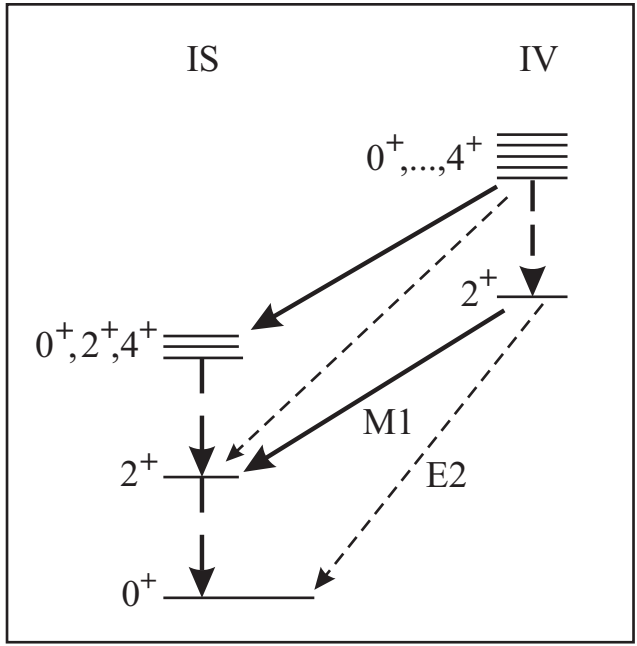}
\caption{Schematic representation of the $M1$
and associated $E2$ transitions of states with
symmetric (l.h.s.) and mixed-symmetric (r.h.s.) wave functions in
vibrational nuclei near closed shells.}\label{Fig3}
\end{figure}

In the spherical, lighter $fp$-shell nuclei, but also in the region of
lighter rare-earth nuclei (nuclei with neutron number $N$ near to 82,
and proton number $Z$ between 50 to 66), a somewhat different and
slightly more simple structure emerges (see Fig.~\ref{Fig2}): \\
(i) On the lower excitation energy side, one observes a rather stable
(in energy) predominantly orbital $1^{+}$ excitation which is produced
through coupling the lowest proton $2^{+}_{\pi}$ excitation with the
lowest neutron  $2^{+}_{\nu}$ excitation. Depending on the precise
shell structure, one can find also some higher-lying $4^{+}_{\pi}
\otimes 4^{+}_{\nu}$, $6^{+}_{\pi} \otimes 6^{+}_{\nu}$ fragments, in
general, with rapidly decreasing $M1$ excitation strength. This feature
is particularly evident in the light nuclei with protons and neutrons
filling the spherical $1f_{7/2}$ shell-model orbital. \\
(ii) Removed from this lower state by several MeV and near to 10 MeV of
excitation energy for the nuclei in the mass $A=50$ region, one finds
the spin Gamow-Teller giant resonance. Here, the repulsive spin-isospin
force component is responsible for a coherent state made of $1p-1h$
excitations, mainly. For nuclei around mass $A=50$, this will
predominantly be a $1f_{7/2}\rightarrow1f_{5/2}$ excitation, whereas in
the somewhat heavier mass $A=90$ region, one will encounter
$1g_{9/2}\rightarrow1g_{7/2}$ transitions. \\
The particular magnetic dipole response shown schematically in
Fig.~\ref{Fig2} forms the salient features amply discussed in the
present article when highlighting $J^{\pi} = 1^{+}$ states and the
associated $M1$ strength for light and medium-heavy even-even nuclei.
Besides the extensive exploration of the magnetic dipole response in
deformed nuclei and also in the region of light to medium-heavy nuclei,
uncovering the presence of a scissors mode (Sec.~\ref{intromagdipol}),
during the last decades, much progress has been made lately in the
study of proton-neutron $2^+$ excited states in vibrational and
transitional nuclei, corresponding with mixed-symmetry wave functions
in the proton and neutron building blocks. Here, small amplitude
quadrupole oscillations (phonons) dominate the low-energy nuclear
structure properties. Coupling the proton $2^{+}_{\pi}$ and neutron
$2^{+}_{\nu}$ phonons can result in multiphonon states where the
phonons move in phase (characterized by symmetric wave functions with
shorthand notation S), but also with out-of-phase motion (with
mixed-symmetry wave functions and shorthand notation MS). This is
illustrated in Fig.~\ref{Fig3}. Recent review articles
\cite{Kneissl:2006,Pietralla:2008} extensively discuss the regions in
which such $2^+$ states have been observed as well as present the
experimental techniques needed to characterize unambiguously those
isovector excitations (photon scattering, electron scattering, Coulomb
excitation, $\beta$-decay, inelastic neutron scattering and light-ion
fusion reactions as the major probes).

In view of the isoscalar and isovector components of the magnetic
dipole operator (see also Secs.~\ref{sec:th-mdr} and
\ref{sec:isovector} for its precise structure and a more detailed
discussion), the experimental identification of these isovector
excitations, shown schematically in Fig.~\ref{Fig3}, is
characterized by \\
(i) strong $M1$ transitions (matrix element of $\simeq 1\mu_N$) between
the MS and S states (thick full lines), \\
(ii) weakly-collective $E2$ transitions ( a few \% of the large
isoscalar $B(E2;2^+_1 \rightarrow 0^+_1)$ transition probability,
typical for vibrational and transitional nuclei) between the MS
and S states (thin dashed lines), and \\
(iii) strongly collective $E2$ transitions in between states with MS
character, with a strength typical for the collective E2 transitions
between symmetric states (thick dashed lines). \\
These particular fingerprints properties result from a systematic
exploration of isovector excitations in vibrational and transitional
regions during the last decade \cite{Pietralla:2008}, and clear-cut
evidence for the observation of the lowest $2^+_{ms}$ (both of a
one-phonon and two-phonon nature) as well as two-phonon $3^+_{ms}$ and
$1^+_{ms}$ isovector excitations exists now for the mass regions with
$Z\sim40,N\sim50$ and $Z\sim50,50<N<82$, rare-earth nuclei in the
region $54<Z\leq82,72<N\leq82$ (i.e., Xe,Ba,Ce,Nd and Sm nuclei), as
well as nuclei in the $A\sim60$ mass region, and heavy nuclei near
$^{208}$Pb.

\subsection{Scissors modes in other many-body systems}
\label{intromanybody}

In the introductory part to this review, we have pointed out that the
magnetic dipole response in strongly deformed atomic nuclei, is
characterized  by a clearly separated orbital scissors mode
(small-angle vibration of neutrons versus protons) at the lower
energies and at the higher energy side by a resonance-like structure
comprised of proton and neutron spin-flip excitations. In nuclei with
vibrational and transitional character, near to closed shells,
mixed-symmetry excitations result from the isovector coupling of the
lowest one-phonon proton and neutron $2^+_1$ excitations. The essential
ingredient in both cases is the presence of two distinguishable
components, proton and neutron fluids, in the atomic nucleus.

It turns out that in other many-body systems very similar rotational
oscillatory scissors motion has been discussed and, in certain cases,
also been observed experimentally. The presence of a two-component or
two-fluid quantum system is essential in this respect. In
Sec.~\ref{sec:manybody}, we address the main results obtained in the
study of (i) metallic clusters \cite{Lipparini:1989b,Nesterenko:1999},
(ii) elliptically deformed quantum dots
\cite{Serra:1999,Alhassid:2000}, and (iii) the oscillatory behavior
induced by the rotation of the atomic cloud in a deformed trap and the
corresponding superfluid effects caused by Bose-Einstein condensation
\cite{Guery-Odelin:1999,Marago:2000,Marago:2001}, as well as their
connections with the study of magnetic dipole excitations in atomic
nuclei. The underlying common mechanism as well as some typical
illustrations will be presented too.

\section{EARLY THEORETICAL SUGGESTIONS FOR THE MAGNETIC SCISSORS MODE
AND ITS EXPERIMENTAL DISCOVERY}
\label{sec:earlyth}

\subsection{Overview: a piece of history}
\label{sec:overview}

Low-energy collective modes in atomic nuclei, for both spherical and
deformed nuclei, displaying  nuclear density oscillations or more
permanently deformed structures of the density, have been rather well
described within the Bohr-Mottelson model \cite{Bohr:1975}. In those
excitations, both the proton density $\rho_{\pi}(\vec{r})$ and the
neutron density $\rho_{\nu}(\vec{r})$ exhibit variations that act in
phase, i.e. isoscalar collective modes are obtained. It was rather soon
realized that, besides these symmetric collective modes, non-symmetric
density variations might show up too. The latter excitations are
expected to occur at much higher excitation energies though because of
the symmetry energy term, coupling the proton and neutron density
oscillations preferentially in a symmetric way. The giant
electric-dipole mode - which in even-even nuclei excites negative
parity states - is the best known and well-documented example for such
isovector excitations, the mode in which the centre-of-mass for the
charge and mass distributions do not coincide but perform an
out-of-phase motion around the equilibrium value.

Non-symmetric collective modes were considered quite early by
\textcite{Greiner:1965,Greiner:1966} and \textcite{Faessler:1966},
independently, at the end of the 60's for spherical nuclei and this on
the basis of isovector quadrupole collective excitations. Somewhat
later but again, about at the same time and independently,
\textcite{Hilton:1976}, \textcite{Suzuki:1977}, and
\textcite{LoIudice:1978,LoIudice:1979} suggested an extension of the
Bohr-Mottelson description of rotational motion. The latter treated the
nucleus in terms of a geometrical two-rotor model (TRM), in which a
collective magnetic dipole ($M1$) mode could be formed by a rotational
oscillation of the proton versus the neutron deformed density
distribution (or fluids giving a rotational flow that is strongly
excited as an orbital magnetic excitation). The name 'scissors mode'
was suggested much later after the experimental discovery of this mode
originating in the peculiar nature of its geometrical picture
\cite{Richter:1983,Bohle:1984a}.

At that time, these non-symmetric excitations, in particular the
scissors mode in deformed systems, were a mere theoretical suggestion
built on the strong fundament in describing nuclear collective motion
(both for spherical and deformed nuclei). A determining factor in order
to estimate the excitation energy and the strength was of course the
knowledge of the symmetry energy connected to the non-symmetric motion.
In the early calculations, the full symmetry energy term, known from a
liquid-drop model treatment of global nuclear structure properties, was
considered giving rise to energies and $B(M1)$ values much too large.
It was only when the excitation energy of the scissors mode (having
both a mass parameter and a restoring force strength) was adjusted to
the observed experimental low-lying $B(E2)$ values including also
pairing correlation in the protons and neutrons participating in the
scissors motion that excitation energies of about $3-5$~MeV
\cite{Suzuki:1977,DeFranceschi:1983,DeFranceschi:1984,Lipparini:1983}
but still fairly high $B(M1)$ values of $9-18$~$\mu^{2}_{N}$ were
obtained \cite{Lipparini:1983,DeFranceschi:1983,DeFranceschi:1984}.

Strong support for the above idea came from a different way of treating
collective modes of motion in the nuclear many-body problem. Working in
an algebraic framework and using the concepts of symmetry,
\textcite{Arima:1975a,Arima:1975b} formulated a model in which the
interacting fermion problem is replaced by an interacting boson
problem, only considering $s$ ($L=0$) and $d$ ($L=2$) boson degrees of
freedom (interacting boson model, IBM). It was known from standard
shell-model two-body interactions that the $0^{+}$ coupled pair state
and, to a much lesser extent, the $2^{+}$ coupled pair dominate the
binding of two-nucleon systems. By studying the symmetries of such an
interacting boson model with the U(6) symmetry (the symmetry describing
an interacting system of $s$ and $d$ bosons) and at the same time
incorporating the proton and neutron degrees of freedom (IBM-2), a
class of states with non-symmetric spatial and also non-symmetric
charge structure (mixed-symmetry states) showed up naturally. By
fitting the parameters in the IBM-2 model to known fermionic
properties, a $B(M1)$ value for a scissors like mode of about
$2-3$~$\mu^{2}_{N}$ has been predicted \cite{Iachello:1981}.

\subsection{Experimental discovery}

Both the predictions of \textcite{LoIudice:1978,LoIudice:1979} within
the TRM and of  \textcite{Iachello:1981} and \textcite{Dieperink:1983}
within the IBM-2 formed the essential basis for high-resolution
inelastic electron scattering experiments at the Darmstadt Electron
Linear Accelerator (DALINAC) to search for this mode. And indeed after
great experimental efforts it has first been detected in the strongly
deformed nucleus $^{156}$Gd with essentially all the properties that
were predicted \cite{Richter:1983}. \label{sec:expdis}
\begin{figure}[hb]
\includegraphics[angle=90,width=8.5cm]{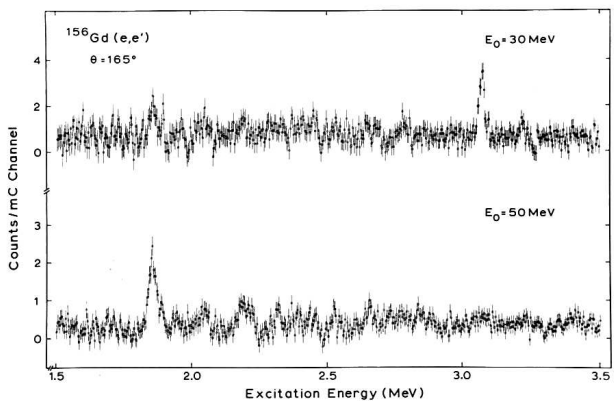}
\caption{Backward angle $^{156}${Gd}$(e,e')$ spectra indicating the
almost uniform excitation of many-low spin states in the measured
energy region except for a strongly excited $J^{\pi} = 1^{+}$ state
seen in the 30 MeV spectrum at $E_{x}$ = 3.075 MeV and a known
$J^{\pi} = 3^{-}$ state at $E_{x}$ = 1.825 MeV in the 50 MeV
spectrum. The excitation of the state at $E_{x}$ = 3.075 MeV is due
to the scissors mode \cite{Bohle:1984a}.}\label{Fig4}
\end{figure}
Figure \ref{Fig4} from the work of \textcite{Bohle:1984a} displays two
of the original spectra measured at a backward angle of $\theta =
165^{\circ}$ where magnetic excitations are expected to show up
strongly. The spectrum, taken at the low bombarding energy E$_{0}$ = 30
MeV (corresponding to a low momentum transfer), reveals a rich fine
structure of excited low-spin states with known experimental levels up
to about 2.5 MeV of excitation energy. The only strong transition is to
a state at E$_{x}$ = 3.075 MeV. This state is almost absent in the
E$_{0}$ = 50 MeV spectrum (i.e. at a higher momentum transfer), in
which, however, a well known collective $J^{\pi} = 3^{-}$ state at
E$_{x}$ = 1.852 MeV is strongest. The state at E$_{x}$ = 3.075 MeV
dominates all measured spectra at low momentum transfer, has a form
factor behaviour consistent with a $J^{\pi} = 1^{+}$ assignment and a
transition strength B(M1)$\uparrow$ of about $1.5$~$\mu_{N}^{2}$.

Immediately after its discovery in $^{156}$Gd it has been shown that
the newly found $M1$ mode of course is not unique to this particular
nucleus but is a general property of heavy deformed nuclei. This has
been proven by inelastic electron scattering measurements at the
DALINAC on the deformed nuclei $^{154}$Sm, $^{158}$Gd, $^{164}$Dy,
$^{168}$Er and $^{174}$Yb \cite{Bohle:1984b}, where also first evidence
for the fragmentation of the scissors mode strength has been presented
in the nucleus in which it has been discovered, $^{156}$Gd.
Furthermore, the new mode has also immediately been verified in nuclear
resonance fluorescence experiments with real photons
\cite{Berg:1984a,Berg:1984b,Berg:1984c}.

Since then, the experimental evidence for such scissors-like
excitations in strongly deformed nuclei but also in vibrational,
transitional and gamma-soft nuclei has become compellingly large.
Moreover, the scissors mode has been explored besides the purely
electromagnetic probes (electrons, photons) also with hadronic probes
(proton scattering and low-energy neutron-induced compound reactions)
all over the nuclear mass table. A rather complete summary of the
experimental data from electron and photon scattering can be found in
\textcite{Enders:2005}. Several articles covering this work both from
the experimental and the theoretical point of view appeared in the past
\cite{Richter:1985,Richter:1990a,Richter:1991,Richter:1993a,%
Richter:1993b,Richter:1995,Heyde:1989,Berg:1987,Kneissl:1996,%
LoIudice:1997,LoIudice:2000a,Zawischa:1998}. In the next section
(Sec.~\ref{sec:expevi}), we will discuss this experimental evidence for
magnetic excitation of states of mixed-symmetric character (spatially
and in the proton-neutron charge coordinates) and present all of the
salient features in depth although not exhaustively. We put the accent
mainly on the overall and systematic features rather than on each
individual case, separately. Thereby, we are able to focus on the
essential properties of this mode of motion.

\section{MAGNETIC DIPOLE EXCITATIONS IN HEAVY NUCLEI}
\label{sec:expheavy}

\subsection{Low-energy scissors mode}
\label{sec:lowscissor}

\subsubsection{Experimental evidence}
\label{sec:expevi}

\paragraph{Overview} The early systematics of the scissors mode in
heavy deformed nuclei known mainly from high-resolution electron and
photoexcitation experiments at the DALINAC and the S-DALINAC in
Darmstadt and the DYNAMITRON accelerator in Stuttgart
has been described by \textcite{Richter:1995} and
\textcite{Kneissl:1996}. The majority of nuclei studied are even-even
ones but the scissors mode has also been detected in a number of heavy
odd-mass nuclei (Sec.~\ref{sec:magnetic-dipol-heavy}). Very often the
transition strength of the scissors mode is fragmented. In order to
detect weak transitions, highly efficient gamma-ray detector systems
with proper background suppression have to be used in photon scattering
(also called nuclear resonance fluorescence, NRF) experiments as
comprehensively discussed by \textcite{Kneissl:1996}. In present-day
nuclear resonance fluorescence experiments reduced transition strengths
of $B(M1)\uparrow \approx 0.01 \mu_{N}^2$ for magnetic dipole
transitions at an energy of about 3 MeV can be detected. The scissors
mode has been studied in vibrational and rotational nuclei, in chains
of isotopic nuclei. Advances in gamma spectroscopy - e.g.\ by the use
of EUROBALL detector modules \cite{vonNeumann-Cosel:1997a} - have
extended our knowledge of the scissors mode to $\gamma$-soft nuclei,
$^{194,196}$Pt \cite{vonBrentano:1996,Linnemann:2003} and the chain of
Xe isotopes \cite{Garrel:2006}. For example, in the rare-earth region a
total of 42 isotopes ranging from Nd to Pt have been studied providing
detailed experimental information on the systematics of the scissors
mode. Furthermore, it has been verified in the actinide mass region
\cite{Heil:1988,Margraf:1990}.

\begin{figure}[tbh]
\includegraphics[angle=0,width=8.5cm]{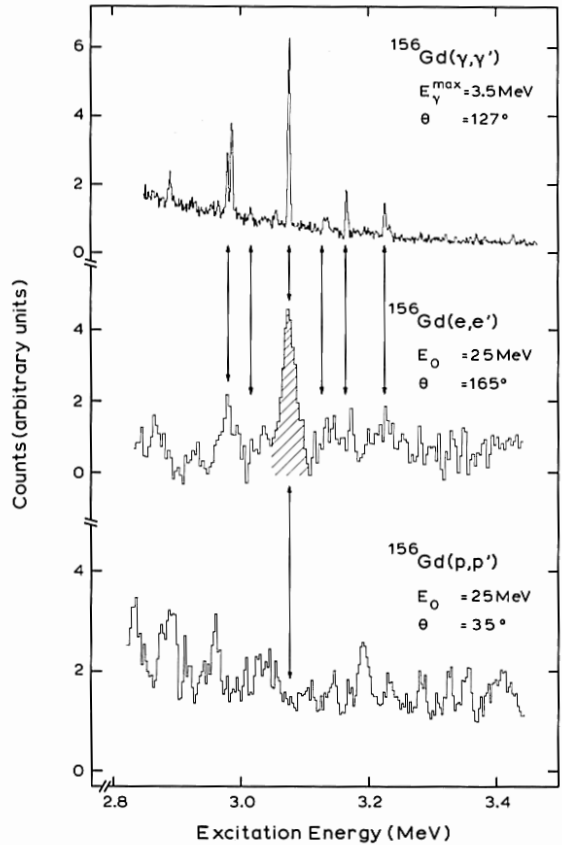}
\caption{High-resolution nuclear fluorescence (upper part) and
inelastic electron scattering (middle part) spectra showing a
strongly excited $J^{\pi} = 1^{+}$ state (hatched) and several weaker
$1^{+}$ states, all marked by arrows. These states are conspicuously
absent in the inelastic proton scattering (lower part) spectrum
\cite{Richter:1990a}.}\label{Fig6}
\end{figure}

A comparison of different probes for the case of $^{156}$Gd  is made in
Fig.~\ref{Fig6} \cite{Richter:1990a}. A combined analysis of the
(e,e$^{\prime}$) and ($\gamma,\gamma^{\prime}$) experiments
\cite{Bohle:1986} revealed five more weakly excited $J^{\pi} = 1^+$
states in the vicinity of the strongest one, bringing the total $M1$
strength up to about 2.4~$\mu_{N}^{2}$. The strongest state might be
viewed acting as a doorway for the others. A high-resolution inelastic
proton scattering spectrum \cite{Wesselborg:1986} is also shown in
Fig.~\ref{Fig6} and the absence of any of the states seen in the
(e,e$^{\prime}$) and ($\gamma,\gamma^{\prime}$) reactions is already a
strong hint that the scissors mode is excited through the orbital part
of the magnetic dipole operator.

The salient features of the scissors mode in heavy deformed nuclei
unraveled in high-resolution electron, photon and proton scattering
experiments \cite{Richter:1995} are the following: \\
(i) Its mean excitation energy scales roughly as  $66 \delta
A^{-1/3}$~MeV with $\delta$ being the deformation
parameter.\footnote{In this article, the quadrupole deformation
parameter used is mainly denoted by $\delta$. Slightly different
parameterizations ($\beta_2$, $\epsilon_2$) exist, which can all be
related with each other \cite{Loebner:1970,Hasse:1988}. To lowest order
$\epsilon_2 \approx \delta \approx
\frac{3}{4}\sqrt{\frac{5}{\pi}}\beta_2$.}
This puts the center of gravity of the orbital $M1$ strength
distribution in rare earth nuclei at E$_x \approx$ 3 MeV. \\
(ii) The total transition strength from the ground state into the
$J^{\pi} = 1^+$ states is $\sum B(M1)\!\uparrow \approx 3~\mu_N^2$ and
the maximum strength that is carried in the transition to an individual
state is roughly 1.5~$\mu_N^2$. \\
(iii) In the nuclear transition current the orbital part dominates over
the spin part and one has typically $B_l(M1)/B_{\sigma}(M1) \approx
10$. \\
(iv) The summed experimental transition strength up to E$_x \approx$ 4
MeV varies quadratically with the quadrupole ground state deformation
parameter.

Before these observations are compared with various nuclear model
predictions in Sec.~\ref{sec:th-col-mic} we concentrate briefly
on a few more experimental characteristics of the mode.

\begin{figure}[tbh]
\includegraphics[angle=0,width=8.5cm]{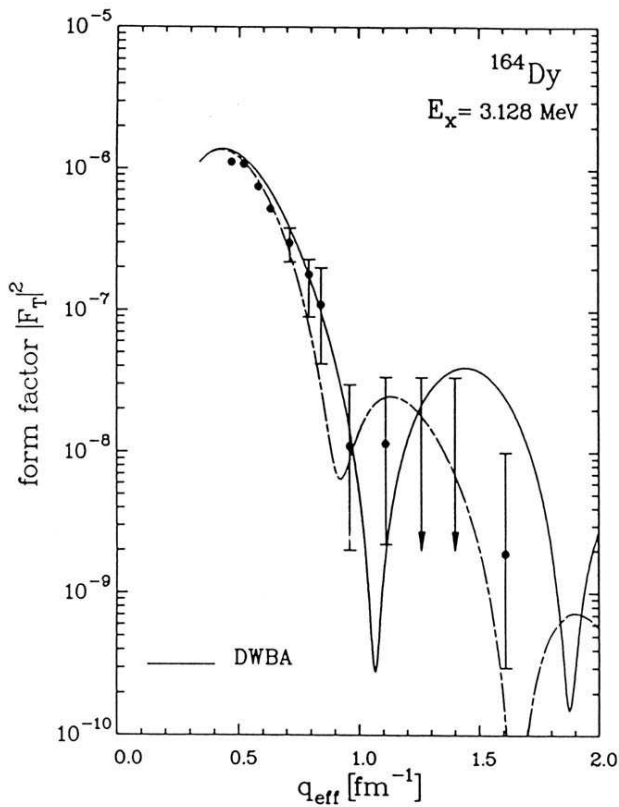}
\caption{Comparison of the experimental form factor for the
transition into one of the $J^{\pi} = 1^{+}$ scissors mode states in
$^{164}${Dy} with form factors predicted in QRPA
\cite{Scholtz:1989b}. The dashed curve denotes the orbital part only
\cite{Richter:1990a}.}\label{Fig7}
\end{figure}

\paragraph{Form factor} One of the first strong hints that indeed $J^{\pi}
= 1^+$ states were excited from the $J^{\pi} = 0^+$ ground state came
from the measurements of form factors in inelastic electron scattering
at low momentum transfers \cite{Bohle:1984a,Bohle:1984b}. In all cases
a shape characteristic for a magnetic dipole form factor has been
found. As an example, one such form factor \cite{Bohle:1987a} is shown
in Fig.~\ref{Fig7} for the transition into a scissors mode state in
$^{164}$Dy at E$_x$ = 3.11 MeV. It is compared to a form factor
calculated \cite{Scholtz:1989b} in distorted wave Born approximation
(DWBA) with a QRPA transition density to a state at E$_x$ = 3.128 MeV,
i.e.\ very close to the experimental state. The full curve is the total
form factor (orbital plus spin) and the dashed one its orbital part
alone. It is evident that the experimental form factor is well
described at and above its first maximum by the model, considering the
sizable uncertainty due to the smallness of the form factor values at
momentum transfers larger than 1~fm$^{-1}$. Similarly to the QRPA
model, the Interacting Boson Model-2 (IBM-2)- to be discussed later on
- describes the data as well and points in particular also to the
orbital nature of the transition.

\begin{figure}[tbh]
\includegraphics[angle=0,width=8.5cm]{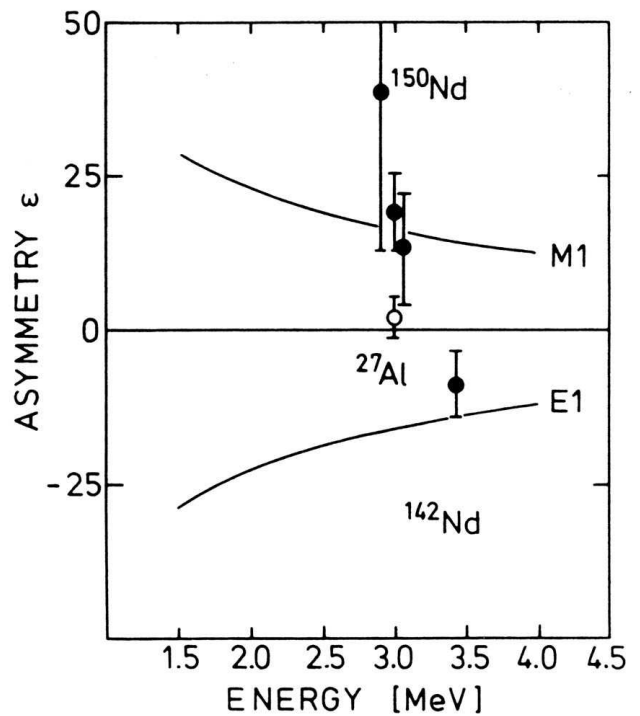}
\caption{Experimental asymmetries $\epsilon = (N_{ \perp} -
N_{\parallel})/ (N_{ \perp} + N_{\parallel})$ of Compton-scattered
photons determined by a five-detector polarimeter. The experimental
data are compared with the calculated polarization sensitivity of
the setup given by the solid lines. Three strongly excited $M1$
transitions into states around 3 MeV excitation energy in the
deformed nucleus $^{150}$Nd have been identified. For comparison an
E1 transition into a state at 3.4 MeV in the spherical nucleus
$^{142}$Nd is also shown. (Reprinted with permission from
\textcite{Kasten:1989}. \copyright (2009) Am.~Phys.~Soc.)}\label{Fig9}
\end{figure}

\paragraph{Photon polarization and parity assignments}
Although the excitation of the scissors mode states by inelastic
electron scattering and the measurement of the respective form factors
has already provided a clear indication of the magnetic nature of the
transition, it can be established beyond any doubt by using polarized
photons either in the entrance or in the exit channel in NRF
experiments (see \textcite{Kneissl:1996} for a detailed discussion). In
the former method of ($\vec{\gamma},\gamma^{\prime}$), which has been
successfully used at the photon scattering facilities at Gent
\cite{Govaert:1994} and Rossendorf \cite{Schwengner:2007}, the linearly
polarized bremsstrahlung causes a positive azimuthal asymmetry for pure
electric and a negative one for pure magnetic dipole excitations of the
detected photon in the exit channel. In the latter method of
($\gamma,\vec{\gamma}^{\prime}$), which has been employed vigorously in
recent years at Stuttgart \cite{Margraf:1995}, the parity assignments
for the excited states in the nucleus result from the measurement of
the linear polarization of the scattered photons with a Compton
polarimeter. In a pioneering experiment \cite{Kasten:1989} in the field
of the scissors mode this technique has been used to measure directly
(and model-independently) the parity of three strongly excited states
near 3 MeV in the deformed nucleus $^{150}$Nd. They were shown to be of
positive parity, i.e. $J^{\pi} = 1^+$ states excited through the
scissors mode (Fig.~\ref{Fig9}).

However, for transitions into states at excitation energies above 4
MeV, the application of this technique will be difficult, simply
because the already small analyzing power of the Compton polarimeter at
low energies becomes even smaller at higher energies. Here the use of
polarized photons in the entrance channel of the reaction and the
subsequent  measurement of the intensity distribution of the scattered
photon is preferable for the parity determination of nuclear dipole
transitions. The analyzing power of this process is 100 \% and
independent of the energy of the scattered photon. In a conceptually
simple experiment, \textcite{Pietralla:2002} recently measured
unambiguously the parity of a number of dipole excitations in
$^{138}$Ba between $E_x = 5.5 - 6.5$ MeV using photons from the
Duke/OK-4 Storage Ring Free Electron Laser backscattered from
relativistic electrons, now called the High Intensity Gamma-Ray Source
(HI$\gamma$S) facility \cite{Weller:2009}. The produced photon beam has
been intense (10$^7$ photons/s) and nearly monochromatic ($\Delta
E_{\gamma}/E_{\gamma} \approx$ 3.8 \%). This very efficient technique
indeed will have a future in the study of elementary dipole transitions
below threshold for particle emission.

\paragraph{Branching ratios of spin-one states in deformed nuclei
and the K quantum number}

\begin{figure}[tbh]
\includegraphics[angle=0,width=8.5cm]{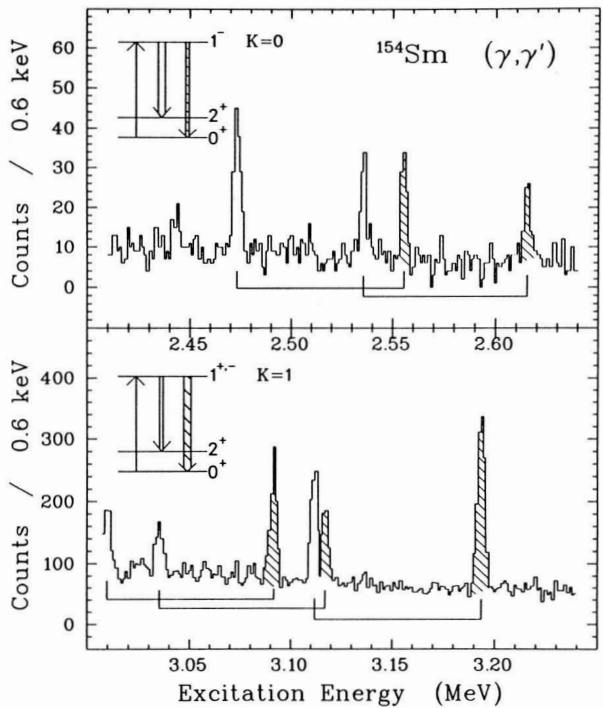}
\caption{Two parts of a NRF spectrum of
$^{154}$Sm. The ratio of the areas of peaks linked by brackets,
corresponding to the ground state and the $2^{+}$ transitions, allows
the assignment of the $K$ quantum number to the excited spin-1
state. In the upper part two examples for $J^{\pi};K = 1^{-};0$, and
in the lower part two examples for $J^{\pi};K = 1^{+,-};1$
assignments are shown \cite{Richter:1991}.}\label{Fig10}
\end{figure}

In NRF measurements, spin-one levels of both parities are selectively
exited which decay either to the $0^+$ ground state or to low-lying
excited states. Figure \ref{Fig10} displays two parts of a NRF spectrum
taken at a bremsstrahlung endpoint energy slightly above 4.6 MeV for
the strongly deformed ($\delta$ = 0.27) nucleus $^{154}$Sm. A number of
transitions clustering around 3~MeV, i.e.\ the location of the states
excited by the scissors mode, are observed \cite{Ziegler:1993}.
Multipolarities of individual transitions were ascertained by
simultaneous two-point angular distribution measurements at
90$^{\circ}$ and 127$^{\circ}$ with respect to the direction of the
incoming bremsstrahlung beam. Those data sufficed to clearly
distinguish between quadrupole and dipole transitions. In the case of
$^{154}$Sm it has been possible to determine the nature of the dipole
transitions ($M1$ or $E1$) by supplementing the present
($\gamma,\gamma^{\prime}$) data with (e,e$^{\prime}$) form factor
measurements \cite{Ziegler:1993}. However, there is still another
property attached to the $J = 1$ levels excited from the $J^{\pi} =
0^+$ ground state, i.e. its $K$ quantum number. The pairs of the lines
connected by brackets in the spectrum of Fig.~\ref{Fig10} correspond to
the decay of the $J=1$ levels into the ground state or the first
excited $J^{\pi} = 2^+$ state of $^{154}$Sm. From the observed
branching ratio one obtains information about the $K$ quantum number of
the $J=1$ state as demonstrated in Fig.~\ref{Fig10}.
The branching ratios for the deexcitation of levels in deformed nuclei
to various states of a rotational band are governed by the so-called
Alaga rules \cite{Alaga:1955} which yield for the ratio of transition
strengths $B(1\rightarrow 2)/B(1\rightarrow 0) = 0.5$ for $\Delta K =
1$ and 2 for $\Delta K = 0$ transitions, respectively.
\begin{figure}[tbh]
\includegraphics[angle=0,width=8.5cm]{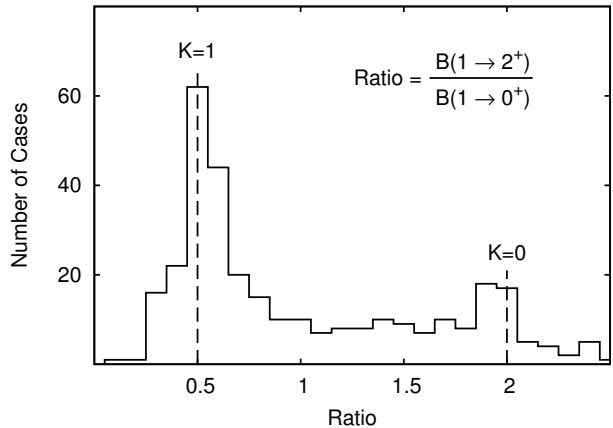}
\caption{Frequency distribution of experimental branching ratios for
about 320 transitions observed in NRF in
deformed rare-earth and actinide nuclei. One notices two distinct maxima at 0.5
for $K=1$ and at $2.0$ for K=0 spin-one levels
(adapted from \textcite{Zilges:1990b}).}\label{Fig11}
\end{figure}

The examination of the decay of about 200 \ $J = 1$ levels in a number
of rare earth nuclei \cite{Zilges:1990b} indeed showed maxima at
experimental branching ratios of 0.5 (as expected for states with $K =
1$) and of 2.0 (as expected for states with $K = 0$), respectively.
Figure~\ref{Fig11} presents an update including additional data which
appeared since the original publication \cite{Enders:2005}. The number
of $K = 1$ states is at least twice the number of $K = 0$ (note that
there are no positive parity states with $K = 0$). Thus, the majority
of the branching ratios are in good agreement with the Alaga rules.
Furthermore, recent parity measurements of strong dipole transitions in
$^{172,174}$Yb confirm the $E1/M1$ assignments based on the $K$ quantum
numbers \cite{Savran:2005}.

But what is the reason for a number of cases with branching ratios in
between the limits set by the Alaga rules? Some of them are known to
result from rather strong $E1$ transitions of $J^{\pi} = 1^-$ states
with a strength hitherto still unexplained \cite{Zilges:1996}. The
deviations, however, may also be taken as evidence for possible $K$
mixing. In fact $K$ mixing matrix elements have been calculated from
the measured energies, branching ratios and absolute transition
strengths and are about 50~keV \cite{vonBrentano:1994a}.

\paragraph{Evidence for quasi-particle excitations at low energy}
The high-resolution $(e,e')$ data, obtained by \textcite{Bohle:1984b}
as well as NRF results \cite{Wesselborg:1988} on $^{164}$Dy have shown,
in addition to the strongly excited $J^{\pi} = 1^+$ states around E$_x
\approx$ 3 MeV, a second group of $J^{\pi} = 1^+$ states about 0.5 MeV
lower in excitation energy which carry a much weaker strength . Similar
results have been obtained in the
$^{168}$Er($e,e^{\prime}$) spectrum proving that the occurrence of a
lower-lying weakly excited group of $J^{\pi} = 1^+$ states seems to be
a general phenomenon in heavy deformed nuclei. On the basis of $(e,e')$
form factor measurements, albeit difficult for the lower group of
states because of the  smallness of the experimental cross sections and
the corresponding large uncertainty, it has been speculated upon that
the two groups of states might be of very different structure
\cite{Richter:1990a}.

\begin{figure}[tbh]
\includegraphics[angle=0,width=8.5cm]{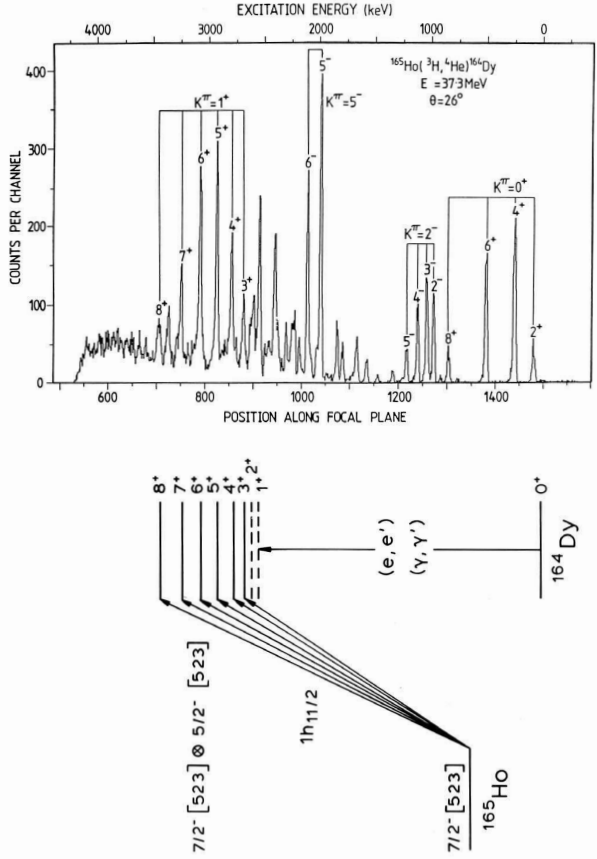}
\caption{Members of a $K^{\pi} = 1^{+}$ rotational band observed in
the single-particle transfer reaction
$^{165}$Ho($^3$H,$^4$He)$^{164}$Dy. The (not populated) $1^{+}$ band
head lies at an excitation energy of 2.543(13)~MeV which can be
identified with the $1^{+}$ state energy of 2.539~MeV, observed in
nuclear resonance fluorescence and inelastic electron scattering
experiments \cite{Richter:1990a}.}\label{Fig12}
\end{figure}

That this is indeed the case has been proven independently in a study
of the proton pick-up reaction $^{165}$Ho ($\alpha$,t) $^{164}$Dy
\cite{Freeman:1989}. As can be seen from the top part of
Fig.~\ref{Fig12} the $J^{\pi} = 3^+$ to 8$^+$ members of a $K^{\pi} =
1^+$ rotational band have been identified in this single-particle
transfer reaction. The reconstruction of the (not populated) $1^+$ band
head energy resulted in $E_x = 2.543(13)$~MeV which can be safely
identified with the $J^{\pi} = 1^+$ state energy of $E_x = 2.539$~MeV
observed in $(e,e')$ scattering \cite{Bohle:1987a} and the NRF
experiments \cite{Wesselborg:1988}. This $1^+$ state thus corresponds
most likely to a two-quasiproton configuration (bottom part of
Fig.~\ref{Fig12}) of the form
$\frac{7}{2}^-$[523]$\otimes\frac{5}{2}^-$[523], and such a
configuration is of course {\it{not}} consistent with a collective
magnetic dipole transition from the ground state into it, as  has been
pointed out earlier by \textcite{Hamamoto:1984} and later also by
\textcite{Otsuka:1990b}. Furthermore, no rotational states built upon
the $J^{\pi} = 1^+$ scissors mode states around E$_x \approx$ 3.1 MeV
are seen in the spectrum supporting the collective interpretation of
the latter.

\begin{figure}[tbh]
\includegraphics[angle=0,width=8.5cm]{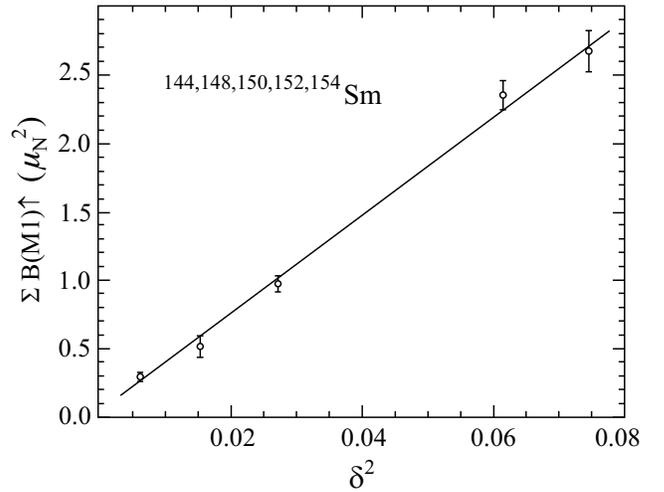}
\caption{The summed orbital $M1$ strengths observed in even-even Sm
isotopes and plotted vs.\ the square of the deformation parameter
$\delta$ \cite{Ziegler:1990}. See also Fig.~\ref{Fig28} for the
systematics of rare-earth nuclei including the present data.}
\label{Fig13}
\end{figure}

\paragraph{The scissors mode and deformation} So far deformation has already
been mentioned alongside the discussion of the experimental data on the
scissors mode. The certainly most important observation made since the
discovery of the mode itself has been, that the measured orbital
magnetic dipole strength increases linearly with the square of the
deformation parameter $\delta$. This is shown in Fig.~\ref{Fig13},
where the summed $M1$ strength is plotted vs. $\delta^2$, for a chain
of even-even Sm isotopes \cite{Ziegler:1990}. This striking result has
later been verified in corresponding experiments on a series of
even-even Nd isotopes \cite{Margraf:1993}. Those observations have been
anticipated in a systematic theoretical study \cite{DeCoster:1989c} of
$M1$ strength in nuclei of the rare-earth region within the Nilsson
model. Immediately after the experimental results displayed in
Fig.~\ref{Fig13}, it was further realized that the orbital $M1$
strength in a given nucleus is also proportional to the strength of the
$E2$ transition to the lowest $J^{\pi} = 2^+$ state
\cite{Rangacharyulu:1991}. The empirical relation
\begin{equation}
\label{eqn:sumB} \sum_f B_f(M1)\!\uparrow~\sim~B(E2; 0^+_{1}
\rightarrow 2^+_{1})~\sim~\delta^2 ,
\end{equation}
thus represents - as expressed by \textcite{LoIudice:2000a} - the most
spectacular manifestation of the scissors nature of the low-lying
magnetic dipole transitions and their collectivity. Since the
neutron-proton interaction is mainly responsible for the nuclear
quadrupole deformation the experimental observation of the strong
$M1/E2$ correlation is of great interest for a test of nuclear models
of deformation as will be shown below.

\paragraph{Summary} In the subsections {\it{a}}-{\it{f}} above, experimental
evidence has been presented for the existence of strong magnetic dipole
transitions of orbital character into states at an excitation energy of
E$_x \approx$ 3 MeV in even-even heavy deformed nuclei. The total
{\it{orbital}} strength into these states amounts in strongly deformed
nuclei to about $\sum B(M1) \approx 3~\mu_N^2$. It hardly moves at all
with excitation energy, i.e.\ it remains low-lying, is
{\it{scissors-like}} and weakly collective, but strong on the
single-particle scale. Its observability is a strong effect as a
consequence of the fact that the nuclear particle-hole force has swept
the competing stronger spin-flip strength up to higher excitation
energy (see Fig.~\ref{Fig1} and Sec.~\ref{sec:th-spinflip}).
Furthermore the existence of weakly excited two-quasiparticle $J^{\pi}
= 1^+$ excitations at about twice the pairing gap, i.e.\ below the
states associated with the scissors mode, has been shown. Thus,
returning to the schematic picture of Fig.~\ref{Fig1} above, the
salient experimental features of magnetic dipole excitations below
about 4 MeV of excitation energy in heavy deformed nuclei have been
demonstrated in this subsection.

The weakly collective scissors mode excitation has become an ideal test
of models - especially microscopic models - of nuclear vibrations. Most
models are usually calibrated to reproduce properties of strongly
collective excitations (e.g.\ of $J^{\pi} = 2^+$ or $3^-$ states, giant
resonances, ...). Weakly-collective phenomena, however, force the
models to make genuine predictions, and the fact that the transitions
in question are strong on the single-particle scale makes it impossible
to dismiss failures as a mere detail, especially in the light of the
overwhelming experimental evidence for them in many nuclei
\cite{Richter:1995,Kneissl:1996}. This should be kept in mind in the
assessment of the wide variety of nuclear models which the scissors
mode has inspired after its discovery about two and a half decades ago.

\subsubsection{Theoretical description: from collective to microscopic models}
\label{sec:th-col-mic}

\begin{figure}[tbh]
\includegraphics[angle=0,width=8.5cm]{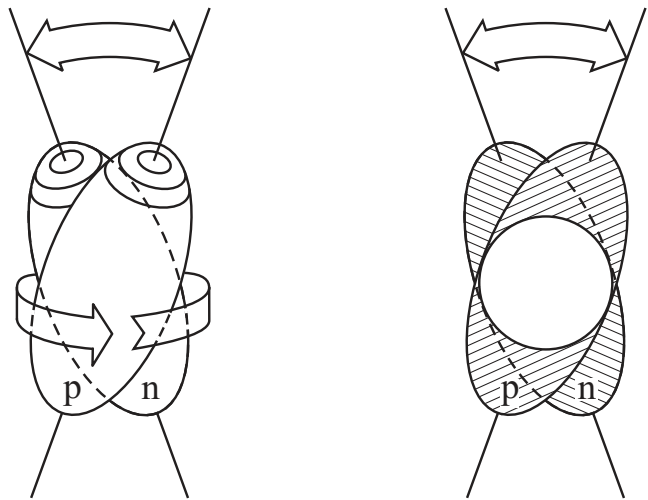}
\caption{Pictorial representation of the scissors mode in a
proton-neutron two-fluid model (left part) and using a presentation
where an inert core exists and only a small part of the proton and
neutron fluids take part in the scissors motion (right part). The
proton-neutron rotational oscillation is the basis of the two-rotor
model which is the rotational analogon of the semiclassical model of
the electric dipole giant resonance.}\label{Fig14}
\end{figure}

\paragraph{Geometric collective models} \label{sec:th-coll-geo} As a
special case of the generalized isovector Bohr-Mottelson model, a
two-rotor model considering both proton ($\pi$) and neutron ($\nu$)
degrees of freedom was worked out in detail by
\textcite{LoIudice:1978,LoIudice:1979}. Describing these two systems as
axially symmetric rigid rotors that are able to perform rotational
oscillations around a common axis orthogonal to their symmetry axes
(Fig.~\ref{Fig14}, l.h s.), the following Hamiltonian was set up
\begin{equation}
\label{eqn:H} H=\frac{(\hat{I}_{\pi}+\hat{I}_{\nu})^{2}}{2J_{intr}}+
\frac{(\hat{I}_{\pi}-\hat{I}_{\nu})^{2}}{2J_{intr}}+\frac{1}{2}C\theta^{2},
\label{eq:hamiltonian}
\end{equation}
in which the restoring force constant $C$ is related to the symmetry
energy constant in the semi-empirical mass formula. Using known
properties of deformed nuclei: the moment of inertia $J_{intr}$ of the
ground-state band, the symmetry energy and the
$B(E2;0^{+}_{1}\rightarrow2^{+}_{1}$) transition strength, one can
determine both the excitation energy $E_{sc} =\hbar\sqrt{C/J_{intr}}$
and the $B(M1)$ transition strength
\begin{equation}
\label{eqn:B} B(M1;0^{+}\rightarrow1^{+})=\frac{3}{16\pi}
\frac{J_{intr}}{\hbar^2}
E_{sc}(g_{p}-g_{n})^{2}~\mu^{2}_{N}. \label{eq:m1-moi}
\end{equation}
Here $g_{p}$ and $g_{n}$ denote the orbital gyromagnetic factors
associated with the rotation of the deformed proton and neutron
systems, respectively. It turned out that the early calculations of
both the excitation energy and the $M1$ strength gave too large values
compared with the first experimental observations of $1^{+}$ scissors
excitations in deformed nuclei (the $B(M1)$ value exceeds experimental
values for the strongest $1^{+}$ states by a factor of almost $7$).
However, it cannot be emphasized enough how important the seminal work
of \textcite{LoIudice:1978,LoIudice:1979} has been in the experimental
search which finally led to the discovery of the scissors mode in
$^{156}$Gd \cite{Bohle:1984a}. Moreover, the confrontation of the
conceptually simple two-rotor model (TRM) with the wealth of
experimental data having accumulated rapidly in the late eighties and
early nineties of the last century has led to a steady improvement of
the model \cite{LoIudice:1997,LoIudice:2000a} which is still the one
allowing a good first insight in the dynamics which causes the scissors
mode to show up at all in deformed nuclei.

A geometrical model for strongly deformed nuclei (assuming axial
symmetry) with separate proton and neutron deformations has been
formulated by \textcite{Rohozinski:1985}. There, the scissors mode was
explained as a relative vibration of the proton and neutron collective
surface. Moreover, rotational bands are obtained on top of K$^{\pi}$ =
0$^+$, 1$^+$ and 2$^+$ vibrational excitations.

Through the experimental observations, we now know the restoring force
constant $C$ in Eq.~(\ref{eqn:H}) much better which determines largely
the excitation energy $E_{sc}$ of the mode. Furthermore, introducing
pairing correlations amongst the participating protons and neutrons and
thus effectively reducing the number of protons and neutrons
participating in the scissors motion (Fig.~\ref{Fig14}, r.h.s.)
resulted in reduced $M1$ strengths, but still too big by factors $4-5$
as compared to experiment. In this purely collective approach, all $M1$
strength is concentrated in a single state whereas the data are much
more fragmented.

As a consequence of the too high excitation energy and the too large M1
strength for the $1^+$ scissors excitation, \textcite{Faessler:1986a}
concentrated on a more detailed study of the restoring force for this
isovector mode \cite{Faessler:1986b}. Considering the fact that the
symmetry energy coefficient exhibits a strong $\rho^{2/3}$ density
dependence, the isovector symmetry energy becomes
\cite{Nojarov:1986,Faessler:1987}
\begin{equation}
\label{eqn:density}
E^{iv}_{sym}= D\int (\rho_p -\rho_n)^2/(\rho_p+\rho_n)^{1/3} d\vec{r},
\end{equation}
with $D = 91.6$ MeVfm$^2$. This symmetry energy has been calculated
microscopically, using proton and neutron densities $\rho_p,\rho_n$
derived from a deformed Woods-Saxon potential, also including pairing.
Identifying the isovector symmetry energy with the scissors potential
energy $1/2C\theta^2$, a much improved restoring force constant $C$
could be derived. This reduces the scissors excitation energy by a
factor of about 2 but still a too large $B(M1)$ value is obtained.
Isovector motion of protons and neutrons has also been discussed
\cite{Faessler:1966,Faessler:1987} for systems performing harmonic
small-amplitude vibrations around a spherical equilibrium shape, using
an extended Bohr-Mottelson model. The implications will be studied
extensively in Sec.~\ref{sec:isovector}.

The collective two-fluid model, both for rotational as well as for
vibrational excitations, has been studied and refined over the years in
great detail. The results have been summarized in review articles by
\textcite{LoIudice:1997,LoIudice:2000a} and partly in an article by
\textcite{Zawischa:1998}. We refer the reader to these for interesting
details.

\paragraph{Algebraic collective models} \label{sec:th-coll-algebra}
When discussing the nuclear structure aspects of an interacting
fermion system, it is striking that for the low-energy collective
modes to develop, the nucleon-nucleon correlations acting in the $L
= 0$ (paired state) and also in the $L = 2$ configuration are
particularly important. It has been shown that the $L = 0$
correlations amongst identical nucleons lead to a generalized
seniority classification while the addition of the $L = 2$ pair
component gives rise to the possibility to develop strong collective
excitations when both proton and neutron valence particles are
present. Considering those two-nucleon pair configurations, it is
possible to formulate a model in which these pairs are now treated
as genuine bosons: the $L = 0$ pair is mapped onto the $s$ boson and
the $L = 2$ pair onto the $d$ boson \cite{Arima:1975a,Arima:1975b}.
This system of interacting bosons (IBM concept) has been studied in
great detail, in particular emphasizing the group structure (which
is the group U(6)) and its reductions, by Arima and Iachello in a
series of papers \cite{Iachello:1987}.

\begin{figure}[tbh]
\includegraphics[angle=0,width=8.5cm]{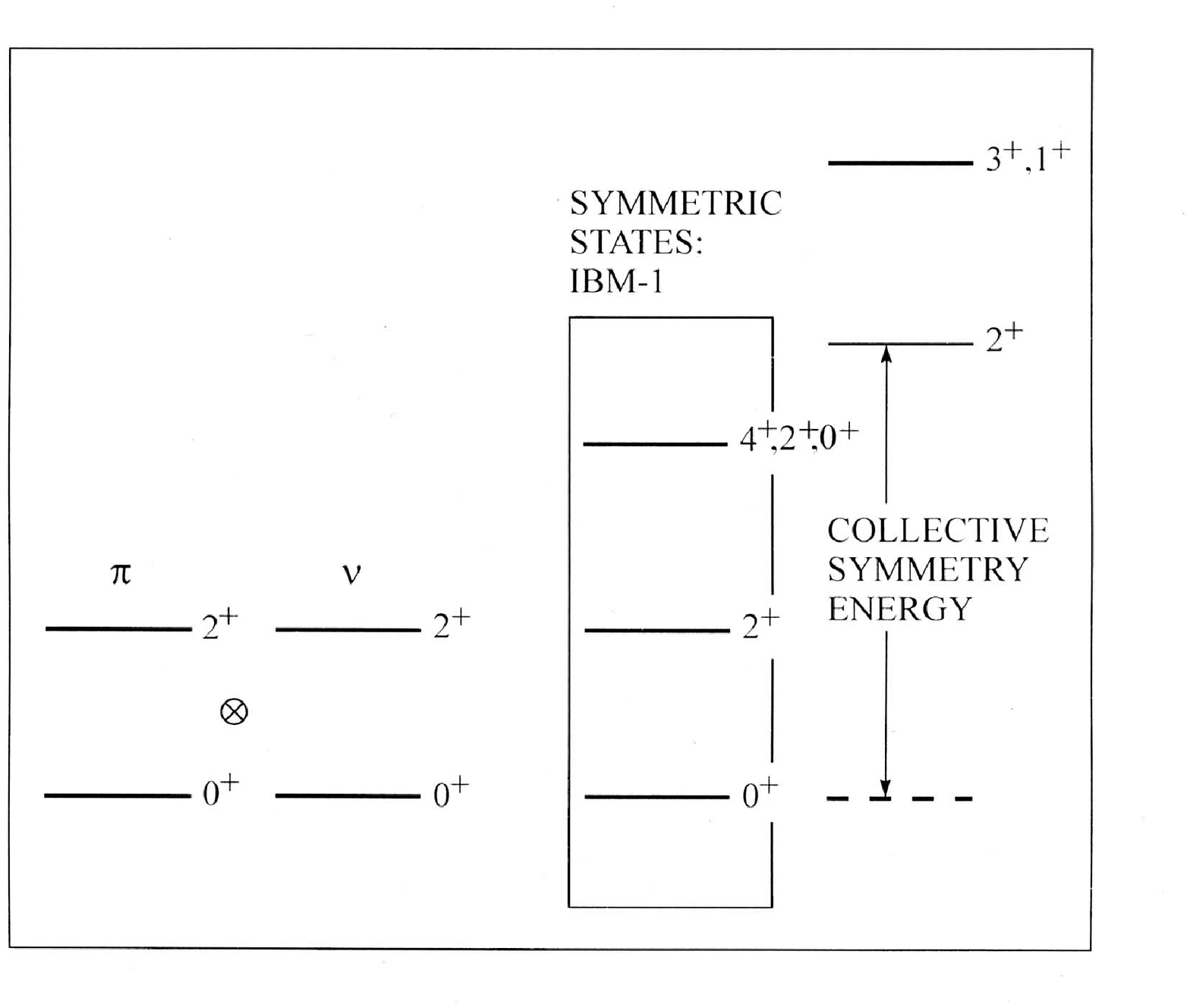}
\caption{Schematic representation of a coupled proton-neutron system
with boson numbers $N_{\pi} = 1, N_{\nu} =1$. The symmetric states
(in the box) and the antisymmetric states are drawn at the
right-hand side of the figure (adapted from \cite{Heyde:1989}).}\label{Fig15}
\end{figure}

In making the model more closely-looking to the shell model with
protons and neutrons interacting, the charge degree of freedom was
introduced in order to distinguish between $s$ and $d$ proton and
neutron bosons, doubling the space of independent degrees of freedom.
The fact that for bosons the total wave function needs to be symmetric
under the interchange of any two bosons, it is still possible to
construct wave functions that have mixed-symmetry character in both the
spatial and the charge part, separately. Using the group-theoretical
formulation, the product irrep.\ (irreducible representations) of
U$_{\pi}(6)$$\otimes$U$_{\nu}(6)$ contains, besides the one-row, also
two-row irrep., or even more explicitly,
\begin{equation}
\label{eqn:ncrossn} [N_{\pi}]\otimes [N_{\nu}] =
[N_{\pi}+N_{\nu},0]\oplus [N_{\pi}+N_{\nu} -1,1]\oplus ...
\end{equation}
The physics of these mixed-symmetry U(6) irrep.\ becomes most clear
when studying the energy eigenvalues for a total IBM-2 Hamiltonian
which contains, besides the pure proton and neutron parts, also the
coupling term between the proton-neutron combined parts. This IBM-2
Hamiltonian can be written as
\begin{equation}
\label{eqn:ibm2} H=\epsilon_{{d}_{\pi}}\hat{n}_{{d}_{\pi}} +
\epsilon_{{d}_{\nu}}\hat{n}_{{d}_{\nu}} + \kappa ( \hat{Q}_{\pi}
+\hat{Q}_{\nu} )\cdot( \hat{Q}_{\pi} + \hat{Q}_{\nu}) +
\lambda\hat{M}_{\pi\nu}.
\end{equation}
We can show the essential results easily in the case of only two bosons
(Fig.~\ref{Fig15}) where one $s_{\pi}$ and $d_{\pi}$ and one $s_{\nu}$
and $d_{\nu}$    boson are considered.
The symmetric coupling [2,0] corresponds to the 0, 1 and 2 quadrupole
phonon structure of the well-known symmetric quadrupole vibrator; the
[1,1] irrep. gives rise to the non-symmetric $1^{+}$, $2^{+}$ and
$3^{+}$ levels. Here too, the energy separation between the $0^{+}$ and
the $2^{+}_{ms}$ states is related to the collective symmetry energy in
the interacting boson model, an energy which is governed by the
strength $\lambda$ of the Majorana term $\hat{M}_{\pi\nu}$ in the IBM-2
Hamiltonian of Eq.~(\ref{eqn:ibm2}), see \textcite{Scholten:1985b}.
This is very much the same physics underlying the splitting of the
various isospin $T$ components resulting from combining protons and
neutrons in fermion space, as discussed in the introduction
(Fig.~\ref{Fig3}). An equivalent two-valued variable (called $F$-spin)
has thus been introduced \cite{Iachello:1984,Arima:1977} to
characterize the charge (or spatial) part of the boson wave function.
If the proton and neutron bosons are characterized with the help of
their $F$-spin quantum number $F=1/2, F_{z}= +1/2$ and $F=1/2, F_{z}=
-1/2$, respectively, a system of $N_{\pi}$ proton and $N_{\nu}$ neutron
bosons can be classified according to its total $F$-spin. The totally
symmetric orbital ($sd$ boson space) states have maximal $F$-spin,
i.e.\ $F_{max} = 1/2(N_{\pi} + N_{\nu})$ while the mixed-symmetric
states are labeled by decreasing $F$-spin values down to $F_{min} =
1/2|N_{\pi}-N_{\nu}|$. The class of mixed-symmetry states with $F =
F_{max} - 1$ are the lowest-lying and can be excited from the ground
state of an even-even nucleus by a $\Delta F = 1$ transition.

The usual $M1$ operator in fermion space
\begin{equation}
\label{eqn:m1op-fermion}
T^{F}(M1) = \sqrt{\frac{3}{4\pi}}\sum_{i}[g_{l}(i)\hat{l}_{i} +
g_{s}(i)\hat{s}_{i}]\mu_{N} ,
\end{equation}
has its image in boson space
\begin{equation}
T^{B}(M1) = \sqrt{\frac{3}{4\pi}}(g_{\pi}\hat{L}_{\pi} +
g_{\nu}\hat{L}_{\nu})\mu_{N} ,
\end{equation}
with $g_l$, $g_s$ being the fermion orbital and spin $g$-factors and
$g_{\pi}$, $g_{\nu}$ the respective proton and neutron boson $g$
factors. The quantities $\hat{L}_{\pi}$ and $\hat{L}_{\nu}$ are the
corresponding orbital angular momentum operators of the proton and
neutron boson system.

In IBM-2 studies, concentrating on deformed nuclei, one is using the
SU(3) reduction of the U(6) group structure and, for those nuclei, it
was shown that the lowest-lying states of the family of mixed-symmetry
character were characterized by the $F_{max} -1, J^{\pi} = 1^{+}$
quantum numbers \cite{Iachello:1981,Iachello:1984}. These findings
corroborate results obtained from a totally different starting point,
viz.\ the TRM. The IBM-2 $J^{\pi} = 1^{+}$ states are also called
scissors states although there is no immediate reference in the
algebraic formulation to specific coordinate forms and thus also not of
shapes and shape dynamics. Using a coherent-state formalism,
\textcite{Dieperink:1983} was able to show the correspondence
explicitly and, moreover, found indeed that only the valence nucleons
are contributing to the strength of the scissors mode thus leading in a
natural way to a much lower $B(M1)$ strength compared to the early TRM
calculations.

In studying the $M1$ excitation properties within the IBM-2, because of
the specific difference in magnetization properties for proton and
neutron bosons, it was clear that $M1$ transitions could appear
naturally now, in contrast to the former IBM-1. Using mapping from
fermion magnetic properties onto boson ones, it was possible to also
determine the analogous boson $g_{\pi}$  and $g_{\nu}$ factors
\cite{Sambataro:1981,Sambataro:1984,Allaart:1988}. This item has been a
topic of much discussion because the mapping calculations all seem to
come up more or less with values $g_{\pi}\simeq 1$ $\mu_{N}$ and
$g_{\pi}\simeq 0$ $\mu_{N}$ but empirical fits in various mass regions
have indicated quite important deviations
\cite{Wolf:1987,Mizusaki:1991,Kuyucak:1995}. For the pure SU(3) limit,
though, an analytically closed form could be derived for the transition
strength
\begin{equation}
B(M1) =
\frac{3}{4\pi}\frac{8N_{\pi}N_{\nu}}
{2N-1}(g_{\pi}-g_{\nu})^{2}~\mu^{2}_{N},
\end{equation}
an expression when applied to $^{156}$Gd and using the above boson
factors $g_{\pi},g_{\nu}$ gives the result of $B(M1)\simeq
2.8$~$\mu^{2}_{N}$, quite close to the experimentally observed value.
Thus the realistic estimate of \textcite{Iachello:1981}, in the IBM-2,
for the transition strength has been equally important in the search
for the scissors mode \cite{Bohle:1984a} as the estimate of
\textcite{LoIudice:1978,LoIudice:1979} in the TRM for the excitation
energy.

\begin{figure}[tbh]
\includegraphics[angle=0,width=8.5cm]{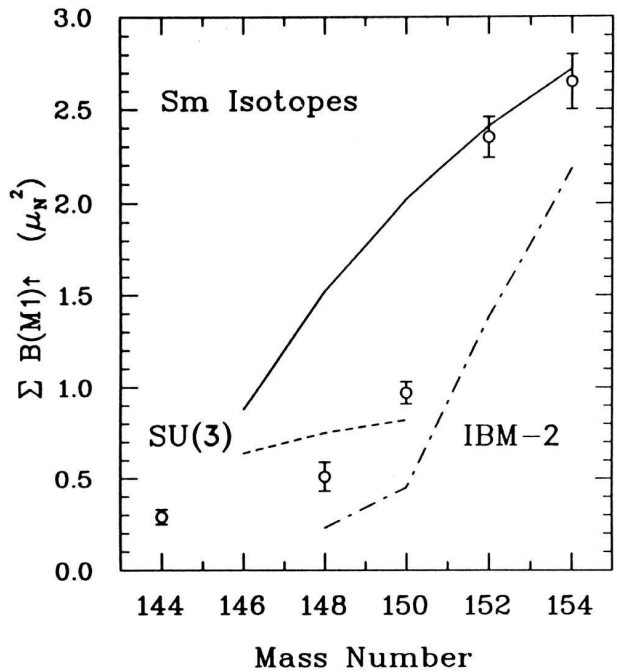}
\caption{Orbital $M1$ strength versus mass number of the Sm
isotopes. The results from full IBM-2 calculations (dash-dotted
line) of \textcite{Scholten:1985b} are shown, together with results
in the SU(3) limit using a $Z=50$ subshell closure (solid line) and
a $Z=64$ subshell closure (dashed line)
\cite{Richter:1991}.}\label{Fig17}
\end{figure}

The subject of mixed-symmetry states appearing as a new class of states
in the IBM-2 has been investigated afterwards in much detail. There has
been an investigation of the various limiting cases that appear if
dynamical symmetries hold \cite{VanIsacker:1986} but also rather
extensive numerical studies have been carried out
\cite{Scholten:1985b}. As an illustration, we compare in
Fig.~\ref{Fig17} the experimental $M1$ transition strengths in Sm
isotopes with the results of the pure SU(3) limit and of the numerical
IBM-2 calculations \cite{Scholten:1985b}.

\paragraph{Microscopic descriptions} \label{sec:th-mmstudies}
On the opposite side from the collective model concepts, the nuclear
shell-model allows for possibilities to describe nuclear coherent
motion from first principles using a Hartree-Fock basis and a
self-consistent procedure in order to determine both global and local
nuclear structure properties. The study of scissors motion starting
from a microscopic shell-model basis can be separated into two parts.
For light and medium-heavy nuclei, regular shell-model calculations
have been performed and also $M1$ excitation properties been studied.
Once entering the region of heavy and deformed nuclei, the model space
to be considered becomes prohibitively large and approximations to the
shell model have been used, mainly the QRPA.

Magnetic dipole excitations have, by now, been measured over a large
part of the nuclear mass table (Sec.\ \ref{sec:expevi}). In contrast to
most of the purely collective models, the low-lying M1 strength is
spread out over an energy interval in the region of $2.5-4$~MeV,
depending on the specific nucleus and thus depending on its proximity
to the closed shells.

\begin{figure}[tbh]
\includegraphics[angle=0,width=8.5cm]{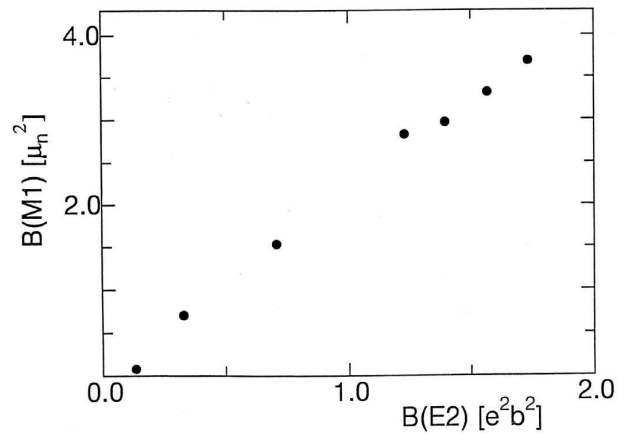}
\caption{Summed orbital $B(M1)$ strength from the ground state
calculated in a Monte-Carlo shell-model approach versus the
corresponding $ B(E2; 0^{+}_{1} \rightarrow 2^{+}_{1})$ values in
the Ba isotopes with mass numbers $A = 138 - 150$. (Reprinted with permission from
\textcite{Shimizu:2001}. \copyright (2009) Am.\ Phys.\ Soc.)}
\label{Fig18}
\end{figure}

Large-scale shell-model studies would be an ideal way to probe the
presence of concentration and fragmentation of $M1$ strength but
this has not been possible for heavy nuclei until recently. Within
the context of a Monte-Carlo shell-model approach, worked out by
Otsuka and collaborators \cite{Honma:1995,Honma:1996,Mizusaki:1996,%
Mizusaki:1999,Utsuno:1999,Otsuka:1998,Otsuka:1999,Shimizu:2001},
large-scale shell-model studies have been performed in order to study
the transition from spherical to deformed shapes with an application to
the Ba isotopes. Starting from a given Hamiltonian and for a given
single-particle energy spectrum that remains fixed through the Ba
isotopes, such microscopic calculations beyond mean-field approaches
have given first evidence, that shape changes indeed do occur due to a
change of the number of interacting protons and neutrons. The $M1$
strength of the scissors mode serves as a good measure of the
deformation of the ground state as has been discussed in
Sec.~\ref{sec:expevi} in particular in the context of Fig.~\ref{Fig13}
and Eq.~(\ref{eqn:sumB}). \textcite{Shimizu:2001} have calculated the
$M1$ sum rule for the ground state with the orbital g-factors having
the free nucleon values. The spin contributions have been omitted for
reasons of simplicity, at present. In Fig.~\ref{Fig18}, the $B(M1)$ sum
rule values are plotted versus the corresponding $ B(E2;0^{+}_{1}
\rightarrow 2^{+}_{1}) $ values. One notices a nearly perfect linearity
between these two quantities first observed experimentally
\cite{Ziegler:1990}. As pointed out in the previous subsection,
starting from symmetries within the IBM and in the nuclear shell model,
such a relation also exists. The Monte-Carlo shell-model calculation,
however, presents a first microscopic underpinning of the connection
between $M1$ and $E2$ properties in nuclei and shows a validity that
does not rely any longer on particular symmetries of the nuclear
many-body system, i.e.\ it holds for the whole isotopic series in the
Ba nuclei.

Besides the nuclear shell-model explicitly, the QRPA or the
Quasi-Particle Tamm-Dancoff Approximation (QDTA), where no ground-state
correlations are considered, present an alternative to study the
properties and the internal orbital and spin character of the magnetic
dipole transitions involving the specific $1^{+}$ states under study.
The QRPA and QDTA models in itself will not be discussed in the present
review and we refer the reader to the literature for a concise
discussion \cite{Rowe:1970,Ring:1980,Eisenberg:1987,Soloviev:1992}. In
the early calculations when applying the QRPA to study the scissors
mode excitations, most often rather schematic forces have been used:
quadrupole proton-neutron forces and spin-spin and spin-isospin
separable interactions, including pairing (both monopole and quadrupole
multipoles) in order to study the role of these components
\cite{Iwasaki:1984,Bes:1984,Kurasawa:1984,Hamamoto:1971}. Quite
recently \cite{Balbutsev:2004,Balbutsev:2007}, a separable
quadrupole-quadrupole residual interaction has been used within a
time-dependent Hartree-Fock (TDHF) framework and its Wigner transform.
Here, the aim was to derive a set of equations describing different
multipole moments, in particular the scissors and isovector giant
quadrupole resonance and their coupling. Pairing, however, is not
included in the formalism. The results obtained are closely related to
the earlier work of \textcite{Lipparini:1989a}.

The pairing force is particularly important in generating the necessary
correlations that relate the overall summed orbital $M1$ strength to
the nuclear quadrupole deformation characterizing a given nucleus.
Moreover, the spin-isospin parts are determining factors in placing the
spin-flip part of the $M1$ response (Fig.~\ref{Fig1}) at the excitation
energy between 5 and 10~MeV. These studies were instrumental in finding
out the first and foremost important physics elements that are at the
origin of coherent orbital magnetic excitations with a scissors
character. It has been pointed out \cite{Ikeda:1993,Heyde:1993a} that a
correct treatment of the full Coriolis term and considering the
coupling of 2qp excitations to a rotational core, induces very specific
correlations that concentrate the independent $M1$ components into a
single peak.

The realistic application of the QRPA to the study of the scissors
mode exciting $J^{\pi} = 1^+$ states has met with some initial
problems because the overall rotational motion of an intrinsically
deformed nucleus carries the same angular momentum as the scissors
mode states themselves. Very much like in case of the electric
dipole mode leading to the $J^{\pi} = 1^-$ states, one has to
separate the spurious rotational motion of the whole nucleus from
the relevant intrinsic $1^{+}$ excitations. This was the origin of
the fact that various QRPA calculations gave rise to rather large
differences in the $1^{+}$ excitation energy and, more importantly,
in the $M1$ response. Removing this spurious rotational motion,
however, using various techniques such as (i) constructing a basis
orthogonal to the spurious rotational motion, (ii) adding a specific
symmetry restoration term to the Hamiltonian, (iii) using the Pyatov
prescription \cite{Baznat:1975} of replacing the quadrupole field in
the Hamiltonian in such a way that rotational invariance is imposed,
the results on $J^{\pi} = 1^{+}$ energies and on $M1$ strengths have
all been converging with similar conclusions.
\begin{figure}[tbh]
\includegraphics[angle=0,width=8.5cm]{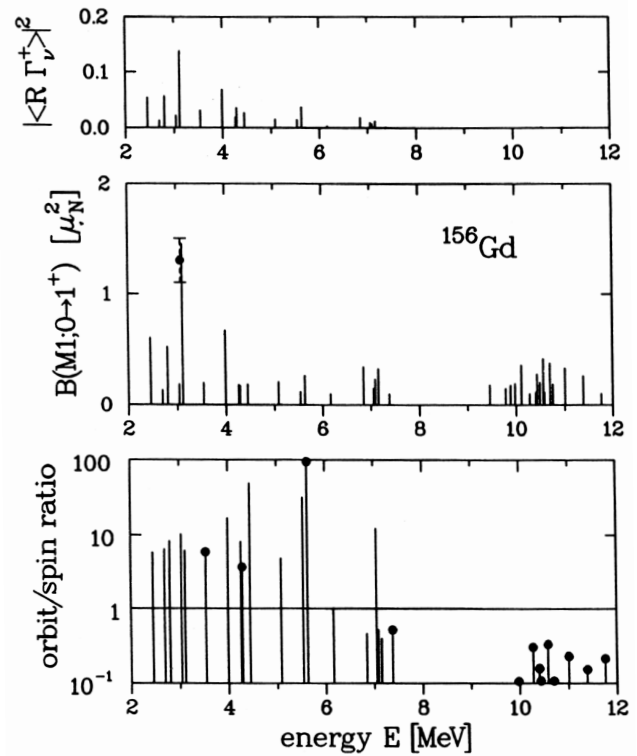}
\caption{Calculated QRPA magnetic dipole strength distribution
compared to the experimental data point of the strongest transition
observed in $^{156}Gd$ (middle part) and orbit-to-spin ratios of the
$M1$ transition matrix elements (bottom part). The magnitude of
these ratios is shown and negative ratios are indicated by a dot on
the top of the bar. The average overlap of the calculated $1^+$
states with the scissors mode state (denoted $R$) is displayed in the
upper part. (From \textcite{Nojarov:1990}, reprinted with permission
from Springer Service + Business Media.)}\label{Fig19}
\end{figure}

\begin{figure}[tbh]
\includegraphics[angle=0,width=8.5cm]{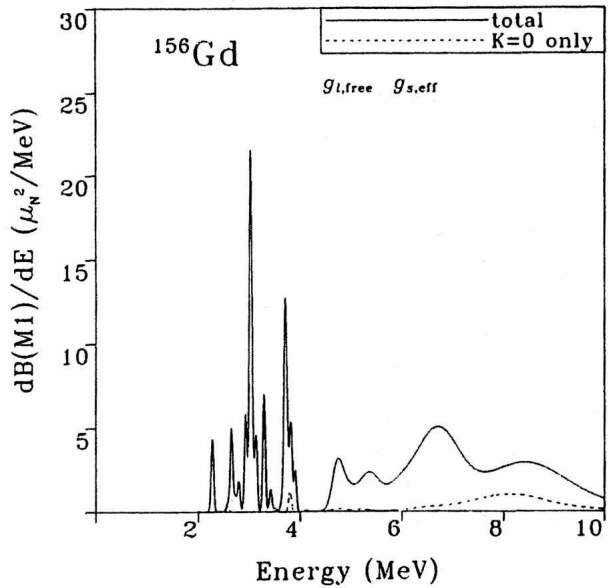}
\caption{The complete magnetic dipole response of $^{156}$Gd up to
10 MeV excitation energy, calculated using the QRPA approach. The
individual levels have been folded with Gaussian functions. (Reprinted with
permission from \textcite{Zawischa:1990a,Zawischa:1990b}. \copyright (2009)
Am.~Phys.~Soc.)}\label{Fig20}
\end{figure}

\begin{figure}[tbh]
\includegraphics[angle=0,width=8.5cm]{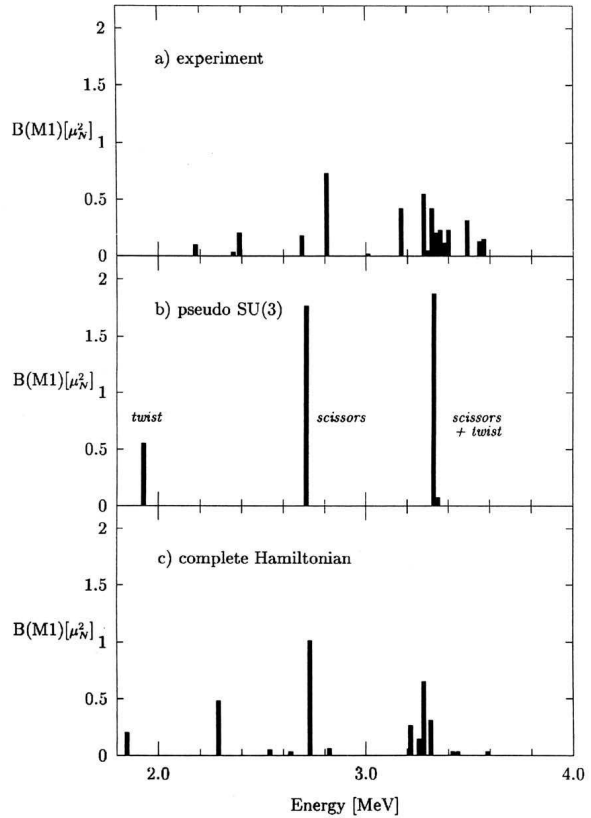}
\caption{$M1$ transition strengths in $^{160}$Gd. (a) Experimental
data, (b) pure pseudo-SU(3) scheme, and (c) complete pseudo-SU(3)
calculation. The twist mode results when considering triaxial proton and
neutron distributions which allow additional rotation around their principal
axes. It can be combined with the scissors mode
into a scissors plus twist mode. (Reprinted with permission from \textcite{Beuschel:2000}.
\copyright (2009) Am.~Phys.~Soc.)}\label{Fig21}
\end{figure}

So, the various QRPA studies are now very close with respect to a
number of key issues characterizing the observation of a magnetic
scissors-type of excitation
\cite{Nojarov:1990,Zawischa:1990a,Zawischa:1990b,Soloviev:1997a,%
Soloviev:1997b,Soloviev:1997c,Beuschel:2000}. Examples are shown in
Figs.~\ref{Fig19}-\ref{Fig21}. Work using variants of QRPA starting
from a deformed Woods-Saxon potential, HFB calculations, and schematic
forces has also been carried out
\cite{Hamamoto:1987,Sugawara-Tanabe:1989,Hilton:1995,Hilton:1998,%
LoIudice:1996a,LoIudice:1996b}.

(i) A concentration of low-lying $M1$ strength is found close to the
energy of $3$ MeV, whose overlap with the scissors mode is as big as
$\simeq$ 40 \% if summed up to 4~MeV(see Fig.~\ref{Fig19}, upper part).
The low-energy $M1$ strength is of dominant orbital character with
orbit/spin ratios of the order of 10 or larger (cf.~Fig.~\ref{Fig19},
bottom part as an example). Furthermore, all calculations predict the
existence of a higher-lying scissors part. This issue, however, is not
closed at all since various authors come to largely different
conclusions concerning its mean energy and, more importantly, its
fragmentation.

(ii) There is common agreement on the very importance of the pairing
correlations in establishing a strong relationship between the summed
$M1$ strength and nuclear deformation. These at first quite different
observables have a deeper connection which is born out from the
microscopic QRPA studies too as will be pointed out below in some more
detail.

(iii) An important spin-flip $M1$ mode is observed in the strongly
deformed rare-earth nuclei in the energy region $5-10$ MeV. There is,
also here, an open debate on the specific way in which the spin-flip
strength is distributed in energy. A two-bump picture shows up but in
some calculations the bumps are mainly isoscalar and isovector in
character, whereas other calculations comply with a rather definite
separation between proton and neutron spin-flip modes. The solution
here should come from selective reaction studies (see
Sec.~\ref{sec:th-spinflip} for a deeper discussion on this issue).

We cannot provide a complete discussion of the multitude of QRPA and
QTDA studies in both the rare-earth and actinide (and even light)
nuclei. We refer to \textcite{LoIudice:1997} for a detailed but still
succinct presentation of the major results. We note, however, that very
recently relativistic QRPA calculations within a self-consistent
relativistic mean-field (RMF) framework have been carried out for
axially deformed nuclei \cite{Arteaga:2008}. Spin, orbital and total
$M1$ strengths were derived for $^{160}$Gd and $^{20}$Ne, with clear
evidence for a scissors mode in $^{160}$Gd \cite{Arteaga:2007}.

\paragraph{Relationship between collective and microscopic models}
\label{sec:relations}
Starting from collective models, the proton and neutron degrees of
freedom form the essential ingredients to generate mixed-symmetry
charge (and spatial) wave function of the $J^{\pi} = 1^+$ states. This
was the case in the two-rotor geometrical model and also within the
proton-neutron interacting boson model (IBM-2).

Because the building blocks constituting the bosons are nucleon pairs,
the IBM-2 approach is rooted closely in the nuclear shell-model (see
e.g. the studies in the light $1f_{7/2}$ nuclei by
\textcite{McCullen:1964,Zamick:1986a,Zamick:1986b,Liu:1987a,Liu:1987b,Liu:1987c}).
In the QRPA, on the other hand, the building blocks are highly
correlated particle-hole (or 2 qp) excitations that make up for a
microscopic description to the collective phonon modes in the nucleus.
It has been pointed out explicitly \cite{LoIudice:1997} that in
schematic models, the connection between the microscopic QRPA and the
collective two-fluid approaches (TRM) can be shown in detail. An
approximate relation for the scissors total summed $M1$ strength is
then
\begin{equation}
B(M1)\uparrow\simeq \frac{3}{16\pi}\frac{J_{intr}}{\hbar^2}E_{sc}~\mu^{2}_{N},
\label{eqn:m1_sum_LoIudice}
\end{equation}
if all the $1^{+}$ states could be approximated by a single centroid
energy $E_{sc}$ like in Eq.~(\ref{eqn:B}).

In all of the more detailed microscopic calculations, be it large-scale
shell-model studies or QRPA calculations, the immediate connection is
not so obvious anymore. Still, the observed spreading (or
fragmentation) of the population of $1^{+}$ states is rather well
reproduced. The QRPA calculations mainly start from different
single-particle structures (Nilsson or Woods-Saxon deformed mean-field,
self-consistent energy spectra,..) and also have been using differing
two-body interactions and so, a large sensitivity of the $1^{+}$ energy
and strength distribution in realistic cases is expected.

There exists an interesting avenue in trying to give a microscopic
support of collective magnetic dipole excitations starting from the
SU(3) for light nuclei and pseudo-SU(3) for the heavy deformed nuclei.
This model goes back to the seminal work of
\textcite{Elliott:1958,Elliott:1963} indicating that collective model
aspects are inherent in the proton-neutron shell model structure for
light $sd$ shell nuclei \cite{Elliott:1985, Elliott:1990}. Calculations
within this approximation have been carried out in $sd$- and $fp$-shell
nuclei for the orbital $M1$ properties and compared to the experimental
values \cite{Chaves:1986,Poves:1989,Retamosa:1990}. The extension to
heavier nuclei was hampered for some time as SU(3) is largely broken
because of the strong spin-orbit splitting. A suggestion has, however,
been made in order to reorganize the level structure conform with a
pseudo SU(3) scheme, explicitly incorporating proton and neutron
degrees of freedom. In former studies of this type, one had to rely on
very simple Hamiltonians; this has been overcome and one can now handle
general one- and two-body Hamiltonians. So this model, deeply rooted in
the shell-model, serves as an interesting bridge to connect the
underlying microscopic structure to the collective building blocks.
Results for Sm, Gd and Dy nuclei have been obtained
\cite{Beuschel:1998,Beuschel:2000,Rompf:1998} and even a reasonable
fragmentation of $M1$ strength is achieved as illustrated in
Fig.~\ref{Fig21} for $^{160}$Gd.

\subsubsection{Fragmentation of orbital dipole strength and sum rules}
\label{sec:th-frag}
\paragraph{Fragmentation of the orbital strength}
\label{sec:th-mmstudies-$M1$strength}

An important element in discussing relationships, similarities but also
complementary aspects of the collective model approaches and
shell-model or QRPA studies is the amount of fragmentation resulting
from these various models. As discussed in Sec.~\ref{sec:th-col-mic},
in collective models, the $M1$ scissors strength is concentrated in a
single or very few strongly excited $1^{+}$ states
\cite{Scholten:1985b} while in the microscopic models, strength needs
to become concentrated in fewer strong states compared with the
unperturbed spectrum of $1^{+}$ states.
\begin{figure}[tbh]
\includegraphics[angle=0,width=8.5cm]{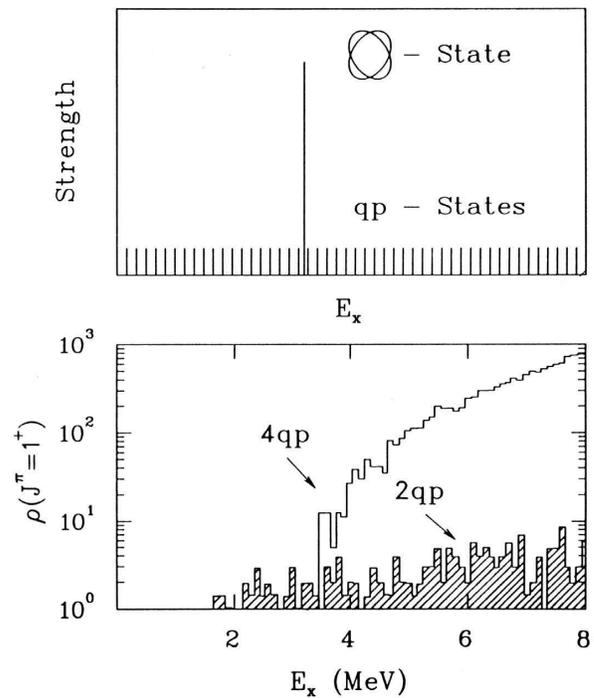}
\caption{Upper part: schematic picture of the scissors mode embedded
into the background of dense qp states (no coupling). Lower part:
level density for 2qp and 4qp $1^{+}$ states for Gd isotopes
calculated using the Nilsson model (adapted from \cite{Heyde:1989}).}\label{Fig22}
\end{figure}

\begin{figure}[tbh]
\includegraphics[angle=0,width=8.5cm]{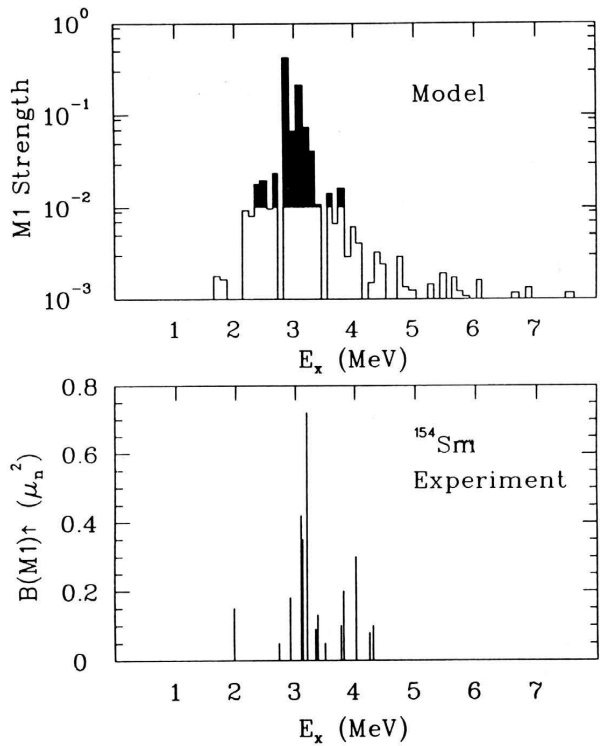}
\caption{Upper part: scissors $M1$ strength damped into the
background of 2qp and 4qp states shown in lower part of
Fig.~\ref{Fig22}, averaged over intervals of 100~keV. The black part
indicates transitions with a relative strength larger than 1\%
(adapted from \textcite{Heyde:1989}). Lower part: experimentally measured orbital $M1$
strength distribution \cite{Ziegler:1993}.}\label{Fig23}
\end{figure}

What are the major issues here? We present the above elements of
fragmenting collective $M1$ strength into a background of microscopic
configurations (2qp,4qp,..) much in the same way as fragmentation is
generally described in nuclear reaction theory (Fig.~\ref{Fig22}, upper
part). There might remain some structure in the fragmented strength due
to interactions with states that are intermediate in complexity between
the strongly collective states on one side and the regular shell-model
configurations on the other side. Such states could be due to
hexadecapole configurations (in certain regions of the nuclear mass
table this degree of freedom in the upper part of the rare-earth mass
region, particularly, can be important), triaxial shape
configurations,.. that first split the $M1$ scissors strength in the
manner of a doorway-state before it gets fragmented into the
microscopic background of $1^{+}$ states \cite{Scholten:1985b}. These
strength function phenomena have been discussed e.g.\ in the Appendix
D2 of \textcite{Bohr:1969}. If the average coupling strength $\langle V
\rangle$ between a single collective state and the background
configurations is larger than the distance between the discrete levels
$D =$ 1/$\rho$ with $\rho$ denoting the level density, then a
Breit-Wigner damping of collective strength over the microscopic
background results, given by a width of $\Gamma = 2\pi \langle V
\rangle^{2}\rho$. The strength function, i.e.\ the probability of
finding a simple, collective state in a unit energy interval of the
spectrum, can subsequently be derived. If, on the other hand, the level
density becomes too low for the above conditions to be valid, one has
to resort to diagonalizing the coupled system of collective and
shell-model configurations. A model where this has been worked out
explicitly for the scissors $1^{+}$ state, as obtained in the IBM-2,
coupling to the underlying background of $1^{+}$ 2qp and 4qp
configurations was described by  \textcite{Heyde:1996}. The background
structure, as calculated using a Nilsson model for the Gd nuclei, is
shown in the lower part of Fig.~\ref{Fig22}. The damping and
fragmentation of the scissors mode into this background has been
derived using a constant coupling matrix element between the states in
the two model spaces (Fig.~\ref{Fig23}, upper part). For the
fragmentation down to the 1\% level, a coupling matrix element of about
50 keV is obtained between the scissors mode and the 2qp states. The
experimental $M1$ strength plotted in the lower part of
Fig.~\ref{Fig23} indicates a similar fragmentation.

Starting from the opposite side, i.e., a shell-model or QRPA approach,
the diagonalization within the unperturbed basis of 2qp, 4qp,... states
will eventually result in the building up of collective correlations
that are reminiscent of a scissors magnetic dipole mode. Examples have
been shown in Figs.~\ref{Fig19}-\ref{Fig21} in the heavy deformed
rare-earth nuclei. The discussion in Sec.~\ref{sec:th-col-mic} has
precisely concentrated on this issue with references to a number of
calculations of that type. One can make use of very similar arguments
as those used at the time by \textcite{Brown:1959} in the discussion of
how individual $1p-1h$ $1^{-}$ configurations, interacting through a
zero-range force, could built a single collective, isovector electric
dipole mode. In contrast to the study of this giant electric dipole
resonant state in which the collective state is mainly built from
1$\hbar\omega$ excitations, the orbital magnetic dipole strength is
mainly of 0$\hbar\omega$ nature. \textcite{Richter:1980} have shown
that the separable characteristics of the two-body matrix elements,
which is essential for the schematic Brown-Bolsterli model, does not
hold any more for the magnetic multipole excitations, i.e. the
particle-hole matrix elements do not scale with the $M1$ transition
amplitudes. As a result, rather small energy shifts show up compared to
the unperturbed energy spectrum. Therefore, there remains a
concentration of mainly orbital $M1$ strength of $0\hbar\omega$ origin
in the energy interval $2.5-4$ MeV. An example of the fragmentation
process, starting from the pseudo SU(3) description, but treating a
more general Hamiltonian, has been discussed by the group of Draayer
and a particularly interesting case is {$^{196}$Pt \cite{Beuschel:2000}
which is a prime example of a $\gamma$-soft nucleus
\cite{vonBrentano:1996}}.

\paragraph{Level spacing distribution of scissors mode states}
\label{sec:th-levelspacing}

As we have noticed in the foregoing subsection, the complexity of
the nuclear many-body problem is clearly manifest in the
fragmentation of the experimental transition strength, which is
distributed over several levels of the same spin and parity. This
complexity has led Wigner, more than forty years ago, to the
introduction of Random Matrix Theory (RMT), reviewed in detail by
\textcite{Guhr:1998} and \textcite{Weidenmueller:2009}. This
statistical approach models spectral fluctuation properties: if the
levels are correlated, one expects a linear repulsion between them
and Wigner-Dyson statistics for the nearest neighbor spacing
distribution (NNSD). However, if correlations are absent, there is
no level repulsion and the NNSD is of Poisson-type, i.e.\ an
exponential distribution. The validity of this ansatz has been
confirmed in various data analyses and has been summarized in the
two review articles cited.

In heavy nuclei, the picture emerges that high-lying single-particle
states containing many complex configurations show Wigner-Dyson
statistics \cite{Haq:1982}, whereas low-lying collective states of
simple structure lack correlations and yield a Poisson distribution
\cite{Shriner:1991,Garrett:1997}. This in turn allows to use RMT to
conclude from spectral statistics if excitations are mainly of
single-particle or of collective character. This idea has also been
applied to the states which belong to the scissors mode
\cite{Enders:2000}. As has been pointed out in Sec.~\ref{sec:expevi}
above, an unprecedented data set is now available covering doubly even
nuclei in the $N = 82 - 126$ major shell. By combining the data sets
from 13 heavy deformed nuclei, a data ensemble has been constructed
with a total number of 152 states in the excitation energy window of
about $2.5 < E_x < 4.0$ MeV (Fig.~\ref{Fig1}).

\begin{figure}[tbh]
\includegraphics[angle=0,width=8.5cm]{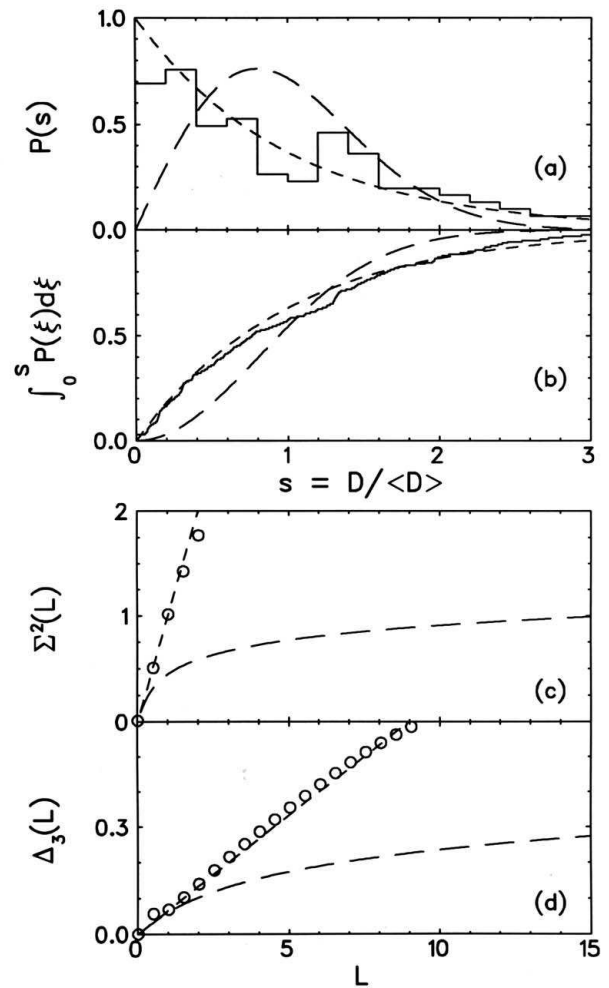}
\caption{Level spacing distribution of scissors mode states in heavy
deformed nuclei (152 states from 13 nuclei). (a) nearest-neighbor
spacing distribution, (b) cumulative nearest-neighbor distribution,
(c) number variance ${\Sigma}^{2} (L)$, and (d) spectral rigidity
$\Delta_{3} (L)$, where $L$ denotes the length of the sequence. The
histograms and open circles display the data. Poissonian
behavior and expectations from the Gaussian Orthogonal Ensemble
(Wigner) are shown as short and long-dashed lines, respectively
\cite{Enders:2000}.}\label{Fig24}
\end{figure}

After unfolding the experimental spectra, i.e. removing the energy
dependence of the average level spacing, the NNSD, the cumulative NNSD,
the number variance $\Sigma^2$ and the spectral rigidity $\Delta_3$
\cite{Bohigas:1989,Guhr:1998} were extracted from the data ensemble.
The results are shown in Fig.~\ref{Fig24}. All evaluated statistical
measures agree very well with the Poissonian behavior of uncorrelated
levels. Although the individual level sequences are rather short (of
order ten only), the functions $\Sigma^2$ and $\Delta_3$ clearly show
the lack of long-range correlations. It has also been shown by
\textcite{Enders:2000} that the influence of missing levels due to the
experimental conditions is negligible. Consequently the remarkable
conclusion from the statistics of the level spacings distribution may
be drawn that the scissors-mode states all have the same structure and
are excited collectively by the same mode.

\paragraph{Sum rules and relation to other observables}
\label{sec:th-sumrule-rel}

Even though the magnetic dipole strength appears rather fragmented in
the energy region $2.5$ to $4.0$ MeV, experimental methods have been
set up in order to distinguish those $1^{+}$ states that carry mainly
orbital $M1$ strength (see also the discussion by
\textcite{Kneissl:1996}). The summed experimental strengths show
remarkable correlations to collective observables of the low-energy
spectrum.

\begin{figure}[tbh]
\includegraphics[angle=0,width=8.5cm]{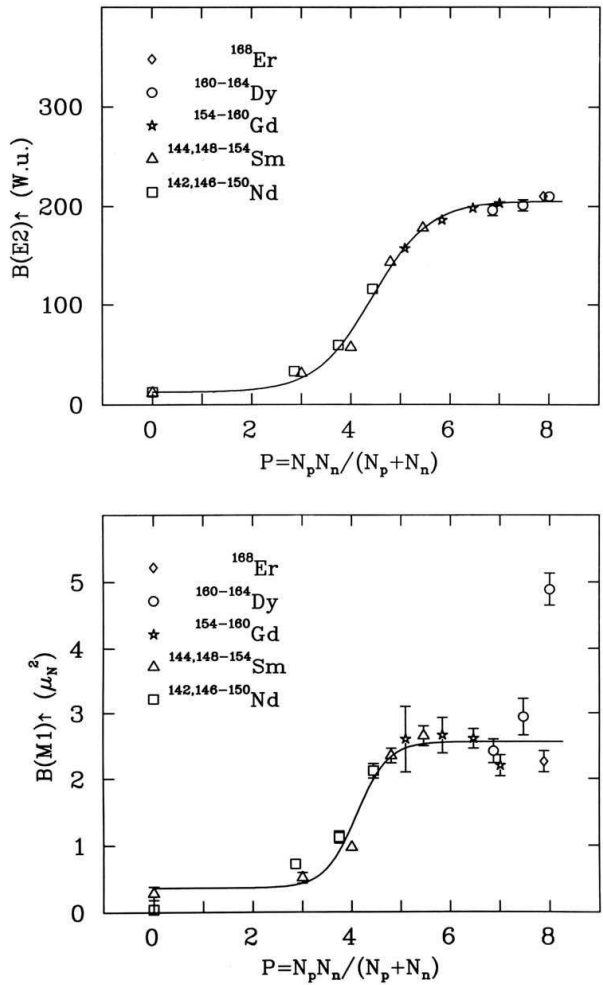}
\caption{Upper part: $E2$ transition strengths of the $2^{+}_{1}$
states in the even-even rare-earth nuclei indicated in the figure
versus $P$. Lower part: The same for the summed $M1$ strength in the
energy range $E_{x} = 2.5 - 4$ MeV. The solid lines correspond to
fits explained in \textcite{Rangacharyulu:1991}.}\label{Fig26}
\end{figure}

Firstly, as pointed out already above, the summed $M1$ strength
correlates with the square of the equilibrium quadrupole deformation
value for series of isotopes like the Nd and Sm nuclei
(Fig.~\ref{Fig13}, taken from \textcite{Ziegler:1990}). Secondly, the
summed $M1$ strength also correlates linearly with the
$B(E2;0^{+}_{1}\rightarrow2^{+}_{1}$) value for most nuclei in this
mass region \cite{Richter:1995}. Furthermore, both the summed $M1$
strength and the particular $B(E2)$ value scale in the same way and
saturate \cite{Rangacharyulu:1991} when plotted, not as a function of
proton(neutron) number (Fig.~\ref{Fig26}), but using the variable $P =
N_{p}N_{n}/(N_{p} + N_{n})$ with $\frac{N_p}{2}, \frac{N_n}{2}$, the
number of proton and neutron pairs, respectively, counted from the
nearest closed shells (the number of bosons in the IBM-2 model),
introduced by \textcite{Casten:1987}. Thirdly, it has been shown that
the summed $M1$ strength even scales linearly with the isotopic shift
for those nuclei (Nd, Sm, Dy) where both sets of data are available
\cite{Heyde:1993c}. These three important observations all point
towards close interconnections between the $M1$, $E2$ and the $E0$
electromagnetic properties for the rare-earth region.

Before going into some more detail discussing what theoretical
models predict for the behavior of the summed $M1$ strength, the
deformation dependence of the $B(E2;0^{+}_{1}\rightarrow2^{+}_{1})$
strength and the radius are rather obvious from a collective
geometrical approach \cite{Bohr:1975}. The connection of the latter
quantities to the summed magnetic dipole strength, however, was much
less expected to appear, in particular the quadratic dependence on
deformation and the same saturation behavior when passing through
the rare-earth region.

\begin{figure}[tbh]
\includegraphics[angle=0,width=8.5cm]{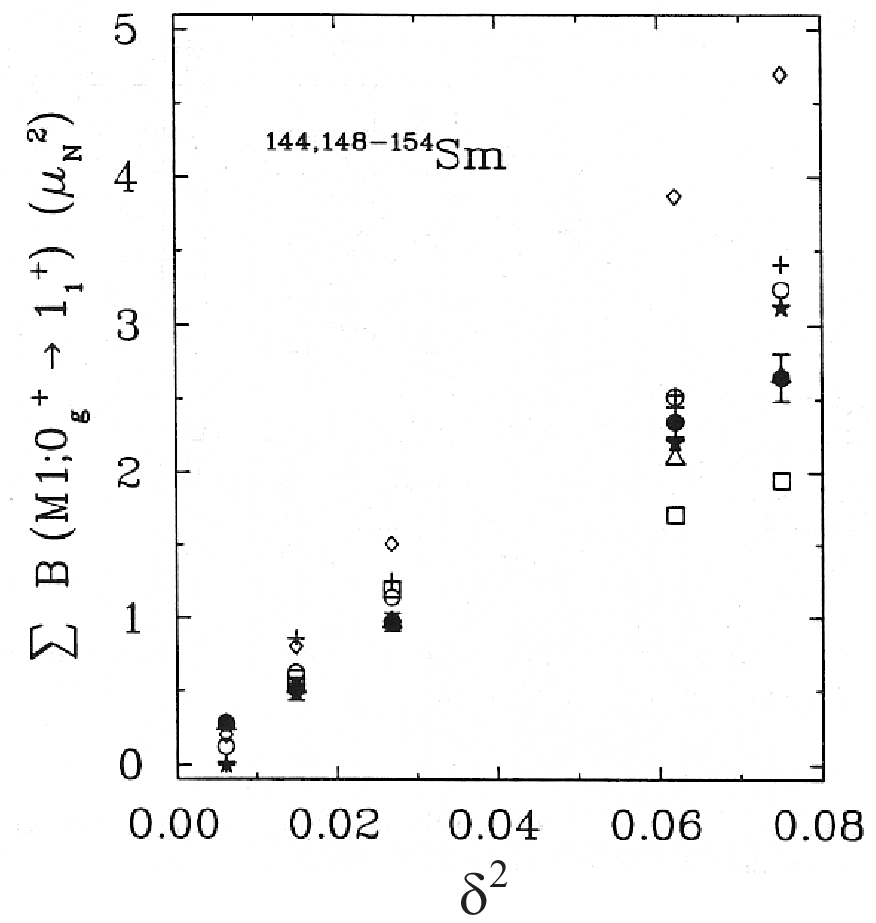}
\caption{Orbital $M1$ strength versus the square of deformation in
the Sm isotopes. The experimental points (black circles with error
bars) are compared to six theoretical predictions from
\textcite{Hamamoto:1991} (circles), \textcite{Garrido:1991}
(crosses), \textcite{Hilton:1993} (diamonds), \textcite{Heyde:1991}
(squares), \textcite{LoIudice:1993b} (triangles), and
\textcite{Sarriguren:1994} (stars), respectively
\cite{Richter:1995}. Note that the experimental points do overlap
with the triangles in a number of cases, making the distinction
difficult.}\label{Fig27}
\end{figure}

Starting from the generalized Bohr-Mottelson model,
\textcite{LoIudice:1993b} have worked out a sum rule that holds very
generally and is essentially model independent and parameter free. The
resulting expression
\begin{equation}
B(M1) \uparrow \simeq 0.0042 E_{sc}A^{\frac{5}{3}}\delta^{2}
(g_{p}-g_{n})^{2}~\mu^{2}_{N} ,
\end{equation}
directly contains the dependence on deformation as well as on the
gyromagnetic factors associated with the collective motion of the
deformed proton and neutron systems. In the region covering
transitional and strongly deformed rare-earth nuclei, the ratio of the
experimental over the theoretical summed $M1$ strength is close to one
over a large mass span, with a drop off towards the heavier masses. A
specific example in which the sum rule is compared to a number of other
models and the data for the Sm nuclei is given in Fig.~\ref{Fig27}
\cite{Hamamoto:1991,Garrido:1991,Hilton:1993,Heyde:1991,LoIudice:1993b}.
While all of them roughly reproduce the quadratic dependence, the
theoretical model results exhibit rather different slopes, in some
cases with serious deviations from the data for the more deformed $A
=152,154$ nuclei. \textcite{LoIudice:1997} has also shown that the TRM
and IBM-2 sum rules are very closely related and all can be brought
back to the basic sum rule structure as deriving from the QRPA study of
collective magnetic dipole excitations.

After the above experimental observations were well established,
various theoretical ideas in deriving closed expressions for this
summed $M1$ strength, aiming at establishing at the same time a
relation to the $E2$ and the $E0$ nuclear properties, have been
explored. Here, we discuss  implications of both non-energy weighted
and energy-weighted sum rules, using collective model as well as
shell-model approaches.

\begin{figure}[tbh]
\includegraphics[angle=0,width=8.5cm]{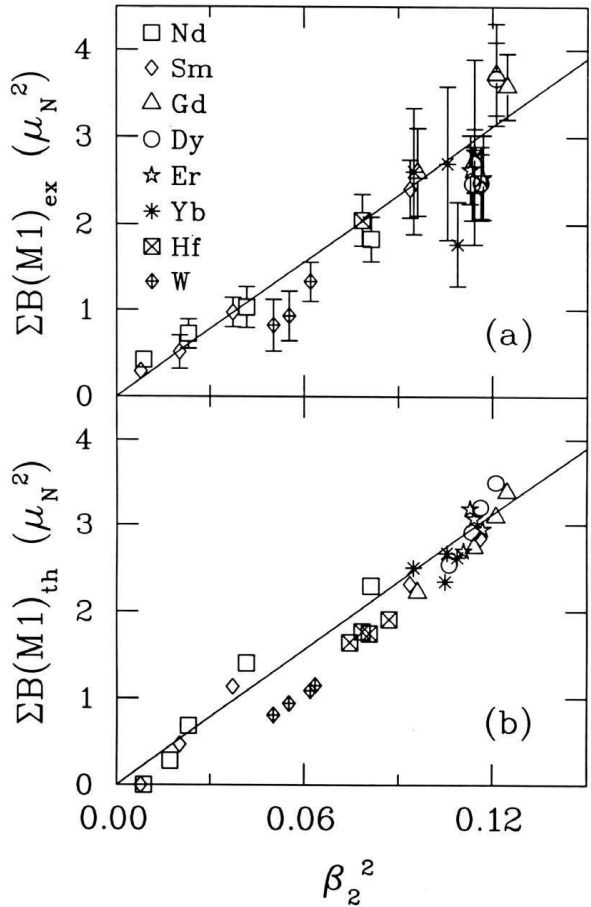}
\caption{Experimental $M1$ scissors mode strengths as a function of
the deformation parameter $\beta_{2}^2$. (a) The straight line is a
least-squares fit assuming intercept zero. (b) Prediction of an
IBM-2 sum rule for the scissors mode strength of all even-even
stable nuclei from Nd to W
\cite{vonNeumann-Cosel:1995}.}\label{Fig28}
\end{figure}

\textcite{Ginocchio:1991} proposed a non-energy weighted $M1$ sum rule
within the IBM for an $N$-boson system
%
\begin{eqnarray}
\label{eqn:ibm-newsr}
\sum B(M1)=\frac{9}{4\pi}
\left(g_{\pi}-g_{\nu}\right)^{2}\frac{P}{N-1}\left\langle
0^{+}\left|\hat{n}_{d}\right|0^{+}\right\rangle,
\end{eqnarray}
%
an expression which connects the summed $M1$ strength with the
expectation value of the number of $d$ bosons  in the nuclear ground
state. This latter quantity is also a measure of deformation because
the average number of $d$ bosons in the ground-state $\left\langle
0^{+}\left|\hat{n}_{d}\right|0^{+}\right\rangle / N$ can be expressed
by a deformation parameter $\beta_{IBM}$. A relation was derived
\cite{vonNeumann-Cosel:1995} between this specific IBM quantity and the
corresponding geometrical definition of deformation like the
Bohr-Mottelson parameter $\beta_{2}$, as
\begin{equation}
\beta_{IBM} =
\frac{3\lambda}{2\sqrt{\pi}}\frac{Z}{Z_{val}}\beta_{2} .
\end{equation}
Here, $Z_{val}$ describes the number of protons in the valence shell
and $\lambda$  - not to be confused with Majorana strength parameter in
Eq.~(\ref{eqn:ibm2}) - is a measure of how much of the $E2$ sum rule is
exhausted by the transition to the $2^{+}_{1}$ state. The original sum
rule of Ginocchio has thereby been extended and applied to the full
range of nuclei spanning the Nd to W region providing an excellent
description of the quadratic quadrupole deformation dependence. This is
illustrated in Fig.~\ref{Fig28}.

Making use of the IBM-2, also other sum rules have been derived
(energy-weigthed $M1$ sum rule and even sum rules for other
multipoles). The energy-weighted sum rule indicates a direct
relation between summed $M1$ strength and the
$B(E2;0^{+}_{1}\rightarrow2^{+}_{1}$) value when the assumption is
made that most $E2$ strength remains in the first excited $2^{+}$
state (which is a rather good approximation for transitional and
definitely so for deformed nuclei) and gives rise to the following
relations
%
\begin{eqnarray}
&\left\langle
0^{+}_{1}\left|[[\hat{H},\hat{T}(M1)],\hat{T}(M1)]\right|
0^{+}_{1}\right\rangle \\
&= \frac{2}{\sqrt{3}} \sum B(M1)
\left\{E_{x}(1^{+})-\lambda N \right\}, \nonumber
\end{eqnarray}
%
or,
%
\begin{equation}
\label{eqn:ibm-m1e2}
\sum B\left(M1\right)
\left\{E_{x}\left(1^{+}\right) - \lambda N\right\} = c
\sum B\left(E2\right)
\end{equation}
%
when using $F$-spin symmetry in evaluating the quadrupole expectation
value. Here, $\lambda$ denotes the strength of the Majorana term in
Eq.~(\ref{eqn:ibm2}). This expression indeed relates the
energy-weighted $M1$ sum rule with the non-energy weighted $E2$ sum
rule. The quantity $c$ is introduced to match dimensions of the left-
and right-hand side of this equation. This relation is discussed and
illustrated more explicitly in \textcite{Heyde:1992} in which it is
shown under what approximations the above relation reduces to the
non-energy weighted $M1$ sum rule. There, it is also shown that the
effect of the Majorana term can be incorporated to a large extent.

It is important though to study analogous sum rules for the magnetic
dipole strength but now starting from a shell-model formulation of the
problem. Using protons and neutrons, explicitly, \textcite{Zamick:1992}
were able to derive an energy-weighted magnetic sum rule which was
refined subsequently \cite{MoyadeGuerra:1993} with the more general
result
%
\begin{eqnarray}
\sum B\left(M1\right)E_{x} \left(1^{+}\right)=
\frac{9\chi}{16\pi}\left[B\left(E2\right)_{IS}-
B\left(E2\right)_{IV}\right],
\label{eq:m1-ewsr}
\end{eqnarray}
%
which is indeed very close to the form of the IBM-2 result in
Eq.~(\ref{eqn:ibm-m1e2}), except for the additional isovector (IV)
contribution. Here, $\chi$ denotes the strength of the
quadrupole-quadrupole interaction and the $B(E2)$ values are in units
of e$^{2}$fm$^{4}$. For many of the transitional and definitely for the
strongly deformed nuclei, the second term is very small and thus one
recovers the IBM-2 result exactly. For a number of cases, though,
\textcite{Zamick:1992} explicitly show the need for the isovector term
in order to have the correct physics in connecting $M1$ and $E2$
electromagnetic properties. It is important to note that, similar to
the collective IBM-2 formulation and the TRM treatment, the residual
two-body forces may contain besides the strong quadrupole forces,
pairing interactions amongst identical nucleons. The sum rule is not
affected by the addition of the latter term.

Moreover, one can relate the non-energy weighted $M1$ sum rule to the
nuclear monopole properties \cite{Heyde:1993c}. Starting from the IBM
expression for the monopole operator
\begin{equation}
T(E0)=\gamma_{0}\hat{n}_{s}+\beta_{0}\hat{n}_{d}=\gamma_{0}\hat{N}+\beta'_{0}
\hat{n}_{d} ,
\end{equation}
with $\hat{n}_{s},\hat{n}_{d},\hat{N}$ representing the $s$, $d$ and
total boson number operators, respectively, and $\beta'_{0} \equiv
\beta_{0}-\gamma_{0}$, one derives the mean-square radius as
\begin{equation}
\langle r^{2} \rangle = \gamma_{0}N + \beta'_{0} \langle \hat{n}_{d}
\rangle.
\end{equation}
Thus, the $M1$ sum rule of Eq.~(\ref{eqn:ibm-newsr}) can be recast in
the form
%
\begin{eqnarray}
\sum B(M1)=\frac{9}{4\pi}
\left(g_{\pi}-g_{\nu}\right)^{2}
\frac{1}{\beta'_{0}}\frac{P}{N-1}\left[ \langle r^{2} \rangle -
\gamma_{0} N \right].
\end{eqnarray}

\begin{figure}[tbh]
\includegraphics[angle=90,width=8.5cm]{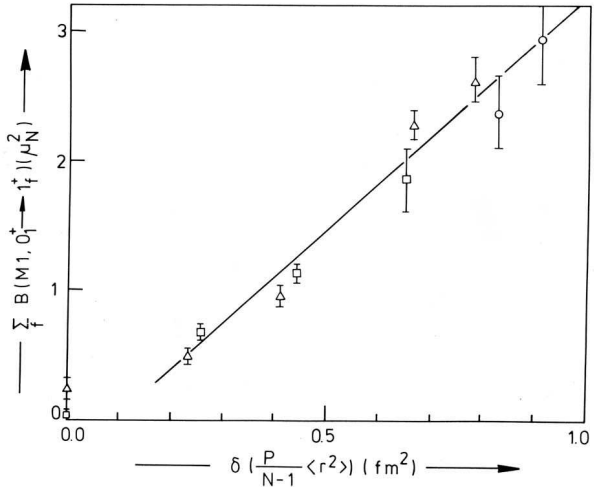}
\caption{Relation between the experimental summed $M1$ strength of
all $1^{+}$ states below 4 MeV for $^{142,146-150}$Nd (squares),
$^{144,148-154}$Sm (triangles), $^{160-162}$Dy (circles ) and the
variation of the quantity $ \delta (\frac{P}{N-1}
\langle{r}^2\rangle)$  related to the isotope shift
\cite{Heyde:1993c}.}\label{Fig29}
\end{figure}

This latter expression (see Fig.~\ref{Fig29} for a comparison with
data) shows the connection between summed $M1$ strength and nuclear
radial properties (the latter taken from the compilation of
\textcite{Otten:1989}). Similar connections have also been suggested by
\textcite{Iachello:1981} and by \textcite{Otsuka:1992a}, separately
because of the connection between the nuclear radial variation $\Delta
\langle r^{2} \rangle$ for a liquid drop and the variation in
quadrupole deformation $\Delta \langle \beta^{2}_2 \rangle$.

\paragraph{Deformation dependence and saturation}
\label{th-deform-sat}

Starting from a microscopic approach (deformed mean-field with residual
pairing and quadrupole interactions and even using Hartree-Fock
calculations using Skyrme effective forces
\cite{Garrido:1991,Smith:1995,LoIudice:1998}), the very same dependence
between the summed M1 strength and nuclear deformation results, as
discussed before. Remarkably, it is the presence of pairing forces that
modifies the dependence of the summed $M1$ strength from linear to
quadratic. This was shown to be the case by \textcite{Hamamoto:1991} in
a Nilsson model study, and also by \textcite{LoIudice:1993b} using an
RPA study of magnetic scissors motion.

The fact that the summed $M1$ strength shows this striking
collective behavior immediately leads to saturation because
the equilibrium quadrupole deformation (in passing through the
rare-earth region from $A=140$ towards the mass $A=180$ region)
stabilizes at a value of $\delta \simeq 0.25$ in the region $A \geq
160$ of strongly deformed even-even nuclei. The precise origin of this
saturation stems from the specific single-particle structure in the
deformed mean field and from the balancing effects of shell- and
pairing corrections to the liquid-drop energy \cite{Heyde:1992}. The
saturation arises after a steep increase in deformation, which is
reflected in a steep rise in both the summed $M1$ strength and in the
$B(E2;0^{+}_{1}\rightarrow2^{+}_{1})$ value when entering the region of
deformation, starting from closed-shell nuclei (see Fig.~\ref{Fig26}).
This strong correlation between the summed $M1$ strength and the
ground-state equilibrium quadrupole deformation has been discussed by
\textcite{DeCoster:1989c}.

\begin{figure}[tbh]
\includegraphics[angle=0,width=8.5cm]{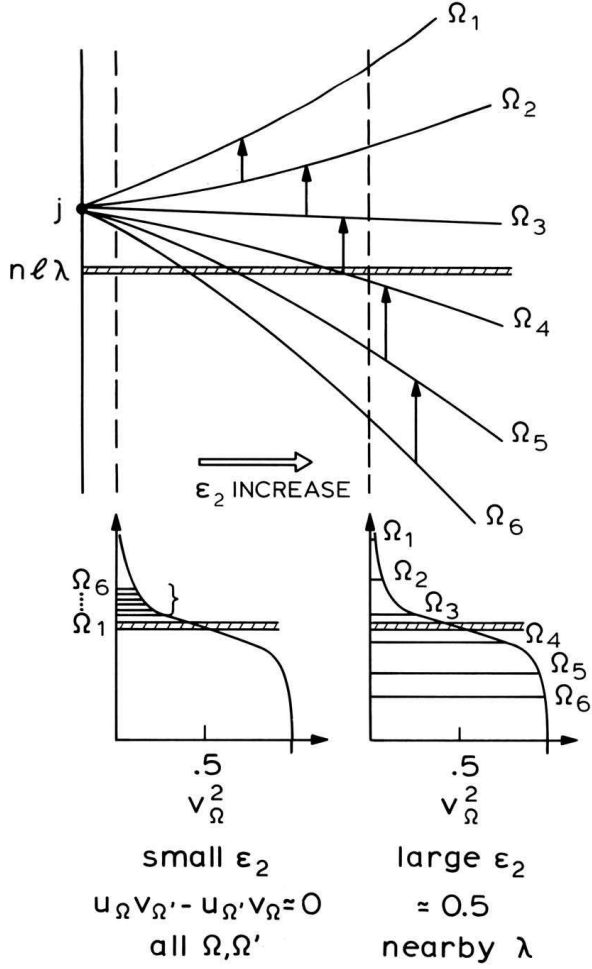}
\caption{Schematic representation of the effect of deformation on
the orbital $ (lj) \rightarrow (lj)$ transition strength. The Fermi
level is denoted by the hatched lines \cite{Richter:1990a}.}\label{Fig30}
\end{figure}

The dominant role of pairing to obtain the correct deformation
dependence for the magnetic summed strength is not straightforward but
comes in indirectly. In a quadrupole deformed potential, the strength
of $M1$ transitions between Nilsson orbitals, characterized by $\Omega$
(the projection of $j$ on the symmetry axis) and illustrated by the
arrows in the top part of Fig.~\ref{Fig30}, can be expressed as
\begin{equation}
\label{eqn:nilsson}
B(M1)= \frac{3}{4\pi}(u_{1}v_{2}-u_{2}v_{1})^{2} {\mid \langle
\Omega_{1}|g_{l}\hat{l}_{+} + g_{s}\hat{s}_{+}|\Omega_{2}
\rangle \mid}^{2}.
\end{equation}
The occupation probabilities $v_{i}^2$ (with $u_{i}^2 = 1 - v_{i}^2$)
of the Nilsson orbitals $\Omega_{i}$ are schematically drawn in the
bottom part of Fig.~\ref{Fig30} for both small and large quadrupole
deformations. The doubly-hatched lines indicate the position of the
Fermi level (top part) and for an occupation number 0.5 (lower part).
For small deformation the pairing factor $(u_{1}v_{2}-u_{2}v_{1})^2$
quenches the $M1$ strength and vanishes for zero deformation. With
increasing deformation, the Nilsson single-particle orbitals $\Omega_i$
originating from a single $j$ shell-model orbital are more spread out
and the corresponding occupation probabilities become different
resulting in a rather large pairing factor. Then, as deformation does
not change much for increasing mass numbers, the pairing factors remain
roughly constant, causing saturation before they start to decrease
again towards the end of the shell. Using the approximation that the
energy of the Nilsson orbitals vary linearly with deformation for not
too large values of deformation, it can even be shown that the pairing
factor $(u_{1}v_{2}-u_{2}v_{1})^2$ becomes proportional to
$\delta^{2}$. This result comes very close to what
\textcite{Hamamoto:1991} obtained too.

A particularly interesting example of this deformation dependence has
resulted from the study of magnetic dipole strength in superdeformed
nuclei \cite{Hamamoto:1992,Hamamoto:1994}. The summed $B(M1)$ strength
was found to be much larger than in nuclei at normal deformation. This
can be understood from the growing proton orbital contribution with
increasing deformation and the fact that for the weak pairing present
in the superdeformed configuration, the pairing factor in
Eq.~(\ref{eqn:nilsson}) becomes maximal. Applications for nuclei in the
proton-rich deformed Kr-Zr nuclei show similar results
\cite{Nakatsukasa:1994}.

Whereas QRPA calculations for strongly deformed nuclei give rise to the
$\delta^{2}$ dependence of the M1 strength and saturation for both the
$B(E2;0^+_1 \rightarrow 2^+_1)$ and summed M1 strength due to the
importance of pairing correlations amongst the interacting nucleons,
the IBM does not exhibit this characteristic behavior. In both the
SU(3) and in the O(6) limit of the IBM-2, the $B(M1)$ transition into
the scissors $1^{+}$ mixed-symmetric state is proportional to the $P$
factor \cite{Casten:1987} which implies an almost linear rise towards
mid-shell before dropping off towards the end of the major shell
\cite{Scholten:1985a}. This sheds light on the way in which to count
the boson number when highly deformed systems are being described.
\textcite{Casten:1988} have shown that one should consider in this
region an effective boson number, following arguments by
\textcite{Otsuka:1990a}, through which Pauli blocking is taken into
account.

\paragraph{A comprehensive analysis} \label{sec:th-comprehen}
Even though it was shown in the foregoing sections that one can
obtain a qualitative understanding of the observed behavior of the
summed magnetic dipole strength at low energies (deformation
dependence, saturation, relation to other multipoles), a
quantitative agreement between the large body of experimental data
and theory has still been lacking. Part of the problem is related to the
fact that most model approaches use still too idealized assumptions
concerning the way in which nucleons behave inside the atomic
nucleus: moments of inertia, gyromagnetic ratios, etc.. Therefore, a
global study in heavy even-even nuclei was carried out in order to
obtain as accurate a description of the scissors mode when one uses
as input the realistic physical parameters in the calculations
\cite{Enders:1999,Enders:2005}.

It makes use of the sum rule method as described by
\textcite{Lipparini:1983,Lipparini:1989a}, LS in shorthand notation,
and starts from the energy-weighted ($S_{+1}$) and the inverse
energy-weighted ($S_{-1}$) sum-rule expressions
\begin{eqnarray}
S_{+1} & = & E_{sc}B(M1)  \nonumber \\
& = & \frac{3}{20\pi}r^{2}_{0}A^{\frac{5}{3}}\delta^{2}
E^{2}_{D}\frac{m_{N}}{\hbar^2}\left(g_{p}-g_{n}\right)^2~{\mu}^{2}_{N}
{\rm MeV},\label{eq:s+1}
\end{eqnarray}
and (cf.\ Eq.~(\ref{eqn:m1_sum_LoIudice}))
\begin{eqnarray}
S_{-1}=\frac{B(M1)}{E_{sc}}=\frac{3}{16\pi}\frac{J_{sc}}{\hbar^2}
\left(g_{p}-g_{n}\right)^2~\frac{{\mu}^{2}_{N}}{\rm MeV},
\label{eq:s-1}
\end{eqnarray}
%
%
%
%
with $r_{0} = 1.15$ fm, $A$ the nuclear mass number, $\delta$ the
nuclear deformation parameter, $E_{D}$ the isovector giant electric
dipole resonance (IVGDR) excitation energy, m$_{N}$ the nucleon mass
and g$_{p}$ (g$_{n}$) the $g$ factors for protons (neutrons). With
$E_{sc}$ we denote the excitation energy of the scissors mode and
$J_{sc}$ describes the moment of inertia associated with the scissors
mode vibrations, which are of isovector type. These expressions are
rather general and express e.g.\ the fact that the scissors mode and
the IVGDR are both of isovector nature and strongly related through the
restoring force acting on the deformed proton and neutron bodies and
also that the major contribution to $S_{-1}$ comes from the low-lying
scissors mode ($0 \hbar \omega$ strength in Fig.~\ref{Fig1}) whereas
the high-lying scissors mode ($2 \hbar \omega$ strength in
Fig.~\ref{Fig1}) mainly contributes to $S_{+1}$.

First we discuss $S_{-1}$ because the low-lying scissors mode, after
all, is very well studied by now in the heavy nuclei. To start with,
we use the common relative $g$-factor values $g_{rel} = g_{p} -
g_{n} = 2Z/A$ \cite{Bohr:1975}, and so deduce the moment of inertia
(taking the tacit assumption that the scissors $M1$ strength resides
in the energy region $2.5-4$~MeV). It is an interesting observation
to see \cite{Enders:1999,Enders:2005} that the ground-band and
scissors motion moments of inertia are very close to each other
(except for a systematic deviation in heavier nuclei).

In a next step, we take these two moments of inertia to be
equal\footnote{More precisely the moment of inertia of isoscalar and
isovector motion differ by a factor $4NZ/A^2 \simeq 0.96$
\cite{LoIudice:1993b}.}, i.e.\ $J_{sc}= J_{gb}$, and evaluate the $g$
factors starting from the sum-rule value of $S_{-1}$. There appears a
striking agreement between these values with only small deviations (of
the order of 10\%). When recalculating the moment of inertia related to
the scissors motion in Eqs.~(\ref{eq:s+1},\ref{eq:s-1}), using the
experimentally deduced $g(2^{+}$) values and comparing these moments
with the ground-state moments of inertia, an almost perfect overlap
between the two sets results. So, we can draw the conclusion that the
$g$ factors acting in the scissors mode are the same as the ones for
the ground band.

Since all quantities in the sum rules (\ref{eq:s+1},\ref{eq:s-1}) are
fixed, we are now in a position to rederive the energies and the
strengths for the scissors modes. However, when dealing with $S_{+1}$
one has to take into account that contributions from the $K=1$
component of the isovector giant quadrupole resonance (IVGQR) will
dominate, which have to be removed in order to compare with the
experimental data. This is achieved with a procedure described by
\textcite{Lipparini:1983}, which leads to a correction factor $\xi =
E_Q^2/(E_Q^2 + 2E_D^2)$ to Eq.~(\ref{eq:s+1}) where $E_Q$ denotes the
centroid energy of the isoscalar giant quadrupole resonance (ISGQR).

When putting all low-lying (high-lying) $M1$ strength
$\sum_{low}B(M1)=B_{l}$ $(\sum_{high}B(M1)=B_{h})$ into a single state
with energy $E_{l}(E_{h})$, one obtains the relations
\begin{equation}
S_{-1}\approx \frac{B_{l}}{E_{l}},~~S_{+1}\approx B_{h}\cdot E_{h},
\end{equation}
and using the expressions for $S_{-1}$ and $S_{+1}$, we can derive an
average energy
\begin{equation}
\bar{\omega} = \frac{2}{\sqrt{15}} \sqrt{\frac{m_N}{\hbar^2}}  r_0 \sqrt{\frac{4 N
Z}{A^2}} A^{5/6} E_D \sqrt{E_{2^+} \xi} \delta
\label{eq:omega},
\end{equation}
where use was made of the relation between scissors mode and
ground-state band moment of inertias established above and the latter
is expressed through the energy of the first excited state of the
rotational band $E_{2^+}$ by $J_{gb} = 3 \hbar^2 / E_{2^+}$. The
centroid energies of the IVGDR and ISGQR are taken from mass-dependent
systematics \cite{Harakeh:2001}.

\begin{figure}[tbh]
\includegraphics[angle=0,width=8.5cm]{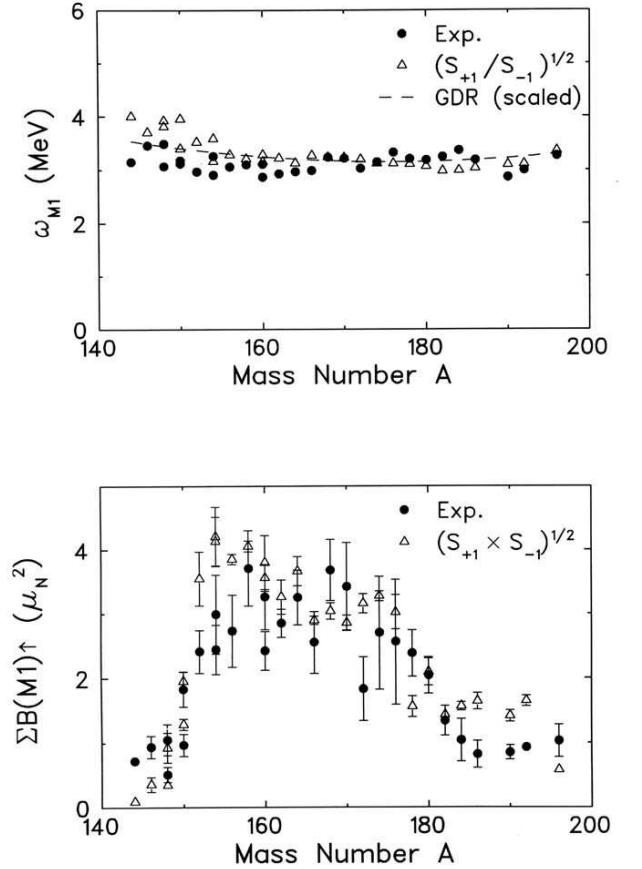}
\caption{Excitation energy (upper part) and transition strength systematics
(lower part) from a sum-rule analysis of the scissors mode in rare-earth nuclei.
Experimental values (solid circles) and parameter-free predictions
(open triangles) are shown for the mean excitation energy (upper
part) and the summed $M1$ strength (lower part). The
deformation dependence of the moment of inertia leads to
the proportionality of the excitation energies of the scissors mode
and the IVGDR as indicated by the dashed line \cite{Enders:2005}.
}\label{Fig32}
\end{figure}

The values resulting from Eq.~(\ref{eq:omega}) are drawn as triangles
in Fig.~\ref{Fig32} (upper part). Excellent agreement with the
experimental data is obtained.  When scaled by the ratio of the
scissors-mode to the liquid-drop moment of inertia, the energy of the
isovector giant dipole resonance (dashed line) also shows the
proportionality to the energy of the scissors mode predicted in
Eq.~(\ref{eq:omega}). We note that the observed near constancy of
$\omega$ could also be explained by \textcite{Pietralla:1998b} within a
schematic RPA approach after inclusion of the deformation dependence of
paring effects analogous to the discussion in the previous subsection.

In the same way, one obtains for the low-lying scissors strength
\begin{equation}
B_l = \frac{3}{\pi} \sqrt{\frac{3}{20}} r_0  \sqrt{\frac{4 N
Z}{A^2}} A^{5/6} E_D \sqrt{\frac{m_N \xi}{\hbar^2 E_{2^+}}} \delta
g_{gb}^2. \label{eq:strength}
\end{equation}
with $g_{gb}= \frac{1}{2}(g_p + g_n)$, the $g$ factor of the $2^+$
level of the ground-state band. The comparison to the experimental
scissors mode strength is depicted in the lower part of
Fig.~\ref{Fig32}. The agreement is very satisfactory except for some
nuclei with $A > 180$. The strong deformation dependence is generated
by the interplay of $E_{2^+}$ and $\delta$. Indeed the experimentally
established quadratic dependence of the scissors-mode strength on the
ground-state deformation is easily derived from Eq.~(\ref{eq:strength})
recalling \cite{Bohr:1975} that the moment of inertia $J_{gb}$ is
roughly proportional (albeit much larger) to the superfluid moment of
inertia $J_{liq}$
\begin{equation}
J_{gb} \propto J_{liq} \simeq J_{rig}
\delta^2, \label{eq:ISliqrig}
\end{equation}
where $J_{rig}$ stands for the moment of inertia of a rigid rotor.

\paragraph{Sum rule relation between magnetic dipole and octupole strength}
\label{sec:th-sumrule-relation}

In the spirit of the IBM-2 non-energy weighted $M1$ sum rule
\cite{Ginocchio:1991}, a relation between magnetic dipole and
octupole strength could be derived starting from energy-weighted sum
rules \cite{Heyde:1994}. It has been possible to obtain an
approximate, yet very simple relation
\begin{eqnarray}
\frac{\sum_{f}B(M1;0^{+}_{1} \rightarrow 1^{+}_{f})E_{x}(1^{+}_{f})}
{\sum_{f}B(M3;0^{+}_{1} \rightarrow 3^{+}_{f})E_{x}(3^{+}_{f})}
\cong {\left [ \frac{6(g_{\pi} -
g_{\nu})}{7(\Omega_{\pi}-\Omega_{\nu})} \right ]}^{2}.
\end{eqnarray}

This relation establishes a link between different parameters of the
IBM-2, the gyromagnetic boson factors and octupole boson moments, and
hence imposes constraints on choosing them when fitting to other
spectroscopic data. One can, on the other hand, also use the relation
starting from common values for these parameters
$(g_{\pi},g_{\nu},\Omega _{\pi},\Omega_{\nu})$ as derived from a
phenomenological and/or microscopic starting point and deduce an
estimate for the summed $M3$ strength, whenever information on the
summed $M1$ dipole strength is available. This has been discussed by
\textcite{DeCoster:1995}, where an estimate of the summed $M3$ strength
is presented in the mass region $144 \leq A \leq 164$. Even though it
contains a number of approximations, in the absence of systematics on
the $M3$ strength, the above method might be a first guide for further
experimental studies.

There are very few studies on $M3$ transitions carried out by now, both
theoretically and experimentally. On the theoretical side, $M3$
transitions were investigated within the framework of the IBM-2 by
\textcite{Scholten:1984} and within the context of a schematic RPA
study for heavy deformed nuclei by \textcite{LoIudice:1988}. There has
been an early experimental search for $M3$ strength in $^{164}$Dy by
\textcite{Bohle:1987b} that made use of both the electron accelerators
at Darmstadt and Amsterdam. Only an upper limit for such strength has
been derived. A more systematic search for magnetic octupole strength
is called for in the light also of the above approximate connection
with the summed $M1$ strength.

\subsection{Spin-flip mode: experimental evidence and theoretical description}
\label{sec:th-spinflip}

\subsubsection{Qualitative nature of the magnetic dipole response}
\label{sec:th-mdr}


The magnetic dipole operator, for a system of protons and neutrons
reads
\begin{equation}
T(M1) =\sqrt{\frac{3}{4\pi}}\sum_{i}\left\{g_{l}(i)\hat{l_{i}} +
g_{s}(i)\hat{s_{i}}\right\}\mu_{N},
\end{equation}
with the usual orbital and spin $g_{l},g_{s}$ factors for neutrons
and protons (see also Eq.~(\ref{eqn:m1op-fermion})). Using the
isospin labels $t_{z}(i)= \pm\frac{1}{2}$ for neutron and proton,
respectively, the magnetic dipole operator can be split into an
isoscalar and isovector term in the following way
\begin{equation}
T(M1)= \sqrt{\frac{3}{4\pi}}(g_J\hat{J} + g_S \hat{S})\mu_{N} + T(M1,IV).
\end{equation}
Since $\hat{J}$ denotes the total angular momentum operator, this
term does not induce any $M1$ transitions and because $g_S =
[(g^{\pi}_{s}+g^{\nu}_{s})-1]/2$,
the isoscalar spin part only contributes in a minor way to $M1$
transitions. This is a consequence of the opposite signs in the proton
and neutron spin $g_{s}$ factors resulting in a value of $g_S$= 0.38.

The isovector part of the $M1$ operator $T(M1,IV)$
%
\begin{eqnarray}
T(M1,IV) & = &\sqrt{\frac{3}{4\pi}}\ \left
\{\frac{1}{2}(\hat{L}_{\pi}-\hat{L}_{\nu}) \right. \nonumber \\
&& \left. + \frac{1}{2} (g^{\pi}_{s}-g^{\nu}_{s})(\hat{S}_{\pi}-\hat{S}_{\nu})\
\right \}~\mu_{N} \label{iv-m1-op}
\end{eqnarray}
%
%
splits into two pieces: the first part, describing the relative angular
momentum between protons and neutrons, generates the scissors orbital
motion whereas the second part, a spin-flip part, is nothing else but
the $\Delta T_{z}$=0 component of the Gamow-Teller operator. This term
can strongly enhance spin-flip $M1$ transitions because of the large
factor $\frac{1}{2}(g^{\pi}_{s}-g^{\nu}_{s})$ in front (with a
numerical value 4.72 using free $g_s$ factors).

\begin{figure}[tbh]
\includegraphics[angle=0,width=8.5cm]{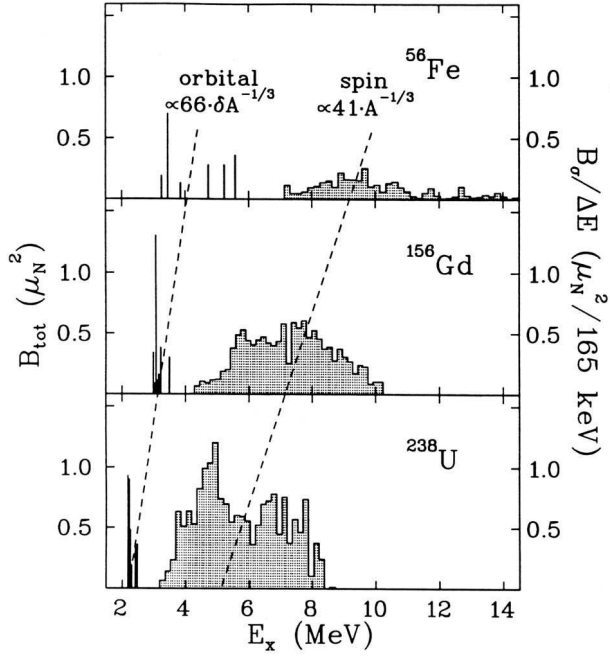}
\caption{The nuclear orbital and spin magnetic dipole response in a
medium-heavy, a heavy and a very heavy nucleus, derived from
experiments with electromagnetic and hadronic probes, respectively
\cite{Richter:1993b}.}\label{Fig33}
\end{figure}

That the simple picture of the nuclear magnetic dipole response is
approximately correct is shown in Fig.~\ref{Fig33} by using the three
nuclei $^{56}$Fe, $^{156}$Gd and $^{238}$U as examples. As pointed out
above, the mean excitation energy of the orbital mode scales
approximately with deformation as $E_{x} \simeq 66 \delta A^{-1/3}$
MeV. The spin $M1$ strength obtained in inelastic proton scattering
\cite{Frekers:1990} lies at $E_{x} \simeq 41A^{-1/3}$ MeV and thus
exhibits a shell-model like excitation energy dependence. As seen in
Fig.~\ref{Fig33}, the spin strength represents the largest fraction of
the $M1$ strength.

It is the residual particle-hole interaction acting in the spin-isospin
channel that shifts this spin $M1$ strength to higher energies as can
be seen e.g.\ in schematic model
\cite{DeCoster:1991a,DeCoster:1992,Zawischa:1990a,Zawischa:1994a} and
in RPA studies \cite{Sarriguren:1993,Sarriguren:1994,Sarriguren:1996}.
The orbital part of the $M1$ strength is hardly moved by this force
component and so one expects to observe in the experiments a rather
good overall separation of the orbital dipole magnetic excitations, at
the lower energy end of $2.5-4$~MeV, from the higher-lying spin
magnetic dipole excitations.

\begin{figure}[tbh]
\includegraphics[angle=0,width=8.5cm]{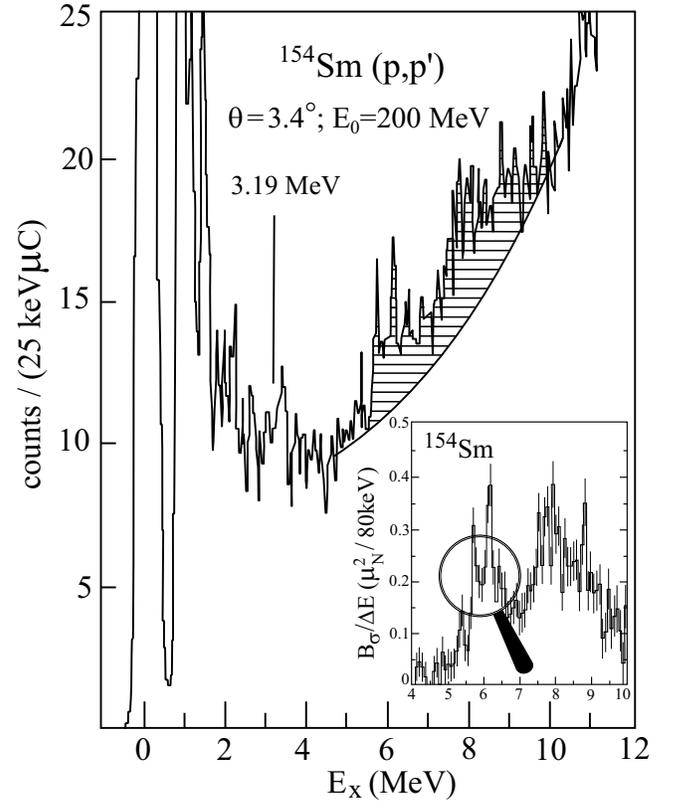}
\caption{Forward angle inelastic proton scattering spectrum taken at
200 MeV incident energy on $^{154}$Sm. The shaded area constitutes
the spin magnetic dipole giant resonance. In the insert, the extracted
B(M1) strength (in units $\mu_N^2/80$ keV) distribution is shown
(adapted from \textcite{Frekers:1990}).}\label{Fig34}
\end{figure}

The experimental detection of spin strength in the energy region above
4~MeV needs a probe that is particularly sensitive to the spin part of
the nuclear current. Intermediate-energy scattering of (polarized)
protons at small forward angles should be the optimal selective
reaction to carry out such a search. First experiments performed by a
Darmstadt/M\"unster/TRIUMF collaboration on $^{154}$Sm, $^{158}$Gd, and
$^{168}$Er used 200 MeV protons at an angle of $3.4^\circ$ covering
final states up to 12 MeV. Analyzing those data clearly showed in all
nuclei the presence of extra strength sitting on the tail of the IVGDR
(Fig.~\ref{Fig34}) with a double-hump structure
\cite{Frekers:1990,Richter:1995}. The double-hump has centroids around
6 and 8.5 MeV and widths of about 1.5 and 2 MeV, respectively. Even
more detailed substructure becomes visible (see insert in
Fig.~\ref{Fig34}). Such a pronounced splitting and fragmentation of
magnetic dipole strength has not been observed as yet in spherical
nuclei. The selectivity of the $(p,p')$ reaction could be demonstrated
at the same time: the orbital $M1$ tranition at $3.19$ MeV strongly
excited in $(e,e')$ and ($\gamma,\gamma'$) reactions was not observed
with an upper limit  $B(M1)< 0.1$ $\mu^2_N$.

\begin{figure}[tbh]
\includegraphics[angle=0,width=8.5cm]{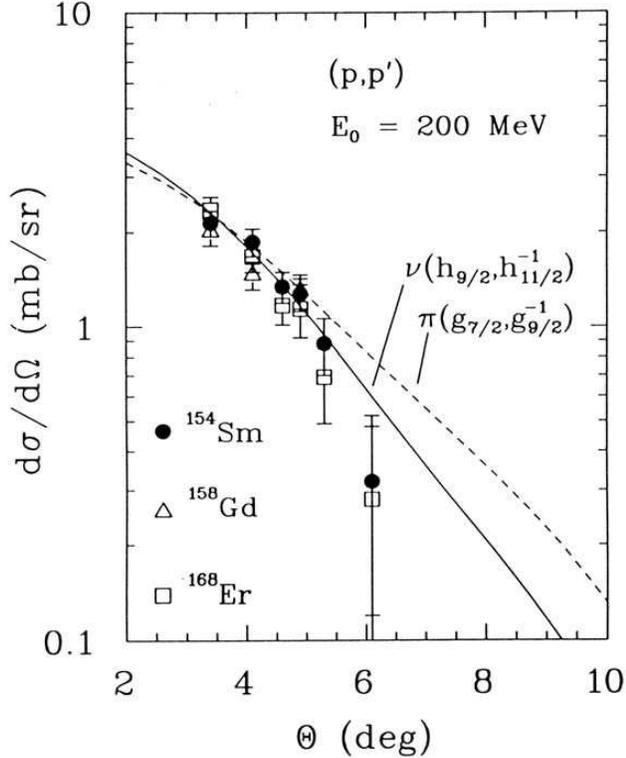}
\caption{Angular distribution of the summed double-humped structures
observed in $^{154}$Sm, $^{158}$Gd, and $^{168}$Er. The dashed and full curves
result from DWBA calculations based on the indicated particle-hole
excitations \cite{Richter:1991}.}\label{Fig36}
\end{figure}

In order to obtain an even more complete insight in the structure of
these excitations, angular distributions have been taken for these
three nuclei (Fig.~\ref{Fig36}). No strong $Z$ nor $A$ dependence shows
up. A DWBA fit for a $\Delta S=1$, $\Delta L=0$ transition considering
a neutron spin-flip $1h_{11/2}\rightarrow 1h_{9/2}$ or a proton
$1g_{9/2}\rightarrow 1g_{7/2}$ transition has been performed. These
orbitals are clearly the dominant ones in this mass region and for
deformed nuclei; even the Nilsson states are dominated by these
particular spherical components. A strength  $B(M1) =
10.5(2.0)$~$\mu_{N}^2$ could be extracted, a value in line with
expectations for the theoretical spin-flip strength (see e.g.\
Fig.~\ref{Fig20}).

\begin{figure}[tbh]
\includegraphics[angle=0,width=8.5cm]{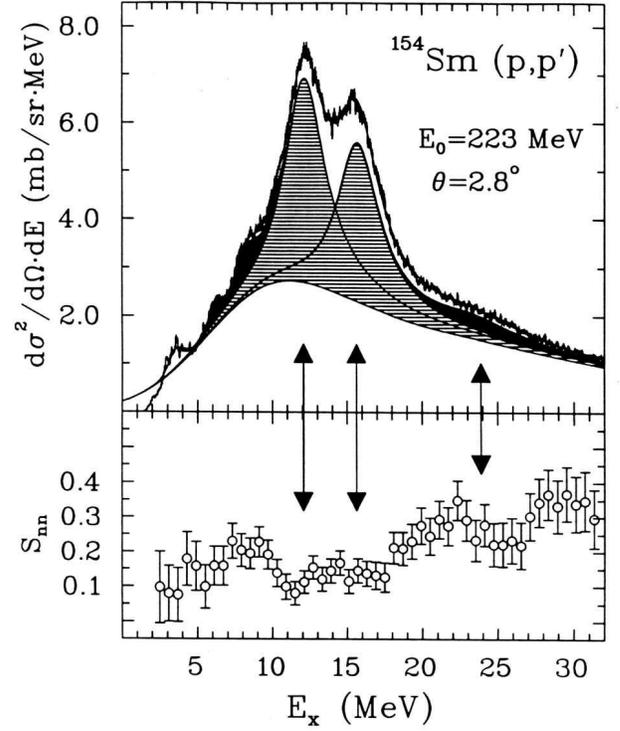}
\caption{Differential cross-section and transverse spin-flip
probability for inelastic polarized proton scattering on $^{154}$Sm.
The hatched areas show the double-humped GDR. Visible on the
low-energy side of the the GDR is the spin-flip $M1$ resonance
between $5-12$ MeV excitation energy and at $E_{x}=23.4$~MeV for the IVGQR.
The arrows visualize the connection between the electric resonances
and dips in the spin-flip probability \cite{Richter:1995}.}\label{Fig37}
\end{figure}

Complementary experiments detecting the transverse spin-flip
probability $S_{nn}$ have been carried out \cite{Woertche:1994}.
Thereby the probability that an incoming proton, interacting with the
target nucleus, will leave with its spin flipped ($\Delta S=1$ process)
is measured. The results for $^{154}$Sm are displayed in
Fig.~\ref{Fig37} where, besides the cross-section in the energy
interval $4-32$ MeV, the corresponding spin-flip probability $S_{nn}$
is given. Here one notices the presence of increased $S_{nn}$ values in
the region of the observed spin $M1$ strength (6 to 8.5 MeV),
confirming the spin-flip character of the structures located on the
low-energy tail of the GDR, the latter being split in two fragments,
due to nuclear deformation. This figure also shows the presence of an
isovector giant quadrupole resonance (IVGQR) derived from a multipole
decomposition around 23 MeV, an excitation energy where it is has also
been detected in other heavy nuclei \cite{Harakeh:2001}. The analysis
of the $S_{nn}$ values is carried out consistently with the DWBA
analysis of the angular distributions.

In recent years it has become clear that the $E1$ response in heavy
nuclei generally exhibits a local concentration of strength - called
the pygmy dipole resonance (PDR) - well  below the IVGDR overlapping
with the excitation energy region of the spin-flip resonance. The PDR
has been observed at a variety of shell closures (see
e.g.~\textcite{Kneissl:2006} and references therein) but not yet in
heavy deformed nuclei. This raises the question whether part of the
strength attributed to the spin-flip $M1$ resonance is in fact of $E1$
nature. While this problem needs further experimental investigation,
there are immediately two arguments in favor of the present
interpretation: the angular distributions shown in Fig.~\ref{Fig36} are
distinct from those of Coulomb-excited $E1$ transitions, and the
$S_{nn}$ values in the bottom part of Fig.~\ref{Fig37} display a local
maximum of the spin-flip strength.

\begin{figure}[tbh]
\includegraphics[angle=0,width=8.5cm]{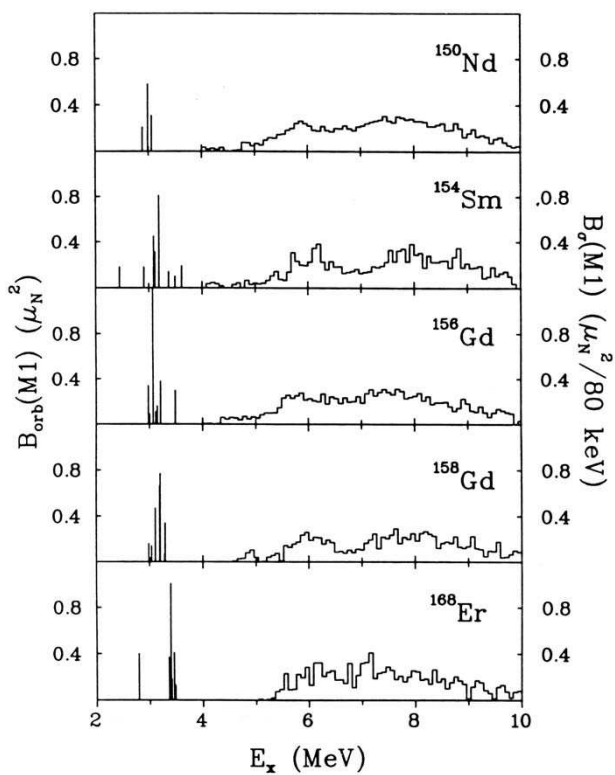}
\caption{Magnetic dipole response of several deformed rare-earth
nuclei determined by inelastic electron, photon and proton
scattering \cite{Richter:1995}.}\label{Fig38}
\end{figure}

\begin{figure}[tbh]
\includegraphics[angle=0,width=8.5cm]{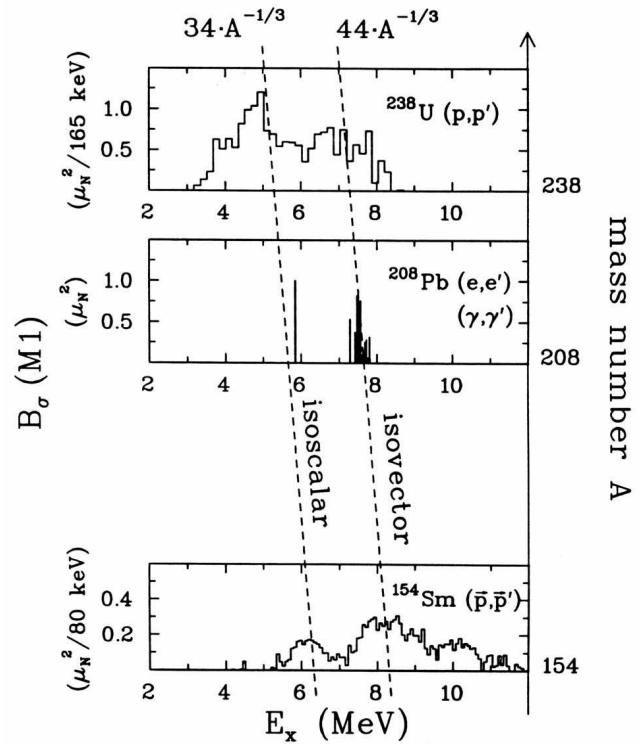}
\caption{Spin magnetic dipole strength distributions in $^{238}$U,
$^{208}$Pb and $^{154}$Sm. The center of gravity of the excitation
energy of the two peaks representing the main strength follows a
simple $A^{-1/3}$ law, characteristic for spin-flip excitations
between spin-orbit partners. The experimental strength distribution
for $^{208}$Pb has been combined from inelastic electron and photon
scattering experiments \cite{Richter:1995}.}\label{Fig39}
\end{figure}

Before concentrating on the theoretical description, we show finally
the $M1$ response for a set of nuclei spanning a wide region of
deformed rare-earth nuclei (Fig.~\ref{Fig38}). In all of these nuclei,
a particularly stable pattern is emerging: at the lower energy side, at
energies $2.5-4$ MeV, a concentration of orbital magnetic dipole
strength shows up with a ratio $\sqrt{B_{l} / B_{\sigma}} \approx 4$,
where $B_l$ and $B_\sigma$ denote the reduced transition strength of
the orbital and spin part of the magnetic dipole operator
(Eq.~\ref{iv-m1-op}) and $B(M1) = (\sqrt{B_l} \pm \sqrt{B_\sigma})^2$
\cite{Willis:1989}. Higher up, starting at 5.5 MeV up to almost 10 MeV,
a rather broad and extended region with a clear double-hump structure
in most of these nuclei appears in which $\sqrt{B_{l} / B_{\sigma}}
\leq 1$. Figure \ref{Fig38} therefore reflects the magnetic dipole
response for strongly deformed nuclei in the rare-earth region to
electromagnetic and hadronic probes. The low-energy part corresponds to
an orbital magnetic dipole structure, the scissors mode, and the higher
part is the spin-flip part mainly caused by proton and neutron
single-particle transitions between spin-orbit partners. Whereas the
energy of the spin-flip $M1$ mode will be localized at the energy of
the gap in closed shells for spherical nuclei, in the region where
deformation sets in one expects splitting of the various Nilsson energy
levels causing a spreading of the spin-flip strength around the
spherical centroid energy. If this argument is correct, one should
obtain an $A^{-1/3}$ excitation-energy dependence for the observed
peaks of $M1$ strength throughout the whole mass region (cf.\
Fig.~\ref{Fig33}). This seems indeed the case and is demonstrated in
Fig.~\ref{Fig39}, in which the double-hump $M1$ spin strength in
deformed rare-earth and actinide nuclei is connected
\cite{Richter:1995} to the detailed knowledge of the spin-flip strength
in the doubly-magic $^{208}$Pb nucleus \cite{Laszewski:1988}.
Indications of a similar splitting is observed in medium-mass nuclei
\cite{Djalali:1982} except for the stable Zr isotopes
\cite{Crawley:1982}, where the forward-angle $(p,p')$ cross sections
exhibit a single bump at $E_x \approx 9$~MeV identified as a spin-flip
$M1$ resonance. This may be related to the special shell structure at
$Z = 40$, where the high-$j$ orbital near the Fermi surface
($\pi1g_{9/2}$) is essentially unoccupied and the corresponding
$1g_{9/2}\rightarrow 1g_{7/2}$ transition suppressed.

\subsubsection{Theoretical description}
\label{sec:th-spinflip-des}

When trying to study the systematics of centroid energy and strength of
the spin-flip transitions in rare-earth nuclei, a first approximation
is to look at the unperturbed $M1$ strength originating from a deformed
single-particle model. In carrying out this procedure,
\textcite{DeCoster:1991a} have studied the summed spin $M1$ strength
throughout the whole rare-earth region from $^{140}$Ce up to
$^{198}$Pt. The strongest values are obtained at the end of the major
shell near $Z=82$ and $N=126$ through proton
$1h_{11/2}\rightarrow1h_{9/2}$ and neutron
$1i_{13/2}\rightarrow1i_{11/2}$ transitions. The unperturbed energy of
these transitions is situated in the energy region $4-10$ MeV. With the
residual interaction switched on, the $M1$ strength will be
redistributed but the total strength should not change much from the
unperturbed case.

\begin{figure}[tbh]
\includegraphics[angle=0,width=8.5cm]{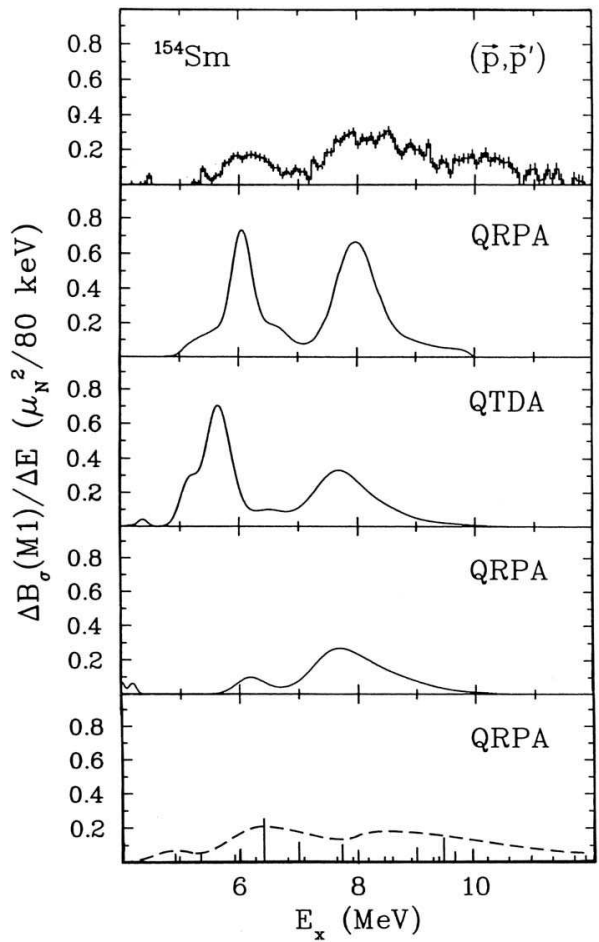}
\caption{Experimental and theoretical spin magnetic dipole strength
distributions in $^{154}$Sm. Underneath the experimental data,
theoretical predictions from various calculations are given (in
descending order): QRPA from
\textcite{Zawischa:1990a,Zawischa:1990b}, QTDA from
\textcite{DeCoster:1991a,DeCoster:1992}, QRPA from \textcite{Sarriguren:1993}, and
QRPA from \textcite{Hilton:1998}.}\label{Fig40}
\end{figure}

A comparison of the experimental strength distribution in $^{154}$Sm
with a number of QRPA and QTDA studies has been carried out
\cite{Zawischa:1990a,Zawischa:1990b,Sarriguren:1993,Hilton:1998,DeCoster:1991a},
see\ Fig.~\ref{Fig40}. The theoretical results have been folded with a
Gaussian of variable width in order to facilitate comparison. One has
to conclude that the agreement between experiment and theory is still
on a qualitative level. The position of the two peaks does not vary a
lot but the relative strength of the peaks is changing in a rather
important way pointing out the sensitivity of the calculations to both
the underlying single-particle structure as well as to the residual
interactions used. The QRPA calculations that come closer to the data
\cite{Sarriguren:1993,Hilton:1998} have been carried out after the
experiments were performed. One also notices that the difference
between QRPA and QTDA
\cite{Zawischa:1990a,Zawischa:1990b,DeCoster:1991a} are not dramatic
pointing out that ground-state correlations do not seem to play a major
role in determining both the energy and the strength for these
spin-flip transitions.

The specific double structure of the strength distributions is related
to the residual spin-spin interaction which changes the unperturbed
picture in an important way \cite{DeCoster:1992}. Besides a shift of
the spin strength to higher energies, as expected from schematic $p-h$
models studying isovector excitations, a rather clear separation into a
proton-like and a neutron-like collective spin mode remains. As a
result, in $^{154}$Sm, the lower peak mainly originates from the proton
$1g_{9/2} \rightarrow 1g_{7/2}$ and the $1h_{11/2} \rightarrow
1h_{9/2}$ excitations whereas the second, higher-lying peak is mainly
due to the $1h_{11/2} \rightarrow 1h_{9/2}$ and the $1i_{13/12}
\rightarrow 1i_{11/2}$ spherical components. These simple $p-h$
configurations act as doorway states for the fragmentation of the
resonance in analogy to the discussion for the scissors mode (cf.\
Fig.~\ref{Fig23}).

For larger strengths of the spin-spin proton-neutron interaction as
used by \textcite{Sarriguren:1993,Sarriguren:1994,Sarriguren:1996}, the
spin strength becomes concentrated more into a full isoscalar and
isovector part with proton and neutron configurations strongly mixed.
\textcite{Zawischa:1990b} obtain results somewhat intermediate between
the two more extreme cases of very weak coupling and strong coupling
between the individual proton and neutron spin-flip $M1$
configurations. The higher peak shows a structure that is reminiscent
of a genuine giant spin-flip (or Gamow-Teller) mode of isovector
character. In comparing both the incoherent sum of the separate proton
and neutron contributions with the actual calculation where
interference effects do play an important role, it seems like the lower
part is mainly of proton character but also an isoscalar part is
present. They come to the conclusion that for the higher peak in
$^{154}$Sm, using a Landau-Migdal residual interaction, the neutron
contributions play the dominant role and come close to the results of
\textcite{DeCoster:1991a}.

It is beyond discussion that the starting points, i.e.\ different
single-particle deformed potentials (Nilsson, deformed Woods-Saxon,
deformed Hartree-Fock mean field) and different residual interactions,
lead to results that differ in an important way in their interpretation
of the nature of the double-peak structure (proton and neutron vs.\
isoscalar and isovector), see also \textcite{Lipparini:1984}.
Experiments that are sensitive to the proton-to-neutron content in
exciting those states - like inelastic $\pi^{\pm}$ scattering -  can
most probably solve this issue and, at the same time, give invaluable
information concerning the proton-neutron part of the spin-spin
component in the effective residual two-body interaction.

In conclusion, the general structure and evolution of spin-flip $M1$
strength can be studied using schematic models too
\cite{DeCoster:1991c,DeCoster:1992,Zawischa:1990b,Zawischa:1994a} and
these results are in general consistent with those from the more
detailed QRPA studies. For not too strong spin-spin proton-neutron
coupling, while considering a two-level model (or a four level model)
in the rare-earth region, the strength becomes concentrated,
separately, into a pure proton and neutron collective spin-flip state.
With increasing strength, all components eventually contribute into an
isovector mode at the higher energy and an isoscalar part at the lower
energy side, albeit with the neutron configurations and proton
configurations dominating in these two modes, respectively.

\subsection{Magnetic dipole strength at higher excitation energy:
prediction and experimental hints} \label{sec:mds-exp}

Experiments have succeeded in studying the response of the nucleus to
medium energy protons towards much higher excitation energies. This was
illustrated already in Fig.~\ref{Fig37} in which the cross section and
the $S_{nn}$ transverse spin-flip probability in $^{154}$Sm up to an
excitation energy of 32 MeV has been shown. Besides the dominant giant
electric dipole resonance state, split by deformation, on the lower
side, the spin-flip $M1$ strength has been detected and is discussed in
detail in Sec.~\ref{sec:th-spinflip}. On the higher energy tail though,
excess strength is observed, which can be described by a Lorentzian
centered at $E_{x} = 23.4$~MeV with a width of 6.8(6) MeV. These
parameters as well as an exhaustion of the corresponding
energy-weighted sum rule of  $76(11)$\% agree well with the (scarce)
systematics \cite{Harakeh:2001} of the isovector giant electric
quadrupole resonance (IVGQR). The arrows indicate dips in the $S_{nn}$
behaviour which are directly related to the electric character of the
strong states as compared to the other spin magnetic excitations.

This is probably the $K^{\pi} = 1^{+}$ component of the IVGQR, which
is split into various $K$ components for strongly deformed nuclei,
and taken as the genuine manifestation of a classical scissors
motion. \textcite{LoIudice:1989} have pointed out that the lower RPA
$1^{+}$ solution does not collect the whole $M1$ strength. They have
shown that a non-negligible fraction is obtained at higher energy.
For $A=164$ ($N=Z$), they obtain a value of 25 MeV for the
excitation energy of a high-energy mode with a corresponding
strength of $B(M1)= 4.5~\mu_{N}^{2}$. The quadrupole component then
acquires, through the $M1$ transition, an additional scissors
characteristic.

\begin{figure}[tbh]
\includegraphics[angle=270,width=8.5cm]{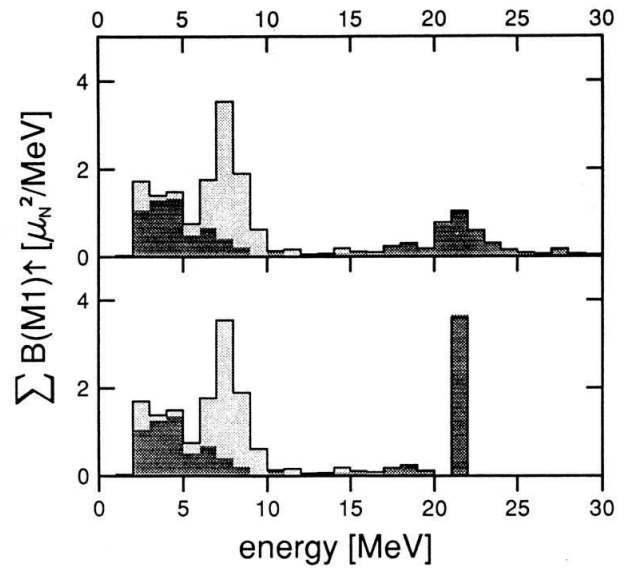}
\caption{Prediction of the full $M1$ strength distribution (summed
in bins of 1 MeV) including the high-energy part in $^{160}$Gd. The
dark-hatched zone represents spin strength, the light-hatched one
orbital strength. In the lower part, only 2qp configurations up to
20 Mev are taken into account. In the upper part, a much higher
energy cut-off for these 2qp configurations is imposed. (Reprinted
with permission from \textcite{Nojarov:1995b}. \copyright (2009)
Am.\ Phys.\ Soc.)}\label{Fig42}
\end{figure}

As was discussed in Sect.~\ref{sec:th-frag}, making a comprehensive
analysis of $M1$ excitations in atomic nuclei, a relation between the
energy and strength of a low- and high-energy scissors state was
indicated i.e. $\omega_{h} B_{h} = 4\omega_{l} B_{l}$ which gives
interesting information on both the expected excitation energy and the
$M1$ strength for a high-energy mode. Such a mode will be mainly built
out of $2\hbar\omega$ quasiparticle excitations for which the $M1$
strength becomes concentrated into a single strong state above 20 MeV.
The issue of how well such a strong state at that high energy will
remain intact is not clearly settled. A number of calculations
\cite{Hamamoto:1992,Hamamoto:1994,Nojarov:1995b,Zawischa:1994a,Zawischa:1998},
using schematic or more realistic forces come to different conclusions.
\textcite{Nojarov:1995b} and \textcite{Hamamoto:1992} obtain a large
concentration in a strong peak if they truncate the 2qp model space up
to 20 MeV, in line with calculations within a schematic picture in
which the higher-lying $M1$ strength also remains concentrated (see
Fig.~\ref{Fig42}). Using a more extended space spreads out this
strength considerably. Anyhow, at this energy, the resonance is highly
unbound which will induce further spreading making it a difficult task
to unambiguously detect and measure the amount of $M1$ strength. Thus,
in view of the largely different theoretical results, the question of a
still observable concentrated strong state will be difficult to solve.
Moreover, as illustrated in Fig.~\ref{Fig37}, there are considerable
experimental difficulties arising from a strong background mainly due
to quasifree scattering, whose exact shape and strength are unknown and
had be approximated by a semiempirical approach \cite{Lisantti:1984}.

The identification of the IVGQR permits an interesting test of the
energy-weighted $M1$ sum rule given in Eq.~(\ref{eq:m1-ewsr}) relating
the summed $M1$ strength on the l.h.s.\ to the difference of the
isoscalar and isovector $E2$ summed strengths on the r.h.s. of
Eq.~(\ref{eq:m1-ewsr}). With the experimental numbers for $^{154}$Sm
\cite{Ziegler:1993} one obtains for the energy-weighted $M1$ sum rule a
value of $7.71 \pm 0.44$~$\mu^{2}_{N}$~MeV whereas the right-hand part
(the difference of summed $E2$ strength) becomes $9.32 \pm 0.31
$~$\mu^{2}_{N}$~MeV. This is a rather good indication that below 4 MeV
- the region where the sum of $M1$ strength was carried out - the $M1$
sum rule is exhausted already by 80\%, leaving room for about 20\% at
the high energy part of $M1$ strength.

\section{MAGNETIC DIPOLE EXCITATIONS IN HEAVY ODD-MASS NUCLEI}
\label{sec:magnetic-dipol-heavy}

\subsection{Experimental results and systematics}
\label{sec:md-exp-heavy}

Naturally, the issue of what will happen in odd-mass nuclei when a
single nucleon (proton or neutron) is coupled to the scissors mode in
the even-even underlying core system comes up. From the concept of
particle-core coupling and considering the low-lying isoscalar
quadrupole and octupole vibrational excitations in spherical and
transitional nuclei, ample evidence for all states and a sharing of
electromagnetic strength amongst the particle-core coupled multiplet
members has been shown (Bohr and Mottelson, 1975). In the situation
where the scissors mode $M1$ strength at the low-energy region is
already spread out over a large energy span ($2.5-4$ MeV), it is clear
that an experimental verification of (i) the presence of
particle(hole)-scissors coupled configurations, and more compelling,
(ii) if the total $M1$ strength summed up in the appropriate energy
interval in the odd-mass nucleus is consistent with the observed $M1$
strength in the even-even adjacent nuclei, will not be easy.
\begin{figure}[tbh]
\includegraphics[angle=0,width=8.5cm]{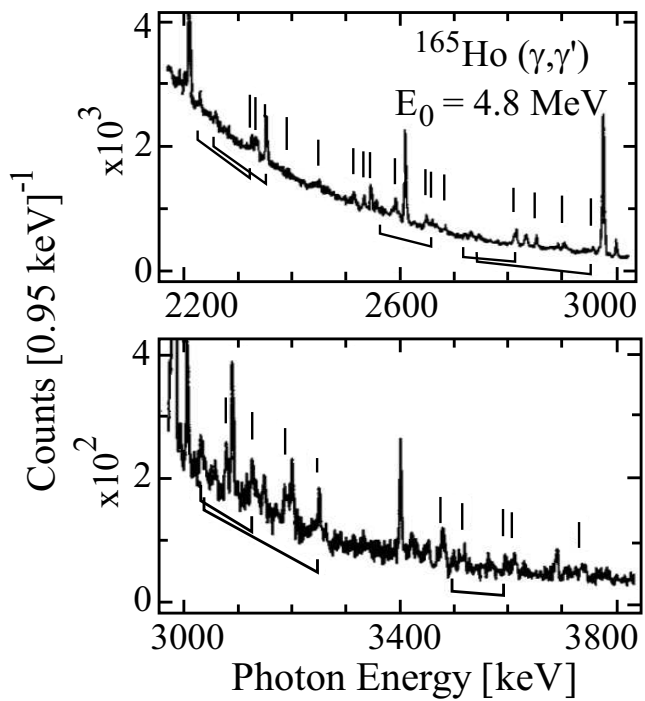}
\caption{Spectra of the $^{165}$Ho$(\gamma,\gamma')$ reaction in
the energy region $E_\gamma = 2.2 - 3.8$~MeV observed
with a Euroball Cluster detector placed under $130^\circ$ with
respect to the incident beam with an endpoint energy of 4.8~MeV.
Only the strongest transitions assigned to $^{165}$Ho are marked
with lines; brackets connecting two peaks indicate decay branches to
low-lying excited states. Other transitions are due to the
calibration standard $^{27}$Al or result from background sources
\cite{Huxel:1999}.}\label{Fig52}
\end{figure}

The problem is clearly one of detecting all the $M1$ strength, in
particular the $M1$ strength residing in the background of many and
complex configurations. On the other hand, also the challenge for a
good description from the theoretical side is not an easy one too. From
a more phenomenological approach and taking the core $M1$ strength to
be concentrated in one state, one will clearly not be able to correctly
reproduce the strong fragmentation, however, the summed strength puts a
constraint on this kind of model studies. From a more microscopic
approach, odd-mass nuclei can be studied using a Quasiparticle-Phonon
Nuclear Model (QPNM) \cite{Soloviev:1992}. Here,one needs to take into
account the fact that the QRPA phonons themselves are partly
constructed from the quasi-particle configurations one is coupling to.
This so-called Pauli blocking has been treated for deformed nuclei
\cite{Soloviev:1992}. For spherical odd-mass nuclei, a detailed study
of the transition strength from core-coupled configurations provided
quantitative evidence for Pauli blocking \cite{Scheck:2008}.
Ultimately, one aims at exact shell-model calculations but for the
strongly deformed rare-earth region, this is at present outside reach.

The experimental work covering a large part of the deformed rare-earth
region has mainly been carried out by groups at Stuttgart, K\"oln and
Darmstadt, using inelastic photon scattering through the excited states
in these odd-mass nuclei. The first search in the odd-proton $^{165}$Ho
nucleus, using photon scattering with an endpoint of about 2.5 MeV did
result in appreciable amounts of $M1$ strength \cite{Huxel:1992} albeit
to be associated with transitions amongst single one-quasiparticle
proton excitations in this particular nucleus. Partly due to the low
endpoint energy, no clear evidence for the presence of $M1$ strength
into the mixed-symmetric configurations, to be expected beyond 2.5~MeV,
was detected. Using a higher endpoint energy of 4.8 MeV and an EUROBALL
cluster module \cite{vonNeumann-Cosel:1997a}, spectra of much higher
quality could be measured at the S-DALINAC \cite{Huxel:1999} and we
give an illustration for the case of $^{165}$Ho (Fig.~\ref{Fig52}).

The first and quite clear indication for such $M1$ excitations was
obtained in the $^{163}$Dy odd-neutron nucleus \cite{Bauske:1993}.
Here, a concentration of $M1$ strength near to 3 MeV excitation energy
was detected and the summed strength fits with the observed $M1$
strength in the neigboring even-even $^{162,164}$Dy nuclei. Subsequent
experiments on $^{161}$Dy and $^{157}$Gd \cite{Margraf:1995}, however,
showed a huge fragmentation of strength in the latter nucleus. This is
quite difficult to understand  in the light of the proximity to
$^{161}$Dy (just a difference of 2 protons and 2 neutrons). Further
experiments on some key nuclei in order to build up systematics in this
region of the rare-earth nuclei were performed on $^{155}$Gd and
$^{159}$Tb (the latter an odd-proton nucleus), and on the heavier
$^{167}$Er \cite{Schlegel:1996} and $^{165}$Ho, $^{169}$Tm
\cite{Huxel:1999} nuclei.
\begin{figure}[tbh]
\includegraphics[angle=0,width=8.5cm]{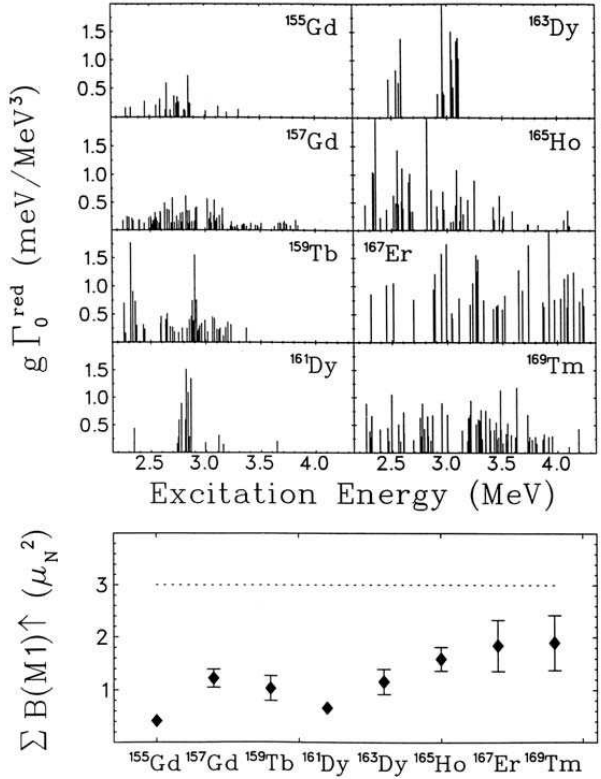}
\caption{Comparison of the magnetic dipole strength in the odd-mass
rare-earth nuclei \cite{Enders:1997}. Upper part: distribution of reduced ground-state
decay widths. Lower part: summed $B(M1)$ strengths assuming magnetic
dipole character of all observed transitions in the energy range between
2.5 and 3.7~MeV. Large differences in total strength, fragmentation and the number of
detected ground-state transitions is observed. Data are from
\textcite{Bauske:1993,Margraf:1995,Schlegel:1996,Nord:1996,Huxel:1999}.}
\label{Fig53}
\end{figure}

\begin{figure}[tbh]
\includegraphics[angle=0,width=8.5cm]{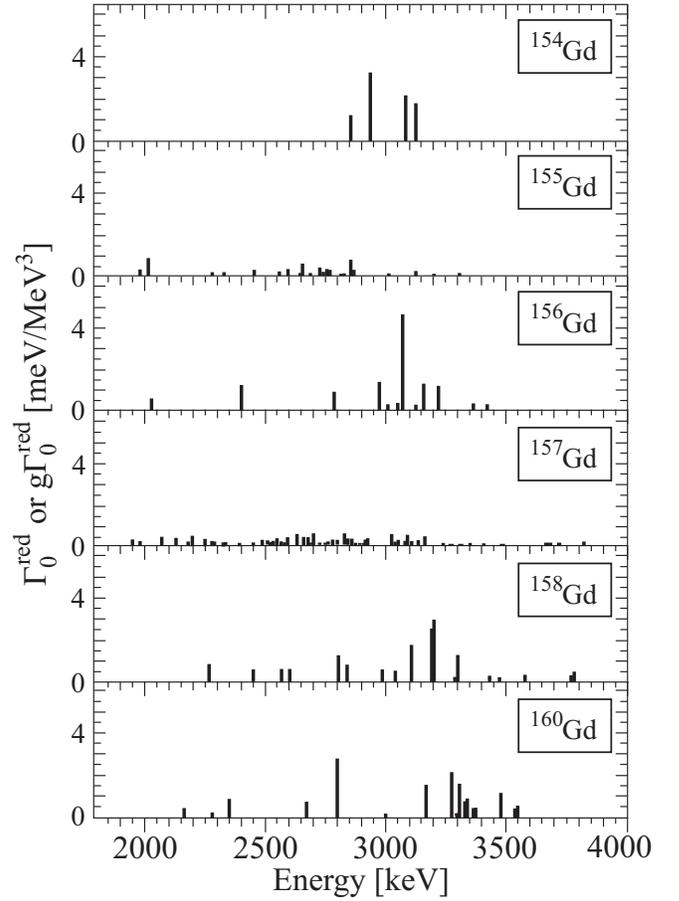}
\caption{Ground-state decay width distributions for
$^{154,155,156,157,158,160}$Gd extracted from photon scattering
experiments. For the even-even nuclei all $\Delta K= 1$ transitions
are shown. In a number of cases, the $M1$ character has been
determined by Compton polarimetry. Information from $(e,e')$ form
factor measurements has also been considered (Reprinted
with permission from \textcite{Nord:1996}. \copyright (2009)
Am.\ Phys.\ Soc.)}\label{Fig54}
\end{figure}

Bringing these data together in Fig.~\ref{Fig53}, one observes that
starting from $^{155}$Gd, passing over the odd-mass Dy nuclei and
progressing towards heavier nuclei, $M1$ strength seemed to become more
concentrated, precisely in those regions that were expected from the
knowledge of the $M1$ scissors mode strength in the nearby even-even
nuclei, \cite{Enders:1997}. In Fig.~\ref{Fig54} the full systematics of
the Gd nuclei combining mass-even and -odd isotopes is shown
\cite{Kneissl:1996,Nord:1996}. More recent experimental data on
$^{151,153}$Eu, and with increased sensitivity, on $^{163}$Dy and
$^{165}$Ho, have been provided by \textcite{Nord:2003}. Conclusions
from all these data are (i) the fragmentation pattern is at best rather
badly understood, and (ii) even worse, only about half to one-third of
the $M1$ strength observed in the even-even nuclei (when summing the
strength in the odd-mass nuclei in the interval $2.5-3.7$~MeV) could be
detected firmly (see lower part of Fig.~\ref{Fig53}).

One notable exception is the study of $^{167}$Er
\cite{Schlegel:1996}, where experiments have been carried out with
endpoint energies going up to 5.8 MeV at the S-DALINAC. The summed
strength reaches 3.49(1.15)~$\mu_{N}^2$, a value that is at variance
with many of the former experiments by a factor of about three. This
experiment gave a first hint that one needed to look for $M1$
fragments at energies higher than was first thought. As we shall
discuss below, the calculated $M1$ strength using the
interacting-boson fermion model \cite{Iachello:1991} very well
accounts for this total strength although in the theoretical study
only two major peaks are obtained below 4 MeV.

\subsection{Missing strength: experimental problem and its solution}
\label{sec:md-exp-heavy-miss}

\begin{figure}[tbh]
\includegraphics[angle=270,width=8.5cm]{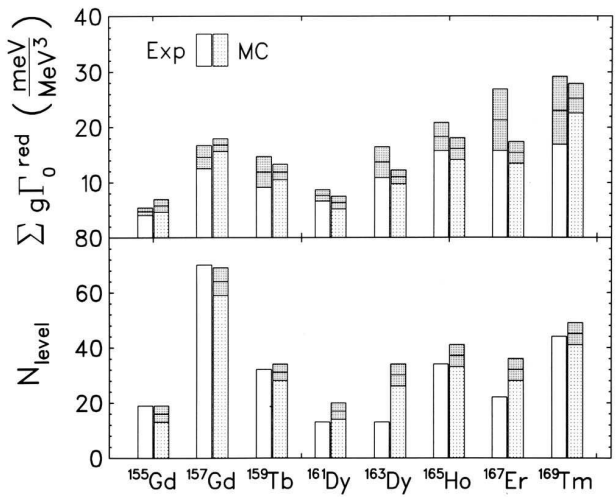}
\caption{Number of ground-state dipole transitions and their summed
reduced width in deformed odd-mass rare-earth nuclei. Open bars
refer to the data. Dotted bars stand for Monte-Carlo generated
strength distributions based on a statistical approach.
Uncertainties of the random spectra have been estimated by 1000-fold
repetition of the calculation. Error bars denote a $1 \sigma$
deviation \cite{Huxel:1999}.}\label{Fig56}
\end{figure}

The comparison between odd-mass and even-even nuclei immediately poses
the question:  where has the $M1$ strength gone in the odd-mass nuclei?
The search was on for the observation of a large part of $M1$ strength
residing in a very large number of complex states but with very small
$B(M1)$ values and hidden in the background of the spectra. A detailed
statistical model analysis of the high-quality data in $^{165}$Ho and
$^{169}$Tm, obtained with an EUROBALL cluster module, indeed revealed
that a significant part of the $M1$ strength is carried by the
background states \cite{Enders:1997,Enders:1998,Huxel:1999}. It has
been shown \cite{Enders:1997} that the statistical assumptions
underlying the fluctuation analysis approach are also capable to
explain the large variations in the measured dipole distributions seen
in Fig.~\ref{Fig53}. Monte-Carlo distributions have been generated
taking into account the properties of $M1$ and $E1$ distributions in
the even-even neighboring nuclei and allowing for the energy dependence
of the experimental sensitivity limits. These results are presented in
Fig.~\ref{Fig56} and compared to the data
\cite{Enders:1997,Huxel:1999}. Overall, the large variations of the
total number of observed levels and the summed dipole strengths can be
simultaneously reproduced in a very satisfactory manner.

\begin{figure}[tbh]
\includegraphics[angle=0,width=8.5cm]{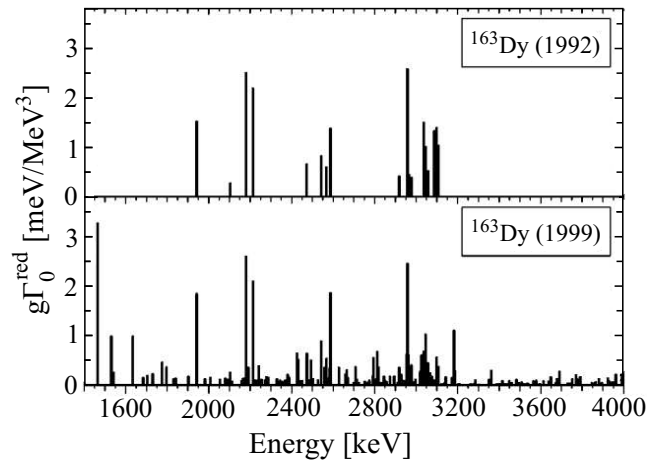}
\caption{Ground-state decay width distributions for
$^{163}$Dy from \textcite{Bauske:1993} and \textcite{Nord:2003}.
The latter experiment had an order-of-magnitude improved sensitivity.
While transition strength agree well for prominent excitations, many
previously unresolved weak transitions are visible. }\label{FigSc51}
\end{figure}

Subsequent experiments with unrivaled sensitivity confirmed these
results for the cases of $^{163}$Dy and $^{165}$Ho \cite{Nord:2003}.
The experiments showed a wealth of previously unresolved weak
transition as demonstrated in Fig.~\ref{FigSc51}, where the strength
distribution deduced for $^{163}$Dy is compared to the first
measurement \cite{Bauske:1993}. The sum of the reduced dipole strength
is roughly doubled. However, the fragmentation pattern into a few
rather strong and many very weak transitions in $^{163}$Dy seems to be
peculiar, since a fluctuation analysis cannot explain the still missing
strength. On the other hand, for $^{165}$Ho good agreement with the
previous work \cite{Huxel:1999} was obtained when combining the
strength of resolved and unresolved transitions. It may also be noted
that a very recent NRF measurement on $^{235}$U also shows good
agreement of the total $M1$ strength deduced from a comparable
statistical analysis with that in the even-mass neighbor $^{236}$U
\cite{Yevetska:2009}.

So, as a conclusion for the present-day situation on scissors states
and scissors $M1$ strength in odd-mass nuclei one can say that in the
deformed odd-mass rare-earth (and probably also actinide) nuclei, the
mode seems to be present with a strength expected from the even-even
systematics but a significant part - which can change quite importantly
from nucleus to nucleus depending on the respective level densities and
the photon scattering end-point energy - escapes detection in the
photon scattering experiments because of the very large fragmentation.

\subsection{Theoretical description}
\label{sec: oddatheo}

As remarked above, coupling an odd particle or hole (proton or neutron)
to the collective modes of the even-even core nucleus generally results
in the observation of core-coupled multiplets \cite{Bohr:1975}. Because
of the subsequent fragmentation of M1 scissors strength, it is clear
that in comparing theoretical results with data, at most  indications
for the total summed $M1$ strength will be the guiding principle to
judge the level of agreement.

Within the context of the interacting boson-fermion model (IBFM), a sum
rule has been derived by \textcite{Ginocchio:1997} in which they have
studied the coupling of a single $j$-shell particle (the unnatural
parity orbital in fact) to the underlying scissors mode. In the limit
of good $F$ spin and large boson number $N$, the resulting new sum
rule, has been compared with a similar sum rule derived for even-even
nuclei \cite{Ginocchio:1991,vonNeumann-Cosel:1995}. These results yield
upper limits and in the case of the odd-neutron $1i_{13/2}$ particle
coupled to the scissors for the nuclei $^{161}$Dy and $^{167}$Er as
some extremes (see the data), it is observed that the theoretical sum
rule in $^{167}$Er is consistent with the data but for $^{161}$Dy
definitely largely overestimates them.
\begin{figure}[tbh]
\includegraphics[angle=0,width=8.5cm]{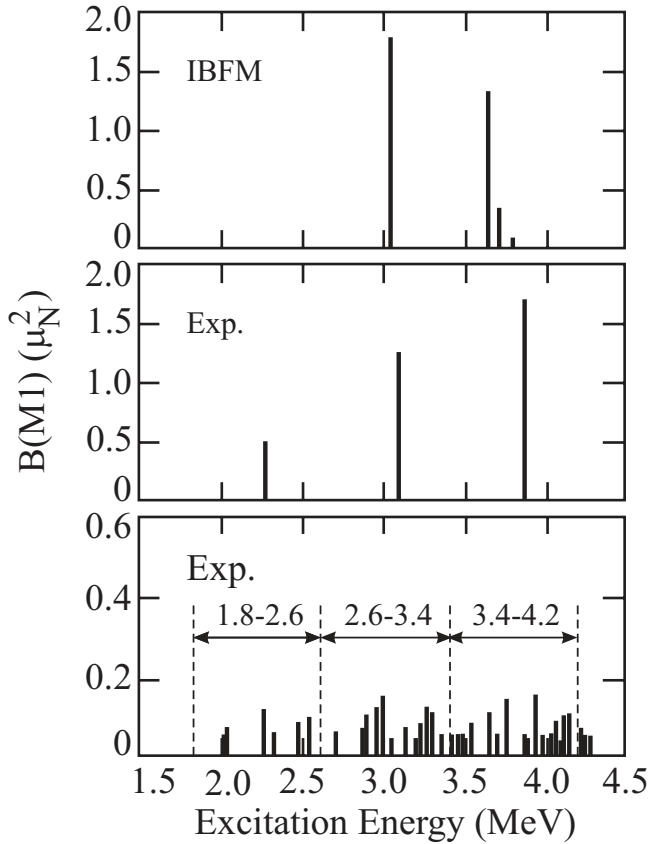}
\caption{Bottom to top: experimental $M1$ strength distribution in
$^{167}$Er, strength distribution summed up in the indicated energy
bins, comparison with an IBFM calculation, and expected splitting of
the energy spectrum through coupling of the unpaired particle
\cite{Schlegel:1996}.}\label{Fig57}
\end{figure}

A number of interesting results have been derived by
\textcite{VanIsacker:1989} and by \textcite{Frank:1991} using the more
general IBFM. Analytical results could still be derived under the
assumption of good $F$ spin and considering a single-$j$ shell.
Calculations were performed for $^{169}$Tm and $^{165}$Ho
\cite{Huxel:1992}. A more detailed study has also been carried out for
$^{163}$Dy and $^{167}$Er. In the case of $^{167}$Er, where the
$1i_{13/2}$ odd-neutron determines the ground-state structure, this
orbital only is considered and again comparisons with summed strengths
are possible, as depicted in Fig.~\ref{Fig57}. The multiplet structure
resulting from coupling the odd-particle to the scissors excitation
always underestimates fragmentation by far. Within the algebraic
formulation, \textcite{Devi:1992a,Devi:1992b,Devi:1996} have been
studying group-theoretical reductions for odd-$A$ nuclei now including
the $g$ boson. Because of the very fact of introducing an extra boson
degree of freedom, the effect of fragmentation of course increases.
\begin{figure}[tbh]
\includegraphics[angle=0,width=8.5cm]{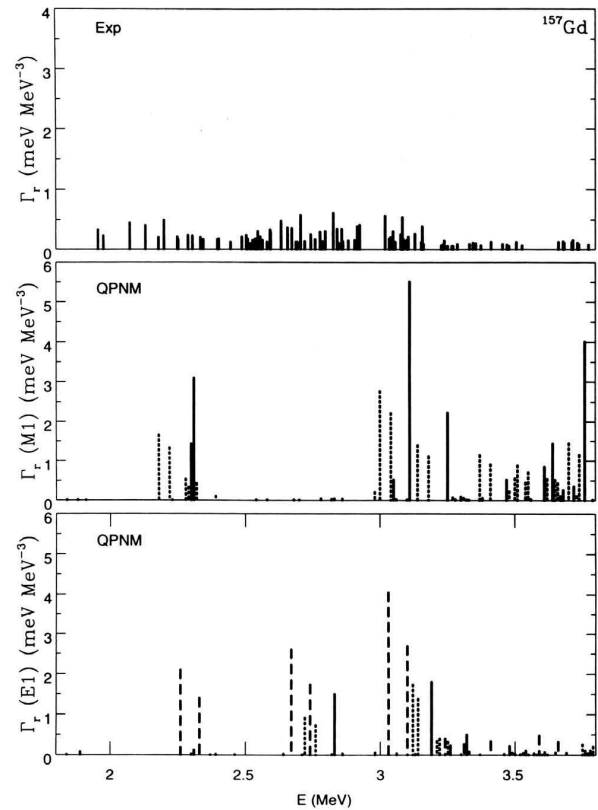}
\caption{Experimental ground-state reduced width distribution in
$^{157}$Gd (top), together with the QPNM predictions
\cite{Soloviev:1992} for $M1$ (middle) and $E1$ (bottom)
transitions to $K=1/2,3/2,5/2$ final states displayed by full, dashed and dotted
lines, respectively \cite{Soloviev:1997c}.}\label{Fig58}
\end{figure}

Microscopic studies in the deformed rare-earth region have also been
carried out using the coupling of a 1qp excitation with the underlying
collective phonon structure. Calculations have been performed by
\textcite{Raduta:1989c} and \textcite{Raduta:1990a}, using a
coherent-state formalism to describe the collective phonons. A more
extensive calculation of this kind exists also for a number of odd-mass
nuclei \cite{Soloviev:1996b,Soloviev:1997a,Soloviev:1997c}. In the
particular case of $^{157}$Gd illustrated in Fig.~\ref{Fig58}, $M1$
strength is localized mainly between 2 and 2.5 and between 3 and 3.7
MeV, clearly overestimating the observed $M1$ strength. The strongest
concentration of $M1$ strength is situated near 3 MeV with a subsequent
large fragmentation (i) due to the fact that from the $J;K$ ground
state $M1$ excitations are possible into $J+1,J,J-1;K-1$ and $J+1;K+1$
excited states, and (ii) the 1qp-phonon coupling mechanism
redistributes strength too. The calculations indeed give support to the
presence of scissors particle-core coupled configurations in the
odd-mass strongly-deformed rare-earth nuclei, albeit with a varying
character of splitting and further fragmentation over a background of
complex microscopic configurations.

\section{MAGNETIC DIPOLE EXCITATIONS IN LIGHT AND MEDIUM-HEAVY NUCLEI}
\label{sec:magnetic-dipol-light}

\subsection{Experimental data}
\label{sec:md-exp-light}

Like in heavier nuclei, most experimental information on ground-state
isovector magnetic dipole transitions comes from inelastic photon and
electron scattering experiments. A particularly interesting technique
for the study of even-mass nuclei is electron scattering at $180^\circ$
\cite{Fagg:1975} because of the dramatic suppression of the background
due to the radiative tail of elastic scattering. If combined with a
large-acceptance spectrometer it represent a powerful tool for the
study of the $M1$ response \cite{Luettge:1995,Luettge:1996a}. Data from
$(\gamma,\gamma')$ in lighter nuclei are limited. Summaries are
provided e.g.\ by \textcite{Berg:1987} and \textcite{Raman:1991}. The
most exhaustive studies of $M1$ strength distributions are available
from $(e,e')$ data covering all stable nuclides in the $p$-shell, $N =
Z$ and $Z+2$ nuclides in the $sd$-shell, the stable Ca isotope chain,
the $N =28$ isotones, the open-shell nuclei $^{46,48}$Ti, $^{50}$Cr,
$^{56}$Fe, and finally $^{58}$Ni.

\begin{figure}[tbh]
\includegraphics[angle=270,width=8.5cm]{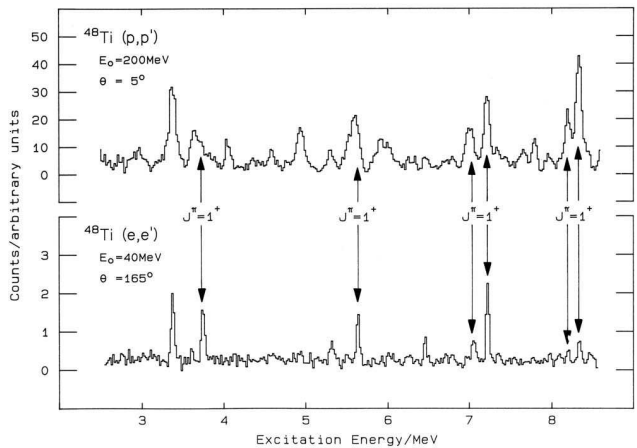}
\caption{High-resolution $(p,p')$ and $(e,e')$ spectra in $^{48}Ti$.
The $J^{\pi} = 1^{+}$ states are marked by arrows
\cite{Richter:1990a}.}\label{Fig43}
\end{figure}

\begin{figure}[tbh]
\includegraphics[angle=0,width=8.5cm]{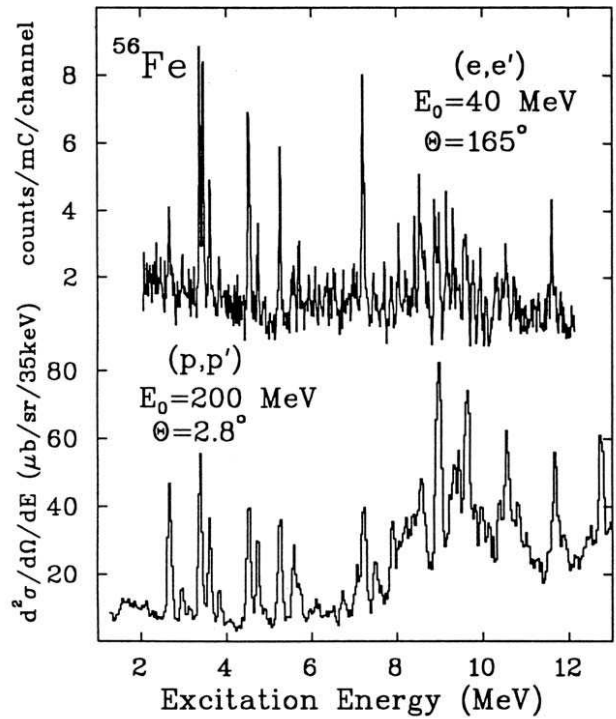}
\caption{High-resolution inelastic electron and proton scattering
spectra in $^{56}$Fe measured at TRIUMF. Lines above 6~MeV correspond
to the excitation of $J^{\pi} = 1^{+}$ states under the kinematic
conditions of the two experiments \cite{Richter:1993b}.}\label{Fig44}
\end{figure}

Complementary information on the spin part of the $M1$ strength stems
from inelastic proton scattering experiments. As an example, in
Fig.~\ref{Fig43} spectra of the $(p,p')$ and $(e,e')$ reactions off
$^{48}$Ti under kinematics favoring $M1$ excitations are compared
\cite{Richter:1990a}. All transitions identified to have $M1$ character
(marked by arrows) are seen in both spectra although with different
relative intensities due to the interference of orbital and spin
strength in the latter reaction. Another case, $^{56}$Fe, highlighting
the close resemblance of spectra obtained with both probes
\cite{Richter:1993b} is displayed in Fig.~\ref{Fig44}. Here, all
transitions observed above $E_x = 6$ MeV possess $M1$ character and
represent the spin-$M1$ Gamow-Teller resonance. At lower excitation
energies mostly $2^+$ states are populated except for the prominent
transition seen in the $(e,e')$ spectrum at about 3.5 MeV, which again
carries most of the orbital $M1$ strength.

\subsection{Theoretical description: the shell model and random phase
approximation}
\label{sec:th-md-light}

Contrary to the problems encountered to describe the magnetic dipole
excitation modes in strongly deformed heavy nuclei within a shell-model
framework, for the light nuclei large-scale shell-model calculations
have been performed and used to study $M1$ excitation properties in
detail. Concentrating on mixed-symmetry states, it were in particular
Zamick and coworkers who studied such excitations for nuclei in the
$1f_{7/2}$ shell-model region
\cite{Zamick:1985,Zamick:1986a,Zamick:1986b,Liu:1987a,Liu:1987b,Liu:1987c}.
Let us treat $^{48}$Ti with 2 proton particles and 2 neutron holes
outside of the closed shell nucleus $^{48}$Ca as an example (analogous
arguments hold for other nearby nuclei). The wave functions within the
$1f_{7/2}$ model space solely are expanded as
%
%
\begin{equation}
\Psi(J^{\pi})=\sum_{L_{p},L_{n}}D_{J}(L_{p},L_{n})|
(1f_{7/2})^{2}_{L_{p}},(1f_{7/2})^{-2}_{L_{n}};JM\rangle,
\end{equation}
in which the notation is self-explanatory and where the coefficients
$D_{J}(L_{p},L_{n})$ denote the amplitudes that the two protons couple
to $L_{p}$ and the two neutron holes to $L_{n}$. The $M1$ operator now
induces transitions from the $0^{+}$ ground state into the $1^{+}$
states with a corresponding $B(M1)$ value of
%
\begin{eqnarray}
B(M1)& = & \frac{3}{4\pi}\left(g_j^{\pi}-g_j^{\nu}\right)^{2} \times\\
&&\left|\sum_{L}D_{0}(L,L)D_{1}(L,L) \right|^{2} L(L+1)~\mu_{N}^{2} \nonumber,
\end{eqnarray}
%
in which $g_j^{\pi}$ and $g_j^{\nu}$ are the conventional
single-particle gyromagnetic factors $g_j^{\rho}=[(2j-1)g_l^{\rho} +
g_s^{\rho}]/2j$, and $g_j^{\rho}=[(2j+3)g_l^{\rho} -
g_s^{\rho}]/2(j+1)$, for $j=l+\frac{1}{2}$ and $l-\frac{1}{2}$,
respectively, and $\rho = \pi,\nu$ \cite{Brussaard:1977}. One can even
derive a sum rule for the strength into all possible final $1^{+}$
states
%
\begin{eqnarray}
\sum B(M1)&=&\frac{3}{4\pi}
\left(g_j^{\pi}-g_j^{\nu}\right)^{2} \times \\
&&\left|\sum_{L}D_{0}(L,L)\right|^{2}L(L+1)~\mu_{N}^{2}.  \nonumber
\end{eqnarray}

\begin{figure}[tbh]
\includegraphics[angle=0,width=8.5cm]{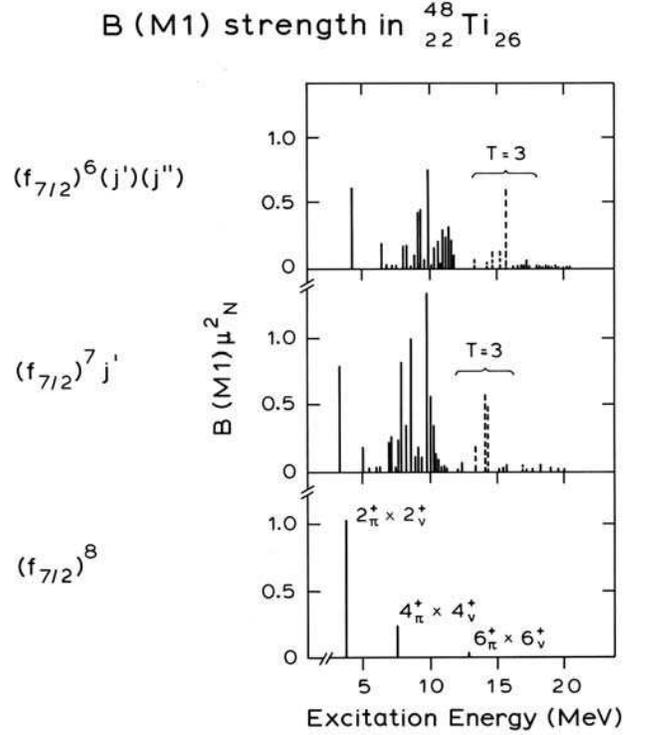}
\caption{Magnetic dipole strength distributions for $^{48}$Ti as
calculated for the model spaces given on the l.h.s.\ of the figure.
(Reprinted with permission from \textcite{Liu:1987b}. \copyright (2009)
Am.\ Phys.\ Soc.)}\label{Fig46}
\end{figure}

For the case of $^{48}$Ti the mixed-symmetric $2^{+}$ state results
from the mixed-symmetric combination of the $L_{p}=0,L_{n}=2$ and
$L_{p}=2,L_{n}=0$ components whereas the $1^{+}$ scissors counterpart
originates from the combinations $L_{p}=2,L_{n}=2;L_{p}=4,L_{n}=4;~{\rm
etc.}$ While the use of a single $1f_{7/2}$ orbital keeps the $M1$
strength very much concentrated in a single strong excitation,
gradually increasing the shell-model space with the inclusion of the
$1f_{5/2},2p_{3/2}$ and $2p_{1/2}$ orbitals opens the way into many new
states and fragmentation starts to set in, as illustrated in
Fig.~\ref{Fig46} \cite{Liu:1987b}. One notices that the low-lying
$J^{\pi}$ = $1^{+}$ state is always well separated from the other
$1^{+}$ states at higher excitation energy. It is mainly of orbital
nature and can be associated with an experimental state at $E_{x}$ =
3.74 MeV which carries a strength of $B(M1)\! \uparrow =
0.52(8)~\mu_{N}^{2}$ \cite{Guhr:1990}. We also emphasize that within
this single $1f_{7/2}$ shell-model space, the orbital $M1$ strength is
a measure of dynamical quadrupole correlations in the ground state
since it depends on the ($2^{+}_{p} 2^{+}_{n}$) configuration admixture
in the ground state.

An important issue in the study of scissors mode excitations in the
open-shell nuclei situated in the Ti, Cr, Fe region is the
orbital-to-spin ratio for the low-lying $1^{+}$ states. In contrast to
heavy nuclei, only a few orbitals are determining the structure of the
wave functions and therefore, a non-negligible spin contribution will
be present, even in the lowest $1^{+}$ state. This can be studied in a
most illustrative way by comparing $(e,e')$ and $(p,p')$ experiments
(cf.\ Figs.~\ref{Fig43} and \ref{Fig44}). However, a quantitative
analysis of the $(p,p')$ data is hampered by the dependence of the
extracted spin-$M1$ strengths on the choice of the effective projectile
target interaction, which can lead to variations up to about 40\%
\cite{Hofmann:2007}.

\begin{figure}[tbh]
\includegraphics[angle=270,width=8.5cm]{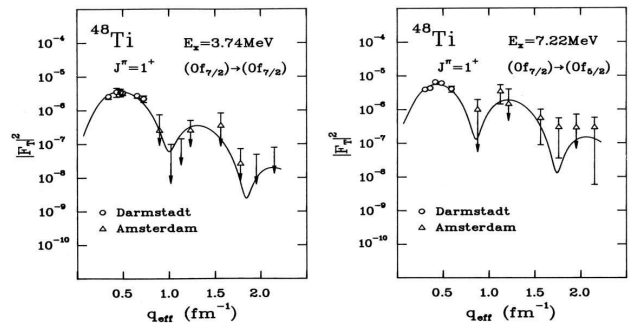}
\caption{Inelastic electron scattering form factors for a
predominantly orbital (left-hand side) and predominantly spin-flip
(right-hand side) $M1$ transition \cite{Richter:1990a}.}\label{Fig50}
\end{figure}

Thus, it is important to find other means to disentangle the spin and
orbital parts. It has been discussed e.g.\ by \textcite{Abdelaziz:1987}
that the GT matrix element in $\beta$ decay might be used to estimate
the spin contribution to a collective isovector $M1$ transition. Such
matrix elements can also be measured in charge-exchange reactions
populating analog states in the odd-odd neighboring nuclei. A wealth of
high-resolution data on the GT strength distributions has recently
become available \cite{Frekers:2006,Fujita:2008} and the dependence on
the effective projectile-target interaction in hadronic reactions can
partly be circumvented by normalizing to $\beta$-decay results.
However, isospin selection rules limit the applicability to special
cases (some of which are discussed below). Electron scattering form
factors present another method to derive bounds on the relative
importance of orbital versus spin magnetism in a number of transitions.
We illustrate in Fig.~\ref{Fig50} two form factors in $^{48}$Ti for
transitions to the 3.74 MeV and 7.22 MeV $1^{+}$ states, respectively
\cite{Guhr:1990}. Whereas the first transition seems to proceed through
the $1f_{7/2}$ orbital only (recoupling), the second transition is
consistent with a spin-flip transition $1f_{7/2} \rightarrow 1f_{5/2}$.

A systematic study of the orbital-to-spin ratio was carried out by
\textcite{Oda:1987} in which a dependence on both the model space and
the use of effective $g$ factors was explored. The result is that
including excitations from the $1f_{7/2}$ orbital into the higher-lying
$1f_{5/2},2p_{3/2},2p_{1/2}$ orbitals increases the orbital part over
the spin part in a systematic way for a number of Ti, Cr, and Fe
isotopes. Moreover, quenching the $g$ factors from the free-nucleon
values further reduces the spin strength and reinforces the model space
extension. The latter result was also derived independently by
\textcite{Heyde:1984}; see also \textcite{Heyde:1989} for an
illustration. Studies of the $M1$ response for the light Ti nuclei have
been carried out using the QRPA approach with very similar results and
conclusions concerning the orbital-to-spin ratio
\cite{Nojarov:1987,Nojarov:1991,Nojarov:1992,Faessler:1988,Faessler:1989}.

On the other hand, truncations of the large-scale shell-model space and
choosing a specific proton-neutron force may lead to a symmetry-based
approach to study $M1$ properties. For light nuclei, the SU(3)
shell-model has been used and applied to both $sd$- and $fp$-shell
nuclei \cite{Chaves:1986,Poves:1989,Retamosa:1990}. A comparison of
shell-model and IBM calculation was performed for the light Sc, Ti, V
$fp$-shell nuclei \cite{Abdelaziz:1988}. A particular symmetry-dictated
truncation to realistic shell-model calculations, emphasizing the
importance of $S$,$D$, and $G$ pairs, was used for $^{54,56}$Cr and
$^{56-60}$Fe \cite{Halse:1990,Halse:1991a}. Moreover, a pseudo SU(3)
model was suggested to describe rotational properties in this mass
region \cite{Halse:1991b}.

\begin{figure}[tbh]
\includegraphics[angle=0,width=8.5cm]{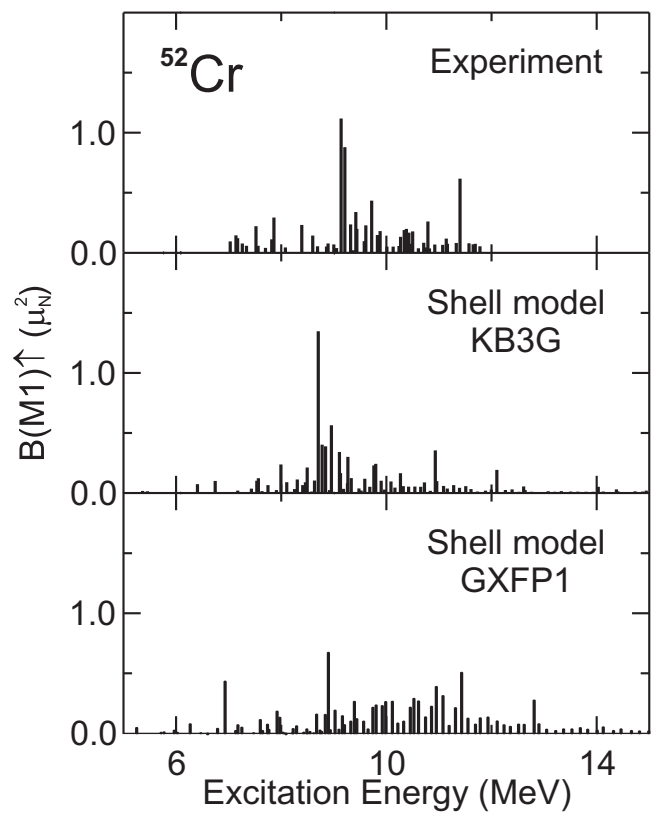}
\caption{$M1$ strength distribution in $^{52}$Cr from (top to
bottom) high-resolution $(e,e')$ experiments \cite{Sober:1985} and
large-scale shell-model calculations using the KB3G
\cite{Poves:2001} and GXPF1 \cite{Honma:1994} interactions,
respectively.}\label{Fig48}
\end{figure}

Recent computational progress allows shell-model studies of the $M1$
strength in large model spaces to describe details of the fragmentation
of the mode. For example, unrestricted calculations in the full $fp$
model space are possible now for $^{46,48}$Ti \cite{Fearick:2006}. As
an example of the state-of-the-art, a study of the stable $N =28$
isotones $^{48}$Ca, $^{50}$Ti, $^{52}$Cr, $^{54}$Fe
\cite{Langanke:2004} is discussed, whose experimental $M1$ strength
distributions have been measured in great detail
\cite{Steffen:1980,Sober:1985}. In Fig.~\ref{Fig48}, the results for
$^{52}$Cr are shown together with calculations based on two widely used
shell-model interactions called KB3G \cite{Poves:2001} and GXPF1
\cite{Honma:1994} derived in a $G$-matrix approach from nucleon-nucleon
interaction potentials. Because of the $N = 28$ shell closure a
spherical ground state can be expected. Correspondingly, no low-lying
orbital transitions are observed. In the energy region above 6~MeV,
presented in Fig.~\ref{Fig48}, a resonance structure arising from
spin-flip transitions is visible. The shell-model results are quite
successful in reproducing the features of the strength distribution
qualitatively and also quantitatively when analyzing the resonance
centroid and total strength, but in detail differences remain.

The comparison made in Fig.~\ref{Fig48} raises an important and
nontrivial question: how can one quantify the degree of correspondence
between data and calculation? One possible way may be the extraction of
scales characterizing the fine structure of the mode by means of a
wavelet analysis \cite{Shevchenko:2008}. Fine structure has been shown
to be a global phenomenon of giant resonances
\cite{Shevchenko:2004,Shevchenko:2009,Kalmykov:2006}. A recent
application of this method to $M1$ strength distributions in $fp$-shell
nuclei \cite{Petermann:2008} indeed reveals considerable differences in
the characteristic scales extracted from the $M1$ strength functions
obtained with different effective interactions including those shown in
Fig.~\ref{Fig48}, and in the case of $^{52}$Cr the results from the
KB3G calculation are found to be closer to the data.

The capability of large-scale shell model calculations to describe the
interference of spin and orbital parts has been investigated in a
detailed study of the electron scattering form factor of the prominent
$M1$ transition at $E_x =3.449$ MeV in $^{56}$Fe, which contains spin
and orbital matrix elements of comparable size \cite{Fearick:2003}.
While different effective shell-model interactions describe the
strength reasonably well, the predicted $q$ dependence differs
considerably. Clearly, the spin-orbit interplay remains a challenge to
shell-model studies even in very large model spaces.

It also became clear in the above studies that a reduction of the $g$
factors from the free-nucleon values is generally increasing the
orbital-to-spin matrix element ratio. An independent approach to
understand the $g$-factor quenching has been carried out in this region
of medium-heavy and light nuclei, concentrating on the comprehensively
studied $N = 28$ nuclei. Shell-model calculations require in all cases
a reduction of the spin part of the magnetic dipole operator. A
consistent description for the stable $N = 28$ isotones can be reached
using a value $g^{eff}_{s} = 0.75(2) g^{free}_{s}$
\cite{vonNeumann-Cosel:1998}. The required reduction is remarkably
close to the quenching factor 0.744(15) obtained from a recent
shell-model analysis of GT $\beta$-decay transitions in the lower
$fp$-shell region \cite{Martinez-Pinedo:1996}. Indeed, the most
important mechanism responsible for the quenching is, viz.\ the mixing
with two-particle two-hole configurations at high excitation energies,
is expected to be the same as in the GT case
\cite{Bertsch:1982,Ichimura:2006}.

\subsection{Some astrophysical implications}
\label{sec:md-astro-impl}

\begin{figure}[tbh]
\includegraphics[angle=0,width=8.5cm]{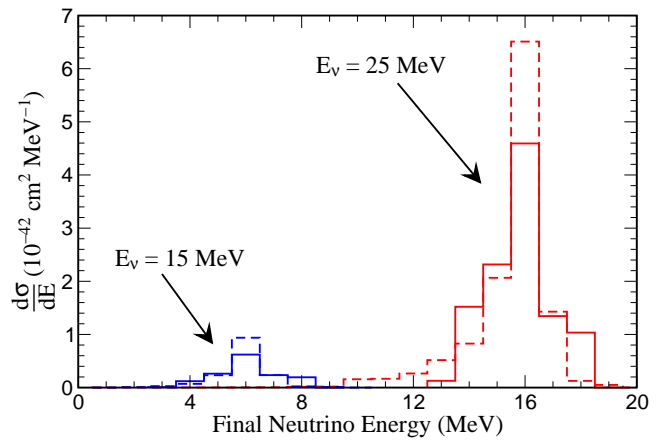}
\caption{ (Color online) Differential inelastic neutrino cross sections for
$^{52}$Cr and initial neutrino energies $E_\nu=15$ MeV and 25 MeV.
The solid histograms are obtained from the $M1$ data, the dashed ones
from shell-model calculations.  The bumps represent the GT$_0$
strength shifted by the centroid energy of the resonance. The final
neutrino energies are given by $E_f = E_\nu- E_x$
\cite{Langanke:2004}.}\label{Fig47}
\end{figure}

The knowledge of the magnetic dipole strength in $fp$-shell nuclei is
also crucial in supernova modeling. It permits to determine cross
sections of inelastic neutrino-nucleus scattering, a process whose
importance for supernova dynamics was recognized only recently
\cite{Hix:2003}. Under the conditions of a supernova type II in massive
stars, neutrino-nucleus reactions are dominated by GT transitions
\cite{Langanke:2003}. The description of inelastic scattering processes
requires knowledge of the $T_0 \rightarrow T_0$ isospin component of
the GT strengths, where $T_0$ denotes the ground-state isospin. Except
for an overall factor relating the weak and electromagnetic interaction
this is nothing but the spin part of the $M1$ strength. Orbital
strength is negligible in the stable $N = 28$ isotones due to the shell
closure; thus, the measured $M1$ distribution represents to a good
approximation the needed GT$_0$ strength. Figure~\ref{Fig47} displays
differential cross sections of inelastic neutrino scattering on
$^{52}$Cr calculated under this assumption for two typical neutrino
energies $E_\nu = 15$ and 25 MeV (solid line).

Since the reaction network calculations in a supernova require
information on the GT$_0$ strength in many nuclei, one has to rely
on model calculations. The set of highly precise data in the $N =
28$ isotones \cite{Sober:1985} was used to demonstrate the
capability of large-scale shell-model calculations to describe $M1$
strength distributions \cite{Langanke:2004}. If applied to the
present problem, differential neutrino scattering cross sections
shown by the dashed line in Fig.~\ref{Fig47} result. These are
indeed in good agreement with the data, and it was concluded that
present-day shell-model calculations can provide the necessary
GT$_0$ strengths needed as input to the supernova simulations.
Inclusion of inelastic neutrino-nucleus scattering increases the
neutrino opacities noticeably and strongly reduces the high-energy
tail of the neutrino spectrum emitted in the neutrino burst at shock
breakout. Relatedly, the expected event rates for the observation of
such neutrinos by earthbound detectors are reduced by up to about
60\% \cite{Langanke:2008}.

\subsection{Selected problems in light and medium-heavy nuclei}
\label{sec:md-very-light}

In light nuclei, magnetic dipole transitions have been measured in
almost all stable nuclei and detailed theoretical studies (mostly
shell-model) have been performed. We do not attempt an exhaustive
discussion of these results. Rather, we highlight a few topics
accessible only in light nuclei or extending beyond the range of
physics questions discussed so far.

\subsubsection{Quenching of the spin $M1$ strength: the case of
$^{48}$Ca} \label{sec:md-very-light-1}

A classical example of a clear-cut isovector $M1$ transition with a
remarkably large transition strength $B(M1)\!\uparrow = 3.9(3)~\mu_N^2$
was found in $^{48}$Ca in electron scattering experiments
\cite{Steffen:1980}. It was interpreted as a rather pure neutron
$1f_{7/2} \rightarrow 1f_{5/2}$ excitation \cite{Richter:1985}. The
$(e,e')$ form factor of this transition serves as a another prime
example to study the problem of quenching of the spin part of the
isovector $M1$ response. Since GT and $M1$ strengths are quenched by
comparable amounts one can expect that the responsible processes are
the same. Two major mechanisms are expected to contribute
\cite{Ericson:1988}. On the one hand, a polarization of the nuclear
core by the nucleon undergoing a spin-isospin transition occurs, which
leads to a virtual excitation of high-lying states by the tensor force
in second order; hence it is named second-order core polarization.
Alternatively, the GT and $M1$ operators are modified by virtual
$\Delta$-hole excitations, analogous to the Lorenz-Lorentz correction
in dielectrics \cite{Delorme:1976}. Both mechanisms lead to different
redistributions of the strength. Thus, by measuring the response up to
high excitation energies, the former mechanism was shown to dominate
for GT transitions \cite{Ichimura:2006}. The study of the $(e,e')$ form
factor permits to delineate the contributions in the case of the $M1$
response \cite{Richter:1991}. The most refined calculations
\cite{Takayanagi:1988}, although still not able to fully explain the
amount of quenching, confirm the dominant role of core polarization
leading to mixing with $2p2h$ states over the $\Delta$-hole part, which
contributes about 10\% to the strength reduction. This value is in
agreement with an analysis of the $\Delta$-hole contributions to the
quenching of GT strength \cite{Ichimura:2006}.

\subsubsection{Cross-shell transitions in $^{36,38,40}$Ar}
\label{sec:md-very-light-3}

Magnetic dipole strength distributions in $^{36}$Ar and $^{38}$Ar
deduced from electron scattering experiments \cite{Foltz:1994} reveal
marked differences. Calculations in a $0 \hbar \omega$ model space
($sd$ shell) with a phenomenological interaction, generally successful
in the description of $M1$ and GT strengths \cite{Brown:1988}, work
well for $^{36}$Ar (with a closed neutron $sd$ shell) but fail
completely for $^{38}$Ar (with two neutrons in the $fp$ shell),
indicating the importance of $sd$$\rightarrow$$fp$ cross-shell
contributions. This problem has recently been studied by
\textcite{Lisetskiy:2007} based on an effective interaction including
the coupling of $sd$- to the $1f_{7/2}2p_{3/2}$- orbitals
\cite{Caurier:2001}. Such cross-shell calculations present a limit of
present-day computational capabilities and still require a significant
truncation of the model space. The first results support the importance
of an inclusion of cross-shell transitions and make specific
predictions for the even more neutron-rich $^{40}$Ar. Some information
on $M1$ strength in $^{40}$Ar has recently been reported
\cite{Li:2006}, but a measurement of the full $M1$ strength
distribution would be important.

\subsubsection{$l$-forbidden transitions}\label{sec:md-very-light-5}

In $sd$-shell nuclei, an effective shell-model $M1$ operator has been
determined by an empirical fit to the large body of data on magnetic
and $M1$ transitions \cite{Brown:1987}. The deviations from the bare
operator are incorporated in correction factors for the spin and
orbital parts and an induced-tensor term. Microscopic calculations
\cite{Towner:1987,Arima:1987} are in good agreement except for an
isovector tensor correction. Tensor corrections are generally weak and
therefore buried in the dominant spin strength for most $M1$
transitions. However, experimental information on the tensor correction
terms can be obtained from $l$-forbidden transitions ($1d_{3/2}
\leftrightarrow 2s_{1/2}$ in the $sd$ shell). The term '$l$-forbidden'
refers to a selection rule for the one-body operator of $M1$ or GT
transitions which does not allow a change of the orbital quantum
number. The higher-order corrections to the $l$-forbidden transitions
are theoretically expected to be dominated by $\Delta$ resonance
admixtures into the nuclear wave functions
\cite{Towner:1987,Arima:1987} and they are a unique observable in this
respect. The problem has been studied extensively in $1d_{3/2}
\rightarrow 2s_{1/2}$ single-hole transitions in $A =39$ nuclei. One
finds an order of magnitude larger $M1$ strength \cite{Grundey:1981}
relative to the GT strength \cite{Hagberg:1994}, while the microscopic
results predict the tensor correction governing the strength to be the
same. However, the interpretation could be blurred by weak cross-shell
admixtures of the type discussed in Sec.~\ref{sec:md-very-light-3}.
Therefore, data away from the end of the $sd$ shell are important. One
such example is an $(e,e')$ study of the $l$-forbidden transition to
the $1^+$ state in $^{32}$S at $E_x = 7.003$ MeV \cite{Reitz:1999}. The
form factor exhibits an anomalous momentum-transfer dependence compared
to allowed $M1$ transitions because its finite strength results from
higher-order terms only. The shell-model analysis reconfirms the
discrepancy between empirical and microscopic approaches to determine
the tensor correction, and the problem remains unresolved so far.

Of course, $l$-forbidden transitions are not restricted to the case
$1d_{3/2} \leftrightarrow 2s_{1/2}$ but can appear between all pairs of
shell-model orbitals with quantum numbers $(n,l,j=l+1/2)$ and
$(n-1,l+2,j'=(l+2)-1/2)$, where $n$ is the radial quantum number, and
$l$ and $j$ are the orbital and total angular momenta, respectively.
Experimentally, the single-particle energies of corresponding pairs of
states show a near degeneracy in many nuclei. This led to the concept
of pseudospin symmetry, where the doublet structure is expressed in
terms of a `pseudo'-orbital angular momentum $\tilde{l} = l + 1$, in
which the two levels represent spin-orbit partners with a `pseudo'spin
$\tilde s = 1/2$. While pseudospin symmetry was empirically established
40 years ago \cite{Hecht:1969,Arima:1969}, a deeper understanding has
been lacking. Relativistic corrections \cite{Bohr:1969} have been
suggested as a possible source and applied with some success to
describe magnetic moments of pseudospin partners near the $N = 82$
shell closure \cite{Heyde:1977}. Recently, pseudospin symmetry has been
interpreted as a relativistic SU(2) symmetry of the Dirac-Hamiltonian
which occurs when the attractive scalar and repulsive vector nuclear
mean fields cancel \cite{Ginocchio:2005}. Evaluating this concept,
\textcite{Ginocchio:1999} derived a relation between the magnetic
moments of the pseudospin partners and the strength of the
$l$-forbidden $M1$ transition between them. Application to data near a
variety of magic numbers reveals overall good correspondence with a few
marked deviations \cite{vonNeumann-Cosel:2000a}.

\subsubsection{Enhancement of magnetic dipole strength by meson exchange currents}
\label{sec:md-very-light-4}

Direct signatures of mesonic exchange currents (MEC) in experimental
observables are usually restricted to few-body nuclear systems
\cite{Ericson:1988} with the exception of magnetic dipole properties
(moments and transition strengths) as discussed in the previous
subsection. One such example is discussed here. It is well established
in $sd$-shell nuclei that full $0 \hbar \omega$ shell model
calculations with an effective operator are able to describe the $M1$
and GT matrix elements \cite{Brown:1987,Towner:1987,Arima:1987}. In
selfconjugate even-even nuclei with ground-state spin and isospin $J;T
= 0^+;0$ the set of final states populated by isovector $M1$, GT$_-$
and $GT_+$ transitions forms a triplet of isobaric analogue states.
Their transition strengths are directly related, if spin-orbital
interference effects are negligible. This is certainly not the case for
individual transitions but holds on the level of 10\% when studying
full strength distributions of $sd$-shell nuclei \cite{Hino:1988}
because of the sign variations of the mixing term.

When comparing $M1$ and GT strength distributions in $^{24}$Mg,
excellent agreement of the GT strengths amongst each other and with the
shell model result. However, the same calculations significantly
underpredict the $M1$ strength. Such an enhancement of the experimental
$M1$ strength can be traced back \cite{Richter:1990b} to MEC
contributions. To make this clear it is convenient to describe the $M1$
and GT strengths schematically in the following form
\begin{eqnarray}
{\mathrm B(M1)} & = &C
[ M_{\sigma} + M_l + M_{\Delta} + M_V^{MEC}]^2 \;\; , \label{eq:m1me} \\
\label{eq:gtme} {\mathrm B(GT)} & = & [ M_{\sigma} + M_{\Delta} +
M_A^{MEC}]^2 \;\; .
\end{eqnarray}
Here, $M_l$ and $M_{\sigma}$ are the orbital and spin matrix elements,
and $M_{\Delta}$ stands for the contribution of $\Delta$-isobar
admixtures to the strength. The numerical factor $C$ before the square
brackets in Eq.~(\ref{eq:m1me}) equals to 2.643~$\mu_N^2$ using free
nucleon $g$ factors. Neglecting the orbital part, the main difference
between $M1$ and GT excitations lies in the MEC contributions, which
are of vector type for the former and of axial-vector type for the
latter. Since axial vector currents are strongly suppressed because of
the conservation of $G$ parity \cite{Towner:1987}, deviations of the
ratio $R(M1/GT) = \sum B(M1)/2.643 \sum B(\mathrm{GT})$ from unity
point towards an enhancement of the $M1$ strength by vector-type MEC
contributions. Besides $^{24}$Mg, a clear enhancement was also observed
in $^{28}$Si \cite{Luettge:1996b,vonNeumann-Cosel:1997b}. In $^{32}$S
the situation is less clear \cite{Hofmann:2002} because $(e,e')$ form
factors indicate significant orbital admixtures in some of the
strongest transitions, and the experimental information on the $M1$
strength distribution is limited to an excitation energy of 12 MeV and
therefore incomplete. Another problem noted in \textcite{Richter:1990b}
was that spin-$M1$ strengths in selfconjugate $sd$-shell nuclei deduced
from forward-angle $(p,p')$ data \cite{Crawley:1982} are systematically
about 20\% larger than the corresponding GT$\pm$ strengths. However,
this discrepancy can probably be resolved utilizing the latest
experimental developments allowing true $0^\circ$ measurements combined
with high energy resolution \cite{Tamii:2009}.

\subsubsection{Isoscalar and isovector $M1$ transitions in $^{12}$C and isospin mixing}

Isospin is an approximate symmetry in nuclei broken by the long-range
Coulomb force but also by small charge-dependent components of the
nuclear interaction. In light nuclei, Coulomb effects are weak and
excited states possess a well defined isospin quantum number $T$
experimentally known in many cases. This allows to study isospin mixing
between states of the same $J^\pi$ but different $T$. Evidence for
isospin mixing beyond the Coulomb force has been claimed from the
observation of very large isospin mixing matrix elements but later it
was realized that the predictions exhibit a strong dependence on the
poorly known radial wave functions of the involved single-particle
states \cite{Auerbach:1983}.

A unique testing ground are the $M1$ transitions to the pair of
{$J^\pi; T = 1^+; 0$} and $J^\pi; T = 1^+; 1$ states in $^{12}$C at
12.71 and 15.11~MeV, respectively. These are of $1p_{3/2} \rightarrow
1p_{1/2}$ spin-flip character. Form factors of both transitions (albeit
weak for the isoscalar case) at low momentum transfer have been
measured with high precision in inelastic electron scattering
\cite{vonNeumann-Cosel:2000b}. Analysis in a two-state model determines
not only the mixing amplitudes but also the relative sign through the
$q$ dependence of the form factors. The resulting Coulomb matrix
element $\langle H_c \rangle = 118(8)$ keV, determined with unequaled
precision, is large but can be fully explained by Coulomb mixing
\cite{Harney:1986}.

\section{ISOVECTOR MAGNETIC DIPOLE TRANSITIONS IN VIBRATIONAL NUCLEI}
\label{sec:isovector}

\subsection{Introduction}
\label{subsec-vibr}

In most of this review we have concentrated on the magnetic dipole
orbital and spin response in stable, deformed nuclei, which has been
studied using both electromagnetic and hadronic scattering off the
nuclear ground state (Sec.~\ref{intromagdipol}). In particular, the
scissors $1^+$ mode was shown to be excited with a summed strength that
scales with the square of the nuclear deformation. Therefore, with
decreasing deformation entering the region near closed shells, the
scissors mode as well as a stable intrinsic quadrupole deformation will
cease to be formed. This mass region is characterized by small
amplitude quadrupole vibrational oscillations as the major degree of
freedom, with typical $B(E2;2^+_1 \rightarrow 0^+_1)$ strength of the
order a few tens of Weisskopf units (W.u.). The low-energy nuclear
structure properties result in a first excited $2^+_1$ phonon
excitation which is an in phase motion (or symmetric mode) in the
proton and neutron collective motion, also called isoscalar (IS) mode.
Multiphonon states can then be constructed, as shown in
Fig.~\ref{Fig3}, in which the proton and neutron motion combines into
symmetric (or IS) excitations . However, multiphonon states can also
arise rise from proton and neutron motion combining into non-symmetric
(or IV) excitations. Besides the $1^+$ states, which is the counterpart
of the scissors mode in the vibrational nuclei, also $0^+,2^+,3^+,4^+,$
states result.

The appearance of IV proton-neutron excitations has been proposed in
the context of the IBM-2
\cite{Arima:1977,Iachello:1987,Iachello:1984,Otsuka:1985}. This
approach points out that the isovector excitations appear in a natural
way by combining the lowest-lying proton and neutron $2^+_1$ $d$-boson
configurations \cite{Iachello:1984,VanIsacker:1986} into states of
mixed-symmetry (MS) character \cite{Iachello:1984}. Besides, the shell
model constitutes a microscopic framework in order to describe
excitations that are non-symmetric in its proton and neutron
coordinates \cite{Heyde:1986,Lisetskiy:2000,Holt:2007,Boelaert:2007b}.
Moreover, it is possible to describe isoscalar and isovector
excitations within the framework of a quasiparticle-phonon model which
defines RPA phonons, and then to construct states in a basis of one-,
two- and three-phonon components. Since this approach has a microscopic
(QRPA) underpinning, it allows to bridge the gap between a fully
microscopic shell-model approach and the algebraic IBM-2
\cite{LoIudice:2000b,LoIudice:2002,LoIudice:2004,LoIudice:2006,LoIudice:2008,LoIudice:2009}.

In order to locate these mixed-symmetry (MS) states, one can use the
particular structure of the $M1$ and $E2$ operators. We have already
shown that the $M1$ magnetic dipole operator can be separated into its
isoscalar and isovector parts (see Sec.~\ref{sec:th-mdr}). In view of
the structure of the isovector part, one expects strong magnetic dipole
transitions in the decay of the mixed-symmetry states into the
low-lying symmetric states. Likewise, one can separate the electric
quadrupole operator $T(E2)$
\begin{equation}
T(E2)=e_{\pi}\sum_{i=1}^{Z} r_{i,\pi}^2 Y_2(\hat{r}_{i,\pi}) +
e_{\nu}\sum_{i=Z+1}^{A} r_{i,\nu}^2 Y_2(\hat{r}_{i,\nu}),
\end{equation}
with $e_{\pi}$ and $e_{\nu}$ the proton and neutron effective charges
into an isoscalar and isovector part
\begin{equation}
T(E2)=\frac{e_{\pi}+e_{\nu}}{2}T(E2,IS)+
      \frac{e_{\pi}-e_{\nu}}{2}T(E2,IV) .
\end{equation}
Here, $T(E2,IS)$ and $T(E2,IV)$ are the symmetric and antisymmetric
combinations of the proton and neutron parts of the $E2$ operator.
Because of the specific symmetry character of the IS and IV
excitations, strong $E2$ transitions are expected between S and MS
states, separately, but rather weak $E2$ transitions from MS to S
states. These characteristics are highlighted in Fig.~\ref{Fig3}.

In view of the above discussion, the key signature, in order to assign
mixed-symmetry character to a state, derives from the $E2$ and $M1$
decay properties:
(i) strong $M1$ transitions ({$B(M1)$ of the order of $\simeq 1
\mu_N^2$}) to low-lying symmetric states restricting to transitions
between states with equal number of phonons, mainly,
(ii) weak collective $E2$ transitions (with transition probabilities
about $10 \% $ of the strong $E2$ transitions such as $2^+_{1}
\rightarrow 0^+_{1}$) to low-lying symmetric states, and
(iii) strong collective $E2$ transitions amongst the MS states
themselves.

\subsection{Experimental results and theoretical description}
\label{subsec-exp-vibr}

>From an experimental point of view, the study of MS states in nuclei of
vibrational and transitional structure is rather different from the
mapping of scissors mode $1^+$ excitations in deformed nuclei. In the
latter case (see Sec.~\ref{intromagdipol}), electron, photon and hadron
scattering starting from the $0^+$ ground state in deformed nuclei
allowed to determine orbital and spin-flip $M1$ strength. In the
present situation, the identifying elements are both a strong $M1$
transition into the isoscalar (mostly the $2^+_1$ state) state
accompanied by a weak $E2$ transition into the $0^+$ groundstate (with
a magnitude of the order of a few \% of the $B(E2;2^+_1 \rightarrow
0^+_1)$ reduced transition probability). Thus, in order to obtain a
unique characterization of the mixed-symmetry states, different types
of experiments have to be carried out and combined. Typical experiments
will need to probe the lifetime of a given level, the determination of
$J^{\pi}$ values, $\gamma$-decay branching ratios and the
$\delta(E2/M1)$ mixing ratios and this for as complete a set of states
with given $J^{\pi}$ value. Classical $\gamma$-ray spectroscopic
methods have to be used extensively as well photon-scattering
experiments as e.g.\ $\gamma$ decay following $\beta$ decay that
populates levels in nuclei under study. Determining the nuclear
lifetimes needs typically the study of Doppler-shifted attenuation
techniques (DSAM) in ($n,n'\gamma$), light-ion induced reactions and
Coulomb excitation in inverse kinematics. Direct excitation as used in
electron scattering, photon scattering and Coulomb excitation on stable
nuclei gives rise to lifetimes in a rather straightforward way. A
detailed review of these techniques is presented by
\textcite{Kneissl:2006} and \textcite{Pietralla:2008}.

\subsubsection{The Z $\sim$ 40, N $\sim$ 50 mass region}
\label{subsubsec-Mo}

\begin{figure}[tbh]
\includegraphics[angle=0,width=8.5cm]{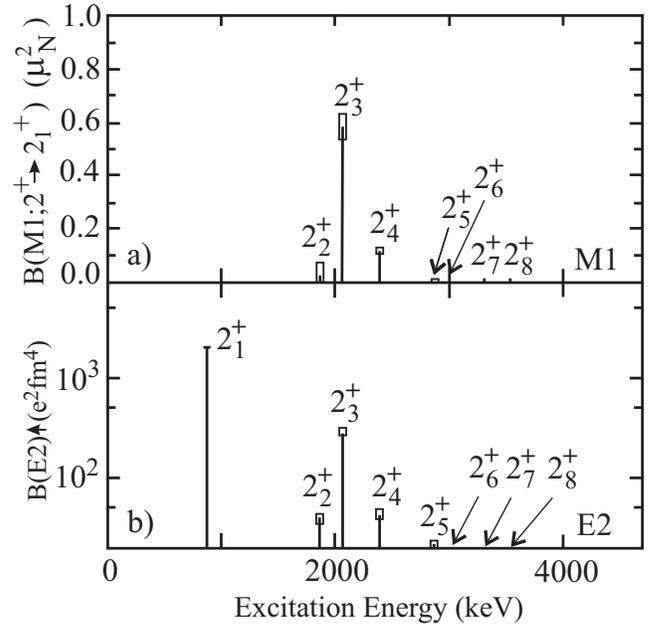}
\caption{Measured $E2$ and $M1$ strengths in order to identify the
$2^+_{1,ms}$ state in $^{94}$Mo. (a) $B(M1;2^+ \rightarrow 2^+_1)$
values for the seven lowest-lying identified nonyrast $2^+$ states.
(b) Corresponding $B(E2;0^+_1 \rightarrow 2^+_i)$ values.
The error bars are displayed as boxes.
(Reprinted with permission from \textcite{Fransen:2003}. \copyright
(2009) Am.\ Phys.\ Soc.)}\label{Fig60}
\end{figure}

\begin{figure}[tbh]
\includegraphics[angle=0,width=8.5cm]{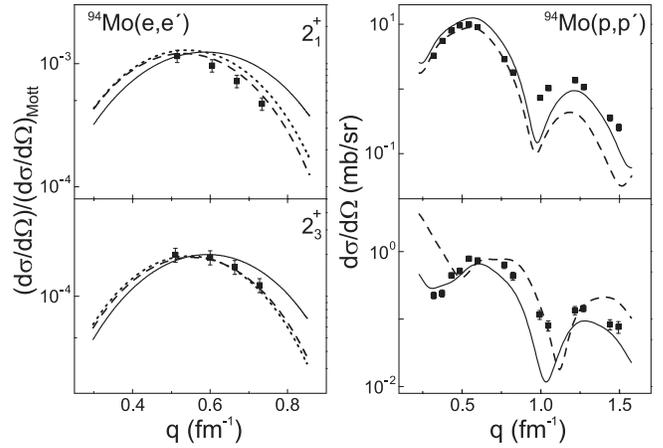}
\caption{Momentum-transfer dependence of the symmetric
($2^+_1$, upper part) and mixed-symmetric ($2^+_3$, lower part)
one-phonon excitation cross-sections in $^{94}$Mo in inelastic electron (left side) and
proton (right side) scattering. The data (full squares) are compared
to QPM (solid lines), shell-model(dashed lines) and IBM-2 (dotted
lines) predictions \cite{Burda:2007}.}\label{Fig61}
\end{figure}

The nucleus $^{94}$Mo forms a particularly suitable test case of the
above schematic picture because it has 4 protons outside of the $Z=38$
subshell closure and 2 neutrons outside of the $N=50$ closed shell.
With these building blocks, ideal conditions show up to form both
symmetric and antisymmetric couplings of these pairs. Detailed
experimental studies have identified proton-neutron mixed symmetry
$2^+$ and $1^+$ states \cite{Pietralla:1999}. In Fig.~\ref{Fig60}, the
specific signatures of a MS $2^+$ state (strong $M1$ transition into
the $2^+_1$ state, weakly collective B(E2) transition from the ground
state) are shown. The data point towards the $2^+_3$ state as the ideal
MS candidate.
Experiments in $^{94}$Mo have furthermore shown \cite{Pietralla:2000}
evidence for two-phonon MS states built from combining a symmetric
($2^+_1$) and anti-symmetric ($2^+_3$) state, thus forming states with
spins in the range $0^+$- $4^+$. Clear-cut identification of the $2^+$
\cite{Fransen:2001} and $3^+$ \cite{Pietralla:2000} members could be
achieved and candidates for the other spins were identified
\cite{Fransen:2003} based on the above discussed signatures of MS
states.
Independent evidence for one-phonon symmetric and mixed-symmetric
states has been recently demonstrated in a combined study of
high-resolution inelastic electron and proton scattering off MS $2^+$
states in $^{94}$Mo \cite{Burda:2007} when comparing with theoretical
results derived from quasiparticle-phonon, shell-model and IBM-2 as
shown in Fig.~\ref{Fig61}. Multi-phonon MS states have also been
observed in the N=54 $^{96}$Mo nucleus \cite{Lesher:2007}.

The study of the variation of the MS states, keeping the neutron number
fixed at $N=52$ but changing the proton number, is quite interesting.
In this spirit, experiments have been carried out in $^{96}$Ru with
$Z=44$ \cite{Pietralla:2001,Klein:2002} pointing out a strong
similarity with $^{94}$Mo. Moreover, evidence for a MS $1^+$ state has
been shown by \textcite{Linnemann:2005}. A study of the $^{92}$Zr
nucleus, with the same number of neutrons i.e., $N=52$, is quite
different because in the core nucleus,  $^{90}$Zr, two $0^+$ states
appear resulting from  the presence of both $(1g_{9/2})^2_{0^+}$ and
$(2p_{1/2})^2_{0^+}$ configurations. The $0^+$ ground state therefore
acquires extra binding energy which distorts the vibrational spectra as
compared with the $N=52$ Mo and Ru nuclei. Still, photon scattering
\cite{Werner:2002} and ($n,n'\gamma$) inelastic neutron scattering have
enabled to observe $2^+$ and $1^+$ states with a MS character. Recent
experiments \cite{Elhami:2007,Elhami:2008,Werner:2008} have
concentrated on the special situation in the Zr nuclei.

The nuclei with proton number $40 \leq Z \leq 50$ and neutron number
close to $N=50$ (here, $N=52$), with in particular the nucleus
$^{94}$Mo, form an ideal testing ground for both the IBM-2, the QPM as
well as the nuclear shell model.  It was shown \cite{Iachello:1984}
that the IBM-2 framework naturally predicts a class of states with MS
character in the proton and neutron contributions. Detailed selection
and intensity rules have been derived for $E2$ and $M1$ transitions in
the various limits of the IBM-2 \cite{VanIsacker:1986,Scholten:1985b}.
A comparison of the IBM-2 results has been carried out in $^{94}$Mo as
well as the adjacent $N=52$ isotones Ru and Zr with a consistent
description of MS states with both a one ($2^+$) and two-phonon
($1^+,2^+,3^+,4^+$) character as well as of their decay properties to
lower-lying symmetric zero and one-phonon states (see e.g.\
\textcite{Pietralla:2008} for more details). Keeping within the context
of the IBM-2, a particular scheme (called $Q$-phonon scheme) has been
set up by \textcite{Otsuka:1994b}, which allows for the description of
these symmetric and MS excitations as phonons, applicable in the U(5),
O(6) symmetry limits and in the transitional nuclei between these two
limits.

The IBM-2 has a drawback because the operator only addresses the
orbital part and specific spin contributions are only considered in an
average way. Therefore, microscopic techniques are needed such as the
standard shell model and quasi-particle phonon (QPM) approaches. The
QPM approach has been applied with considerable success in the region
of vibrational nuclei by
\textcite{LoIudice:2000b,LoIudice:2002,LoIudice:2004,LoIudice:2006} and
\textcite{LoIudice:2008}. In the $Z=40$, $N=50$ region, it turns out
that the $2^+_1$ RPA phonon has mainly a symmetric structure in the
interchange of proton and neutron labels (or is $F$-spin symmetric in
the IBM-2 language) whereas the $2^+_2$ RPA phonon is antisymmetric (or
of $F$-spin MS nature) to a good approximation. The QPM eigenvalue
problem is then solved in a basis including up to three-phonon states
and considering many phonons of different $J^{\pi}$ nature. These
results give support to the IBM-2 calculations carried out in which
these $s$ and $d$ bosons are the only building blocks. Detailed results
are given for $^{94}$Mo \cite{LoIudice:2000b,LoIudice:2002} and for
$^{92}$Zr \cite{LoIudice:2004,LoIudice:2006}.

\begin{figure}[tbh]
\includegraphics[angle=0,width=8.5cm]{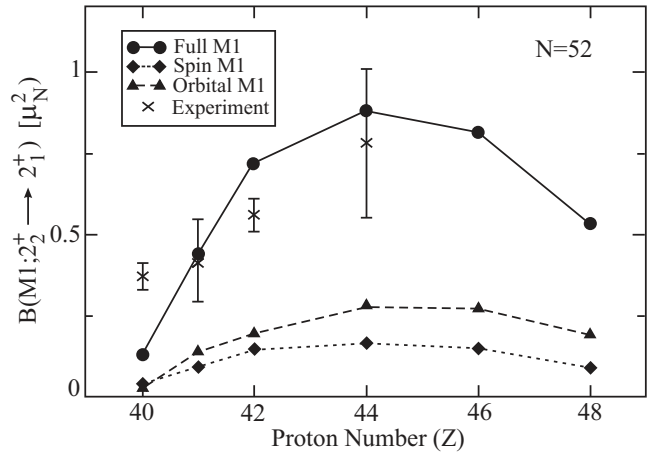}
\caption{Evolution of the total, orbital, and spin $B(M1;2^+_2
\rightarrow 2^+_1)$ values for the $N=52$ isotones. The results of
recent shell-model calculations are compared with
the data. (Reprinted with permission from \textcite{Holt:2007}. \copyright
(2009) Am.\ Phys.\ Soc.)}\label{Fig63}
\end{figure}

Large-scale shell-model calculations have been carried out for
$^{94}$Mo \cite{Lisetskiy:2000}, $^{96}$Ru \cite{Werner:2002} and
$^{92}$Zr \cite{Klein:2002} in the $Z=40$, $N=50$ mass region using a
surface delta interaction and treating all valence protons and neutrons
outside of the $^{88}$Sr core with $Z=38$ and $N=50$. Here, both the
orbital and spin matrix elements contribute, the latter part being
non-negligible for $M1$ transitions and moments. Like in the IBM-2 and
QPM, very much the same structure shows up. The specific
characteristics of $M1$ and $E2$ decay characterizing excited states
give a microscopic underpinning to the concepts of symmetric and mixed
symmetric excitations as used in the Q-phonon classification. The
fingerprints that characterize the decay of MS states (see Section
\ref{subsec-vibr}) are clearly observed in the results
\cite{Lisetskiy:2000}. Shell-model calculations for the $N=52$ nuclei
have been performed within the same model space but now using matrix
elements derived from the low-momentum $V_{low~k}$ nucleon-nucleon
interaction \cite{Holt:2007} giving a rather good reproduction of the
experimentally observed results. In Fig.~\ref{Fig63}, besides the
calculated total M1 strength the orbital and spin contributions, which
interfere constructively, are given separately.

\subsubsection{Nuclei near other doubly-closed shell regions}
\label{subsubsec-other}

\paragraph{The region near Z $\sim$ 50}
\label{paragraph-Sn}

The Cd and  Te nuclei with 2 proton holes (2 proton particles) away
from the $Z=5$0 Sn closed core, combined with the neutron filling of
the $50<N<82$ neutron shell form an interesting region to expect
mixed-symmetric states. In this mass region, a variety of experiments
have been carried out, including the  ($n,n'\gamma$) reaction
\cite{Garrett:1996,Bandyopadhyay:2003} for $^{112,114}$Cd, photon
scattering, eventually combined with Compton polarimetry to deduce the
parity unambiguously \cite{Lehmann:1999,Gade:2003,Kohstall:2005} for
$^{108-116}$Cd, $\beta$-decay \cite{Linnemann:2007} for $^{106}$Cd, and
recoil-distance Doppler-shift measurement (RDDS) after
fusion-evaporation reactions \cite{Boelaert:2007a} for $^{102,104}$Cd.
Evidence for the presence of MS $2^+$ states, slightly above 2 MeV
excitation energy, as well as for MS $1^+$ states, at the higher
excitation energy near 3 MeV, have been obtained in almost all of these
nuclei.

Shell-model calculations have also been reported for the light Cd
nuclei with $A=98-106$, using the same core as before ($^{88}$Sr),
treating all the available valence protons (10 in the case of Cd) and
neutrons \cite{Boelaert:2007b}. A detailed mapping of shell-model
states onto MS states for both $2^+$ and $1^+$ states was performed
using as criterium, strong $M1$ transitions to the $0^+_1$ and $2^+_1$
symmetric states combined with weak $E2$ transitions to these same
states, and strong $E2$ transitions in between the MS $2^+,1^+$ states.
Likewise, in the even-even $^{122-130}$Te nuclei, an experimental
search for MS $1^+$ and $2^+$ states was performed by
\textcite{Schwengner:1997a,Schwengner:1997b} using photon inelastic
scattering, and in $^{124}$Te by \textcite{Georgii:1995} using a
variety of reactions. This resulted in the detection of candidates for
mixed-symmetric $2^+$ states, slightly above 2 MeV excitation energy.
More recently, \textcite{Hicks:2008} have provided detailed results on
fragmentation of MS $2^+$ states in the $^{122-130}$Te nuclei. The
deformation dependence of the g.s.\ scissors mode strength in these
isotopes could by successfully reproduced by QRPA calculations
\cite{Guliyev:2002}.

\paragraph{The rare-earth region: $54<Z\leq60$ and $72<N\leq82$}
\label{paragraph-rare}

\begin{figure}[tbh]
\includegraphics*[angle=0,width=8.5cm]{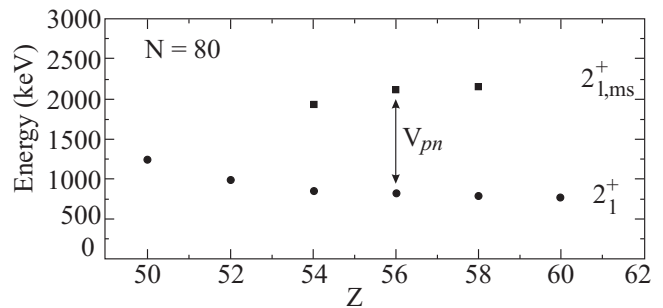}
\caption{Excitation energies of the $2^+_1$ and $2^+_{ms}$ states in
$N = 80$ isotones. The splitting is a measure of the residual
proton-neutron interaction \cite{Ahn:2009}. \label{Fig1new} .}
\end{figure}

Early evidence for the presence of MS $2^+$ states in rare-earth nuclei
resulted from an analysis by \textcite{Hamilton:1984}. He showed that
in nuclei with two neutrons outside of the $N=82$ closed neutron shell,
with an even number of protons filling the $50<Z<82$ proton shell,
$2^+$ states near 2 MeV would show up. Experiments at the ILL Grenoble
using $\gamma-\gamma$ directional correlation experiments (from
$\delta$-mixing ratios and branching ratios) allowed to find in
$^{140}$Ba, $^{142}$Ce and $^{144}$Nd a $2^+_3$ state with the typical
$M1$ and $E2$ branching into the lower-lying symmetric excitations.
More recently, these nuclei have been studied using NRF and
$(n,n'\gamma)$ scattering, even extending up to $^{148}$Sm
\cite{Gade:2000,Gade:2004,Vanhoy:1995,Hicks:1998,Li:2005,Mukhopadhyay:2008}
showing evidence for MS $2^+$ states, in most cases exhibiting
fragmentation over a number of $2^+$ states near or just above to 2 MeV
excitation energy. Likewise, experiments have been carried out in the
$N=80$ nuclei $^{134}$Xe,$^{136}$Ba and $^{138}$Ce using the same
techniques and Coulomb excitation
\cite{Pietralla:2008,Pietralla:1998a,Rainovski:2006,Scheck:2004,Williams:2009}.
They again show the presence of MS $2^+$ states and fragmentation, the
latter changing quite drastically with the proton number increasing
from $Z=56$ to $Z=58$. As demonstrated in Fig.~\ref{Fig1new} it is even
possible to extract the strength of the residual proton-neutron
interaction from the energy splitting between lowest symmetric and
mixed-symmetric $2^+$ state in these $N=80$ nuclei \cite{Ahn:2009}.
The $N=80$ nuclei have been extensively studied by
\textcite{LoIudice:2008} within the QPM approach and in a large-scale
shell-model study \cite{Sieja:2009}. These calculations describe rather
well both the variation in excitation energy as well as the changing
fragmentation pattern, moving from $^{132}$Xe to $^{140}$Nd.

A number of nuclei in which both the proton number and neutron number
is steadily increasing moving away from the $Z=50$ and $N=82$ shell
closure, such as $^{126,128,130}$Xe, $^{134}$Ba, $^{136}$Ce, have been
studied using techniques as discussed before
\cite{Gade:2000,Wiedenhoever:1997,Fazekas:1992,Betterman:2009,Ahn:2007}.
In these nuclei, the signature of a MS $2^+$ state shows up
consistently slightly above 2 MeV excitation energy.

\paragraph{The $A\approx60$ region}
\label{paragraph-60}

The region near the doubly-magic nucleus $^{56}Ni$ with proton and
neutron  hole or particle pairs outside of the $Z = N = 28$ core may
well give rise to MS couplings of proton and neutron building blocks.
The nucleus $^{56}$Fe forms an ideal testing case, and has been studied
by \textcite{Eid:1986},using $\gamma$-decay studies, and by
\textcite{Hartung:1989} in electron scattering. Clear-cut evidence for
fragmentation of MS $2^+$ strength has been observed around
$2.6-2.9$~MeV. In nearby nuclei such as $^{54}$Cr and $^{66}$Zn,
candidates for MS $2^+$ strength have been detected near 3~MeV
excitation energy \cite{Lieb:1988,Gade:2002}. It is interesting to note
the increase in energy from the heavier nuclei, where the typical
energies are closer to 2~MeV.

\paragraph{Heavy nuclei in the vicinity of $^{208}Pb$}
\label{paragraph-Pb}

The idea of low-lying $2^+_{ms}$ excitations appearing in regions where
the number of protons and neutrons forms a stable, closed shell, has
been shown to be a general property all through the nuclear mass
region. Therefore, the region around $Z=82$ and $N=126$ should be a
most interesting region one in order to explore the appearance of
states which exhibit a MS character in the protons and neutrons. Early
evidence was shown by \textcite{Ahmad:1989} for $2^+$ states near 1.5
MeV in $^{200}$Hg. Likewise a $2^+$ state near 2.2 MeV was observed in
$^{196}$Pt \cite{Jewell:1997,vonBrentano:1996}. This region has by now
not been studied in a systematic way but the Hg, Pt and also the Po, Rn
nuclei with neutron numbers close to $N=126$ should form an ideal
testing ground but require radioactive beams.

\subsection{Summary}
\label{vib-sum}

To sum up this chapter, we have presented evidence for isovector
proton-neutron excitations in which small amplitude quadrupole
oscillations form the basic building blocks. These isovector
excitations result in a natural way in both shell-model, collective
(geometrical and algebraic) and quasi-particle-phonon theoretical
approaches. The characteristic fingerprint of strong $M1$ transitions
between MS and symmetric collective states, associated with weak E2
transitions from MS states into the symmetric collective states, has
allowed to identify the MS states. They are most clearly observed when
a given nucleus only contains a few proton particles (holes) and
neutron particle (holes)outside of closed shells. The $Z \approx40$, $N
\approx 50$ region is one of the best studied regions showing besides
the lowest MS $2^+$ state, more complex MS $1^+,2^+,3^+$ and $4^+$
states (Zr, Mo, Ru nuclei). More recently, the presence of MS states
has been accumulated in nuclei near $Z \approx 50$ (such as the Cd, Te
nuclei), in the rare-earth region ($54<Z \leq 60$, $72 < N \leq 82$)
and also for lighter nuclei near the $Z=N=28$ closed shells. In view of
the very general characteristic of these excitations, one can expect
them to appear in nuclei adjacent to any doubly-closed shell nucleus.

\section{SCISSORS MODES IN OTHER MANY-BODY SYSTEMS}
\label{sec:manybody}

\subsection{Rotational magnetic excitations in other fermion systems}
\label{fermionic}

\subsubsection{Deformed metallic clusters}
\label{metallic}

Mass spectra for a particular class of metallic clusters were produced
more than 20 years ago \cite{Knight:1984}. They exhibit large abundance
peaks at $N=8,20,40,58,92,138, \dots$ These clusters were produced  in
the expansion of an inert gas (typically Argon or Xenon) through a
$0.1-0.2$ mm wide nozzle, in which the random thermal motion of the
atoms is converted into a uniform translational motion, thereby also
causing a cooling of the inert gas. Introducing atomic Na vapor into
this system results into large clusters  with an overall broad size
distribution. Details of the mechanism are described by de
\textcite{deHeer:1993}. The peculiar observation for Na clusters, and
later shown to exist also for Ag, Au, and Cs, points towards extra
stability associated with the delocalized motion of atomic $3s$
electrons, bound in a spherical potential (see
\textcite{deHeer:1987,Bjornholm:1990,Bjornholm:1992} for a detailed and
extensive discussion). The observation of quantal effects in clusters
of atoms established a deep connection between various fields in
physics such as electronic motion in atoms and nucleonic motion in
nuclei.

The Hamiltonian describing the neutral cluster consisting of $N$
nuclei with $Z$ electrons each, is fully determined through the
Coulomb force, but is generally too complex to be solved exactly.
In simple metals, though, such as Ag, Al, Na, \dots, the separation
into valence electrons and core electrons (well bound and localized)
leads to the simplification of $xN$ interacting electrons ($x$ the
number of valence electrons per atom) moving in the field caused by
the $N$ ions. A further step results in completely ignoring the
nuclear motion, thus leading to an electronic Hamiltonian of the
form
\begin{equation}
H_{\it el}= \sum_{k=1}^{{\it x}N} \{\frac{{\bf p}_k^2}{2m} + V_I({\bf
r}_k)\} + \frac{1}{2} \sum_{{\it l}\neq{\it k}=1}^{{\it x}N}
\frac{{\it e}^2}{\mid {\bf r}_k -{\bf r}_l \mid} ,
\end{equation}
with the ionic potential $V_I({\bf r}_k)$ defined as
\begin{equation}
V_I({\bf r}_k) = - \sum_{{\it i}=1}^N \frac{{\it xe^2}}{\mid {\bf
r}_k -{\bf R}_i \mid} .
\end{equation}
The latter potential most often is replaced by some
pseudo-potential. At the end, a rather drastic but efficient
approximation consists of averaging out the ionic structure and
replacing the corresponding charge distribution by a ``constant''
background charge in a finite (spherical, deformed, vibrating and/or
rotating) volume. This defines the so-called {\it jellium} model as
used in the description of metallic bulk and surface properties
\cite{Brack:1991a,Brack:1991b,Brack:1992}.

\begin{figure}[tbh]
\includegraphics[angle=0,width=8.5cm]{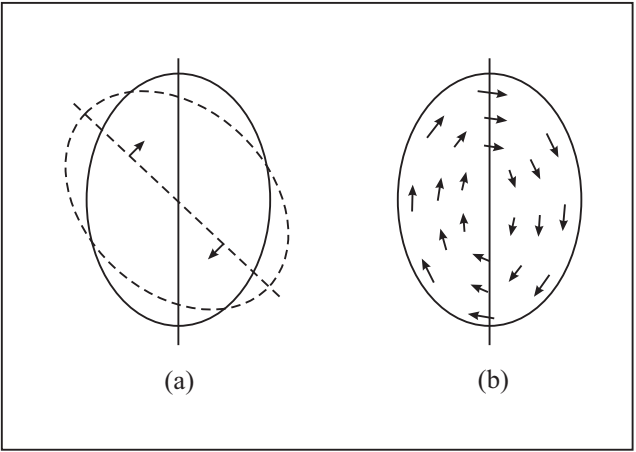}
\caption{Displacement field for the magnetic $M1$ low-lying
rotational state corresponding to (a) a rigid rotation of the
electrons with respect to the jellium background and (b) a rotation
within a rigid surface.The direction of the unit vector $\bf{\hat
\omega}$ ($z$ direction) points out of the plane. (Reprinted with
permission from \textcite{Lipparini:1989b}. \copyright
(2009) Am.\ Phys.\ Soc.)}\label{Fig64}
\end{figure}

Collective dipole excitations are well known in alkali-metal clusters
\cite{deHeer:1987,deHeer:1993} which correspond to the classical
surface-plasmon oscillations \cite{Ashcroft:1976} of the electron cloud
against the positively charged ions forming the cluster. The relative
motion of protons versus neutrons in atomic nuclei, giving rise to the
electric giant dipole mode is equivalent to the electric dipole mode
that results in electron motion in atoms. For deformed metallic
clusters, a magnetic excitation of orbital nature was predicted by
\textcite{Lipparini:1989a} at an energy much lower than the classical
plasmon frequency. A macroscopic illustration of this new magnetic
excitation derives from a displacement field ${\bf u}$ of the electron
motion in the valence cloud. The suggested form is
\begin{equation}
{\bf u} = {\bf \hat{\omega} x r} + \frac{\delta}{1+ \delta/3}{\bf
\nabla}(xy) ,
\end{equation}
with {\bf $\hat{\omega}$} the unit vector in the z-direction. The cluster
is described with a deformed electron density profile
\begin{equation}
\rho_e = \rho_0(\frac{x^2}{R^2_x} + \frac{y^2}{R^2_y} + \frac{z^2}{R^2_z}),
\end{equation}
with $R_x=R_z$ and the deformation $\delta$ defined by the expression
\begin{equation}
\delta = \frac{3}{2}\frac{R^2_y - R^2_x}{R^2_y + 2R^2_x}.
\end{equation}
This displacement field satisfies the condition ${\bf \nabla \cdot
u}$=0. The first term solely corresponds to a rigid rotation of the
electrons with respect to the jellium background and implies a scissors
mode (Fig.~\ref{Fig64}a) with a restoring force that is due to the
Coulomb attraction between the electron cloud and the ion positive
charged background. This is very much identical to the proton-neutron
symmetry term causing the restoring force in case of atomic nuclei.
Including the quadrupole term in the displacement field i.e. ${\bf
\nabla}(xy)$, the corresponding motion (Fig.~\ref{Fig64}b) corresponds
to a rotation within a spheroidal rigid surface with a velocity field
such that ${\bf v \cdot n}\mid_{surface}=0$.
In the limit of small deformation, one can determine the frequency of
the magnetic mode $\omega_{M1} = \sqrt{K/\theta}$ with $K$ the energy
change originating from the displacement field and $\theta$ the
collective mass parameter (related to the moment of inertia)
\begin{equation}
\omega_{M1} = \delta \sqrt{\frac{4\varepsilon_F}{mr^2_s}}N^{-1/3}.
\label{eq:magmode}
\end{equation}
In this section \ref{sec:manybody}, we use throughout the convention
$\hbar = c = 1$. In deriving this result, the approximate relation
\begin{equation}
\langle r^2 \rangle = \frac{3}{5} r^2_s N^{2/3},
\end{equation}
was used. In the particular case of Na clusters,with $r_s = 4$~a.u.\
and $\varepsilon_F=3.1$~eV, the frequency becomes $\omega_{M1} = \delta
4.6 N^{-1/3}$ eV. In the range of clusters with $N = 10-100$ and for
typical deformations $\delta= 0.2 - 0.4$, the frequency amounts to 0.2
- 0.6 eV, much lower than the dipole plasmon frequency of 3.4 eV. This
collective state carries considerable $M1$ strength $B(M1) \approx
\omega_{M1}\theta~\mu_B^2$, or, using explicit expressions for the
frequency and the moment of inertia \cite{Lipparini:2003}, becomes
\begin{equation}
B(M1) = \frac{4}{5}\delta \sqrt{\varepsilon_F mr^2_s}N^{4/3}\mu_B^2,
\end{equation}
which, in the particular case of Na clusters, reduces to the
approximate result $B(M1) \approx \delta N^{4/3}\mu_B^2$ with $\mu_B$
the Bohr magneton. The particular results for both $\omega_{M1}$ and
$B(M1)$ have a very similar form as the expressions derived in
Eqs.~(\ref{eq:hamiltonian}) and (\ref{eq:m1-moi}) for the nuclear case.
\begin{figure}[tbh]
\includegraphics[angle=0,width=8.5cm]{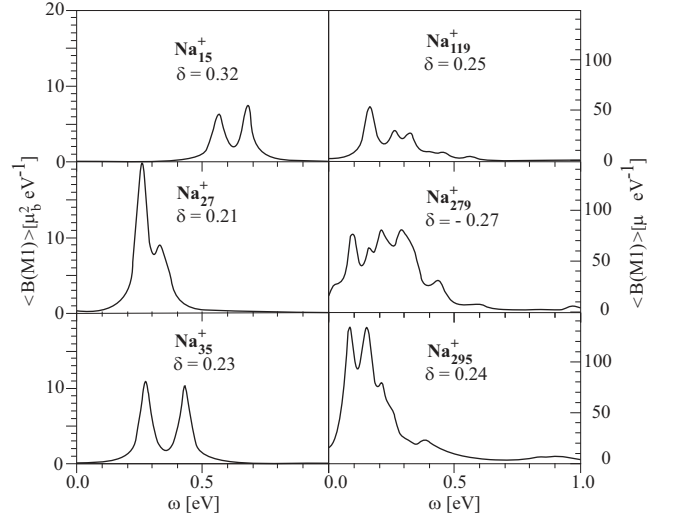}
\caption{Energy distribution of the $M1$ strength for Na clusters
ranging from $N=15$ to $N=295$. The deformation parameter $\delta$
characterizing the specific cluster is also given. (Reprinted with
permission from \textcite{Nesterenko:1999}. \copyright
(2009) Am.\ Phys.\ Soc.)}\label{Fig65}
\end{figure}

The above semi-classical results can also be derived starting from
microcopic calculations. A schematic two-level RPA model has been
studied by \textcite{Lipparini:1989b} with results that corroborate the
macroscopic approach. More detailed studies of the orbital magnetic
dipole mode were carried out by \textcite{Nesterenko:1999}, based on a
self-consistent RPA approach \cite{Nesterenko:1997,Kleinig:1998}. Using
a phenomenological Woods-Saxon potential, the results confirm those
obtained by \textcite{Lipparini:1989a} discussed before. We present in
Fig.~\ref{Fig65} the salient features of the $M1$ strength distribution
as derived for Na clusters, in which the $N$ dependence (moving from
$N=15$ towards $N=295$) becomes particularly clear. In the heavy
clusters with $N \sim 300$, the $M1$ strength reaches very big values
of the order of $350 - 400~\mu_B^2$ with the strength remaining
concentrated in a rather narrow energy interval. Therefore, it is in
the heavy clusters that the collective nature of the $M1$ mode becomes
particularly clear with a $N^{-1/3}$ frequency dependence as pointed
out in Eq.~(\ref{eq:magmode}). This distinguishes the magnetic mode
from the regular electric dipole surface-plasmon mode, which is much
better documented experimentally. As was pointed out before, the
strength in exciting this magnetic dipole mode is so much weaker than
the electric dipole strength that it has not been observed
experimentally as yet, contrary to the nuclear case as discussed in the
major part of the present paper.

\subsubsection{Orbital current modes in deformed quantum dots}
\label{quantumdot}

Recent advances in semiconductor technology have allowed to build
nanostructures with varying shapes. In these systems, electrons are
laterally confined at the semiconductor boundary and form a
two-dimensional (2D) quantum dot (see \textcite{Alhassid:2000}).
Until recently, the vast majority of theoretical and experimental
efforts focused on systems with circular symmetry. Many of the
properties of these dots are very well accounted for when imposing a
parabolic potential as confinement, or, invoking the concept of a
jellium disk (see subsection ~\ref{metallic}). Experiments
addressing deformed nanostructures using Raman scattering and
far-infrared spectroscopy
\cite{Sikorski:1989,Demel:1990,Strenz:1994,Schueller:1996,Austing:1999}
have been at the origin of theoretical studies relaxing on the
circular symmetry \cite{Koskinen:1997,Hirose:1999,Puente:1999}.

With the breaking of the rotational symmetry of the system (metallic
cluster, quantum dot), in particular by introducing a quadrupole
distortion, it turns out that a low-energy orbital current
excitation (OCE) is generated in the 2D dot with an energy
dependence $N^{-1/2}$ in contrast with the $N^{-1/3}$ dependence for
three-dimensional system \cite{Lipparini:2003}. Here, $N$ denotes
the number of electrons confined in the elliptic quantum dot.
Interesting here to remark is the fact that the frequency goes to
zero with $N \rightarrow \infty$.

\begin{figure}[tbh]
\includegraphics[angle=0,width=8.5cm]{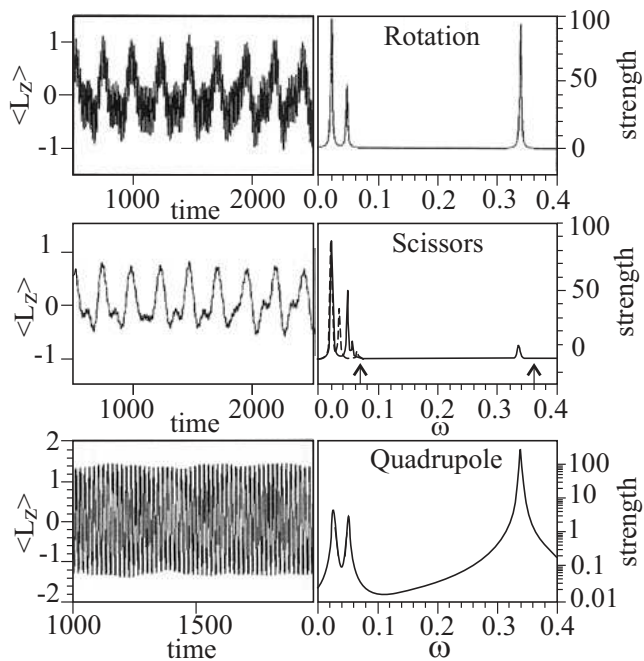}
\caption{Results for the time evolution of an elliptic quantum dot
with $N=20$ electrons, for a deformation value of $\eta$=0.28, and
this using three different initial perturbations: pure rotation,
orbital and quadrupole distortions. In the left-side panels, the
simulated $M1$ signal, expressed as  $ \langle{\it \hat{L}}_z
\rangle$, is plotted as a function of time. The right-side panels
show the corresponding $M1$ excitation strength. The middle
right-side panel also shows the independent electron strength
function (dashed line) and the little arrows indicate the position
of the solutions given by Eqs.~(\ref{eq:omega-oce}) and
(\ref{eq:omega-qcde}). (Reprinted with permission from
\textcite{Serra:1999}. \copyright (2009) Am.\ Phys.\ Soc.)}\label{Fig66}
\end{figure}

One can now perform essentially the same analysis as was used in the
description of deformed metallic clusters, but now  with an elliptic
two-dimensional charge distribution given as
\begin{equation} \label{eq:rhoe}
\rho_e = \rho_0(\frac{x^2}{R^2_x} + \frac{y^2}{R^2_y}),
\end{equation}
with $R_x$ and $R_y$ the ellipse radii and a deformation parameter $\eta$,
defined by the expression
\begin{equation}
\eta = \frac{R^2_y - R^2_x}{R^2_y + R^2_x}.
\end{equation}
The kinetic energy density has an expression analogous to the charge
distribution given in Eq.~(\ref{eq:rhoe}). An OCE follows, using a
displacement field similar to the one discussed in Sec.~\ref{metallic},
but for the 2D case which is expressed as \cite{Serra:1999}
\begin{equation}
{\bf u} = {\bf \hat{\omega} x r} + \eta{\bf \nabla}(xy) \ .
\label{eq:oce}
\end{equation}
The electronic motion becomes a collective rotating flow along
the ellipse contour lines. The frequency for this orbital mode turns
out to be
\begin{equation}
\omega_{OCE}\approx \frac{\eta}{\sqrt{1-\eta^2}}\sqrt{\frac{16
\varepsilon_F}{3 mr_s^2}}N^{-1/2} \ , \label{eq:omega-oce}
\end{equation}
where $r_s$ stands for the Wigner-Seitz radius \cite{Ashcroft:1976}.
Carrying out the same analysis but retaining only the quadrupole
deformation generating displacement term ${\bf u = \nabla}(xy)$, the
frequency of the corresponding quadrupole charge-density excitation
(QCDE) becomes
\begin{equation}
\omega_{QCDE} \approx \sqrt{2}\omega_0, \label{eq:omega-qcde}
\end{equation}
where $\omega_0$ is the average of the frequencies in $x$ and $y$
direction for the confining parabolic potential. This frequency is much
larger than the frequency of the orbital excitation. A most interesting
illustration of the effects of both modes is obtained when evaluating
the magnetic orbital response. This $M1$ response $ \langle {\it
\hat{L}}_z \rangle$ has been calculated \cite{Serra:1999}, as a
function of time, by modifying the electron orbitals using the
displacement operator of Eq.~(\ref{eq:oce}) (left-side part of
Fig.~\ref{Fig66}) as well as the corresponding $M1$ strength
(right-side part of Fig.~\ref{Fig66}). Here, the cases of (i) a pure
rotational perturbation (rotation), (ii) orbital perturbation (OCE or
scissors), and (iii) pure quadrupole perturbation are shown,
respectively. These results are in line with the simple discussion of
the frequencies as presented before. The $M1$ orbital strength is
divided in two distinct regions: one at the higher-energy side, which
is associated with the quadrupole distortion and one at the low-energy
end, associated with the orbital excitation. An in-depth study is
presented by \textcite{Austing:1999}. The distribution of scissors $M1$
strength in these elliptic quantum dots is quite similar to the
situation in strongly deformed nuclei in which the $M1$ strength is
also separated into a low-energy orbital (the scissors mode) part and
the higher-lying $K^{\pi}$=$1^+$ component of the isovector
giant-quadrupole resonance (see Sec.~\ref{sec:mds-exp}).

Besides these orbital charge-density excitations, it is also possible
to describe spin-density oscillations in quantum dots. When the spin
components oscillate in phase, they describe the density modes,
however, when oscillating out of phase, spin modes can be created. A
particular interesting case is obtained as alternating rotation of spin
up and spin down densities in opposite direction, This spin-twist mode
is very soft \cite{Puente:1999}, well below the spin dipole oscillation
modes.

\subsubsection{Other Fermi systems}
\label{fermi}

It turns out that scissor modes can also be realized in a superfluid
Fermi gas \cite{Minguzzi:2001}. Confining the Fermi gas inside a
spherical harmonic trap and solving the equations of motion for the
density and concentration fluctuations gives rise to a single scissors
frequency for the superfluid situation and results in two scissors
frequencies in a normal Fermi gas.

Recently, \textcite{Hatada:2005} have suggested the possibility that
axially symmetric atoms in crystals with ionic bonding can exhibit a
scissors excitation. Its signature in this case is the existence of a
low-lying collective excitation resulting from precessing atoms around
the anisotropy axis of the crystal cells. This excitation has a
magnetic dipole character and could be observed through the absorption
of incoming photons which should exhibit a differential dichroism. An
extension to cover also crystals with cubic symmetry was presented by
\textcite{Hatada:2009b}.

Other ways of observing scissors modes in crystals start from a recent
experiment that studied magnetic properties of rare-earth systems
\cite{vanderLaan:2008}. If one considers crystals in which the internal
electrostatic field is small with respect to the electron spin-orbit
coupling in the atoms, the so-called ``spin-orbit locking'' situation,
an applied external magnetic field will rotate both the spin and charge
density profiles simultaneously. Switching off the magnetic field, the
atoms will start oscillating around the axes of the crystal cells
\cite{Hatada:2009a}. Experiments are proposed that may be sensitive
enough to detect the photons emitted when deexciting the scissors
excitation \cite{Hatada:2010}.

\subsection{Scissors modes of a trapped Bose-Einstein condensate}
\label{bose-einstein}

\begin{figure}[tbh]
\includegraphics[angle=0,width=8.5cm]{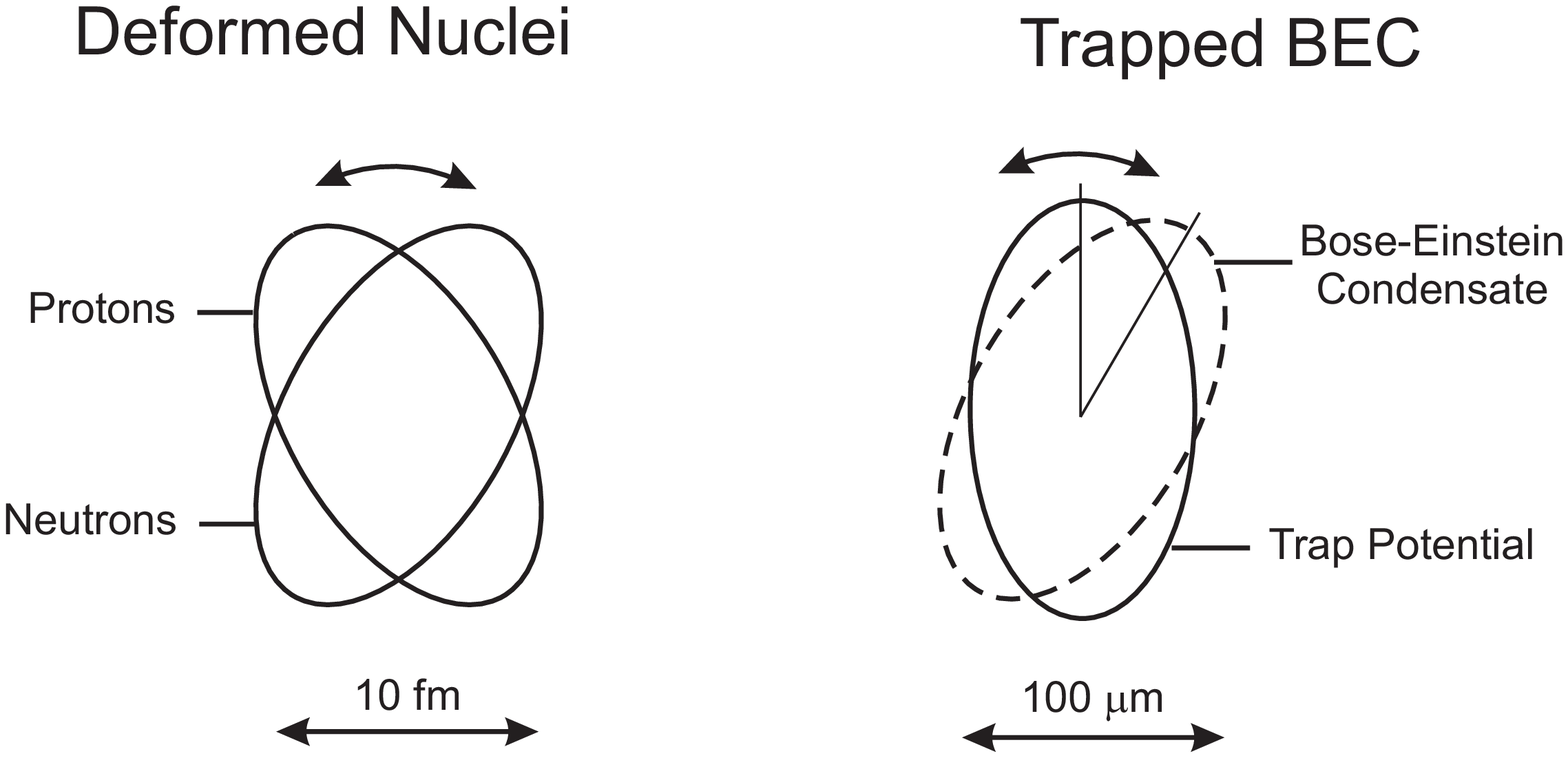}
\caption{Schematic comparison of the scissors motion  in atomic
nuclei with protons and neutrons generating the scissors motion, on
a length scale of 10~fm (left side), and in Bose-Einstein
condensates trapped in an external potential, the latter acting on a
length scale of 100~$\mu$m (right side).}\label{Fig67}
\end{figure}

By now, there exists a vast literature on trapped Bose-Einstein
condensates \cite{Pitaevskii:2003,Giorgini:2008}. Superfluidity in
these condensates is one of the most spectacular consequences. It is,
however, not easy to obtain unambiguous evidence for the superfluid
characteristics. Because superfluidity will affect the moment of
inertia of the trapped condensate  (such as a reduction over the
classical rigid value), one could expect that a study of rotational
properties of such condensates can give rise to experimental evidence
for the existence of superfluidity. \textcite{Guery-Odelin:1999} have
studied the oscillatory behavior caused by rotating a condensate with
respect to the symmetry axis when trapped in a deformed external
potential of parabolic type. They concentrate in particular on the
superfluid effects in the condensate. The restoring force associated
with such a rotation in the $xy$-plane  is proportional to $\delta^2$,
with the trapping potential given by the expression
\begin{equation}
V_{ext}({\bf r})=\frac{m}{2}\omega^2_x x^2 +
\frac{m}{2}\omega^2_y y^2 + \frac{m}{2}\omega^2_z z^2,
\end{equation}
with moreover
\begin{equation}
\omega_x^2=\omega_0^2(1+\delta) ,\omega_y^2=\omega_0^2(1-\delta).
\end{equation}
The mass parameter, determined by the moment of inertia, in the
superfluid case becomes proportional to $\delta^2$ too
\cite{Lipparini:2003,Rowe:1970}. As a result, even when the deformation
of the external potential approaches zero, the frequency keeps a finite
value. It is only in the absence of superfluidity that the moment of
inertia regains its rigid value and therefore, a low-frequency will
characterize the oscillatory motion. The outcome of the theoretical
study (see \textcite{Guery-Odelin:1999,Lipparini:2003}) is that a
sudden rotation of the trap symmetry axis by a small angle $\theta_0$
will perturb the condensate from its equilibrium shape and, if the
angle $\theta_0$ is not too large, will start a scissors-like
motion\footnote{The expression scissors mode has been taken over from
nuclear physics by the atomic physics and BEC community.} in the
$xy$-plane. This is illustrated in a schematic way in Fig.~\ref{Fig67}
(right-hand side) in which both the trapping potential and the
condensate are drawn. An idea of the dimensions of such condensates is
also given. On the left-hand side in Fig.~\ref{Fig67}, we compare with
the analogous situation in strongly deformed atomic nuclei, in which
protons and neutrons can give rise to a scissors motion. Here too, the
dimension of the system is given in order to stress the large
difference in scales but keeping essentially the same physics.
Under the above conditions one obtains
\begin{equation}
\theta(t)=\theta_0 cos(\omega_{sc}t),
\end{equation}
in which $\omega_{sc}=\sqrt{2}\omega_0$, or $\omega_{sc} =
\sqrt{\omega_x^2 + \omega_y^2}$.
%
%
In the absence of a superfluid condensate and entering
a high-temperature regime, analytic solutions become possible
\cite{Guery-Odelin:1999} which, for small rotation angles of the trap
axis, are described by a differential equation for $\theta(t)$. This
equation now allows for different solutions, propagating at high and
low frequencies. In a collisionless regime, the higher frequency
becomes $\omega_+$=$\omega_x+\omega_y$ and can be identified with an
irrotational quadrupole oscillation. The lower frequency
$\omega_-$=$\mid\omega_x-\omega_y\mid$ , however, corresponds to the
rotational mode of the system, a component which is absent in the
superfluid case.

\textcite{Marago:2000,Marago:2001} have reported on a clear observation
of the scissors mode of a Bose-Einstein condensed gas of $^{87}$Rb
atoms brought into a magnetic trap. The authors highlighted the very
importance of the discovery of the scissors mode in atomic nuclei
stating ``The experimental discovery of the scissors mode
\cite{Bohle:1984a}, first predicted in a geometrical model
\cite{LoIudice:1978,Lipparini:1983}, has been one of the most exciting
findings in nuclear physics during the past two decades''.

\begin{figure}[tbh]
\includegraphics[angle=0,width=8.5cm]{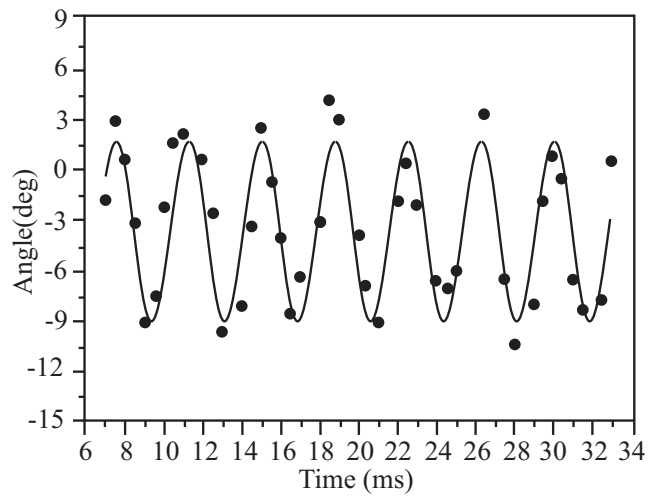}
\caption{The classical frequency and the tilt angle
for a trapped Bose-Einstein condensate of
$^{87}$Rb atoms. (Reprinted with permission from
\textcite{Marago:2000}. \copyright (2009) Am.\ Phys.\
Soc.)}\label{Fig68}
\end{figure}

The scissors mode has been excited by a sudden rotation of the deformed
trapping potential (with the constraints of having $\omega_x \approx
\omega_z > \omega_y$). The condensate, cooled to well below the
critical temperature for the Rb gas contains of the order of $10^4$
atoms. The time evolution of the scissors oscillation
(Fig.~\ref{Fig68}) exhibits a single undamped mode corresponding to
$265.6 \pm 0.8$ Hz, which agrees perfectly with the theoretical value
of $265 \pm 2$ Hz as deduced from $\omega_{sc}$ = $\sqrt{\omega_x^2 +
\omega_y^2}$. This observation thus provides an unambiguous
demonstration of the superfluid nature the Rb condensate. In
Fig.~\ref{Fig68}, we show the classical frequency and the tilt angle
for the trapped Bose-Einstein condensate of $^{87}$Rb atoms, which
correspond to $\nu=265.6 \pm 0.8$ s$^{-1}$ and $\alpha=7.2^{\circ}$,
respectively.

It is instructive to compare the condensate result with the scissors
mode in the deformed nucleus $^{156}$Gd \cite{Bohle:1984a}. We can
extract both the frequency (using the experimental energy of the $1^+$
state at $E_x$=3.075 MeV) and a classical tilt angle identifying the
restoring force with the symmetry energy and making use of the
expression $\alpha = (CJ_{intr}/\hbar^2)^{-1/4}$ deduced from the
proton-neutron collective model \cite{DeFranceschi:1984,Nojarov:1986}
discussed in Sec.~\ref{sec:th-col-mic}. This results in $\nu=7 \times
10^{20}$ s$^{-1}$ and $\alpha \sim 6^{\circ}$, respectively. Thus,
comparable tilt angles are found characterizing the scissors motion,
albeit in different regimes of physics. A striking difference between
the two systems, however, is the fact that in the condensate the
deformation of the trap can be varied by choosing appropriate
frequencies as well as the temperature giving a large range to study
the intricate properties of quantum fluids and their transition from
the irrotational (superfluid case) to the rotational regime. In the
atomic nucleus, the deformation is fixed for a given ($Z,N$)
combination.

Compared to the observation of scissors modes for trapped Bose-Einstein
condensates, scissors-like oscillations have also been observed in a
quantum-degenerate mixture of two such condensates consisting of
different atomic species, i.e., $^{41}$K and $^{87}$Rb
\cite{Modugno:2002}. In this case, the scissors mode is induced by
interactions between the two atomic species present in the trap.
Therefore, this situation is even closer to the nuclear physics case
where proton and neutron fluids are present.

\subsection{Rounding up} \label{similarity}

The above theoretical and experimental studies in both fermionic
(metallic clusters, quantum dots, Fermi gas, crystals) systems in which the electronic
motion in some 'external' potential has been studied and in bosonic
systems (Bose-Einstein condensates in the superfluid phase trapped in
an external anisotropic potential) show the existence of scissors modes
with properties in line of those observed in atomic nuclei. The
arguments can even be turned around: the observation of low-lying
scissors modes in these systems is a proof of the superfluid
characteristic of proton and and neutron fluids in the nuclear
rotational motion.

\section{CONCLUSIONS AND OUTLOOK}
\label{sec:outlook}

\subsection{Conclusions}
\label{conclusions}

The study of the response of nucleons, moving inside the atomic
nucleus, to external (electromagnetic and hadronic) probes in the
magnetic dipole channel, discussed in depth in the present article, can
be crystallized and summarized in a succinct way. Here, we take
Figs.~\ref{Fig1}-\ref{Fig3} as our guide.


(i)
At the lower energy side, orbital nucleonic motion is strongly excited
by the $\hat{L}_{\pi}- \hat{L}_{\nu}$ part of the $M1$ operator
(contra-rotational or scissors-like motion of protons versus neutrons).
This does not show up, however, as a single collective state but the
orbital strength is fragmented within a rather limited energy interval
between 2.5 and 4 MeV. In light nuclei, the strength distribution looks
more simple since it is contained mainly in the $2^{+} (\pi) \otimes
2^{+} (\nu)$, $4^{+} (\pi) \otimes 4^{+} (\nu )$, $\cdots$
configurations. In strongly deformed rare-earth nuclei, there is more
spreading of strength which has been detected using $(e,e')$ and
($\gamma,\gamma'$) reactions and not or very weakly with proton
scattering. A number of interesting results are connected to the fact
that the non-energy weighted $M1$ sum rule strength in the interval
$2.5-4$~MeV exhausts the larger part of the orbital $M1$ strength.
Moreover, it was discovered that this summed $M1$ strength correlates
strongly with deformation and thus with the $B(E2;0^{+}_{1} \rightarrow
2^{+}_{1})$ strengths, indicating saturation when progressing through
the strongly deformed rare-earth region.

(ii)
Spin $M1$ strength is concentrated at higher excitation energies
because the spin-flip part of the $M1$ operator is mainly a
$1\hbar\omega$ excitation in the shell model and has been studied
throughout the region of deformed rare-earth and actinide nuclei. In
light nuclei, the strength is particularly associated with the
spin-flip transition between spin-orbit partners. In the heavier
nuclei, it is the spin-spin isospin dependent part of the residual
interaction ($\vec{\sigma}\cdot\vec{\sigma} \vec{\tau}\cdot\vec{\tau}$)
that rules the concentration or fragmentation of spin $M1$ strength.

(iii)
At still higher excitation energies ($2\hbar\omega$), theoretical
predictions indicate the presence of a $K^{\pi} = 1^{+}$ component of
the isovector giant quadrupole resonance (IVGQR), a state that could
rightly be associated with a collective scissors mode. Because of the
high excitation energy, no systematic experimental studies exist yet
exploring such a mode. Knowledge of the IVGQR strength, connected to
this response, allows the evaluation of an energy-weighted sum rule to
constrain the orbital $M1$ strength, which is exhausted to more than
80\%.

(iv)
In nuclei with just a few valence protons and neutrons outside of
closed shells, it has been experimentally proven that low-lying
mixed-symmetry (isovector) $2^+$ excitations exist. They are
characterized by strong $M1$ decay into the first excited $2^+_1$
state, a symmetric (isoscalar) mode in the proton and neutron motion,
accompanied by weak $E2$ decay into the $0^+$ ground state. The
quadrupole degree of freedom dominates the low-energy structure,
resulting in energy spectra with a vibrational character. However,
isovector combinations can also give rise to a $1^+$ state which is
related to the scissors $1^+$ state as it appears in strongly deformed
nuclei.

(v)
In odd-mass nuclei, the study of the $M1$ strength distribution is
more complicated because of the odd-particle (-hole) coupling to the
$1^{+}$ modes which induces a large fragmentation.
There occurred a number of early problems in accounting for the
observation of the $M1$ strength but the present situation is such
that, through very careful studies of the highly-fragmented background
structures, the full summed $M1$ strength in odd-mass nuclei is
consistent with the summed $M1$ strength obtained in the adjacent
even-even nuclei. Much more work needs to be done from the side of
theoretical studies in order to understand the major mechanisms that
can explain the fragmentation and the sometimes sudden important
changes in the observed fragmentation when going from nucleus to
nucleus.

(vi)
In the even-even nuclei, a number of general features have resulted
from detailed experimental studies of the $M1$ response over many
nuclei, spanning the region from light to very heavy nuclei. One of the
most important observations is
a strong correlation of the orbital magnetic dipole response with other
multipoles, in particular with the $E2$ strength but also, be it in an
indirect way, with the nuclear charge radii, i.e.\ with the $E0$
strength. This connection has been formulated in a more quantitative
way using various $M1$ sum rules that are proportional to the
ground-state expectation value of the quadrupole-quadrupole force and
thus lead to new $E2$ sum rules. A former review article by
\textcite{Lipparini:1989b} already pointed out such a connection. The
microscopic understanding of this intimate relation between the $M1$
properties on one side and the quadrupole and monopole properties on
the other side are more indirect. It has been shown that the quadrupole
deformation of the nucleus spreads the individual single-particle
states (breaking the spherical symmetry) and, with pairing included,
the shell-model description of $M1$ strength is indirectly connected to
the $E2$ ground-state deformation characteristics of the nucleus. This
gives at the same time a natural explanation of the $M1$ orbital summed
strength saturation in the mid-shell region of the rare-earth nuclei.
It is a consequence of the fact that the deformed rare-earth nuclei
also exhibit a saturation in the quadrupole deformation value (about
$\delta\simeq 0.25-0.30$).

(vii)
At the present stage, the magnetic $M1$ strength seems to be rather
well understood and there appears a clear-cut concentration of orbital
strength at lower energies  using various theoretical approaches:
shell-model, QRPA  and QPM calculations, collective geometric and
algebraic models all account rather well with the experimental
situation on a qualitative level. This is a highly comforting situation
with respect to the theoretical understanding of how $M1$ strength
builds up at the lower energy side as mainly orbital in character.
Collective model approaches, however, by their specific nature, fail in
accounting for the detailed fragmentation. This holds for both the
geometrical and algebraic collective models. In order to bridge the gap
between the various theoretical approaches, one will have to address
and explore the possibilities of large-scale shell-model calculations
as a way to extract collective effects starting from a microsocpic
basis.


\subsection{Outlook and future perspectives}
\label{outlook}

Having formulated a number of concluding elements on how the nucleus
responds when excited with electromagnetic and hadronic probes in
the magnetic dipole channel, a number of very clear-cut problems
remain to be solved in the coming years. We also present a number of
topics that will deserve intensive thought and future experimental
efforts such as to bring us closer to a quantitative level of
understanding the magnetic dipole channel (and some related other
multipoles).


(i)
It will be particularly interesting to connect  the scissors $1^+$
mode, which was shown to be strongly excited using electron and photon
scattering off deformed nuclei, to the recent observation of $1^+$
states in nuclei when approaching closed shells. In the latter nuclei,
the quadrupole mode determines the dominant low-lying proton-neutron
isoscalar and isovector excitations. The exploration of a full class of
isovector (also called mixed-symmetry states) excitations with spins
ranging from $0^+$ to $4^+$ has begun but is clearly in need of still
more systematic studies. In this quest, it is of utmost importance to
combine as many complementary probes as possible (Coulomb excitation,
lifetime measurements, detailed $\gamma$- spectroscopy measuring
branching ratios and $\delta(E2/M1)$ mixing ratios,...) so as to be
able to pin down the nuclear wave functions of these states.

\begin{figure}[tbh]
\includegraphics[angle=0,width=8.5cm]{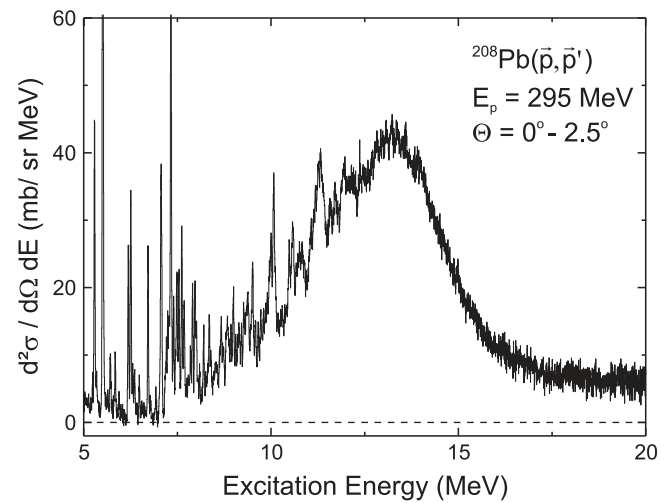}
\caption{Spectrum of the $^{208}$Pb($\vec{\rm p},\vec{\rm p}'$)
reaction at $E = 295$ MeV and $\Theta = 0^\circ$ \cite{vonNeumann-Cosel:2009}
measured with an energy resolution $\Delta E \simeq 25$~keV
(full width at half maximum) .
\label{FigSc13}}
\end{figure}

(ii)
The  spin-flip magnetic dipole response that has been studied using
(polarized) proton scattering off nuclei in the energy region
$5-10$~MeV poses some specific challenges. A particularly important and
unsolved problem, at present, concerns the precise character of the two
humps observed in the spin-flip $M1$ strength distribution in many
deformed nuclei. Theoretical studies come to opposing conclusions and
only a more detailed experimental survey of the precise charge
character of the two humps (proton/neutron versus isoscalar/isovector)
can resolve this issue. From the theoretical side and with the
present-day numerical capacities to perform large-scale shell-model
calculations, a precise survey of the $M1$ response in both the
$sd$-shell nuclei, in particular along the $N=Z$ line with nuclei like
$^{20}$Ne,$^{24}$Mg,$\cdots$,$^{36}$Ar,$^{40}$Ca as well as in the
heavier $fp$-shell nuclei would be very important, also in the light of
existing high-quality data in some of these light and medium-heavy
nuclei. Recent experimental developments \cite{Tamii:2009}, which
combine for the first time high-resolution measurements of the
($\vec{p},\vec{p'}$) reaction at proton energies of several hundred MeV
with measurements at $0^\circ$, should allow to tackle all of these
questions. Indeed, the feasibility of such experiments has been proven
for the heaviest nuclei as demonstrated in Fig.~\ref{FigSc13} for the
example of $^{208}$Pb \cite{vonNeumann-Cosel:2009}.

(iii)
The  spin-flip magnetic dipole response  which is connected to the
axial-vector part of the Gamow-Teller operator, may well be extended
towards higher excitation energies by using the fact that the spin-flip
part probed in proton scattering is also connected to the neutrino
(axial-vector term) scattering contribution. So, proton scattering may
provide valuable information concerning neutrino scattering off nuclei
in the giant resonance region and, subsequently, give insight in
neutrino-nucleus cross-sections and their importance in supernova
processes.

(iv)
There are other magnetic modes to be explored besides the dipole
one, which formed the major part of the present review. There has been
a search for magnetic quadrupole excitations of $J^{\pi}=2^-$ states in
$^{48}$Ca and $^{90}$Zr using high-resolution backward angle inelastic
electron scattering \cite{vonNeumann-Cosel:1999}, extended to $^{58}$Ni
\cite{Reitz:2002}. Macroscopically, the orbital $M2$ mode can be viewed
as a vibrational counter-rotation of different fluid layers in the
upper and lower hemisphere, hence the name ``twist'' mode
\cite{Holzwarth:1977,Holzwarth:1979}. Like for the scissors mode,
occurrence of an orbital $M2$ mode is a general feature of quantum
many-body systems as discussed e.g.\ for the cases of metallic clusters
by \textcite{Nesterenko:2000} and ultracold Fermi gases by
\textcite{Vinas:2001}. Therefore this mode deserves more intensive and
systematic exploration.

(v)
The proportionality of the summed $M1$ strength to the nuclear
deformation, more in particular showing a $\delta^2$ dependence, opens
the possibility to make use of measured magnetic dipole strengths as a
new fingerprint for exploring sudden shape-phase transitions. Since
there appear a number of regions in the nuclear mass table that look
like potential places where sudden changes in nuclear shape may occur,
an in depth study of magnetic dipole excitations in those regions can
lead to extra information.

\begin{figure}[tbh]
\includegraphics[angle=0,width=8.5cm]{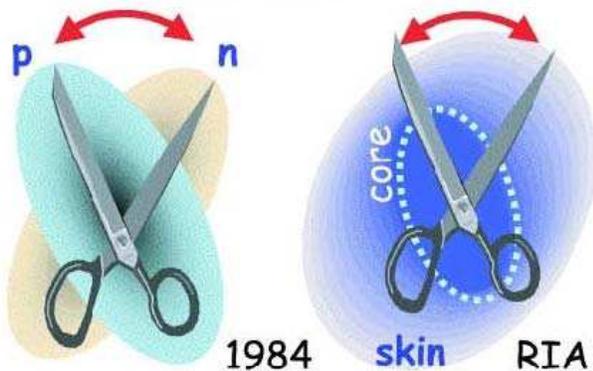}
\caption{(Color online) Schematic picture of the scissors mode in stable deformed
nuclei (left) as compared to exotic nuclei with a weakly-bound
neutron skin (right), taken from \cite{RIA:2003}. \label{FigSc14}}
\end{figure}
(vi)
Last, but not least, a point of current interest is to study the
scissors mode moving away from the valley of $\beta$-stability. Because
the nuclei to be explored are highly unstable, one will have to resort
to reactions in inverse kinematics such as Coulomb excitation, inverse
proton scattering utilizing radioactive ion beam facilities like RIKEN,
FAIR-GSI, SPIRAL II, and further on, FRIB. New phenomena are expected
\cite{RIA:2003} in very neutron-rich nuclei when coupling a weakly
bound neutron skin to a well-bound deformed core (cf.\
Fig.~\ref{FigSc14}).  A soft scissors mode due to the presence of a
neutron skin, analogous to the soft electric dipole mode, has been
proposed within both a geometrical approach by
\textcite{VanIsacker:1992} at an energy approximately half of the
scissors mode energy, and also within the IBM-2
\cite{Warner:1997,Caprio:2004}. Considering the plans at FAIR, one may
even think about electron scattering on unstable nuclei. These and more
exotic possibilities may materialize at future RIB facilities, as
discussed e.g.\ in recent report on upgrading the NSCL facility at MSU
\cite{RIA:2006}. The new generation facilities should also allow to
systematically explore mixed-symmetry states in the $N \approx 50$ and
$N \approx 82$ region, by providing beams of high enough luminosity.

To sum up, in the present review we have shown that (i) the $1^+$
orbital scissors mode at low excitation energy, (ii) the $1^+$
spin-flip mode at higher excitation energy, and, (iii) the $1^+$
component originating from the still higher-lying isovector quadrupole
mode are universal realizations of the nuclear many-body system. It
turns out that these modes are not specific to nuclei only. On the
contrary, the scissors mode also shows up in other many-body systems.
There is compelling evidence for $M1$ excitations in deformed metallic
clusters, in elliptically deformed quantum dots and in Bose-Einstein
condensates of superfluid nature, and we have stressed the fascinating
interplay between nuclear physics and other highly correlated many-body
systems.

\section*{Acknowledgments}
During all the years working on magnetic dipole excitations we have
benefitted much from discussions with many colleagues, in particular
(in alphabetical order) P.~von~Brentano, B.~A.~Brown, R.~Casten,
C.~De~Coster, A.~E.~L.~Dieperink, J.~Enders, A.~Faessler, D.~Frekers,
J.~Ginocchio, I.~Hamamoto, F.~Iachello, P.~Van~Isacker, J.~Jolie,
U.~Kneissl, K.~Langanke, E.~Lipparini, N.~Lo~Iudice, M.~Macfarlane,
G.~Mart\'{\i}nez-Pinedo, B.~Mottelson, E.~Moya~de~Guerra, T.~Otsuka,
F.~Palumbo, N.~Pietralla, A.~Poves, C.~Rangacharyulu, P.~Sarriguren,
O.~Scholten, J.~Speth, S.~Stringari, L.~Zamick, and A.~Zilges. One of
the authors (KH) thanks the ``FWO-Vlaanderen'' and the University of
Ghent, for financial support during the years this review has been
written. This research was also performed in the framework of the BriX
network (IAP P5/07, P6/23) funded by the 'InterUniversity Attraction
Policy Programme - Belgian State - Belgian Science Policy. Furthermore,
the work of the two other authors (PvNC and AR) has been supported
generously by the DFG under contract SFB 634.

\bibliographystyle{apsrmp}

\bibliography{M1_RMP}

\begin{thebibliography}{392}
\expandafter\ifx\csname natexlab\endcsname\relax\def\natexlab#1{#1}\fi
\expandafter\ifx\csname bibnamefont\endcsname\relax
  \def\bibnamefont#1{#1}\fi
\expandafter\ifx\csname bibfnamefont\endcsname\relax
  \def\bibfnamefont#1{#1}\fi
\expandafter\ifx\csname citenamefont\endcsname\relax
  \def\citenamefont#1{#1}\fi
\expandafter\ifx\csname url\endcsname\relax
  \def\url#1{\texttt{#1}}\fi
\expandafter\ifx\csname urlprefix\endcsname\relax\def\urlprefix{URL }\fi
\providecommand{\bibinfo}[2]{#2}
\providecommand{\eprint}[2][]{\url{#2}}

\bibitem[{\citenamefont{Abdelaziz and Elliott}(1987)}]{Abdelaziz:1987}
\bibinfo{author}{\bibnamefont{Abdelaziz}, \bibfnamefont{M.}}, and
  \bibinfo{author}{\bibfnamefont{J.~P.} \bibnamefont{Elliott}},
  \bibinfo{year}{1987}, \bibinfo{journal}{Phys. Lett. B}
  \textbf{\bibinfo{volume}{197}}, \bibinfo{pages}{505}.

\bibitem[{\citenamefont{Abdelaziz} \emph{et~al.}(1988)\citenamefont{Abdelaziz,
  Thompson, Elliott, and Evans}}]{Abdelaziz:1988}
\bibinfo{author}{\bibnamefont{Abdelaziz}, \bibfnamefont{M.}},
  \bibinfo{author}{\bibfnamefont{M.~J.} \bibnamefont{Thompson}},
  \bibinfo{author}{\bibfnamefont{J.~P.} \bibnamefont{Elliott}}, and
  \bibinfo{author}{\bibfnamefont{J.~A.} \bibnamefont{Evans}},
  \bibinfo{year}{1988}, \bibinfo{journal}{J. Phys. G}
  \textbf{\bibinfo{volume}{14}}, \bibinfo{pages}{219}.

\bibitem[{\citenamefont{Ahmad} \emph{et~al.}(1989)\citenamefont{Ahmad,
  Hamilton, Isacker, Ahmad, and Robinson}}]{Ahmad:1989}
\bibinfo{author}{\bibnamefont{Ahmad}, \bibfnamefont{S.~T.}},
  \bibinfo{author}{\bibfnamefont{W.~D.} \bibnamefont{Hamilton}},
  \bibinfo{author}{\bibfnamefont{P.~V.} \bibnamefont{Isacker}},
  \bibinfo{author}{\bibfnamefont{S.~A.} \bibnamefont{Ahmad}}, and
  \bibinfo{author}{\bibfnamefont{S.~J.} \bibnamefont{Robinson}},
  \bibinfo{year}{1989}, \bibinfo{journal}{J. Phys. G}
  \textbf{\bibinfo{volume}{15}}, \bibinfo{pages}{93}.

\bibitem[{\citenamefont{Ahn} \emph{et~al.}(2009)\citenamefont{Ahn, Coqard,
  Pietralla, Rainowski, Costin, Janssens, Lister, Carpenter, Zhu, and
  Heyde}}]{Ahn:2009}
\bibinfo{author}{\bibnamefont{Ahn}, \bibfnamefont{T.}},
  \bibinfo{author}{\bibfnamefont{M.}~\bibnamefont{Coqard}},
  \bibinfo{author}{\bibfnamefont{N.}~\bibnamefont{Pietralla}},
  \bibinfo{author}{\bibfnamefont{G.}~\bibnamefont{Rainowski}},
  \bibinfo{author}{\bibfnamefont{A.}~\bibnamefont{Costin}},
  \bibinfo{author}{\bibfnamefont{R.~F.~V.} \bibnamefont{Janssens}},
  \bibinfo{author}{\bibfnamefont{C.~J.} \bibnamefont{Lister}},
  \bibinfo{author}{\bibfnamefont{M.}~\bibnamefont{Carpenter}},
  \bibinfo{author}{\bibfnamefont{S.}~\bibnamefont{Zhu}}, and
  \bibinfo{author}{\bibfnamefont{K.}~\bibnamefont{Heyde}},
  \bibinfo{year}{2009}, \bibinfo{journal}{Phys. Lett. B}
  \textbf{\bibinfo{volume}{679}}, \bibinfo{pages}{19}.

\bibitem[{\citenamefont{Ahn} \emph{et~al.}(2007)\citenamefont{Ahn, Pietralla,
  Rainovski, Costin, Dusling, Li, Linnemann, and Pontillo}}]{Ahn:2007}
\bibinfo{author}{\bibnamefont{Ahn}, \bibfnamefont{T.}},
  \bibinfo{author}{\bibfnamefont{N.}~\bibnamefont{Pietralla}},
  \bibinfo{author}{\bibfnamefont{G.}~\bibnamefont{Rainovski}},
  \bibinfo{author}{\bibfnamefont{A.}~\bibnamefont{Costin}},
  \bibinfo{author}{\bibfnamefont{K.}~\bibnamefont{Dusling}},
  \bibinfo{author}{\bibfnamefont{T.~C.} \bibnamefont{Li}},
  \bibinfo{author}{\bibfnamefont{A.}~\bibnamefont{Linnemann}}, and
  \bibinfo{author}{\bibfnamefont{S.}~\bibnamefont{Pontillo}},
  \bibinfo{year}{2007}, \bibinfo{journal}{Phys. Rev. C}
  \textbf{\bibinfo{volume}{75}}, \bibinfo{eid}{014313}.

\bibitem[{\citenamefont{Alaga and Mottelson}(1955)}]{Alaga:1955}
\bibinfo{author}{\bibnamefont{Alaga}, \bibfnamefont{K.~A. A.~B., G.}}, and
  \bibinfo{author}{\bibfnamefont{B.}~\bibnamefont{Mottelson}},
  \bibinfo{year}{1955}, \bibinfo{journal}{Dan. Mat. Fys. Medd.}
  \textbf{\bibinfo{volume}{29}}(\bibinfo{number}{9}), \bibinfo{pages}{1}.

\bibitem[{\citenamefont{Alhassid}(2000)}]{Alhassid:2000}
\bibinfo{author}{\bibnamefont{Alhassid}, \bibfnamefont{Y.}},
  \bibinfo{year}{2000}, \bibinfo{journal}{Rev. Mod. Phys.}
  \textbf{\bibinfo{volume}{72}}, \bibinfo{pages}{895}.

\bibitem[{\citenamefont{Allaart} \emph{et~al.}(1988)\citenamefont{Allaart,
  Boeker, Bonsignori, Savoia, and Gambhir}}]{Allaart:1988}
\bibinfo{author}{\bibnamefont{Allaart}, \bibfnamefont{K.}},
  \bibinfo{author}{\bibfnamefont{E.}~\bibnamefont{Boeker}},
  \bibinfo{author}{\bibfnamefont{G.}~\bibnamefont{Bonsignori}},
  \bibinfo{author}{\bibfnamefont{M.}~\bibnamefont{Savoia}}, and
  \bibinfo{author}{\bibfnamefont{Y.~K.} \bibnamefont{Gambhir}},
  \bibinfo{year}{1988}, \bibinfo{journal}{Phys. Rep.}
  \textbf{\bibinfo{volume}{169}}, \bibinfo{pages}{209}.

\bibitem[{\citenamefont{Arima} \emph{et~al.}(1969)\citenamefont{Arima, Harvey,
  and Shimizu}}]{Arima:1969}
\bibinfo{author}{\bibnamefont{Arima}, \bibfnamefont{A.}},
  \bibinfo{author}{\bibfnamefont{M.}~\bibnamefont{Harvey}}, and
  \bibinfo{author}{\bibfnamefont{K.}~\bibnamefont{Shimizu}},
  \bibinfo{year}{1969}, \bibinfo{journal}{Phys. Lett. B}
  \textbf{\bibinfo{volume}{30}}, \bibinfo{pages}{517}.

\bibitem[{\citenamefont{Arima and Iachello}(1975{\natexlab{a}})}]{Arima:1975a}
\bibinfo{author}{\bibnamefont{Arima}, \bibfnamefont{A.}}, and
  \bibinfo{author}{\bibfnamefont{F.}~\bibnamefont{Iachello}},
  \bibinfo{year}{1975}{\natexlab{a}}, \bibinfo{journal}{Phys. Lett. B}
  \textbf{\bibinfo{volume}{57}}, \bibinfo{pages}{39}.

\bibitem[{\citenamefont{Arima and Iachello}(1975{\natexlab{b}})}]{Arima:1975b}
\bibinfo{author}{\bibnamefont{Arima}, \bibfnamefont{A.}}, and
  \bibinfo{author}{\bibfnamefont{F.}~\bibnamefont{Iachello}},
  \bibinfo{year}{1975}{\natexlab{b}}, \bibinfo{journal}{Phys. Rev. Lett.}
  \textbf{\bibinfo{volume}{35}}, \bibinfo{pages}{1069}.

\bibitem[{\citenamefont{Arima} \emph{et~al.}(1977)\citenamefont{Arima, Otsuka,
  Iachello, and Talmi}}]{Arima:1977}
\bibinfo{author}{\bibnamefont{Arima}, \bibfnamefont{A.}},
  \bibinfo{author}{\bibfnamefont{T.}~\bibnamefont{Otsuka}},
  \bibinfo{author}{\bibfnamefont{F.}~\bibnamefont{Iachello}}, and
  \bibinfo{author}{\bibfnamefont{I.}~\bibnamefont{Talmi}},
  \bibinfo{year}{1977}, \bibinfo{journal}{Phys. Lett. B}
  \textbf{\bibinfo{volume}{66}}, \bibinfo{pages}{205}.

\bibitem[{\citenamefont{Arima} \emph{et~al.}(1987)\citenamefont{Arima, Shimizu,
  Bentz, and Hyuga}}]{Arima:1987}
\bibinfo{author}{\bibnamefont{Arima}, \bibfnamefont{A.}},
  \bibinfo{author}{\bibfnamefont{K.}~\bibnamefont{Shimizu}},
  \bibinfo{author}{\bibfnamefont{W.}~\bibnamefont{Bentz}}, and
  \bibinfo{author}{\bibfnamefont{H.}~\bibnamefont{Hyuga}},
  \bibinfo{year}{1987}, \emph{\bibinfo{title}{Advances in Nuclear Physics}},
  volume~\bibinfo{volume}{18} (\bibinfo{publisher}{Plenum},
  \bibinfo{address}{New York}).

\bibitem[{\citenamefont{Pe\~{n}a Arteaga and Ring}(2007)}]{Arteaga:2007}
\bibinfo{author}{\bibnamefont{Pe\~{n}a Arteaga}, \bibfnamefont{D.}}, and
  \bibinfo{author}{\bibfnamefont{P.}~\bibnamefont{Ring}}, \bibinfo{year}{2007},
  \bibinfo{journal}{Prog. Part. Nucl. Phys.} \textbf{\bibinfo{volume}{59}},
  \bibinfo{pages}{314}.

\bibitem[{\citenamefont{Pe\~{n}a Arteaga and Ring}(2008)}]{Arteaga:2008}
\bibinfo{author}{\bibnamefont{Pe\~{n}a Arteaga}, \bibfnamefont{D.}}, and
  \bibinfo{author}{\bibfnamefont{P.}~\bibnamefont{Ring}}, \bibinfo{year}{2008},
  \bibinfo{journal}{Phys. Rev. C} \textbf{\bibinfo{volume}{77}},
  \bibinfo{pages}{034317}.

\bibitem[{\citenamefont{Ashcroft and Mermin}(1976)}]{Ashcroft:1976}
\bibinfo{author}{\bibnamefont{Ashcroft}, \bibfnamefont{E.~W.}}, and
  \bibinfo{author}{\bibfnamefont{N.~D.} \bibnamefont{Mermin}},
  \bibinfo{year}{1976}, \emph{\bibinfo{title}{Solid State Physcis}}
  (\bibinfo{publisher}{Saunders College}, \bibinfo{address}{Philadelphia}).

\bibitem[{\citenamefont{Auerbach}(1983)}]{Auerbach:1983}
\bibinfo{author}{\bibnamefont{Auerbach}, \bibfnamefont{N.}},
  \bibinfo{year}{1983}, \bibinfo{journal}{Phys. Rep.}
  \textbf{\bibinfo{volume}{98}}, \bibinfo{pages}{273}.

\bibitem[{\citenamefont{Austing} \emph{et~al.}(1999)\citenamefont{Austing,
  Sasaki, Tarucha, Reimann, Koskinen, and Manninen}}]{Austing:1999}
\bibinfo{author}{\bibnamefont{Austing}, \bibfnamefont{D.~G.}},
  \bibinfo{author}{\bibfnamefont{S.}~\bibnamefont{Sasaki}},
  \bibinfo{author}{\bibfnamefont{S.}~\bibnamefont{Tarucha}},
  \bibinfo{author}{\bibfnamefont{S.~M.} \bibnamefont{Reimann}},
  \bibinfo{author}{\bibfnamefont{M.}~\bibnamefont{Koskinen}}, and
  \bibinfo{author}{\bibfnamefont{M.}~\bibnamefont{Manninen}},
  \bibinfo{year}{1999}, \bibinfo{journal}{Phys. Rev. B}
  \textbf{\bibinfo{volume}{60}}, \bibinfo{pages}{11514}.

\bibitem[{\citenamefont{Balbutsev and Schuck}(2004)}]{Balbutsev:2004}
\bibinfo{author}{\bibnamefont{Balbutsev}, \bibfnamefont{E.~B.}}, and
  \bibinfo{author}{\bibfnamefont{P.}~\bibnamefont{Schuck}},
  \bibinfo{year}{2004}, \bibinfo{journal}{Nucl. Phys. A}
  \textbf{\bibinfo{volume}{731}}, \bibinfo{pages}{256}.

\bibitem[{\citenamefont{Balbutsev and Schuck}(2007)}]{Balbutsev:2007}
\bibinfo{author}{\bibnamefont{Balbutsev}, \bibfnamefont{E.~B.}}, and
  \bibinfo{author}{\bibfnamefont{P.}~\bibnamefont{Schuck}},
  \bibinfo{year}{2007}, \bibinfo{journal}{Ann. Phys.}
  \textbf{\bibinfo{volume}{322}}, \bibinfo{pages}{489}.

\bibitem[{\citenamefont{Bandyopadhyay}
  \emph{et~al.}(2003)\citenamefont{Bandyopadhyay, Reynolds, Fransen,
  Boukharouba, McEllistrem, and Yates}}]{Bandyopadhyay:2003}
\bibinfo{author}{\bibnamefont{Bandyopadhyay}, \bibfnamefont{D.}},
  \bibinfo{author}{\bibfnamefont{C.~C.} \bibnamefont{Reynolds}},
  \bibinfo{author}{\bibfnamefont{C.}~\bibnamefont{Fransen}},
  \bibinfo{author}{\bibfnamefont{N.}~\bibnamefont{Boukharouba}},
  \bibinfo{author}{\bibfnamefont{M.~T.} \bibnamefont{McEllistrem}}, and
  \bibinfo{author}{\bibfnamefont{S.~W.} \bibnamefont{Yates}},
  \bibinfo{year}{2003}, \bibinfo{journal}{Phys. Rev. C}
  \textbf{\bibinfo{volume}{67}}, \bibinfo{pages}{034319}.

\bibitem[{\citenamefont{Bauske} \emph{et~al.}(1993)\citenamefont{Bauske, Arias,
  von Brentano, Frank, Friedrichs, Heil, Herzberg, Hoyler, Van~Isacker,
  Kneissl, Margraf, Pitz} \emph{et~al.}}]{Bauske:1993}
\bibinfo{author}{\bibnamefont{Bauske}, \bibfnamefont{I.}},
  \bibinfo{author}{\bibfnamefont{J.~M.} \bibnamefont{Arias}},
  \bibinfo{author}{\bibfnamefont{P.}~\bibnamefont{von Brentano}},
  \bibinfo{author}{\bibfnamefont{A.}~\bibnamefont{Frank}},
  \bibinfo{author}{\bibfnamefont{H.}~\bibnamefont{Friedrichs}},
  \bibinfo{author}{\bibfnamefont{R.~D.} \bibnamefont{Heil}},
  \bibinfo{author}{\bibfnamefont{R.-D.} \bibnamefont{Herzberg}},
  \bibinfo{author}{\bibfnamefont{F.}~\bibnamefont{Hoyler}},
  \bibinfo{author}{\bibfnamefont{P.}~\bibnamefont{Van~Isacker}},
  \bibinfo{author}{\bibfnamefont{U.}~\bibnamefont{Kneissl}},
  \bibinfo{author}{\bibfnamefont{J.}~\bibnamefont{Margraf}},
  \bibinfo{author}{\bibfnamefont{H.~H.} \bibnamefont{Pitz}}, \emph{et~al.},
  \bibinfo{year}{1993}, \bibinfo{journal}{Phys. Rev. Lett.}
  \textbf{\bibinfo{volume}{71}}, \bibinfo{pages}{975}.

\bibitem[{\citenamefont{Baznat and Pyatov}(1975)}]{Baznat:1975}
\bibinfo{author}{\bibnamefont{Baznat}, \bibfnamefont{M.~I.}}, and
  \bibinfo{author}{\bibfnamefont{N.~Y.} \bibnamefont{Pyatov}},
  \bibinfo{year}{1975}, \bibinfo{journal}{Sov. J. Nucl. Phys.}
  \textbf{\bibinfo{volume}{21}}, \bibinfo{pages}{365}.

\bibitem[{\citenamefont{Berg}(1984)}]{Berg:1984a}
\bibinfo{author}{\bibnamefont{Berg}, \bibfnamefont{U.~E.~P.}},
  \bibinfo{year}{1984}, \bibinfo{journal}{J. Phys. (France) C}
  \textbf{\bibinfo{volume}{4}}, \bibinfo{pages}{359}.

\bibitem[{\citenamefont{Berg}
  \emph{et~al.}(1984{\natexlab{a}})\citenamefont{Berg, Ackermann, Bangert,
  Blasing, Naatz, Stock, Wienhard, Brussel, Chapuran, and
  Wildenthal}}]{Berg:1984b}
\bibinfo{author}{\bibnamefont{Berg}, \bibfnamefont{U.~E.~P.}},
  \bibinfo{author}{\bibfnamefont{K.}~\bibnamefont{Ackermann}},
  \bibinfo{author}{\bibfnamefont{K.}~\bibnamefont{Bangert}},
  \bibinfo{author}{\bibfnamefont{C.}~\bibnamefont{Blasing}},
  \bibinfo{author}{\bibfnamefont{W.}~\bibnamefont{Naatz}},
  \bibinfo{author}{\bibfnamefont{R.}~\bibnamefont{Stock}},
  \bibinfo{author}{\bibfnamefont{K.}~\bibnamefont{Wienhard}},
  \bibinfo{author}{\bibfnamefont{M.~K.} \bibnamefont{Brussel}},
  \bibinfo{author}{\bibfnamefont{T.~E.} \bibnamefont{Chapuran}}, and
  \bibinfo{author}{\bibfnamefont{B.~H.} \bibnamefont{Wildenthal}},
  \bibinfo{year}{1984}{\natexlab{a}}, \bibinfo{journal}{Phys. Lett. B}
  \textbf{\bibinfo{volume}{140}}, \bibinfo{pages}{191}.

\bibitem[{\citenamefont{Berg}
  \emph{et~al.}(1984{\natexlab{b}})\citenamefont{Berg, Blasing, Drexler, Heil,
  Kneissl, Naatz, Ratzek, Schennach, Stock, Weber, Wickert, Fischer}
  \emph{et~al.}}]{Berg:1984c}
\bibinfo{author}{\bibnamefont{Berg}, \bibfnamefont{U.~E.~P.}},
  \bibinfo{author}{\bibfnamefont{C.}~\bibnamefont{Blasing}},
  \bibinfo{author}{\bibfnamefont{J.}~\bibnamefont{Drexler}},
  \bibinfo{author}{\bibfnamefont{R.~D.} \bibnamefont{Heil}},
  \bibinfo{author}{\bibfnamefont{U.}~\bibnamefont{Kneissl}},
  \bibinfo{author}{\bibfnamefont{W.}~\bibnamefont{Naatz}},
  \bibinfo{author}{\bibfnamefont{R.}~\bibnamefont{Ratzek}},
  \bibinfo{author}{\bibfnamefont{S.}~\bibnamefont{Schennach}},
  \bibinfo{author}{\bibfnamefont{R.}~\bibnamefont{Stock}},
  \bibinfo{author}{\bibfnamefont{T.}~\bibnamefont{Weber}},
  \bibinfo{author}{\bibfnamefont{H.}~\bibnamefont{Wickert}},
  \bibinfo{author}{\bibfnamefont{B.}~\bibnamefont{Fischer}}, \emph{et~al.},
  \bibinfo{year}{1984}{\natexlab{b}}, \bibinfo{journal}{Phys. Lett. B}
  \textbf{\bibinfo{volume}{149}}, \bibinfo{pages}{59}.

\bibitem[{\citenamefont{Berg and Kneissl}(1987)}]{Berg:1987}
\bibinfo{author}{\bibnamefont{Berg}, \bibfnamefont{U.~E.~P.}}, and
  \bibinfo{author}{\bibfnamefont{U.}~\bibnamefont{Kneissl}},
  \bibinfo{year}{1987}, \bibinfo{journal}{Annu. Rev. Nucl. Part. Sci.}
  \textbf{\bibinfo{volume}{37}}, \bibinfo{pages}{33}.

\bibitem[{\citenamefont{Bertsch and Hamamoto}(1982)}]{Bertsch:1982}
\bibinfo{author}{\bibnamefont{Bertsch}, \bibfnamefont{G.~F.}}, and
  \bibinfo{author}{\bibfnamefont{I.}~\bibnamefont{Hamamoto}},
  \bibinfo{year}{1982}, \bibinfo{journal}{Phys. Rev. C}
  \textbf{\bibinfo{volume}{26}}, \bibinfo{pages}{1323}.

\bibitem[{\citenamefont{Bes and Broglia}(1984)}]{Bes:1984}
\bibinfo{author}{\bibnamefont{Bes}, \bibfnamefont{D.~R.}}, and
  \bibinfo{author}{\bibfnamefont{R.~A.} \bibnamefont{Broglia}},
  \bibinfo{year}{1984}, \bibinfo{journal}{Phys. Lett. B}
  \textbf{\bibinfo{volume}{137}}, \bibinfo{pages}{141}.

\bibitem[{\citenamefont{Bettermann}
  \emph{et~al.}(2009)\citenamefont{Bettermann, Fransen, Heinze, Jolie,
  Linnemann, M\"ucher, Rother, Ahn, Costin, Pietralla, and
  Luo}}]{Betterman:2009}
\bibinfo{author}{\bibnamefont{Bettermann}, \bibfnamefont{L.}},
  \bibinfo{author}{\bibfnamefont{C.}~\bibnamefont{Fransen}},
  \bibinfo{author}{\bibfnamefont{S.}~\bibnamefont{Heinze}},
  \bibinfo{author}{\bibfnamefont{J.}~\bibnamefont{Jolie}},
  \bibinfo{author}{\bibfnamefont{A.}~\bibnamefont{Linnemann}},
  \bibinfo{author}{\bibfnamefont{D.}~\bibnamefont{M\"ucher}},
  \bibinfo{author}{\bibfnamefont{W.}~\bibnamefont{Rother}},
  \bibinfo{author}{\bibfnamefont{T.}~\bibnamefont{Ahn}},
  \bibinfo{author}{\bibfnamefont{A.}~\bibnamefont{Costin}},
  \bibinfo{author}{\bibfnamefont{N.}~\bibnamefont{Pietralla}}, and
  \bibinfo{author}{\bibfnamefont{Y.}~\bibnamefont{Luo}}, \bibinfo{year}{2009},
  \bibinfo{journal}{Phys. Rev. C} \textbf{\bibinfo{volume}{79}},
  \bibinfo{pages}{034315}.

\bibitem[{\citenamefont{Beuschel} \emph{et~al.}(1998)\citenamefont{Beuschel,
  Draayer, Rompf, and Hirsch}}]{Beuschel:1998}
\bibinfo{author}{\bibnamefont{Beuschel}, \bibfnamefont{T.}},
  \bibinfo{author}{\bibfnamefont{J.~P.} \bibnamefont{Draayer}},
  \bibinfo{author}{\bibfnamefont{D.}~\bibnamefont{Rompf}}, and
  \bibinfo{author}{\bibfnamefont{J.~G.} \bibnamefont{Hirsch}},
  \bibinfo{year}{1998}, \bibinfo{journal}{Phys. Rev. C}
  \textbf{\bibinfo{volume}{57}}, \bibinfo{pages}{1233}.

\bibitem[{\citenamefont{Beuschel} \emph{et~al.}(2000)\citenamefont{Beuschel,
  Hirsch, and Draayer}}]{Beuschel:2000}
\bibinfo{author}{\bibnamefont{Beuschel}, \bibfnamefont{T.}},
  \bibinfo{author}{\bibfnamefont{J.~G.} \bibnamefont{Hirsch}}, and
  \bibinfo{author}{\bibfnamefont{J.~P.} \bibnamefont{Draayer}},
  \bibinfo{year}{2000}, \bibinfo{journal}{Phys. Rev. C}
  \textbf{\bibinfo{volume}{61}}, \bibinfo{pages}{054307}.

\bibitem[{\citenamefont{Bj\o{}rnholm}
  \emph{et~al.}(1990)\citenamefont{Bj\o{}rnholm, Borggreen, Echt, Hansen,
  Pedersen, and Rasmussen}}]{Bjornholm:1990}
\bibinfo{author}{\bibnamefont{Bj\o{}rnholm}, \bibfnamefont{S.}},
  \bibinfo{author}{\bibfnamefont{J.}~\bibnamefont{Borggreen}},
  \bibinfo{author}{\bibfnamefont{O.}~\bibnamefont{Echt}},
  \bibinfo{author}{\bibfnamefont{K.}~\bibnamefont{Hansen}},
  \bibinfo{author}{\bibfnamefont{J.}~\bibnamefont{Pedersen}}, and
  \bibinfo{author}{\bibfnamefont{H.~D.} \bibnamefont{Rasmussen}},
  \bibinfo{year}{1990}, \bibinfo{journal}{Phys. Rev. Lett.}
  \textbf{\bibinfo{volume}{65}}, \bibinfo{pages}{1627}.

\bibitem[{\citenamefont{Bj\o{}rnholm}
  \emph{et~al.}(1992)\citenamefont{Bj\o{}rnholm, Borggreen, Hansen, Martin,
  Rasmussen, and Pedersen}}]{Bjornholm:1992}
\bibinfo{author}{\bibnamefont{Bj\o{}rnholm}, \bibfnamefont{S.}},
  \bibinfo{author}{\bibfnamefont{J.}~\bibnamefont{Borggreen}},
  \bibinfo{author}{\bibfnamefont{K.}~\bibnamefont{Hansen}},
  \bibinfo{author}{\bibfnamefont{T.}~\bibnamefont{Martin}},
  \bibinfo{author}{\bibfnamefont{H.~D.} \bibnamefont{Rasmussen}}, and
  \bibinfo{author}{\bibfnamefont{J.}~\bibnamefont{Pedersen}},
  \bibinfo{year}{1992}, in \emph{\bibinfo{booktitle}{Ecole Internationale
  Joliot-Curie de Physique Nucleaire}}, edited by
  \bibinfo{editor}{\bibfnamefont{Y.}~\bibnamefont{Abgrall}}
  (\bibinfo{publisher}{IN2P3}), p. \bibinfo{pages}{311}.

\bibitem[{\citenamefont{Boelaert}
  \emph{et~al.}(2007{\natexlab{a}})\citenamefont{Boelaert, Dewald, Fransen,
  Jolie, Linnemann, Melon, M\"{o}ller, Smirnova, and Heyde}}]{Boelaert:2007a}
\bibinfo{author}{\bibnamefont{Boelaert}, \bibfnamefont{N.}},
  \bibinfo{author}{\bibfnamefont{A.}~\bibnamefont{Dewald}},
  \bibinfo{author}{\bibfnamefont{C.}~\bibnamefont{Fransen}},
  \bibinfo{author}{\bibfnamefont{J.}~\bibnamefont{Jolie}},
  \bibinfo{author}{\bibfnamefont{A.}~\bibnamefont{Linnemann}},
  \bibinfo{author}{\bibfnamefont{B.}~\bibnamefont{Melon}},
  \bibinfo{author}{\bibfnamefont{O.}~\bibnamefont{M\"{o}ller}},
  \bibinfo{author}{\bibfnamefont{N.}~\bibnamefont{Smirnova}}, and
  \bibinfo{author}{\bibfnamefont{K.}~\bibnamefont{Heyde}},
  \bibinfo{year}{2007}{\natexlab{a}}, \bibinfo{journal}{Phys. Rev. C}
  \textbf{\bibinfo{volume}{75}}, \bibinfo{eid}{054311}.

\bibitem[{\citenamefont{Boelaert}
  \emph{et~al.}(2007{\natexlab{b}})\citenamefont{Boelaert, Smirnova, Heyde, and
  Jolie}}]{Boelaert:2007b}
\bibinfo{author}{\bibnamefont{Boelaert}, \bibfnamefont{N.}},
  \bibinfo{author}{\bibfnamefont{N.}~\bibnamefont{Smirnova}},
  \bibinfo{author}{\bibfnamefont{K.}~\bibnamefont{Heyde}}, and
  \bibinfo{author}{\bibfnamefont{J.}~\bibnamefont{Jolie}},
  \bibinfo{year}{2007}{\natexlab{b}}, \bibinfo{journal}{Phys. Rev. C}
  \textbf{\bibinfo{volume}{75}}, \bibinfo{eid}{014316}.

\bibitem[{\citenamefont{Bohigas}(1991)}]{Bohigas:1989}
\bibinfo{author}{\bibnamefont{Bohigas}, \bibfnamefont{O.}},
  \bibinfo{year}{1991}, in \emph{\bibinfo{booktitle}{Les Houches Summer School
  on Theoretical Physics, Session L II: Chaos and Quantum Physics}}, edited by
  \bibinfo{editor}{\bibfnamefont{M.~J.} \bibnamefont{Giannoni}},
  \bibinfo{editor}{\bibfnamefont{A.}~\bibnamefont{Voros}}, and
  \bibinfo{editor}{\bibfnamefont{J.}~\bibnamefont{Zinn-Justin}}
  (\bibinfo{publisher}{North-Holland}, \bibinfo{address}{Amsterdam}),
  p.~\bibinfo{pages}{87}.

\bibitem[{\citenamefont{Bohle}
  \emph{et~al.}(1987{\natexlab{a}})\citenamefont{Bohle, Kilgus, Richter,
  de~Jager, and de~Vries}}]{Bohle:1987a}
\bibinfo{author}{\bibnamefont{Bohle}, \bibfnamefont{D.}},
  \bibinfo{author}{\bibfnamefont{G.}~\bibnamefont{Kilgus}},
  \bibinfo{author}{\bibfnamefont{A.}~\bibnamefont{Richter}},
  \bibinfo{author}{\bibfnamefont{C.~W.} \bibnamefont{de~Jager}}, and
  \bibinfo{author}{\bibfnamefont{H.}~\bibnamefont{de~Vries}},
  \bibinfo{year}{1987}{\natexlab{a}}, \bibinfo{journal}{Phys. Lett. B}
  \textbf{\bibinfo{volume}{195}}, \bibinfo{pages}{326}.

\bibitem[{\citenamefont{Bohle}
  \emph{et~al.}(1984{\natexlab{a}})\citenamefont{Bohle, K\"uchler, Richter, and
  Steffen}}]{Bohle:1984b}
\bibinfo{author}{\bibnamefont{Bohle}, \bibfnamefont{D.}},
  \bibinfo{author}{\bibfnamefont{G.}~\bibnamefont{K\"uchler}},
  \bibinfo{author}{\bibfnamefont{A.}~\bibnamefont{Richter}}, and
  \bibinfo{author}{\bibfnamefont{W.}~\bibnamefont{Steffen}},
  \bibinfo{year}{1984}{\natexlab{a}}, \bibinfo{journal}{Phys. Lett. B}
  \textbf{\bibinfo{volume}{148}}, \bibinfo{pages}{260}.

\bibitem[{\citenamefont{Bohle} \emph{et~al.}(1986)\citenamefont{Bohle, Richter,
  Berg, Drexler, Heil, Kneissl, Metzger, Stock, Fischer, Hollick, and
  Kollewe}}]{Bohle:1986}
\bibinfo{author}{\bibnamefont{Bohle}, \bibfnamefont{D.}},
  \bibinfo{author}{\bibfnamefont{A.}~\bibnamefont{Richter}},
  \bibinfo{author}{\bibfnamefont{U.~E.~P.} \bibnamefont{Berg}},
  \bibinfo{author}{\bibfnamefont{J.}~\bibnamefont{Drexler}},
  \bibinfo{author}{\bibfnamefont{R.~D.} \bibnamefont{Heil}},
  \bibinfo{author}{\bibfnamefont{U.}~\bibnamefont{Kneissl}},
  \bibinfo{author}{\bibfnamefont{H.}~\bibnamefont{Metzger}},
  \bibinfo{author}{\bibfnamefont{R.}~\bibnamefont{Stock}},
  \bibinfo{author}{\bibfnamefont{B.}~\bibnamefont{Fischer}},
  \bibinfo{author}{\bibfnamefont{H.}~\bibnamefont{Hollick}}, and
  \bibinfo{author}{\bibfnamefont{D.}~\bibnamefont{Kollewe}},
  \bibinfo{year}{1986}, \bibinfo{journal}{Nucl. Phys. A}
  \textbf{\bibinfo{volume}{458}}, \bibinfo{pages}{205}.

\bibitem[{\citenamefont{Bohle}
  \emph{et~al.}(1987{\natexlab{b}})\citenamefont{Bohle, Richter, de~Jager, and
  de~Vries}}]{Bohle:1987b}
\bibinfo{author}{\bibnamefont{Bohle}, \bibfnamefont{D.}},
  \bibinfo{author}{\bibfnamefont{A.}~\bibnamefont{Richter}},
  \bibinfo{author}{\bibfnamefont{C.~W.} \bibnamefont{de~Jager}}, and
  \bibinfo{author}{\bibfnamefont{H.}~\bibnamefont{de~Vries}},
  \bibinfo{year}{1987}{\natexlab{b}}, \bibinfo{journal}{Z. Phys. A}
  \textbf{\bibinfo{volume}{328}}, \bibinfo{pages}{463}.

\bibitem[{\citenamefont{Bohle}
  \emph{et~al.}(1984{\natexlab{b}})\citenamefont{Bohle, Richter, Steffen,
  Dieperink, Iudice, Palumbo, and Scholten}}]{Bohle:1984a}
\bibinfo{author}{\bibnamefont{Bohle}, \bibfnamefont{D.}},
  \bibinfo{author}{\bibfnamefont{A.}~\bibnamefont{Richter}},
  \bibinfo{author}{\bibfnamefont{W.}~\bibnamefont{Steffen}},
  \bibinfo{author}{\bibfnamefont{A.~E.~L.} \bibnamefont{Dieperink}},
  \bibinfo{author}{\bibfnamefont{N.~L.} \bibnamefont{Iudice}},
  \bibinfo{author}{\bibfnamefont{F.}~\bibnamefont{Palumbo}}, and
  \bibinfo{author}{\bibfnamefont{O.}~\bibnamefont{Scholten}},
  \bibinfo{year}{1984}{\natexlab{b}}, \bibinfo{journal}{Phys. Lett. B}
  \textbf{\bibinfo{volume}{137}}, \bibinfo{pages}{27}.

\bibitem[{\citenamefont{Bohr and Mottelson}(1969)}]{Bohr:1969}
\bibinfo{author}{\bibnamefont{Bohr}, \bibfnamefont{A.}}, and
  \bibinfo{author}{\bibfnamefont{B.}~\bibnamefont{Mottelson}},
  \bibinfo{year}{1969}, \emph{\bibinfo{title}{Nuclear Structure}},
  volume~\bibinfo{volume}{1} (\bibinfo{publisher}{Benjamin},
  \bibinfo{address}{New York}).

\bibitem[{\citenamefont{Bohr and Mottelson}(1975)}]{Bohr:1975}
\bibinfo{author}{\bibnamefont{Bohr}, \bibfnamefont{A.}}, and
  \bibinfo{author}{\bibfnamefont{B.}~\bibnamefont{Mottelson}},
  \bibinfo{year}{1975}, \emph{\bibinfo{title}{Nuclear Structure}},
  volume~\bibinfo{volume}{2} (\bibinfo{publisher}{Benjamin},
  \bibinfo{address}{New York}).

\bibitem[{\citenamefont{Brack}(1992)}]{Brack:1992}
\bibinfo{author}{\bibnamefont{Brack}, \bibfnamefont{M.}}, \bibinfo{year}{1992},
  in \emph{\bibinfo{booktitle}{Ecole Internationale Joliot-Curie de Physique
  Nucleaire}}, edited by
  \bibinfo{editor}{\bibfnamefont{Y.}~\bibnamefont{Abgrall}}
  (\bibinfo{publisher}{IN2P3}), p. \bibinfo{pages}{325}.

\bibitem[{\citenamefont{Brack}
  \emph{et~al.}(1991{\natexlab{a}})\citenamefont{Brack, Henzken, and
  Hanzen}}]{Brack:1991a}
\bibinfo{author}{\bibnamefont{Brack}, \bibfnamefont{M.}},
  \bibinfo{author}{\bibfnamefont{O.}~\bibnamefont{Henzken}}, and
  \bibinfo{author}{\bibfnamefont{K.}~\bibnamefont{Hanzen}},
  \bibinfo{year}{1991}{\natexlab{a}}, \bibinfo{journal}{Z. Phys. D}
  \textbf{\bibinfo{volume}{19}}, \bibinfo{pages}{51}.

\bibitem[{\citenamefont{Brack}
  \emph{et~al.}(1991{\natexlab{b}})\citenamefont{Brack, Henzken, and
  Hanzen}}]{Brack:1991b}
\bibinfo{author}{\bibnamefont{Brack}, \bibfnamefont{M.}},
  \bibinfo{author}{\bibfnamefont{O.}~\bibnamefont{Henzken}}, and
  \bibinfo{author}{\bibfnamefont{K.}~\bibnamefont{Hanzen}},
  \bibinfo{year}{1991}{\natexlab{b}}, \bibinfo{journal}{Z. Phys. D}
  \textbf{\bibinfo{volume}{21}}, \bibinfo{pages}{65}.

\bibitem[{\citenamefont{von Brentano} \emph{et~al.}(1996)\citenamefont{von
  Brentano, Eberth, Enders, Esser, Herzberg, Huxel, Meise, von Neumann-Cosel,
  Nicolay, Pietralla, Prade, Reif} \emph{et~al.}}]{vonBrentano:1996}
\bibinfo{author}{\bibnamefont{von Brentano}, \bibfnamefont{P.}},
  \bibinfo{author}{\bibfnamefont{J.}~\bibnamefont{Eberth}},
  \bibinfo{author}{\bibfnamefont{J.}~\bibnamefont{Enders}},
  \bibinfo{author}{\bibfnamefont{L.}~\bibnamefont{Esser}},
  \bibinfo{author}{\bibfnamefont{R.-D.} \bibnamefont{Herzberg}},
  \bibinfo{author}{\bibfnamefont{N.}~\bibnamefont{Huxel}},
  \bibinfo{author}{\bibfnamefont{H.}~\bibnamefont{Meise}},
  \bibinfo{author}{\bibfnamefont{P.}~\bibnamefont{von Neumann-Cosel}},
  \bibinfo{author}{\bibfnamefont{N.}~\bibnamefont{Nicolay}},
  \bibinfo{author}{\bibfnamefont{N.}~\bibnamefont{Pietralla}},
  \bibinfo{author}{\bibfnamefont{H.}~\bibnamefont{Prade}},
  \bibinfo{author}{\bibfnamefont{J.}~\bibnamefont{Reif}}, \emph{et~al.},
  \bibinfo{year}{1996}, \bibinfo{journal}{Phys. Rev. Lett.}
  \textbf{\bibinfo{volume}{76}}, \bibinfo{pages}{2029}.

\bibitem[{\citenamefont{von Brentano} \emph{et~al.}(1994)\citenamefont{von
  Brentano, Zilges, Zamfir, and Herzberg}}]{vonBrentano:1994a}
\bibinfo{author}{\bibnamefont{von Brentano}, \bibfnamefont{P.}},
  \bibinfo{author}{\bibfnamefont{A.}~\bibnamefont{Zilges}},
  \bibinfo{author}{\bibfnamefont{N.~V.} \bibnamefont{Zamfir}}, and
  \bibinfo{author}{\bibfnamefont{R.-D.} \bibnamefont{Herzberg}},
  \bibinfo{year}{1994}, in \emph{\bibinfo{booktitle}{Symmetries in Science
  VII}}, edited by \bibinfo{editor}{\bibfnamefont{B.}~\bibnamefont{Gruku}} and
  \bibinfo{editor}{\bibfnamefont{T.}~\bibnamefont{Otsuka}}
  (\bibinfo{publisher}{Plenum Press}, \bibinfo{address}{New York}), p.
  \bibinfo{pages}{123}.

\bibitem[{\citenamefont{Brown and Wildenthal}(1987)}]{Brown:1987}
\bibinfo{author}{\bibnamefont{Brown}, \bibfnamefont{B.~A.}}, and
  \bibinfo{author}{\bibfnamefont{B.~H.} \bibnamefont{Wildenthal}},
  \bibinfo{year}{1987}, \bibinfo{journal}{Nucl. Phys. A}
  \textbf{\bibinfo{volume}{474}}, \bibinfo{pages}{290}.

\bibitem[{\citenamefont{Brown and Wildenthal}(1988)}]{Brown:1988}
\bibinfo{author}{\bibnamefont{Brown}, \bibfnamefont{B.~A.}}, and
  \bibinfo{author}{\bibfnamefont{B.~H.} \bibnamefont{Wildenthal}},
  \bibinfo{year}{1988}, \bibinfo{journal}{Annu. Rev. Nucl. Part. Sci.}
  \textbf{\bibinfo{volume}{38}}, \bibinfo{pages}{29}.

\bibitem[{\citenamefont{Brown and Bolsterli}(1959)}]{Brown:1959}
\bibinfo{author}{\bibnamefont{Brown}, \bibfnamefont{G.~E.}}, and
  \bibinfo{author}{\bibfnamefont{M.}~\bibnamefont{Bolsterli}},
  \bibinfo{year}{1959}, \bibinfo{journal}{Phys. Rev. Lett.}
  \textbf{\bibinfo{volume}{3}}, \bibinfo{pages}{472}.

\bibitem[{\citenamefont{Brussaard and Glaudemans}(1977)}]{Brussaard:1977}
\bibinfo{author}{\bibnamefont{Brussaard}, \bibfnamefont{P.~J.}}, and
  \bibinfo{author}{\bibfnamefont{P.~W.~M.} \bibnamefont{Glaudemans}},
  \bibinfo{year}{1977}, \emph{\bibinfo{title}{Shell-model Applications in
  Nuclear Spectroscopy}} (\bibinfo{publisher}{North-Holland},
  \bibinfo{address}{Amsterdam}).

\bibitem[{\citenamefont{Burda} \emph{et~al.}(2007)\citenamefont{Burda, Botha,
  Carter, Fearick, F\"{o}rtsch, Fransen, Fujita, Holt, Kuhar, Lenhardt, von
  Neumann-Cosel, Neveling} \emph{et~al.}}]{Burda:2007}
\bibinfo{author}{\bibnamefont{Burda}, \bibfnamefont{O.}},
  \bibinfo{author}{\bibfnamefont{N.}~\bibnamefont{Botha}},
  \bibinfo{author}{\bibfnamefont{J.}~\bibnamefont{Carter}},
  \bibinfo{author}{\bibfnamefont{R.~W.} \bibnamefont{Fearick}},
  \bibinfo{author}{\bibfnamefont{S.~V.} \bibnamefont{F\"{o}rtsch}},
  \bibinfo{author}{\bibfnamefont{C.}~\bibnamefont{Fransen}},
  \bibinfo{author}{\bibfnamefont{H.}~\bibnamefont{Fujita}},
  \bibinfo{author}{\bibfnamefont{J.~D.} \bibnamefont{Holt}},
  \bibinfo{author}{\bibfnamefont{M.}~\bibnamefont{Kuhar}},
  \bibinfo{author}{\bibfnamefont{A.}~\bibnamefont{Lenhardt}},
  \bibinfo{author}{\bibfnamefont{P.}~\bibnamefont{von Neumann-Cosel}},
  \bibinfo{author}{\bibfnamefont{R.}~\bibnamefont{Neveling}}, \emph{et~al.},
  \bibinfo{year}{2007}, \bibinfo{journal}{Phys. Rev. Lett.}
  \textbf{\bibinfo{volume}{99}}, \bibinfo{eid}{092503}.

\bibitem[{\citenamefont{Caprio and Iachello}(2004)}]{Caprio:2004}
\bibinfo{author}{\bibnamefont{Caprio}, \bibfnamefont{M.~A.}}, and
  \bibinfo{author}{\bibfnamefont{F.}~\bibnamefont{Iachello}},
  \bibinfo{year}{2004}, \bibinfo{journal}{Phys. Rev. Lett.}
  \textbf{\bibinfo{volume}{93}}, \bibinfo{pages}{242502}.

\bibitem[{\citenamefont{Casten} \emph{et~al.}(1987)\citenamefont{Casten,
  Brenner, and Haustein}}]{Casten:1987}
\bibinfo{author}{\bibnamefont{Casten}, \bibfnamefont{R.~F.}},
  \bibinfo{author}{\bibfnamefont{D.~S.} \bibnamefont{Brenner}}, and
  \bibinfo{author}{\bibfnamefont{P.~E.} \bibnamefont{Haustein}},
  \bibinfo{year}{1987}, \bibinfo{journal}{Phys. Rev. Lett.}
  \textbf{\bibinfo{volume}{58}}, \bibinfo{pages}{658}.

\bibitem[{\citenamefont{Casten} \emph{et~al.}(1988)\citenamefont{Casten, Heyde,
  and Wolf}}]{Casten:1988}
\bibinfo{author}{\bibnamefont{Casten}, \bibfnamefont{R.~F.}},
  \bibinfo{author}{\bibfnamefont{K.}~\bibnamefont{Heyde}}, and
  \bibinfo{author}{\bibfnamefont{A.}~\bibnamefont{Wolf}}, \bibinfo{year}{1988},
  \bibinfo{journal}{Phys. Lett. B} \textbf{\bibinfo{volume}{208}},
  \bibinfo{pages}{33}.

\bibitem[{\citenamefont{Caurier} \emph{et~al.}(2001)\citenamefont{Caurier,
  Langanke, Mart\'{\i}nez-Pinedo, Nowacki, and Vogel}}]{Caurier:2001}
\bibinfo{author}{\bibnamefont{Caurier}, \bibfnamefont{E.}},
  \bibinfo{author}{\bibfnamefont{K.}~\bibnamefont{Langanke}},
  \bibinfo{author}{\bibfnamefont{G.}~\bibnamefont{Mart\'{\i}nez-Pinedo}},
  \bibinfo{author}{\bibfnamefont{F.}~\bibnamefont{Nowacki}}, and
  \bibinfo{author}{\bibfnamefont{P.}~\bibnamefont{Vogel}},
  \bibinfo{year}{2001}, \bibinfo{journal}{Phys. Lett. B}
  \textbf{\bibinfo{volume}{522}}, \bibinfo{pages}{240}.

\bibitem[{\citenamefont{Chaves and Poves}(1986)}]{Chaves:1986}
\bibinfo{author}{\bibnamefont{Chaves}, \bibfnamefont{L.}}, and
  \bibinfo{author}{\bibfnamefont{A.}~\bibnamefont{Poves}},
  \bibinfo{year}{1986}, \bibinfo{journal}{Phys. Rev. C}
  \textbf{\bibinfo{volume}{34}}, \bibinfo{pages}{1137}.

\bibitem[{\citenamefont{Crawley} \emph{et~al.}(1982)\citenamefont{Crawley,
  Anantaraman, Galonsky, Djalali, Marty, Morlet, Willis, Jourdain, and
  Kitching}}]{Crawley:1982}
\bibinfo{author}{\bibnamefont{Crawley}, \bibfnamefont{G.~M.}},
  \bibinfo{author}{\bibfnamefont{N.}~\bibnamefont{Anantaraman}},
  \bibinfo{author}{\bibfnamefont{A.}~\bibnamefont{Galonsky}},
  \bibinfo{author}{\bibfnamefont{C.}~\bibnamefont{Djalali}},
  \bibinfo{author}{\bibfnamefont{N.}~\bibnamefont{Marty}},
  \bibinfo{author}{\bibfnamefont{M.}~\bibnamefont{Morlet}},
  \bibinfo{author}{\bibfnamefont{A.}~\bibnamefont{Willis}},
  \bibinfo{author}{\bibfnamefont{J.~C.} \bibnamefont{Jourdain}}, and
  \bibinfo{author}{\bibfnamefont{P.}~\bibnamefont{Kitching}},
  \bibinfo{year}{1982}, \bibinfo{journal}{Phys. Rev. C}
  \textbf{\bibinfo{volume}{26}}, \bibinfo{pages}{87}.

\bibitem[{\citenamefont{De~Coster and Heyde}(1989)}]{DeCoster:1989c}
\bibinfo{author}{\bibnamefont{De~Coster}, \bibfnamefont{C.}}, and
  \bibinfo{author}{\bibfnamefont{K.}~\bibnamefont{Heyde}},
  \bibinfo{year}{1989}, \bibinfo{journal}{Phys. Rev. Lett.}
  \textbf{\bibinfo{volume}{63}}, \bibinfo{pages}{2797}.

\bibitem[{\citenamefont{De~Coster and
  Heyde}(1991{\natexlab{a}})}]{DeCoster:1991a}
\bibinfo{author}{\bibnamefont{De~Coster}, \bibfnamefont{C.}}, and
  \bibinfo{author}{\bibfnamefont{K.}~\bibnamefont{Heyde}},
  \bibinfo{year}{1991}{\natexlab{a}}, \bibinfo{journal}{Phys. Rev. Lett.}
  \textbf{\bibinfo{volume}{66}}, \bibinfo{pages}{2456}.

\bibitem[{\citenamefont{De~Coster and
  Heyde}(1991{\natexlab{b}})}]{DeCoster:1991c}
\bibinfo{author}{\bibnamefont{De~Coster}, \bibfnamefont{C.}}, and
  \bibinfo{author}{\bibfnamefont{K.}~\bibnamefont{Heyde}},
  \bibinfo{year}{1991}{\natexlab{b}}, \bibinfo{journal}{Nucl. Phys. A}
  \textbf{\bibinfo{volume}{529}}, \bibinfo{pages}{507}.

\bibitem[{\citenamefont{De~Coster} \emph{et~al.}(1992)\citenamefont{De~Coster,
  Heyde, and Richter}}]{DeCoster:1992}
\bibinfo{author}{\bibnamefont{De~Coster}, \bibfnamefont{C.}},
  \bibinfo{author}{\bibfnamefont{K.}~\bibnamefont{Heyde}}, and
  \bibinfo{author}{\bibfnamefont{A.}~\bibnamefont{Richter}},
  \bibinfo{year}{1992}, \bibinfo{journal}{Nucl. Phys. A}
  \textbf{\bibinfo{volume}{542}}, \bibinfo{pages}{375}.

\bibitem[{\citenamefont{De~Coster} \emph{et~al.}(1995)\citenamefont{De~Coster,
  Heyde, Rombouts, and Richter}}]{DeCoster:1995}
\bibinfo{author}{\bibnamefont{De~Coster}, \bibfnamefont{C.}},
  \bibinfo{author}{\bibfnamefont{K.}~\bibnamefont{Heyde}},
  \bibinfo{author}{\bibfnamefont{S.}~\bibnamefont{Rombouts}}, and
  \bibinfo{author}{\bibfnamefont{A.}~\bibnamefont{Richter}},
  \bibinfo{year}{1995}, \bibinfo{journal}{Phys. Rev. C}
  \textbf{\bibinfo{volume}{51}}, \bibinfo{pages}{3510}.

\bibitem[{\citenamefont{De~Franceschi}
  \emph{et~al.}(1983)\citenamefont{De~Franceschi, Palumbo, and
  Lo~Iudice}}]{DeFranceschi:1983}
\bibinfo{author}{\bibnamefont{De~Franceschi}, \bibfnamefont{G.}},
  \bibinfo{author}{\bibfnamefont{F.}~\bibnamefont{Palumbo}}, and
  \bibinfo{author}{\bibfnamefont{N.}~\bibnamefont{Lo~Iudice}},
  \bibinfo{year}{1983}, \bibinfo{journal}{Lett. Nuovo Cimento}
  \textbf{\bibinfo{volume}{37}}, \bibinfo{pages}{61}.

\bibitem[{\citenamefont{De~Franceschi}
  \emph{et~al.}(1984)\citenamefont{De~Franceschi, Palumbo, and
  Lo~Iudice}}]{DeFranceschi:1984}
\bibinfo{author}{\bibnamefont{De~Franceschi}, \bibfnamefont{G.}},
  \bibinfo{author}{\bibfnamefont{F.}~\bibnamefont{Palumbo}}, and
  \bibinfo{author}{\bibfnamefont{N.}~\bibnamefont{Lo~Iudice}},
  \bibinfo{year}{1984}, \bibinfo{journal}{Phys. Rev. C}
  \textbf{\bibinfo{volume}{29}}, \bibinfo{pages}{1496}.

\bibitem[{\citenamefont{Delorme} \emph{et~al.}(1976)\citenamefont{Delorme,
  Ericson, Figureau, and Th\'{e}venet}}]{Delorme:1976}
\bibinfo{author}{\bibnamefont{Delorme}, \bibfnamefont{J.}},
  \bibinfo{author}{\bibfnamefont{M.}~\bibnamefont{Ericson}},
  \bibinfo{author}{\bibfnamefont{A.}~\bibnamefont{Figureau}}, and
  \bibinfo{author}{\bibfnamefont{C.}~\bibnamefont{Th\'{e}venet}},
  \bibinfo{year}{1976}, \bibinfo{journal}{Ann. Phys. (N.Y.)}
  \textbf{\bibinfo{volume}{102}}, \bibinfo{pages}{273 }.

\bibitem[{\citenamefont{Demel} \emph{et~al.}(1990)\citenamefont{Demel,
  Heitmann, Grambow, and Ploog}}]{Demel:1990}
\bibinfo{author}{\bibnamefont{Demel}, \bibfnamefont{T.}},
  \bibinfo{author}{\bibfnamefont{D.}~\bibnamefont{Heitmann}},
  \bibinfo{author}{\bibfnamefont{P.}~\bibnamefont{Grambow}}, and
  \bibinfo{author}{\bibfnamefont{K.}~\bibnamefont{Ploog}},
  \bibinfo{year}{1990}, \bibinfo{journal}{Phys. Rev. Lett.}
  \textbf{\bibinfo{volume}{64}}, \bibinfo{pages}{788}.

\bibitem[{\citenamefont{Devi and Kota}(1992{\natexlab{a}})}]{Devi:1992a}
\bibinfo{author}{\bibnamefont{Devi}, \bibfnamefont{Y.~D.}}, and
  \bibinfo{author}{\bibfnamefont{V.~K.~B.} \bibnamefont{Kota}},
  \bibinfo{year}{1992}{\natexlab{a}}, \bibinfo{journal}{Phys. Lett. B}
  \textbf{\bibinfo{volume}{287}}, \bibinfo{pages}{9}.

\bibitem[{\citenamefont{Devi and Kota}(1992{\natexlab{b}})}]{Devi:1992b}
\bibinfo{author}{\bibnamefont{Devi}, \bibfnamefont{Y.~D.}}, and
  \bibinfo{author}{\bibfnamefont{V.~K.~B.} \bibnamefont{Kota}},
  \bibinfo{year}{1992}{\natexlab{b}}, \bibinfo{journal}{Nucl. Phys. A}
  \textbf{\bibinfo{volume}{541}}, \bibinfo{pages}{173}.

\bibitem[{\citenamefont{Devi and Kota}(1996)}]{Devi:1996}
\bibinfo{author}{\bibnamefont{Devi}, \bibfnamefont{Y.~D.}}, and
  \bibinfo{author}{\bibfnamefont{V.~K.~B.} \bibnamefont{Kota}},
  \bibinfo{year}{1996}, \bibinfo{journal}{Nucl. Phys. A}
  \textbf{\bibinfo{volume}{600}}, \bibinfo{pages}{20}.

\bibitem[{\citenamefont{Dieperink}(1983)}]{Dieperink:1983}
\bibinfo{author}{\bibnamefont{Dieperink}, \bibfnamefont{A.~E.~L.}},
  \bibinfo{year}{1983}, \bibinfo{journal}{Prog. Part. Nucl. Phys.}
  \textbf{\bibinfo{volume}{9}}, \bibinfo{pages}{121}.

\bibitem[{\citenamefont{Djalali} \emph{et~al.}(1982)\citenamefont{Djalali,
  Marty, Morlet, Willis, Jourdain, Anantaraman, Crawley, Galonsky, and
  Kitching}}]{Djalali:1982}
\bibinfo{author}{\bibnamefont{Djalali}, \bibfnamefont{C.}},
  \bibinfo{author}{\bibfnamefont{N.}~\bibnamefont{Marty}},
  \bibinfo{author}{\bibfnamefont{M.}~\bibnamefont{Morlet}},
  \bibinfo{author}{\bibfnamefont{A.}~\bibnamefont{Willis}},
  \bibinfo{author}{\bibfnamefont{J.~C.} \bibnamefont{Jourdain}},
  \bibinfo{author}{\bibfnamefont{N.}~\bibnamefont{Anantaraman}},
  \bibinfo{author}{\bibfnamefont{G.~M.} \bibnamefont{Crawley}},
  \bibinfo{author}{\bibfnamefont{A.}~\bibnamefont{Galonsky}}, and
  \bibinfo{author}{\bibfnamefont{P.}~\bibnamefont{Kitching}},
  \bibinfo{year}{1982}, \bibinfo{journal}{Nucl. Phys. A}
  \textbf{\bibinfo{volume}{388}}, \bibinfo{pages}{1}.

\bibitem[{\citenamefont{Eid} \emph{et~al.}(1986)\citenamefont{Eid, Hamilton,
  and Elliott}}]{Eid:1986}
\bibinfo{author}{\bibnamefont{Eid}, \bibfnamefont{S.~A.~A.}},
  \bibinfo{author}{\bibfnamefont{W.~D.} \bibnamefont{Hamilton}}, and
  \bibinfo{author}{\bibfnamefont{J.~P.} \bibnamefont{Elliott}},
  \bibinfo{year}{1986}, \bibinfo{journal}{Phys. Lett. B}
  \textbf{\bibinfo{volume}{166}}, \bibinfo{pages}{267}.

\bibitem[{\citenamefont{Eisenberg and Greiner}(1987)}]{Eisenberg:1987}
\bibinfo{author}{\bibnamefont{Eisenberg}, \bibfnamefont{J.~M.}}, and
  \bibinfo{author}{\bibfnamefont{W.}~\bibnamefont{Greiner}},
  \bibinfo{year}{1987}, \emph{\bibinfo{title}{Nuclear Theory, Vol.~1: Nuclear
  Models}} (\bibinfo{publisher}{North-Holland}, \bibinfo{address}{Amsterdam}),
  \bibinfo{edition}{3rd} edition.

\bibitem[{\citenamefont{Elhami} \emph{et~al.}(2007)\citenamefont{Elhami, Orce,
  Mukhopadhyay, Choudry, Scheck, McEllistrem, and Yates}}]{Elhami:2007}
\bibinfo{author}{\bibnamefont{Elhami}, \bibfnamefont{E.}},
  \bibinfo{author}{\bibfnamefont{J.~N.} \bibnamefont{Orce}},
  \bibinfo{author}{\bibfnamefont{S.}~\bibnamefont{Mukhopadhyay}},
  \bibinfo{author}{\bibfnamefont{S.~N.} \bibnamefont{Choudry}},
  \bibinfo{author}{\bibfnamefont{M.}~\bibnamefont{Scheck}},
  \bibinfo{author}{\bibfnamefont{M.~T.} \bibnamefont{McEllistrem}}, and
  \bibinfo{author}{\bibfnamefont{S.~W.} \bibnamefont{Yates}},
  \bibinfo{year}{2007}, \bibinfo{journal}{Phys. Rev. C}
  \textbf{\bibinfo{volume}{75}}, \bibinfo{eid}{011301(R)}.

\bibitem[{\citenamefont{Elhami} \emph{et~al.}(2008)\citenamefont{Elhami, Orce,
  Scheck, Mukhopadhyay, Choudry, McEllistrem, Yates, Angell, Boswell, Fallin,
  Howell, Hutcheson} \emph{et~al.}}]{Elhami:2008}
\bibinfo{author}{\bibnamefont{Elhami}, \bibfnamefont{E.}},
  \bibinfo{author}{\bibfnamefont{J.~N.} \bibnamefont{Orce}},
  \bibinfo{author}{\bibfnamefont{M.}~\bibnamefont{Scheck}},
  \bibinfo{author}{\bibfnamefont{S.}~\bibnamefont{Mukhopadhyay}},
  \bibinfo{author}{\bibfnamefont{S.~N.} \bibnamefont{Choudry}},
  \bibinfo{author}{\bibfnamefont{M.~T.} \bibnamefont{McEllistrem}},
  \bibinfo{author}{\bibfnamefont{S.~W.} \bibnamefont{Yates}},
  \bibinfo{author}{\bibfnamefont{C.}~\bibnamefont{Angell}},
  \bibinfo{author}{\bibfnamefont{M.}~\bibnamefont{Boswell}},
  \bibinfo{author}{\bibfnamefont{B.}~\bibnamefont{Fallin}},
  \bibinfo{author}{\bibfnamefont{C.~R.} \bibnamefont{Howell}},
  \bibinfo{author}{\bibfnamefont{A.}~\bibnamefont{Hutcheson}}, \emph{et~al.},
  \bibinfo{year}{2008}, \bibinfo{journal}{Phys. Rev. C}
  \textbf{\bibinfo{volume}{78}}, \bibinfo{eid}{064303}.

\bibitem[{\citenamefont{Elliott}(1985)}]{Elliott:1985}
\bibinfo{author}{\bibnamefont{Elliott}, \bibfnamefont{J.~P.}},
  \bibinfo{year}{1985}, \bibinfo{journal}{Rep. Prog. Phys.}
  \textbf{\bibinfo{volume}{48}}, \bibinfo{pages}{171}.

\bibitem[{\citenamefont{Elliott}(1990)}]{Elliott:1990}
\bibinfo{author}{\bibnamefont{Elliott}, \bibfnamefont{J.~P.}},
  \bibinfo{year}{1990}, \bibinfo{journal}{Prog. Part. Nucl. Phys.}
  \textbf{\bibinfo{volume}{25}}, \bibinfo{pages}{325}.

\bibitem[{\citenamefont{Elliott}(1958)}]{Elliott:1958}
\bibinfo{author}{\bibnamefont{Elliott}, \bibfnamefont{P.~J.}},
  \bibinfo{year}{1958}, \bibinfo{journal}{Proc. Roy. Soc. A}
  \textbf{\bibinfo{volume}{245}}, \bibinfo{pages}{128;562}.

\bibitem[{\citenamefont{Elliott}(1963)}]{Elliott:1963}
\bibinfo{author}{\bibnamefont{Elliott}, \bibfnamefont{P.~J.}},
  \bibinfo{year}{1963}, \bibinfo{journal}{Proc. Roy. Soc. A}
  \textbf{\bibinfo{volume}{272}}, \bibinfo{pages}{557}.

\bibitem[{\citenamefont{Enders} \emph{et~al.}(2000)\citenamefont{Enders, Guhr,
  Huxel, von Neumann-Cosel, Rangacharyulu, and Richter}}]{Enders:2000}
\bibinfo{author}{\bibnamefont{Enders}, \bibfnamefont{J.}},
  \bibinfo{author}{\bibfnamefont{T.}~\bibnamefont{Guhr}},
  \bibinfo{author}{\bibfnamefont{N.}~\bibnamefont{Huxel}},
  \bibinfo{author}{\bibfnamefont{P.}~\bibnamefont{von Neumann-Cosel}},
  \bibinfo{author}{\bibfnamefont{C.}~\bibnamefont{Rangacharyulu}}, and
  \bibinfo{author}{\bibfnamefont{A.}~\bibnamefont{Richter}},
  \bibinfo{year}{2000}, \bibinfo{journal}{Phys. Lett. B}
  \textbf{\bibinfo{volume}{486}}, \bibinfo{pages}{273}.

\bibitem[{\citenamefont{Enders} \emph{et~al.}(1998)\citenamefont{Enders, Huxel,
  Kneissl, von Neumann-Cosel, Pitz, and Richter}}]{Enders:1998}
\bibinfo{author}{\bibnamefont{Enders}, \bibfnamefont{J.}},
  \bibinfo{author}{\bibfnamefont{N.}~\bibnamefont{Huxel}},
  \bibinfo{author}{\bibfnamefont{U.}~\bibnamefont{Kneissl}},
  \bibinfo{author}{\bibfnamefont{P.}~\bibnamefont{von Neumann-Cosel}},
  \bibinfo{author}{\bibfnamefont{H.~H.} \bibnamefont{Pitz}}, and
  \bibinfo{author}{\bibfnamefont{A.}~\bibnamefont{Richter}},
  \bibinfo{year}{1998}, \bibinfo{journal}{Phys. Rev. C}
  \textbf{\bibinfo{volume}{57}}, \bibinfo{pages}{996}.

\bibitem[{\citenamefont{Enders} \emph{et~al.}(1997)\citenamefont{Enders, Huxel,
  von Neumann-Cosel, and Richter}}]{Enders:1997}
\bibinfo{author}{\bibnamefont{Enders}, \bibfnamefont{J.}},
  \bibinfo{author}{\bibfnamefont{N.}~\bibnamefont{Huxel}},
  \bibinfo{author}{\bibfnamefont{P.}~\bibnamefont{von Neumann-Cosel}}, and
  \bibinfo{author}{\bibfnamefont{A.}~\bibnamefont{Richter}},
  \bibinfo{year}{1997}, \bibinfo{journal}{Phys. Rev. Lett.}
  \textbf{\bibinfo{volume}{79}}, \bibinfo{pages}{2010}.

\bibitem[{\citenamefont{Enders} \emph{et~al.}(1999)\citenamefont{Enders,
  Kaiser, von Neumann-Cosel, Rangacharyulu, and Richter}}]{Enders:1999}
\bibinfo{author}{\bibnamefont{Enders}, \bibfnamefont{J.}},
  \bibinfo{author}{\bibfnamefont{H.}~\bibnamefont{Kaiser}},
  \bibinfo{author}{\bibfnamefont{P.}~\bibnamefont{von Neumann-Cosel}},
  \bibinfo{author}{\bibfnamefont{C.}~\bibnamefont{Rangacharyulu}}, and
  \bibinfo{author}{\bibfnamefont{A.}~\bibnamefont{Richter}},
  \bibinfo{year}{1999}, \bibinfo{journal}{Phys. Rev. C}
  \textbf{\bibinfo{volume}{59}}, \bibinfo{pages}{R1851}.

\bibitem[{\citenamefont{Enders} \emph{et~al.}(2005)\citenamefont{Enders, von
  Neumann-Cosel, Rangacharyulu, and Richter}}]{Enders:2005}
\bibinfo{author}{\bibnamefont{Enders}, \bibfnamefont{J.}},
  \bibinfo{author}{\bibfnamefont{P.}~\bibnamefont{von Neumann-Cosel}},
  \bibinfo{author}{\bibfnamefont{C.}~\bibnamefont{Rangacharyulu}}, and
  \bibinfo{author}{\bibfnamefont{A.}~\bibnamefont{Richter}},
  \bibinfo{year}{2005}, \bibinfo{journal}{Phys. Rev. C}
  \textbf{\bibinfo{volume}{71}}, \bibinfo{eid}{014306}.

\bibitem[{\citenamefont{Ericson and Weise}(1988)}]{Ericson:1988}
\bibinfo{author}{\bibnamefont{Ericson}, \bibfnamefont{T.}}, and
  \bibinfo{author}{\bibfnamefont{W.}~\bibnamefont{Weise}},
  \bibinfo{year}{1988}, \emph{\bibinfo{title}{Pions and Nuclei}}, International
  Series of Monographs on Physics, Vol.~74 (\bibinfo{publisher}{Clarendon},
  \bibinfo{address}{Oxford}).

\bibitem[{\citenamefont{Faessler}(1966)}]{Faessler:1966}
\bibinfo{author}{\bibnamefont{Faessler}, \bibfnamefont{A.}},
  \bibinfo{year}{1966}, \bibinfo{journal}{Nucl. Phys.}
  \textbf{\bibinfo{volume}{85}}, \bibinfo{pages}{653}.

\bibitem[{\citenamefont{Faessler} \emph{et~al.}(1986)\citenamefont{Faessler,
  Bochnacki, and Nojarov}}]{Faessler:1986b}
\bibinfo{author}{\bibnamefont{Faessler}, \bibfnamefont{A.}},
  \bibinfo{author}{\bibfnamefont{Z.}~\bibnamefont{Bochnacki}}, and
  \bibinfo{author}{\bibfnamefont{R.}~\bibnamefont{Nojarov}},
  \bibinfo{year}{1986}, \bibinfo{journal}{J. Phys. G}
  \textbf{\bibinfo{volume}{12}}, \bibinfo{pages}{L47}.

\bibitem[{\citenamefont{Faessler and Nojarov}(1986)}]{Faessler:1986a}
\bibinfo{author}{\bibnamefont{Faessler}, \bibfnamefont{A.}}, and
  \bibinfo{author}{\bibfnamefont{R.}~\bibnamefont{Nojarov}},
  \bibinfo{year}{1986}, \bibinfo{journal}{Phys. Lett. B}
  \textbf{\bibinfo{volume}{166}}, \bibinfo{pages}{367}.

\bibitem[{\citenamefont{Faessler and Nojarov}(1987)}]{Faessler:1987}
\bibinfo{author}{\bibnamefont{Faessler}, \bibfnamefont{A.}}, and
  \bibinfo{author}{\bibfnamefont{R.}~\bibnamefont{Nojarov}},
  \bibinfo{year}{1987}, \bibinfo{journal}{Prog. Part. Nucl. Phys.}
  \textbf{\bibinfo{volume}{19}}, \bibinfo{pages}{167}.

\bibitem[{\citenamefont{Faessler and Nojarov}(1988)}]{Faessler:1988}
\bibinfo{author}{\bibnamefont{Faessler}, \bibfnamefont{A.}}, and
  \bibinfo{author}{\bibfnamefont{R.}~\bibnamefont{Nojarov}},
  \bibinfo{year}{1988}, \bibinfo{journal}{Phys. Lett. B}
  \textbf{\bibinfo{volume}{215}}, \bibinfo{pages}{439}.

\bibitem[{\citenamefont{Faessler} \emph{et~al.}(1989)\citenamefont{Faessler,
  Nojarov, and Taigel}}]{Faessler:1989}
\bibinfo{author}{\bibnamefont{Faessler}, \bibfnamefont{A.}},
  \bibinfo{author}{\bibfnamefont{R.}~\bibnamefont{Nojarov}}, and
  \bibinfo{author}{\bibfnamefont{T.}~\bibnamefont{Taigel}},
  \bibinfo{year}{1989}, \bibinfo{journal}{Nucl. Phys. A}
  \textbf{\bibinfo{volume}{492}}, \bibinfo{pages}{105}.

\bibitem[{\citenamefont{Fagg}(1975)}]{Fagg:1975}
\bibinfo{author}{\bibnamefont{Fagg}, \bibfnamefont{L.~W.}},
  \bibinfo{year}{1975}, \bibinfo{journal}{Rev. Mod. Phys.}
  \textbf{\bibinfo{volume}{47}}, \bibinfo{pages}{683}.

\bibitem[{\citenamefont{Fazekas} \emph{et~al.}(1992)\citenamefont{Fazekas,
  Belgya, Molnar, Veres, Gatenby, Yates, and Otsuka}}]{Fazekas:1992}
\bibinfo{author}{\bibnamefont{Fazekas}, \bibfnamefont{B.}},
  \bibinfo{author}{\bibfnamefont{T.}~\bibnamefont{Belgya}},
  \bibinfo{author}{\bibfnamefont{G.}~\bibnamefont{Molnar}},
  \bibinfo{author}{\bibfnamefont{A.}~\bibnamefont{Veres}},
  \bibinfo{author}{\bibfnamefont{R.~A.} \bibnamefont{Gatenby}},
  \bibinfo{author}{\bibfnamefont{S.~W.} \bibnamefont{Yates}}, and
  \bibinfo{author}{\bibfnamefont{T.}~\bibnamefont{Otsuka}},
  \bibinfo{year}{1992}, \bibinfo{journal}{Nucl. Phys. A}
  \textbf{\bibinfo{volume}{548}}, \bibinfo{pages}{249}.

\bibitem[{\citenamefont{Fearick} \emph{et~al.}(2003)\citenamefont{Fearick,
  Hartung, Langanke, Mart\'{\i}nez-Pinedo, von Neumann-Cosel, and
  Richter}}]{Fearick:2003}
\bibinfo{author}{\bibnamefont{Fearick}, \bibfnamefont{R.~W.}},
  \bibinfo{author}{\bibfnamefont{G.}~\bibnamefont{Hartung}},
  \bibinfo{author}{\bibfnamefont{K.}~\bibnamefont{Langanke}},
  \bibinfo{author}{\bibfnamefont{G.}~\bibnamefont{Mart\'{\i}nez-Pinedo}},
  \bibinfo{author}{\bibfnamefont{P.}~\bibnamefont{von Neumann-Cosel}}, and
  \bibinfo{author}{\bibfnamefont{A.}~\bibnamefont{Richter}},
  \bibinfo{year}{2003}, \bibinfo{journal}{Nucl. Phys. A}
  \textbf{\bibinfo{volume}{727}}, \bibinfo{pages}{41}.

\bibitem[{\citenamefont{Fearick} \emph{et~al.}(2006)\citenamefont{Fearick, von
  Neumann-Cosel, Richter, Robinson, and Zamick}}]{Fearick:2006}
\bibinfo{author}{\bibnamefont{Fearick}, \bibfnamefont{R.~W.}},
  \bibinfo{author}{\bibfnamefont{P.}~\bibnamefont{von Neumann-Cosel}},
  \bibinfo{author}{\bibfnamefont{A.}~\bibnamefont{Richter}},
  \bibinfo{author}{\bibfnamefont{S.~J.~Q.} \bibnamefont{Robinson}}, and
  \bibinfo{author}{\bibfnamefont{L.}~\bibnamefont{Zamick}},
  \bibinfo{year}{2006}, \bibinfo{journal}{J. Phys. Soc. Japan}
  \textbf{\bibinfo{volume}{75}}, \bibinfo{pages}{094201}.

\bibitem[{\citenamefont{Foltz} \emph{et~al.}(1994)\citenamefont{Foltz, Sober,
  Fagg, Gr\"af, Richter, Spamer, and Brown}}]{Foltz:1994}
\bibinfo{author}{\bibnamefont{Foltz}, \bibfnamefont{C.~W.}},
  \bibinfo{author}{\bibfnamefont{D.~I.} \bibnamefont{Sober}},
  \bibinfo{author}{\bibfnamefont{L.~W.} \bibnamefont{Fagg}},
  \bibinfo{author}{\bibfnamefont{H.~D.} \bibnamefont{Gr\"af}},
  \bibinfo{author}{\bibfnamefont{A.}~\bibnamefont{Richter}},
  \bibinfo{author}{\bibfnamefont{E.}~\bibnamefont{Spamer}}, and
  \bibinfo{author}{\bibfnamefont{B.~A.} \bibnamefont{Brown}},
  \bibinfo{year}{1994}, \bibinfo{journal}{Phys. Rev. C}
  \textbf{\bibinfo{volume}{49}}, \bibinfo{pages}{1359}.

\bibitem[{\citenamefont{Frank} \emph{et~al.}(1991)\citenamefont{Frank, Arias,
  and Van~Isacker}}]{Frank:1991}
\bibinfo{author}{\bibnamefont{Frank}, \bibfnamefont{A.}},
  \bibinfo{author}{\bibfnamefont{J.~M.} \bibnamefont{Arias}}, and
  \bibinfo{author}{\bibfnamefont{P.}~\bibnamefont{Van~Isacker}},
  \bibinfo{year}{1991}, \bibinfo{journal}{Nucl. Phys. A}
  \textbf{\bibinfo{volume}{531}}, \bibinfo{pages}{125}.

\bibitem[{\citenamefont{Fransen} \emph{et~al.}(2003)\citenamefont{Fransen,
  Pietralla, Ammar, Bandyopadhyay, Boukharouba, von Brentano, Dewald, Gableske,
  Gade, Jolie, Kneissl, Lesher} \emph{et~al.}}]{Fransen:2003}
\bibinfo{author}{\bibnamefont{Fransen}, \bibfnamefont{C.}},
  \bibinfo{author}{\bibfnamefont{N.}~\bibnamefont{Pietralla}},
  \bibinfo{author}{\bibfnamefont{Z.}~\bibnamefont{Ammar}},
  \bibinfo{author}{\bibfnamefont{D.}~\bibnamefont{Bandyopadhyay}},
  \bibinfo{author}{\bibfnamefont{N.}~\bibnamefont{Boukharouba}},
  \bibinfo{author}{\bibfnamefont{P.}~\bibnamefont{von Brentano}},
  \bibinfo{author}{\bibfnamefont{A.}~\bibnamefont{Dewald}},
  \bibinfo{author}{\bibfnamefont{J.}~\bibnamefont{Gableske}},
  \bibinfo{author}{\bibfnamefont{A.}~\bibnamefont{Gade}},
  \bibinfo{author}{\bibfnamefont{J.}~\bibnamefont{Jolie}},
  \bibinfo{author}{\bibfnamefont{U.}~\bibnamefont{Kneissl}},
  \bibinfo{author}{\bibfnamefont{S.~R.} \bibnamefont{Lesher}}, \emph{et~al.},
  \bibinfo{year}{2003}, \bibinfo{journal}{Phys. Rev. C}
  \textbf{\bibinfo{volume}{67}}, \bibinfo{pages}{024307}.

\bibitem[{\citenamefont{Fransen} \emph{et~al.}(2001)\citenamefont{Fransen,
  Pietralla, von Brentanon, Dewald, Gableske, Gade, Lisetsky, and
  Werner}}]{Fransen:2001}
\bibinfo{author}{\bibnamefont{Fransen}, \bibfnamefont{C.}},
  \bibinfo{author}{\bibfnamefont{N.}~\bibnamefont{Pietralla}},
  \bibinfo{author}{\bibfnamefont{P.}~\bibnamefont{von Brentanon}},
  \bibinfo{author}{\bibfnamefont{A.}~\bibnamefont{Dewald}},
  \bibinfo{author}{\bibfnamefont{J.}~\bibnamefont{Gableske}},
  \bibinfo{author}{\bibfnamefont{A.}~\bibnamefont{Gade}},
  \bibinfo{author}{\bibfnamefont{A.}~\bibnamefont{Lisetsky}}, and
  \bibinfo{author}{\bibfnamefont{V.}~\bibnamefont{Werner}},
  \bibinfo{year}{2001}, \bibinfo{journal}{Phys. Lett. B}
  \textbf{\bibinfo{volume}{508}}, \bibinfo{pages}{219}.

\bibitem[{\citenamefont{Freeman} \emph{et~al.}(1989)\citenamefont{Freeman,
  Chapman, Durell, Hotchkis, Khazaie, Lisle, Mo, Bruce, Cunningham, Drumm,
  Warner, and Garrett}}]{Freeman:1989}
\bibinfo{author}{\bibnamefont{Freeman}, \bibfnamefont{S.~J.}},
  \bibinfo{author}{\bibfnamefont{R.}~\bibnamefont{Chapman}},
  \bibinfo{author}{\bibfnamefont{J.~L.} \bibnamefont{Durell}},
  \bibinfo{author}{\bibfnamefont{M.~A.~C.} \bibnamefont{Hotchkis}},
  \bibinfo{author}{\bibfnamefont{F.}~\bibnamefont{Khazaie}},
  \bibinfo{author}{\bibfnamefont{C.~J.} \bibnamefont{Lisle}},
  \bibinfo{author}{\bibfnamefont{J.~N.} \bibnamefont{Mo}},
  \bibinfo{author}{\bibfnamefont{A.~M.} \bibnamefont{Bruce}},
  \bibinfo{author}{\bibfnamefont{R.~A.} \bibnamefont{Cunningham}},
  \bibinfo{author}{\bibfnamefont{P.~V.} \bibnamefont{Drumm}},
  \bibinfo{author}{\bibfnamefont{D.~D.} \bibnamefont{Warner}}, and
  \bibinfo{author}{\bibfnamefont{J.~D.} \bibnamefont{Garrett}},
  \bibinfo{year}{1989}, \bibinfo{journal}{Phys. Lett. B}
  \textbf{\bibinfo{volume}{222}}, \bibinfo{pages}{347}.

\bibitem[{\citenamefont{Frekers}(2006)}]{Frekers:2006}
\bibinfo{author}{\bibnamefont{Frekers}, \bibfnamefont{D.}},
  \bibinfo{year}{2006}, \bibinfo{journal}{Prog. Part. Nucl. Phys.}
  \textbf{\bibinfo{volume}{57}}, \bibinfo{pages}{217}.

\bibitem[{\citenamefont{Frekers} \emph{et~al.}(1990)\citenamefont{Frekers,
  W\"{o}rtche, Richter, Abegg, Azuma, Celler, Chan, Drake, Helmer, Jackson,
  King, Miller} \emph{et~al.}}]{Frekers:1990}
\bibinfo{author}{\bibnamefont{Frekers}, \bibfnamefont{D.}},
  \bibinfo{author}{\bibfnamefont{H.~J.} \bibnamefont{W\"{o}rtche}},
  \bibinfo{author}{\bibfnamefont{A.}~\bibnamefont{Richter}},
  \bibinfo{author}{\bibfnamefont{R.}~\bibnamefont{Abegg}},
  \bibinfo{author}{\bibfnamefont{R.~E.} \bibnamefont{Azuma}},
  \bibinfo{author}{\bibfnamefont{A.}~\bibnamefont{Celler}},
  \bibinfo{author}{\bibfnamefont{C.}~\bibnamefont{Chan}},
  \bibinfo{author}{\bibfnamefont{T.~E.} \bibnamefont{Drake}},
  \bibinfo{author}{\bibfnamefont{R.}~\bibnamefont{Helmer}},
  \bibinfo{author}{\bibfnamefont{K.~P.} \bibnamefont{Jackson}},
  \bibinfo{author}{\bibfnamefont{J.~D.} \bibnamefont{King}},
  \bibinfo{author}{\bibfnamefont{C.~A.} \bibnamefont{Miller}}, \emph{et~al.},
  \bibinfo{year}{1990}, \bibinfo{journal}{Phys. Lett. B}
  \textbf{\bibinfo{volume}{244}}, \bibinfo{pages}{178}.

\bibitem[{\citenamefont{Fujita} \emph{et~al.}(2008)\citenamefont{Fujita, Rubio,
  Adachi, Molina, Algora, Berg, von Brentano, Buscher, Cocolios, Frenne,
  Fransen, Fujita} \emph{et~al.}}]{Fujita:2008}
\bibinfo{author}{\bibnamefont{Fujita}, \bibfnamefont{Y.}},
  \bibinfo{author}{\bibfnamefont{B.}~\bibnamefont{Rubio}},
  \bibinfo{author}{\bibfnamefont{T.}~\bibnamefont{Adachi}},
  \bibinfo{author}{\bibfnamefont{F.}~\bibnamefont{Molina}},
  \bibinfo{author}{\bibfnamefont{A.}~\bibnamefont{Algora}},
  \bibinfo{author}{\bibfnamefont{G.~P.~A.} \bibnamefont{Berg}},
  \bibinfo{author}{\bibfnamefont{P.}~\bibnamefont{von Brentano}},
  \bibinfo{author}{\bibfnamefont{J.}~\bibnamefont{Buscher}},
  \bibinfo{author}{\bibfnamefont{T.}~\bibnamefont{Cocolios}},
  \bibinfo{author}{\bibfnamefont{D.~D.} \bibnamefont{Frenne}},
  \bibinfo{author}{\bibfnamefont{C.}~\bibnamefont{Fransen}},
  \bibinfo{author}{\bibfnamefont{H.}~\bibnamefont{Fujita}}, \emph{et~al.},
  \bibinfo{year}{2008}, \bibinfo{journal}{J. Phys. G}
  \textbf{\bibinfo{volume}{35}}, \bibinfo{pages}{014041}.

\bibitem[{\citenamefont{Gade} \emph{et~al.}(2003)\citenamefont{Gade, Belic, von
  Brentano, Fransen, von Garrel, Jolie, Kneissl, Kohstall, Linnemann, Pitz,
  Scheck, Stedile} \emph{et~al.}}]{Gade:2003}
\bibinfo{author}{\bibnamefont{Gade}, \bibfnamefont{A.}},
  \bibinfo{author}{\bibfnamefont{D.}~\bibnamefont{Belic}},
  \bibinfo{author}{\bibfnamefont{P.}~\bibnamefont{von Brentano}},
  \bibinfo{author}{\bibfnamefont{C.}~\bibnamefont{Fransen}},
  \bibinfo{author}{\bibfnamefont{H.}~\bibnamefont{von Garrel}},
  \bibinfo{author}{\bibfnamefont{J.}~\bibnamefont{Jolie}},
  \bibinfo{author}{\bibfnamefont{U.}~\bibnamefont{Kneissl}},
  \bibinfo{author}{\bibfnamefont{C.}~\bibnamefont{Kohstall}},
  \bibinfo{author}{\bibfnamefont{A.}~\bibnamefont{Linnemann}},
  \bibinfo{author}{\bibfnamefont{H.~H.} \bibnamefont{Pitz}},
  \bibinfo{author}{\bibfnamefont{M.}~\bibnamefont{Scheck}},
  \bibinfo{author}{\bibfnamefont{F.}~\bibnamefont{Stedile}}, \emph{et~al.},
  \bibinfo{year}{2003}, \bibinfo{journal}{Phys. Rev. C}
  \textbf{\bibinfo{volume}{67}}, \bibinfo{pages}{034304}.

\bibitem[{\citenamefont{Gade} \emph{et~al.}(2002)\citenamefont{Gade, Klein,
  Pietralla, and von Brentano}}]{Gade:2002}
\bibinfo{author}{\bibnamefont{Gade}, \bibfnamefont{A.}},
  \bibinfo{author}{\bibfnamefont{H.}~\bibnamefont{Klein}},
  \bibinfo{author}{\bibfnamefont{N.}~\bibnamefont{Pietralla}}, and
  \bibinfo{author}{\bibfnamefont{P.}~\bibnamefont{von Brentano}},
  \bibinfo{year}{2002}, \bibinfo{journal}{Phys. Rev. C}
  \textbf{\bibinfo{volume}{65}}, \bibinfo{pages}{054311}.

\bibitem[{\citenamefont{Gade} \emph{et~al.}(2004)\citenamefont{Gade, Pietralla,
  von Brentano, Belic, Fransen, Kneissl, Kohstall, Linnemann, Pitz, Scheck,
  Smirnova, Stedile} \emph{et~al.}}]{Gade:2004}
\bibinfo{author}{\bibnamefont{Gade}, \bibfnamefont{A.}},
  \bibinfo{author}{\bibfnamefont{N.}~\bibnamefont{Pietralla}},
  \bibinfo{author}{\bibfnamefont{P.}~\bibnamefont{von Brentano}},
  \bibinfo{author}{\bibfnamefont{D.}~\bibnamefont{Belic}},
  \bibinfo{author}{\bibfnamefont{C.}~\bibnamefont{Fransen}},
  \bibinfo{author}{\bibfnamefont{U.}~\bibnamefont{Kneissl}},
  \bibinfo{author}{\bibfnamefont{C.}~\bibnamefont{Kohstall}},
  \bibinfo{author}{\bibfnamefont{A.}~\bibnamefont{Linnemann}},
  \bibinfo{author}{\bibfnamefont{H.~H.} \bibnamefont{Pitz}},
  \bibinfo{author}{\bibfnamefont{M.}~\bibnamefont{Scheck}},
  \bibinfo{author}{\bibfnamefont{N.~A.} \bibnamefont{Smirnova}},
  \bibinfo{author}{\bibfnamefont{F.}~\bibnamefont{Stedile}}, \emph{et~al.},
  \bibinfo{year}{2004}, \bibinfo{journal}{Phys. Rev. C}
  \textbf{\bibinfo{volume}{69}}, \bibinfo{pages}{054321}.

\bibitem[{\citenamefont{Gade} \emph{et~al.}(2000)\citenamefont{Gade,
  Wiedenh\"{o}ver, Gableske, Gelberg, Meise, Pietralla, and von
  Brentano}}]{Gade:2000}
\bibinfo{author}{\bibnamefont{Gade}, \bibfnamefont{A.}},
  \bibinfo{author}{\bibfnamefont{I.}~\bibnamefont{Wiedenh\"{o}ver}},
  \bibinfo{author}{\bibfnamefont{J.}~\bibnamefont{Gableske}},
  \bibinfo{author}{\bibfnamefont{A.}~\bibnamefont{Gelberg}},
  \bibinfo{author}{\bibfnamefont{H.}~\bibnamefont{Meise}},
  \bibinfo{author}{\bibfnamefont{N.}~\bibnamefont{Pietralla}}, and
  \bibinfo{author}{\bibfnamefont{P.}~\bibnamefont{von Brentano}},
  \bibinfo{year}{2000}, \bibinfo{journal}{Nucl. Phys. A}
  \textbf{\bibinfo{volume}{665}}, \bibinfo{pages}{268}.

\bibitem[{\citenamefont{von Garrel} \emph{et~al.}(2006)\citenamefont{von
  Garrel, von Brentano, Fransen, Friessner, Hollmann, Jolie, K\"appeler,
  K\"aubler, Kneissl, Kohstall, Kostov, Linnemann} \emph{et~al.}}]{Garrel:2006}
\bibinfo{author}{\bibnamefont{von Garrel}, \bibfnamefont{H.}},
  \bibinfo{author}{\bibfnamefont{P.}~\bibnamefont{von Brentano}},
  \bibinfo{author}{\bibfnamefont{C.}~\bibnamefont{Fransen}},
  \bibinfo{author}{\bibfnamefont{G.}~\bibnamefont{Friessner}},
  \bibinfo{author}{\bibfnamefont{N.}~\bibnamefont{Hollmann}},
  \bibinfo{author}{\bibfnamefont{J.}~\bibnamefont{Jolie}},
  \bibinfo{author}{\bibfnamefont{F.}~\bibnamefont{K\"appeler}},
  \bibinfo{author}{\bibfnamefont{L.}~\bibnamefont{K\"aubler}},
  \bibinfo{author}{\bibfnamefont{U.}~\bibnamefont{Kneissl}},
  \bibinfo{author}{\bibfnamefont{C.}~\bibnamefont{Kohstall}},
  \bibinfo{author}{\bibfnamefont{L.}~\bibnamefont{Kostov}},
  \bibinfo{author}{\bibfnamefont{A.}~\bibnamefont{Linnemann}}, \emph{et~al.},
  \bibinfo{year}{2006}, \bibinfo{journal}{Phys. Rev. C}
  \textbf{\bibinfo{volume}{73}}, \bibinfo{pages}{054315}.

\bibitem[{\citenamefont{Garrett} \emph{et~al.}(1997)\citenamefont{Garrett,
  Robinson, Foglia, and Jin}}]{Garrett:1997}
\bibinfo{author}{\bibnamefont{Garrett}, \bibfnamefont{J.~D.}},
  \bibinfo{author}{\bibfnamefont{J.~Q.} \bibnamefont{Robinson}},
  \bibinfo{author}{\bibfnamefont{A.~J.} \bibnamefont{Foglia}}, and
  \bibinfo{author}{\bibfnamefont{H.~Q.} \bibnamefont{Jin}},
  \bibinfo{year}{1997}, \bibinfo{journal}{Phys. Lett. B}
  \textbf{\bibinfo{volume}{392}}, \bibinfo{pages}{24}.

\bibitem[{\citenamefont{Garrett} \emph{et~al.}(1996)\citenamefont{Garrett,
  Lehmann, McGrath, Yeh, and Yates}}]{Garrett:1996}
\bibinfo{author}{\bibnamefont{Garrett}, \bibfnamefont{P.~E.}},
  \bibinfo{author}{\bibfnamefont{H.}~\bibnamefont{Lehmann}},
  \bibinfo{author}{\bibfnamefont{C.~A.} \bibnamefont{McGrath}},
  \bibinfo{author}{\bibfnamefont{M.}~\bibnamefont{Yeh}}, and
  \bibinfo{author}{\bibfnamefont{S.~W.} \bibnamefont{Yates}},
  \bibinfo{year}{1996}, \bibinfo{journal}{Phys. Rev. C}
  \textbf{\bibinfo{volume}{54}}, \bibinfo{pages}{2259}.

\bibitem[{\citenamefont{Garrido} \emph{et~al.}(1991)\citenamefont{Garrido,
  Moya~de Guerra, Sarriguren, and Udias}}]{Garrido:1991}
\bibinfo{author}{\bibnamefont{Garrido}, \bibfnamefont{E.}},
  \bibinfo{author}{\bibfnamefont{E.}~\bibnamefont{Moya~de Guerra}},
  \bibinfo{author}{\bibfnamefont{P.}~\bibnamefont{Sarriguren}}, and
  \bibinfo{author}{\bibfnamefont{J.~M.} \bibnamefont{Udias}},
  \bibinfo{year}{1991}, \bibinfo{journal}{Phys. Rev. C}
  \textbf{\bibinfo{volume}{44}}, \bibinfo{pages}{R1250}.

\bibitem[{\citenamefont{Georgii} \emph{et~al.}(1995)\citenamefont{Georgii, von
  Egidy, Klora, Lindner, Mayerhofer, Ott, Schauer, von Neumann-Cosel, Richter,
  Schlegel, Schultz, Khitrov} \emph{et~al.}}]{Georgii:1995}
\bibinfo{author}{\bibnamefont{Georgii}, \bibfnamefont{R.}},
  \bibinfo{author}{\bibfnamefont{T.}~\bibnamefont{von Egidy}},
  \bibinfo{author}{\bibfnamefont{J.}~\bibnamefont{Klora}},
  \bibinfo{author}{\bibfnamefont{H.}~\bibnamefont{Lindner}},
  \bibinfo{author}{\bibfnamefont{U.}~\bibnamefont{Mayerhofer}},
  \bibinfo{author}{\bibfnamefont{J.}~\bibnamefont{Ott}},
  \bibinfo{author}{\bibfnamefont{W.}~\bibnamefont{Schauer}},
  \bibinfo{author}{\bibfnamefont{P.}~\bibnamefont{von Neumann-Cosel}},
  \bibinfo{author}{\bibfnamefont{A.}~\bibnamefont{Richter}},
  \bibinfo{author}{\bibfnamefont{C.}~\bibnamefont{Schlegel}},
  \bibinfo{author}{\bibfnamefont{R.}~\bibnamefont{Schultz}},
  \bibinfo{author}{\bibfnamefont{V.~A.} \bibnamefont{Khitrov}}, \emph{et~al.},
  \bibinfo{year}{1995}, \bibinfo{journal}{Nucl. Phys. A}
  \textbf{\bibinfo{volume}{592}}, \bibinfo{pages}{307}.

\bibitem[{\citenamefont{Ginocchio}(1991)}]{Ginocchio:1991}
\bibinfo{author}{\bibnamefont{Ginocchio}, \bibfnamefont{J.~N.}},
  \bibinfo{year}{1991}, \bibinfo{journal}{Phys. Lett. B}
  \textbf{\bibinfo{volume}{265}}, \bibinfo{pages}{6}.

\bibitem[{\citenamefont{Ginocchio}(1999)}]{Ginocchio:1999}
\bibinfo{author}{\bibnamefont{Ginocchio}, \bibfnamefont{J.~N.}},
  \bibinfo{year}{1999}, \bibinfo{journal}{Phys. Rev. C}
  \textbf{\bibinfo{volume}{59}}, \bibinfo{pages}{2487}.

\bibitem[{\citenamefont{Ginocchio}(2005)}]{Ginocchio:2005}
\bibinfo{author}{\bibnamefont{Ginocchio}, \bibfnamefont{J.~N.}},
  \bibinfo{year}{2005}, \bibinfo{journal}{Phys. Rep.}
  \textbf{\bibinfo{volume}{414}}, \bibinfo{pages}{165}.

\bibitem[{\citenamefont{Ginocchio and Leviatan}(1997)}]{Ginocchio:1997}
\bibinfo{author}{\bibnamefont{Ginocchio}, \bibfnamefont{J.~N.}}, and
  \bibinfo{author}{\bibfnamefont{A.}~\bibnamefont{Leviatan}},
  \bibinfo{year}{1997}, \bibinfo{journal}{Phys. Rev. Lett.}
  \textbf{\bibinfo{volume}{79}}, \bibinfo{pages}{813}.

\bibitem[{\citenamefont{Giorgini} \emph{et~al.}(2008)\citenamefont{Giorgini,
  Pitaevskii, and Stringari}}]{Giorgini:2008}
\bibinfo{author}{\bibnamefont{Giorgini}, \bibfnamefont{S.}},
  \bibinfo{author}{\bibfnamefont{L.~P.} \bibnamefont{Pitaevskii}}, and
  \bibinfo{author}{\bibfnamefont{S.}~\bibnamefont{Stringari}},
  \bibinfo{year}{2008}, \bibinfo{journal}{Rev. Mod. Phys.}
  \textbf{\bibinfo{volume}{80}}, \bibinfo{eid}{1215}.

\bibitem[{\citenamefont{Govaert} \emph{et~al.}(1994)\citenamefont{Govaert,
  Mondelaers, Jacobs, Frenne, Persijn, Pomm\'{e}, Yoneama, Herzberg,
  Lindenstruth, Huber, Jung, Starck} \emph{et~al.}}]{Govaert:1994}
\bibinfo{author}{\bibnamefont{Govaert}, \bibfnamefont{K.}},
  \bibinfo{author}{\bibfnamefont{W.}~\bibnamefont{Mondelaers}},
  \bibinfo{author}{\bibfnamefont{E.}~\bibnamefont{Jacobs}},
  \bibinfo{author}{\bibfnamefont{D.~D.} \bibnamefont{Frenne}},
  \bibinfo{author}{\bibfnamefont{K.}~\bibnamefont{Persijn}},
  \bibinfo{author}{\bibfnamefont{S.}~\bibnamefont{Pomm\'{e}}},
  \bibinfo{author}{\bibfnamefont{M.~L.} \bibnamefont{Yoneama}},
  \bibinfo{author}{\bibfnamefont{R.-D.} \bibnamefont{Herzberg}},
  \bibinfo{author}{\bibfnamefont{S.}~\bibnamefont{Lindenstruth}},
  \bibinfo{author}{\bibfnamefont{K.}~\bibnamefont{Huber}},
  \bibinfo{author}{\bibfnamefont{A.}~\bibnamefont{Jung}},
  \bibinfo{author}{\bibfnamefont{B.}~\bibnamefont{Starck}}, \emph{et~al.},
  \bibinfo{year}{1994}, \bibinfo{journal}{Nucl. Instrum. Methods A}
  \textbf{\bibinfo{volume}{337}}, \bibinfo{pages}{265}.

\bibitem[{\citenamefont{Greiner}(1965)}]{Greiner:1965}
\bibinfo{author}{\bibnamefont{Greiner}, \bibfnamefont{W.}},
  \bibinfo{year}{1965}, \bibinfo{journal}{Phys. Rev. Lett.}
  \textbf{\bibinfo{volume}{14}}, \bibinfo{pages}{599}.

\bibitem[{\citenamefont{Greiner}(1966)}]{Greiner:1966}
\bibinfo{author}{\bibnamefont{Greiner}, \bibfnamefont{W.}},
  \bibinfo{year}{1966}, \bibinfo{journal}{Nucl. Phys.}
  \textbf{\bibinfo{volume}{80}}, \bibinfo{pages}{417}.

\bibitem[{\citenamefont{Grundey} \emph{et~al.}(1981)\citenamefont{Grundey,
  Richter, Schrieder, Spamer, and Stock}}]{Grundey:1981}
\bibinfo{author}{\bibnamefont{Grundey}, \bibfnamefont{T.}},
  \bibinfo{author}{\bibfnamefont{A.}~\bibnamefont{Richter}},
  \bibinfo{author}{\bibfnamefont{G.}~\bibnamefont{Schrieder}},
  \bibinfo{author}{\bibfnamefont{E.}~\bibnamefont{Spamer}}, and
  \bibinfo{author}{\bibfnamefont{W.}~\bibnamefont{Stock}},
  \bibinfo{year}{1981}, \bibinfo{journal}{Nucl. Phys. A}
  \textbf{\bibinfo{volume}{357}}, \bibinfo{pages}{269}.

\bibitem[{\citenamefont{Moya~de Guerra and Zamick}(1993)}]{MoyadeGuerra:1993}
\bibinfo{author}{\bibnamefont{Moya~de Guerra}, \bibfnamefont{E.}}, and
  \bibinfo{author}{\bibfnamefont{L.}~\bibnamefont{Zamick}},
  \bibinfo{year}{1993}, \bibinfo{journal}{Phys. Rev. C}
  \textbf{\bibinfo{volume}{47}}, \bibinfo{pages}{2604}.

\bibitem[{\citenamefont{Gu\'ery-Odelin and
  Stringari}(1999)}]{Guery-Odelin:1999}
\bibinfo{author}{\bibnamefont{Gu\'ery-Odelin}, \bibfnamefont{D.}}, and
  \bibinfo{author}{\bibfnamefont{S.}~\bibnamefont{Stringari}},
  \bibinfo{year}{1999}, \bibinfo{journal}{Phys. Rev. Lett.}
  \textbf{\bibinfo{volume}{83}}, \bibinfo{pages}{4452}.

\bibitem[{\citenamefont{Guhr} \emph{et~al.}(1990)\citenamefont{Guhr, Diesener,
  Richter, de~Jager, de~Vries, and de~Witt~Huberts}}]{Guhr:1990}
\bibinfo{author}{\bibnamefont{Guhr}, \bibfnamefont{T.}},
  \bibinfo{author}{\bibfnamefont{H.}~\bibnamefont{Diesener}},
  \bibinfo{author}{\bibfnamefont{A.}~\bibnamefont{Richter}},
  \bibinfo{author}{\bibfnamefont{C.~W.} \bibnamefont{de~Jager}},
  \bibinfo{author}{\bibfnamefont{H.}~\bibnamefont{de~Vries}}, and
  \bibinfo{author}{\bibfnamefont{P.~K.~A.} \bibnamefont{de~Witt~Huberts}},
  \bibinfo{year}{1990}, \bibinfo{journal}{Z. Phys. A}
  \textbf{\bibinfo{volume}{336}}, \bibinfo{pages}{159}.

\bibitem[{\citenamefont{Guhr} \emph{et~al.}(1998)\citenamefont{Guhr,
  M\"{u}ller-Groeling, and Weidenm\"{u}ller}}]{Guhr:1998}
\bibinfo{author}{\bibnamefont{Guhr}, \bibfnamefont{T.}},
  \bibinfo{author}{\bibfnamefont{A.}~\bibnamefont{M\"{u}ller-Groeling}}, and
  \bibinfo{author}{\bibfnamefont{H.~A.} \bibnamefont{Weidenm\"{u}ller}},
  \bibinfo{year}{1998}, \bibinfo{journal}{Phys. Rep.}
  \textbf{\bibinfo{volume}{299}}, \bibinfo{pages}{189}.

\bibitem[{\citenamefont{Guliyev} \emph{et~al.}(2002)\citenamefont{Guliyev,
  Kuliev, von Neumann-Cosel, and Richter}}]{Guliyev:2002}
\bibinfo{author}{\bibnamefont{Guliyev}, \bibfnamefont{E.}},
  \bibinfo{author}{\bibfnamefont{A.~A.} \bibnamefont{Kuliev}},
  \bibinfo{author}{\bibfnamefont{P.}~\bibnamefont{von Neumann-Cosel}}, and
  \bibinfo{author}{\bibfnamefont{A.}~\bibnamefont{Richter}},
  \bibinfo{year}{2002}, \bibinfo{journal}{Phys. Lett. B}
  \textbf{\bibinfo{volume}{532}}, \bibinfo{pages}{173}.

\bibitem[{\citenamefont{Hagberg} \emph{et~al.}(1994)\citenamefont{Hagberg,
  Alexander, Neeson, Koslowsky, Ball, Dyck, Forster, Hardy, Leslie, Mak,
  Schmeing, and Towner}}]{Hagberg:1994}
\bibinfo{author}{\bibnamefont{Hagberg}, \bibfnamefont{E.}},
  \bibinfo{author}{\bibfnamefont{T.~K.} \bibnamefont{Alexander}},
  \bibinfo{author}{\bibfnamefont{I.}~\bibnamefont{Neeson}},
  \bibinfo{author}{\bibfnamefont{V.~T.} \bibnamefont{Koslowsky}},
  \bibinfo{author}{\bibfnamefont{G.~C.} \bibnamefont{Ball}},
  \bibinfo{author}{\bibfnamefont{G.~R.} \bibnamefont{Dyck}},
  \bibinfo{author}{\bibfnamefont{J.~S.} \bibnamefont{Forster}},
  \bibinfo{author}{\bibfnamefont{J.~C.} \bibnamefont{Hardy}},
  \bibinfo{author}{\bibfnamefont{J.~R.} \bibnamefont{Leslie}},
  \bibinfo{author}{\bibfnamefont{H.-B.} \bibnamefont{Mak}},
  \bibinfo{author}{\bibfnamefont{H.}~\bibnamefont{Schmeing}}, and
  \bibinfo{author}{\bibfnamefont{I.~S.} \bibnamefont{Towner}},
  \bibinfo{year}{1994}, \bibinfo{journal}{Nucl. Phys. A}
  \textbf{\bibinfo{volume}{571}}, \bibinfo{pages}{555}.

\bibitem[{\citenamefont{Halse}(1990)}]{Halse:1990}
\bibinfo{author}{\bibnamefont{Halse}, \bibfnamefont{P.}}, \bibinfo{year}{1990},
  \bibinfo{journal}{Phys. Rev. C} \textbf{\bibinfo{volume}{41}},
  \bibinfo{pages}{2340}.

\bibitem[{\citenamefont{Halse}(1991{\natexlab{a}})}]{Halse:1991b}
\bibinfo{author}{\bibnamefont{Halse}, \bibfnamefont{P.}},
  \bibinfo{year}{1991}{\natexlab{a}}, \bibinfo{journal}{Nucl. Phys. A}
  \textbf{\bibinfo{volume}{526}}, \bibinfo{pages}{152}.

\bibitem[{\citenamefont{Halse}(1991{\natexlab{b}})}]{Halse:1991a}
\bibinfo{author}{\bibnamefont{Halse}, \bibfnamefont{P.}},
  \bibinfo{year}{1991}{\natexlab{b}}, \bibinfo{journal}{Phys. Rev. C}
  \textbf{\bibinfo{volume}{44}}, \bibinfo{pages}{2467}.

\bibitem[{\citenamefont{Hamamoto}(1971)}]{Hamamoto:1971}
\bibinfo{author}{\bibnamefont{Hamamoto}, \bibfnamefont{I.}},
  \bibinfo{year}{1971}, \bibinfo{journal}{Nucl. Phys. A}
  \textbf{\bibinfo{volume}{177}}, \bibinfo{pages}{484}.

\bibitem[{\citenamefont{Hamamoto and {\AA}berg}(1984)}]{Hamamoto:1984}
\bibinfo{author}{\bibnamefont{Hamamoto}, \bibfnamefont{I.}}, and
  \bibinfo{author}{\bibfnamefont{S.}~\bibnamefont{{\AA}berg}},
  \bibinfo{year}{1984}, \bibinfo{journal}{Phys. Lett. B}
  \textbf{\bibinfo{volume}{145}}, \bibinfo{pages}{163}.

\bibitem[{\citenamefont{Hamamoto and Magnusson}(1991)}]{Hamamoto:1991}
\bibinfo{author}{\bibnamefont{Hamamoto}, \bibfnamefont{I.}}, and
  \bibinfo{author}{\bibfnamefont{C.}~\bibnamefont{Magnusson}},
  \bibinfo{year}{1991}, \bibinfo{journal}{Phys. Lett. B}
  \textbf{\bibinfo{volume}{260}}, \bibinfo{pages}{6}.

\bibitem[{\citenamefont{Hamamoto and Nazarewicz}(1992)}]{Hamamoto:1992}
\bibinfo{author}{\bibnamefont{Hamamoto}, \bibfnamefont{I.}}, and
  \bibinfo{author}{\bibfnamefont{W.}~\bibnamefont{Nazarewicz}},
  \bibinfo{year}{1992}, \bibinfo{journal}{Phys. Lett. B}
  \textbf{\bibinfo{volume}{297}}, \bibinfo{pages}{25}.

\bibitem[{\citenamefont{Hamamoto and Nazarewicz}(1994)}]{Hamamoto:1994}
\bibinfo{author}{\bibnamefont{Hamamoto}, \bibfnamefont{I.}}, and
  \bibinfo{author}{\bibfnamefont{W.}~\bibnamefont{Nazarewicz}},
  \bibinfo{year}{1994}, \bibinfo{journal}{Phys. Rev. C}
  \textbf{\bibinfo{volume}{49}}, \bibinfo{pages}{2489}.

\bibitem[{\citenamefont{Hamamoto and Ronstr\"{o}m}(1987)}]{Hamamoto:1987}
\bibinfo{author}{\bibnamefont{Hamamoto}, \bibfnamefont{I.}}, and
  \bibinfo{author}{\bibfnamefont{C.}~\bibnamefont{Ronstr\"{o}m}},
  \bibinfo{year}{1987}, \bibinfo{journal}{Phys. Lett. B}
  \textbf{\bibinfo{volume}{194}}, \bibinfo{pages}{6}.

\bibitem[{\citenamefont{Hamilton} \emph{et~al.}(1984)\citenamefont{Hamilton,
  Irb\"ack, and Elliott}}]{Hamilton:1984}
\bibinfo{author}{\bibnamefont{Hamilton}, \bibfnamefont{W.~D.}},
  \bibinfo{author}{\bibfnamefont{A.}~\bibnamefont{Irb\"ack}}, and
  \bibinfo{author}{\bibfnamefont{J.~P.} \bibnamefont{Elliott}},
  \bibinfo{year}{1984}, \bibinfo{journal}{Phys. Rev. Lett.}
  \textbf{\bibinfo{volume}{53}}, \bibinfo{pages}{2469}.

\bibitem[{\citenamefont{Haq} \emph{et~al.}(1982)\citenamefont{Haq, Pandey, and
  Bohigas}}]{Haq:1982}
\bibinfo{author}{\bibnamefont{Haq}, \bibfnamefont{R.~U.}},
  \bibinfo{author}{\bibfnamefont{A.}~\bibnamefont{Pandey}}, and
  \bibinfo{author}{\bibfnamefont{O.}~\bibnamefont{Bohigas}},
  \bibinfo{year}{1982}, \bibinfo{journal}{Phys. Rev. Lett.}
  \textbf{\bibinfo{volume}{48}}, \bibinfo{pages}{1086}.

\bibitem[{\citenamefont{Harakeh and van~der Woude}(2001)}]{Harakeh:2001}
\bibinfo{author}{\bibnamefont{Harakeh}, \bibfnamefont{M.}}, and
  \bibinfo{author}{\bibfnamefont{A.}~\bibnamefont{van~der Woude}},
  \bibinfo{year}{2001}, \emph{\bibinfo{title}{Giant Resonances: Fundamental
  High-Frequency Modes of Nuclear Excitation}} (\bibinfo{publisher}{Oxford
  University}, \bibinfo{address}{Oxford}).

\bibitem[{\citenamefont{Harney} \emph{et~al.}(1986)\citenamefont{Harney,
  Richter, and Weidenm\"uller}}]{Harney:1986}
\bibinfo{author}{\bibnamefont{Harney}, \bibfnamefont{H.~L.}},
  \bibinfo{author}{\bibfnamefont{A.}~\bibnamefont{Richter}}, and
  \bibinfo{author}{\bibfnamefont{H.~A.} \bibnamefont{Weidenm\"uller}},
  \bibinfo{year}{1986}, \bibinfo{journal}{Rev. Mod. Phys.}
  \textbf{\bibinfo{volume}{58}}, \bibinfo{pages}{607}.

\bibitem[{\citenamefont{Hartung} \emph{et~al.}(1989)\citenamefont{Hartung,
  Richter, Spamer, W\"ortche, Rangacharyulu, de~Jager, and
  de~Vries}}]{Hartung:1989}
\bibinfo{author}{\bibnamefont{Hartung}, \bibfnamefont{G.}},
  \bibinfo{author}{\bibfnamefont{A.}~\bibnamefont{Richter}},
  \bibinfo{author}{\bibfnamefont{E.}~\bibnamefont{Spamer}},
  \bibinfo{author}{\bibfnamefont{H.}~\bibnamefont{W\"ortche}},
  \bibinfo{author}{\bibfnamefont{C.}~\bibnamefont{Rangacharyulu}},
  \bibinfo{author}{\bibfnamefont{C.~W.} \bibnamefont{de~Jager}}, and
  \bibinfo{author}{\bibfnamefont{H.}~\bibnamefont{de~Vries}},
  \bibinfo{year}{1989}, \bibinfo{journal}{Phys. Lett. B}
  \textbf{\bibinfo{volume}{221}}, \bibinfo{pages}{109}.

\bibitem[{\citenamefont{Hasse and Myers}(1988)}]{Hasse:1988}
\bibinfo{author}{\bibnamefont{Hasse}, \bibfnamefont{R.~W.}}, and
  \bibinfo{author}{\bibfnamefont{W.~D.} \bibnamefont{Myers}},
  \bibinfo{year}{1988}, \emph{\bibinfo{title}{Geometrical Relationships of
  Macroscopic Nuclear Physics}} (\bibinfo{publisher}{Springer},
  \bibinfo{address}{Berlin}).

\bibitem[{\citenamefont{Hatada} \emph{et~al.}(2005)\citenamefont{Hatada,
  Hayakawa, and Palumbo}}]{Hatada:2005}
\bibinfo{author}{\bibnamefont{Hatada}, \bibfnamefont{K.}},
  \bibinfo{author}{\bibfnamefont{K.}~\bibnamefont{Hayakawa}}, and
  \bibinfo{author}{\bibfnamefont{F.}~\bibnamefont{Palumbo}},
  \bibinfo{year}{2005}, \bibinfo{journal}{Phys. Rev. B}
  \textbf{\bibinfo{volume}{71}}, \bibinfo{pages}{092402}.

\bibitem[{\citenamefont{Hatada}
  \emph{et~al.}(2009{\natexlab{a}})\citenamefont{Hatada, Hayakawa, and
  Palumbo}}]{Hatada:2009b}
\bibinfo{author}{\bibnamefont{Hatada}, \bibfnamefont{K.}},
  \bibinfo{author}{\bibfnamefont{K.}~\bibnamefont{Hayakawa}}, and
  \bibinfo{author}{\bibfnamefont{F.}~\bibnamefont{Palumbo}},
  \bibinfo{year}{2009}{\natexlab{a}}, \bibinfo{journal}{arXiv:0909.1422} .

\bibitem[{\citenamefont{Hatada}
  \emph{et~al.}(2009{\natexlab{b}})\citenamefont{Hatada, Hayakawa, and
  Palumbo}}]{Hatada:2009a}
\bibinfo{author}{\bibnamefont{Hatada}, \bibfnamefont{K.}},
  \bibinfo{author}{\bibfnamefont{K.}~\bibnamefont{Hayakawa}}, and
  \bibinfo{author}{\bibfnamefont{F.}~\bibnamefont{Palumbo}},
  \bibinfo{year}{2009}{\natexlab{b}}, \bibinfo{journal}{arXiv:0905.2100} .

\bibitem[{\citenamefont{Hatada} \emph{et~al.}(2010)\citenamefont{Hatada,
  Hayakawa, and Palumbo}}]{Hatada:2010}
\bibinfo{author}{\bibnamefont{Hatada}, \bibfnamefont{K.}},
  \bibinfo{author}{\bibfnamefont{K.}~\bibnamefont{Hayakawa}}, and
  \bibinfo{author}{\bibfnamefont{F.}~\bibnamefont{Palumbo}},
  \bibinfo{year}{2010}, \bibinfo{journal}{arXiv:1004.2220} .

\bibitem[{\citenamefont{Hecht and Adler}(1969)}]{Hecht:1969}
\bibinfo{author}{\bibnamefont{Hecht}, \bibfnamefont{K.~T.}}, and
  \bibinfo{author}{\bibfnamefont{A.}~\bibnamefont{Adler}},
  \bibinfo{year}{1969}, \bibinfo{journal}{Nucl. Phys. A}
  \textbf{\bibinfo{volume}{137}}, \bibinfo{pages}{129}.

\bibitem[{\citenamefont{de~Heer}(1993)}]{deHeer:1993}
\bibinfo{author}{\bibnamefont{de~Heer}, \bibfnamefont{W.~A.}},
  \bibinfo{year}{1993}, \bibinfo{journal}{Rev. Mod. Phys.}
  \textbf{\bibinfo{volume}{65}}, \bibinfo{pages}{611}.

\bibitem[{\citenamefont{de~Heer} \emph{et~al.}(1987)\citenamefont{de~Heer,
  Selby, Kresin, Masui, Vollmer, Chatelain, and Knight}}]{deHeer:1987}
\bibinfo{author}{\bibnamefont{de~Heer}, \bibfnamefont{W.~A.}},
  \bibinfo{author}{\bibfnamefont{K.}~\bibnamefont{Selby}},
  \bibinfo{author}{\bibfnamefont{V.}~\bibnamefont{Kresin}},
  \bibinfo{author}{\bibfnamefont{J.}~\bibnamefont{Masui}},
  \bibinfo{author}{\bibfnamefont{M.}~\bibnamefont{Vollmer}},
  \bibinfo{author}{\bibfnamefont{A.}~\bibnamefont{Chatelain}}, and
  \bibinfo{author}{\bibfnamefont{W.~D.} \bibnamefont{Knight}},
  \bibinfo{year}{1987}, \bibinfo{journal}{Phys. Rev. Lett.}
  \textbf{\bibinfo{volume}{59}}, \bibinfo{pages}{1805}.

\bibitem[{\citenamefont{Heil} \emph{et~al.}(1988)\citenamefont{Heil, Pitz,
  Berg, Kneissl, Hummel, Kilgus, Bohle, Richter, Wesselborg, and von
  Brentano}}]{Heil:1988}
\bibinfo{author}{\bibnamefont{Heil}, \bibfnamefont{R.~D.}},
  \bibinfo{author}{\bibfnamefont{H.~H.} \bibnamefont{Pitz}},
  \bibinfo{author}{\bibfnamefont{U.~E.~P.} \bibnamefont{Berg}},
  \bibinfo{author}{\bibfnamefont{U.}~\bibnamefont{Kneissl}},
  \bibinfo{author}{\bibfnamefont{K.~D.} \bibnamefont{Hummel}},
  \bibinfo{author}{\bibfnamefont{G.}~\bibnamefont{Kilgus}},
  \bibinfo{author}{\bibfnamefont{D.}~\bibnamefont{Bohle}},
  \bibinfo{author}{\bibfnamefont{A.}~\bibnamefont{Richter}},
  \bibinfo{author}{\bibfnamefont{C.}~\bibnamefont{Wesselborg}}, and
  \bibinfo{author}{\bibfnamefont{P.}~\bibnamefont{von Brentano}},
  \bibinfo{year}{1988}, \bibinfo{journal}{Nuc. Phys. A}
  \textbf{\bibinfo{volume}{476}}, \bibinfo{pages}{39}.

\bibitem[{\citenamefont{Heyde}(1989)}]{Heyde:1989}
\bibinfo{author}{\bibnamefont{Heyde}, \bibfnamefont{K.}}, \bibinfo{year}{1989},
  \bibinfo{journal}{Int. J. Mod. Phys. A} \textbf{\bibinfo{volume}{4}},
  \bibinfo{pages}{2063}.

\bibitem[{\citenamefont{Heyde and De~Coster}(1991)}]{Heyde:1991}
\bibinfo{author}{\bibnamefont{Heyde}, \bibfnamefont{K.}}, and
  \bibinfo{author}{\bibfnamefont{C.}~\bibnamefont{De~Coster}},
  \bibinfo{year}{1991}, \bibinfo{journal}{Phys. Rev. C}
  \textbf{\bibinfo{volume}{44}}, \bibinfo{pages}{R2262}.

\bibitem[{\citenamefont{Heyde and De~Coster}(1993)}]{Heyde:1993a}
\bibinfo{author}{\bibnamefont{Heyde}, \bibfnamefont{K.}}, and
  \bibinfo{author}{\bibfnamefont{C.}~\bibnamefont{De~Coster}},
  \bibinfo{year}{1993}, \bibinfo{journal}{Phys. Lett. B}
  \textbf{\bibinfo{volume}{305}}, \bibinfo{pages}{322}.

\bibitem[{\citenamefont{Heyde} \emph{et~al.}(1994)\citenamefont{Heyde,
  De~Coster, and Ooms}}]{Heyde:1994}
\bibinfo{author}{\bibnamefont{Heyde}, \bibfnamefont{K.}},
  \bibinfo{author}{\bibfnamefont{C.}~\bibnamefont{De~Coster}}, and
  \bibinfo{author}{\bibfnamefont{D.}~\bibnamefont{Ooms}}, \bibinfo{year}{1994},
  \bibinfo{journal}{Phys. Rev. C} \textbf{\bibinfo{volume}{49}},
  \bibinfo{pages}{156}.

\bibitem[{\citenamefont{Heyde} \emph{et~al.}(1993)\citenamefont{Heyde,
  De~Coster, Ooms, and Richter}}]{Heyde:1993c}
\bibinfo{author}{\bibnamefont{Heyde}, \bibfnamefont{K.}},
  \bibinfo{author}{\bibfnamefont{C.}~\bibnamefont{De~Coster}},
  \bibinfo{author}{\bibfnamefont{D.}~\bibnamefont{Ooms}}, and
  \bibinfo{author}{\bibfnamefont{A.}~\bibnamefont{Richter}},
  \bibinfo{year}{1993}, \bibinfo{journal}{Phys. Lett. B}
  \textbf{\bibinfo{volume}{312}}, \bibinfo{pages}{267}.

\bibitem[{\citenamefont{Heyde} \emph{et~al.}(1992)\citenamefont{Heyde,
  De~Coster, Richter, and W\"ortche}}]{Heyde:1992}
\bibinfo{author}{\bibnamefont{Heyde}, \bibfnamefont{K.}},
  \bibinfo{author}{\bibfnamefont{C.}~\bibnamefont{De~Coster}},
  \bibinfo{author}{\bibfnamefont{A.}~\bibnamefont{Richter}}, and
  \bibinfo{author}{\bibfnamefont{H.~J.} \bibnamefont{W\"ortche}},
  \bibinfo{year}{1992}, \bibinfo{journal}{Nucl. Phys. A}
  \textbf{\bibinfo{volume}{549}}, \bibinfo{pages}{103}.

\bibitem[{\citenamefont{Heyde} \emph{et~al.}(1996)\citenamefont{Heyde,
  De~Coster, Rombouts, and Freeman}}]{Heyde:1996}
\bibinfo{author}{\bibnamefont{Heyde}, \bibfnamefont{K.}},
  \bibinfo{author}{\bibfnamefont{C.}~\bibnamefont{De~Coster}},
  \bibinfo{author}{\bibfnamefont{S.}~\bibnamefont{Rombouts}}, and
  \bibinfo{author}{\bibfnamefont{S.~J.} \bibnamefont{Freeman}},
  \bibinfo{year}{1996}, \bibinfo{journal}{Nucl. Phys. A}
  \textbf{\bibinfo{volume}{596}}, \bibinfo{pages}{30}.

\bibitem[{\citenamefont{Heyde and Sau}(1984)}]{Heyde:1984}
\bibinfo{author}{\bibnamefont{Heyde}, \bibfnamefont{K.}}, and
  \bibinfo{author}{\bibfnamefont{J.}~\bibnamefont{Sau}}, \bibinfo{year}{1984},
  \bibinfo{journal}{Phys. Rev. C} \textbf{\bibinfo{volume}{30}},
  \bibinfo{pages}{1355}.

\bibitem[{\citenamefont{Heyde and Sau}(1986)}]{Heyde:1986}
\bibinfo{author}{\bibnamefont{Heyde}, \bibfnamefont{K.}}, and
  \bibinfo{author}{\bibfnamefont{J.}~\bibnamefont{Sau}}, \bibinfo{year}{1986},
  \bibinfo{journal}{Phys. Rev. C} \textbf{\bibinfo{volume}{33}},
  \bibinfo{pages}{1050}.

\bibitem[{\citenamefont{Heyde} \emph{et~al.}(1977)\citenamefont{Heyde,
  Waroquier, Van~Isacker, and Vincx}}]{Heyde:1977}
\bibinfo{author}{\bibnamefont{Heyde}, \bibfnamefont{K.}},
  \bibinfo{author}{\bibfnamefont{M.}~\bibnamefont{Waroquier}},
  \bibinfo{author}{\bibfnamefont{P.}~\bibnamefont{Van~Isacker}}, and
  \bibinfo{author}{\bibfnamefont{H.}~\bibnamefont{Vincx}},
  \bibinfo{year}{1977}, \bibinfo{journal}{Phys. Rev. C}
  \textbf{\bibinfo{volume}{16}}, \bibinfo{pages}{489}.

\bibitem[{\citenamefont{Hicks} \emph{et~al.}(1998)\citenamefont{Hicks, Davoren,
  Faulkner, and Vanhoy}}]{Hicks:1998}
\bibinfo{author}{\bibnamefont{Hicks}, \bibfnamefont{S.~F.}},
  \bibinfo{author}{\bibfnamefont{C.~M.} \bibnamefont{Davoren}},
  \bibinfo{author}{\bibfnamefont{W.~M.} \bibnamefont{Faulkner}}, and
  \bibinfo{author}{\bibfnamefont{J.~R.} \bibnamefont{Vanhoy}},
  \bibinfo{year}{1998}, \bibinfo{journal}{Phys. Rev. C}
  \textbf{\bibinfo{volume}{57}}, \bibinfo{pages}{2264}.

\bibitem[{\citenamefont{Hicks} \emph{et~al.}(2008)\citenamefont{Hicks, Vanhoy,
  and Yates}}]{Hicks:2008}
\bibinfo{author}{\bibnamefont{Hicks}, \bibfnamefont{S.~F.}},
  \bibinfo{author}{\bibfnamefont{J.~R.} \bibnamefont{Vanhoy}}, and
  \bibinfo{author}{\bibfnamefont{S.~W.} \bibnamefont{Yates}},
  \bibinfo{year}{2008}, \bibinfo{journal}{Phys. Rev. C}
  \textbf{\bibinfo{volume}{78}}, \bibinfo{eid}{054320}.

\bibitem[{\citenamefont{Hilton}(1976)}]{Hilton:1976}
\bibinfo{author}{\bibnamefont{Hilton}, \bibfnamefont{R.~R.}},
  \bibinfo{year}{1976}, \bibinfo{note}{talk at International Conference on
  Nuclear Structure, JINR Dubna, unpublished}.

\bibitem[{\citenamefont{Hilton}(1995)}]{Hilton:1995}
\bibinfo{author}{\bibnamefont{Hilton}, \bibfnamefont{R.~R.}},
  \bibinfo{year}{1995}, in \emph{\bibinfo{booktitle}{Proceedings of the
  International Conference on Perspectives of Nuclear Physics in the Late
  Nineties}}, edited by
  \bibinfo{editor}{\bibfnamefont{N.}~\bibnamefont{Dinh~Dang}},
  \bibinfo{editor}{\bibfnamefont{D.}~\bibnamefont{Feng}},
  \bibinfo{editor}{\bibfnamefont{N.~V.} \bibnamefont{Giai}}, and
  \bibinfo{editor}{\bibfnamefont{N.~D.} \bibnamefont{Tu}}
  (\bibinfo{publisher}{World Scientific}, \bibinfo{address}{Singapore}),
  p.~\bibinfo{pages}{86}.

\bibitem[{\citenamefont{Hilton} \emph{et~al.}(1993)\citenamefont{Hilton,
  H\"ohenberger, and Mang}}]{Hilton:1993}
\bibinfo{author}{\bibnamefont{Hilton}, \bibfnamefont{R.~R.}},
  \bibinfo{author}{\bibfnamefont{W.}~\bibnamefont{H\"ohenberger}}, and
  \bibinfo{author}{\bibfnamefont{H.~J.} \bibnamefont{Mang}},
  \bibinfo{year}{1993}, \bibinfo{journal}{Phys. Rev. C}
  \textbf{\bibinfo{volume}{47}}, \bibinfo{pages}{602}.

\bibitem[{\citenamefont{Hilton} \emph{et~al.}(1998)\citenamefont{Hilton,
  H\"{o}henberger, and Ring}}]{Hilton:1998}
\bibinfo{author}{\bibnamefont{Hilton}, \bibfnamefont{R.~R.}},
  \bibinfo{author}{\bibfnamefont{W.}~\bibnamefont{H\"{o}henberger}}, and
  \bibinfo{author}{\bibfnamefont{P.}~\bibnamefont{Ring}}, \bibinfo{year}{1998},
  \bibinfo{journal}{Eur. Phys. J. A} \textbf{\bibinfo{volume}{1}},
  \bibinfo{pages}{257}.

\bibitem[{\citenamefont{Hino} \emph{et~al.}(1988)\citenamefont{Hino, Muto, and
  Oda}}]{Hino:1988}
\bibinfo{author}{\bibnamefont{Hino}, \bibfnamefont{M.}},
  \bibinfo{author}{\bibfnamefont{K.}~\bibnamefont{Muto}}, and
  \bibinfo{author}{\bibfnamefont{T.}~\bibnamefont{Oda}}, \bibinfo{year}{1988},
  \bibinfo{journal}{Phys. Rev. C} \textbf{\bibinfo{volume}{37}},
  \bibinfo{pages}{1328}.

\bibitem[{\citenamefont{Hirose and Wingreen}(1999)}]{Hirose:1999}
\bibinfo{author}{\bibnamefont{Hirose}, \bibfnamefont{K.}}, and
  \bibinfo{author}{\bibfnamefont{N.~S.} \bibnamefont{Wingreen}},
  \bibinfo{year}{1999}, \bibinfo{journal}{Phys. Rev. B}
  \textbf{\bibinfo{volume}{59}}, \bibinfo{pages}{4604}.

\bibitem[{\citenamefont{Hix} \emph{et~al.}(2003)\citenamefont{Hix, Mezzacappa,
  Messer, and Bruenn}}]{Hix:2003}
\bibinfo{author}{\bibnamefont{Hix}, \bibfnamefont{W.~R.}},
  \bibinfo{author}{\bibfnamefont{A.}~\bibnamefont{Mezzacappa}},
  \bibinfo{author}{\bibfnamefont{O.~E.~B.} \bibnamefont{Messer}}, and
  \bibinfo{author}{\bibfnamefont{S.~W.} \bibnamefont{Bruenn}},
  \bibinfo{year}{2003}, \bibinfo{journal}{J. Phys. G}
  \textbf{\bibinfo{volume}{29}}, \bibinfo{pages}{2523}.

\bibitem[{\citenamefont{Hofmann} \emph{et~al.}(2007)\citenamefont{Hofmann,
  B\"{a}umer, van~den Berg, Frekers, Hannen, Harakeh, de~Huu, Kalmykov, von
  Neumann-Cosel, Ponomarev, Rakers, Reitz} \emph{et~al.}}]{Hofmann:2007}
\bibinfo{author}{\bibnamefont{Hofmann}, \bibfnamefont{F.}},
  \bibinfo{author}{\bibfnamefont{C.}~\bibnamefont{B\"{a}umer}},
  \bibinfo{author}{\bibfnamefont{A.~M.} \bibnamefont{van~den Berg}},
  \bibinfo{author}{\bibfnamefont{D.}~\bibnamefont{Frekers}},
  \bibinfo{author}{\bibfnamefont{V.~M.} \bibnamefont{Hannen}},
  \bibinfo{author}{\bibfnamefont{M.~N.} \bibnamefont{Harakeh}},
  \bibinfo{author}{\bibfnamefont{M.}~\bibnamefont{de~Huu}},
  \bibinfo{author}{\bibfnamefont{Y.}~\bibnamefont{Kalmykov}},
  \bibinfo{author}{\bibfnamefont{P.}~\bibnamefont{von Neumann-Cosel}},
  \bibinfo{author}{\bibfnamefont{V.~Y.} \bibnamefont{Ponomarev}},
  \bibinfo{author}{\bibfnamefont{S.}~\bibnamefont{Rakers}},
  \bibinfo{author}{\bibfnamefont{B.}~\bibnamefont{Reitz}}, \emph{et~al.},
  \bibinfo{year}{2007}, \bibinfo{journal}{Phys. Rev. C}
  \textbf{\bibinfo{volume}{76}}, \bibinfo{eid}{014314}.

\bibitem[{\citenamefont{Hofmann} \emph{et~al.}(2002)\citenamefont{Hofmann, von
  Neumann-Cosel, Neumeyer, Rangacharyulu, Reitz, Richter, Schrieder, Sober,
  Fagg, and Brown}}]{Hofmann:2002}
\bibinfo{author}{\bibnamefont{Hofmann}, \bibfnamefont{F.}},
  \bibinfo{author}{\bibfnamefont{P.}~\bibnamefont{von Neumann-Cosel}},
  \bibinfo{author}{\bibfnamefont{F.}~\bibnamefont{Neumeyer}},
  \bibinfo{author}{\bibfnamefont{C.}~\bibnamefont{Rangacharyulu}},
  \bibinfo{author}{\bibfnamefont{B.}~\bibnamefont{Reitz}},
  \bibinfo{author}{\bibfnamefont{A.}~\bibnamefont{Richter}},
  \bibinfo{author}{\bibfnamefont{G.}~\bibnamefont{Schrieder}},
  \bibinfo{author}{\bibfnamefont{D.~I.} \bibnamefont{Sober}},
  \bibinfo{author}{\bibfnamefont{L.~W.} \bibnamefont{Fagg}}, and
  \bibinfo{author}{\bibfnamefont{B.~A.} \bibnamefont{Brown}},
  \bibinfo{year}{2002}, \bibinfo{journal}{Phys. Rev. C}
  \textbf{\bibinfo{volume}{65}}, \bibinfo{pages}{024311}.

\bibitem[{\citenamefont{Holt} \emph{et~al.}(2007)\citenamefont{Holt, Pietralla,
  Holt, Kuo, and Rainovski}}]{Holt:2007}
\bibinfo{author}{\bibnamefont{Holt}, \bibfnamefont{J.~D.}},
  \bibinfo{author}{\bibfnamefont{N.}~\bibnamefont{Pietralla}},
  \bibinfo{author}{\bibfnamefont{J.~W.} \bibnamefont{Holt}},
  \bibinfo{author}{\bibfnamefont{T.~T.~S.} \bibnamefont{Kuo}}, and
  \bibinfo{author}{\bibfnamefont{G.}~\bibnamefont{Rainovski}},
  \bibinfo{year}{2007}, \bibinfo{journal}{Phys. Rev. C}
  \textbf{\bibinfo{volume}{76}}, \bibinfo{eid}{034325}.

\bibitem[{\citenamefont{Holzwarth and Eckart}(1977)}]{Holzwarth:1977}
\bibinfo{author}{\bibnamefont{Holzwarth}, \bibfnamefont{D.}}, and
  \bibinfo{author}{\bibfnamefont{G.}~\bibnamefont{Eckart}},
  \bibinfo{year}{1977}, \bibinfo{journal}{Z. Phys. A}
  \textbf{\bibinfo{volume}{283}}, \bibinfo{pages}{219}.

\bibitem[{\citenamefont{Holzwarth and Eckart}(1979)}]{Holzwarth:1979}
\bibinfo{author}{\bibnamefont{Holzwarth}, \bibfnamefont{D.}}, and
  \bibinfo{author}{\bibfnamefont{G.}~\bibnamefont{Eckart}},
  \bibinfo{year}{1979}, \bibinfo{journal}{Nucl. Phys. A}
  \textbf{\bibinfo{volume}{325}}, \bibinfo{pages}{1}.

\bibitem[{\citenamefont{Honma} \emph{et~al.}(1995)\citenamefont{Honma,
  Mizusaki, and Otsuka}}]{Honma:1995}
\bibinfo{author}{\bibnamefont{Honma}, \bibfnamefont{M.}},
  \bibinfo{author}{\bibfnamefont{T.}~\bibnamefont{Mizusaki}}, and
  \bibinfo{author}{\bibfnamefont{T.}~\bibnamefont{Otsuka}},
  \bibinfo{year}{1995}, \bibinfo{journal}{Phys. Rev. Lett.}
  \textbf{\bibinfo{volume}{75}}, \bibinfo{pages}{1284}.

\bibitem[{\citenamefont{Honma} \emph{et~al.}(1996)\citenamefont{Honma,
  Mizusaki, and Otsuka}}]{Honma:1996}
\bibinfo{author}{\bibnamefont{Honma}, \bibfnamefont{M.}},
  \bibinfo{author}{\bibfnamefont{T.}~\bibnamefont{Mizusaki}}, and
  \bibinfo{author}{\bibfnamefont{T.}~\bibnamefont{Otsuka}},
  \bibinfo{year}{1996}, \bibinfo{journal}{Phys. Rev. Lett.}
  \textbf{\bibinfo{volume}{77}}, \bibinfo{pages}{3315}.

\bibitem[{\citenamefont{Honma} \emph{et~al.}(2004)\citenamefont{Honma, Otsuka,
  Brown, and Mizusaki}}]{Honma:1994}
\bibinfo{author}{\bibnamefont{Honma}, \bibfnamefont{M.}},
  \bibinfo{author}{\bibfnamefont{T.}~\bibnamefont{Otsuka}},
  \bibinfo{author}{\bibfnamefont{B.~A.} \bibnamefont{Brown}}, and
  \bibinfo{author}{\bibfnamefont{T.}~\bibnamefont{Mizusaki}},
  \bibinfo{year}{2004}, \bibinfo{journal}{Phys. Rev. C}
  \textbf{\bibinfo{volume}{69}}, \bibinfo{pages}{034335}.

\bibitem[{\citenamefont{Huxel} \emph{et~al.}(1999)\citenamefont{Huxel, von
  Brentano, Eberth, Enders, Herzberg, von Neumann-Cosel, Nicolay, Pietralla,
  Prade, Rangacharyulu, Reif, Richter} \emph{et~al.}}]{Huxel:1999}
\bibinfo{author}{\bibnamefont{Huxel}, \bibfnamefont{A.}},
  \bibinfo{author}{\bibfnamefont{P.}~\bibnamefont{von Brentano}},
  \bibinfo{author}{\bibfnamefont{J.}~\bibnamefont{Eberth}},
  \bibinfo{author}{\bibfnamefont{J.}~\bibnamefont{Enders}},
  \bibinfo{author}{\bibfnamefont{R.-D.} \bibnamefont{Herzberg}},
  \bibinfo{author}{\bibfnamefont{P.}~\bibnamefont{von Neumann-Cosel}},
  \bibinfo{author}{\bibfnamefont{N.}~\bibnamefont{Nicolay}},
  \bibinfo{author}{\bibfnamefont{N.}~\bibnamefont{Pietralla}},
  \bibinfo{author}{\bibfnamefont{H.}~\bibnamefont{Prade}},
  \bibinfo{author}{\bibfnamefont{C.}~\bibnamefont{Rangacharyulu}},
  \bibinfo{author}{\bibfnamefont{J.}~\bibnamefont{Reif}},
  \bibinfo{author}{\bibfnamefont{A.}~\bibnamefont{Richter}}, \emph{et~al.},
  \bibinfo{year}{1999}, \bibinfo{journal}{Nucl. Phys. A}
  \textbf{\bibinfo{volume}{645}}, \bibinfo{pages}{239}.

\bibitem[{\citenamefont{Huxel} \emph{et~al.}(1992)\citenamefont{Huxel, Ahner,
  Diesener, von Neumann-Cosel, Rangacharyulu, Richter, Spieler, Ziegler,
  De~Coster, and Heyde}}]{Huxel:1992}
\bibinfo{author}{\bibnamefont{Huxel}, \bibfnamefont{N.}},
  \bibinfo{author}{\bibfnamefont{W.}~\bibnamefont{Ahner}},
  \bibinfo{author}{\bibfnamefont{H.}~\bibnamefont{Diesener}},
  \bibinfo{author}{\bibfnamefont{P.}~\bibnamefont{von Neumann-Cosel}},
  \bibinfo{author}{\bibfnamefont{C.}~\bibnamefont{Rangacharyulu}},
  \bibinfo{author}{\bibfnamefont{A.}~\bibnamefont{Richter}},
  \bibinfo{author}{\bibfnamefont{C.}~\bibnamefont{Spieler}},
  \bibinfo{author}{\bibfnamefont{W.}~\bibnamefont{Ziegler}},
  \bibinfo{author}{\bibfnamefont{C.}~\bibnamefont{De~Coster}}, and
  \bibinfo{author}{\bibfnamefont{K.}~\bibnamefont{Heyde}},
  \bibinfo{year}{1992}, \bibinfo{journal}{Nucl. Phys. A}
  \textbf{\bibinfo{volume}{539}}, \bibinfo{pages}{478}.

\bibitem[{\citenamefont{Iachello}(1981)}]{Iachello:1981}
\bibinfo{author}{\bibnamefont{Iachello}, \bibfnamefont{F.}},
  \bibinfo{year}{1981}, \bibinfo{journal}{Nucl. Phys. A}
  \textbf{\bibinfo{volume}{358}}, \bibinfo{pages}{89c}.

\bibitem[{\citenamefont{Iachello}(1984)}]{Iachello:1984}
\bibinfo{author}{\bibnamefont{Iachello}, \bibfnamefont{F.}},
  \bibinfo{year}{1984}, \bibinfo{journal}{Phys. Rev. Lett.}
  \textbf{\bibinfo{volume}{53}}, \bibinfo{pages}{1427}.

\bibitem[{\citenamefont{Iachello and Arima}(1987)}]{Iachello:1987}
\bibinfo{author}{\bibnamefont{Iachello}, \bibfnamefont{F.}}, and
  \bibinfo{author}{\bibfnamefont{A.}~\bibnamefont{Arima}},
  \bibinfo{year}{1987}, \emph{\bibinfo{title}{The Interacting Boson Model}}
  (\bibinfo{publisher}{Cambridge University}, \bibinfo{address}{New York}).

\bibitem[{\citenamefont{Iachello and Van~Isacker}(1991)}]{Iachello:1991}
\bibinfo{author}{\bibnamefont{Iachello}, \bibfnamefont{F.}}, and
  \bibinfo{author}{\bibfnamefont{P.}~\bibnamefont{Van~Isacker}},
  \bibinfo{year}{1991}, \emph{\bibinfo{title}{The Interacting Boson-Fermion
  Model}} (\bibinfo{publisher}{Cambridge University}, \bibinfo{address}{New
  York}).

\bibitem[{\citenamefont{Ichimura} \emph{et~al.}(2006)\citenamefont{Ichimura,
  Sakai, and Wakasa}}]{Ichimura:2006}
\bibinfo{author}{\bibnamefont{Ichimura}, \bibfnamefont{M.}},
  \bibinfo{author}{\bibfnamefont{H.}~\bibnamefont{Sakai}}, and
  \bibinfo{author}{\bibfnamefont{T.}~\bibnamefont{Wakasa}},
  \bibinfo{year}{2006}, \bibinfo{journal}{Prog. Part. Nucl. Phys.}
  \textbf{\bibinfo{volume}{56}}, \bibinfo{pages}{446}.

\bibitem[{\citenamefont{Ikeda and Shimano}(1993)}]{Ikeda:1993}
\bibinfo{author}{\bibnamefont{Ikeda}, \bibfnamefont{A.}}, and
  \bibinfo{author}{\bibfnamefont{T.}~\bibnamefont{Shimano}},
  \bibinfo{year}{1993}, \bibinfo{journal}{Nucl. Phys. A}
  \textbf{\bibinfo{volume}{577}}, \bibinfo{pages}{573c}.

\bibitem[{\citenamefont{Iwasaki and Hara}(1984)}]{Iwasaki:1984}
\bibinfo{author}{\bibnamefont{Iwasaki}, \bibfnamefont{S.}}, and
  \bibinfo{author}{\bibfnamefont{K.}~\bibnamefont{Hara}}, \bibinfo{year}{1984},
  \bibinfo{journal}{Phys. Lett. B} \textbf{\bibinfo{volume}{144}},
  \bibinfo{pages}{9}.

\bibitem[{\citenamefont{Jewell} \emph{et~al.}(1997)\citenamefont{Jewell,
  Cottle, Kemper, and Riley}}]{Jewell:1997}
\bibinfo{author}{\bibnamefont{Jewell}, \bibfnamefont{J.~K.}},
  \bibinfo{author}{\bibfnamefont{P.~D.} \bibnamefont{Cottle}},
  \bibinfo{author}{\bibfnamefont{K.~W.} \bibnamefont{Kemper}}, and
  \bibinfo{author}{\bibfnamefont{L.~A.} \bibnamefont{Riley}},
  \bibinfo{year}{1997}, \bibinfo{journal}{Phys. Rev. C}
  \textbf{\bibinfo{volume}{56}}, \bibinfo{pages}{2440}.

\bibitem[{\citenamefont{Kalmykov} \emph{et~al.}(2006)\citenamefont{Kalmykov,
  Adachi, Berg, Fujita, Fujita, Fujita, Hatanaka, Kamiya, Nakanishi, von
  Neumann-Cosel, Ponomarev, Richter} \emph{et~al.}}]{Kalmykov:2006}
\bibinfo{author}{\bibnamefont{Kalmykov}, \bibfnamefont{Y.}},
  \bibinfo{author}{\bibfnamefont{T.}~\bibnamefont{Adachi}},
  \bibinfo{author}{\bibfnamefont{G.~P.~A.} \bibnamefont{Berg}},
  \bibinfo{author}{\bibfnamefont{H.}~\bibnamefont{Fujita}},
  \bibinfo{author}{\bibfnamefont{K.}~\bibnamefont{Fujita}},
  \bibinfo{author}{\bibfnamefont{Y.}~\bibnamefont{Fujita}},
  \bibinfo{author}{\bibfnamefont{K.}~\bibnamefont{Hatanaka}},
  \bibinfo{author}{\bibfnamefont{J.}~\bibnamefont{Kamiya}},
  \bibinfo{author}{\bibfnamefont{K.}~\bibnamefont{Nakanishi}},
  \bibinfo{author}{\bibfnamefont{P.}~\bibnamefont{von Neumann-Cosel}},
  \bibinfo{author}{\bibfnamefont{V.~Y.} \bibnamefont{Ponomarev}},
  \bibinfo{author}{\bibfnamefont{A.}~\bibnamefont{Richter}}, \emph{et~al.},
  \bibinfo{year}{2006}, \bibinfo{journal}{Phys. Rev. Lett.}
  \textbf{\bibinfo{volume}{96}}, \bibinfo{eid}{012502}.

\bibitem[{\citenamefont{Kasten} \emph{et~al.}(1989)\citenamefont{Kasten, Heil,
  von Brentano, Butler, Hoblit, Kneissl, Lindenstruth, M\"uller, Pitz, Rose,
  Scharfe, Schumacher} \emph{et~al.}}]{Kasten:1989}
\bibinfo{author}{\bibnamefont{Kasten}, \bibfnamefont{B.}},
  \bibinfo{author}{\bibfnamefont{R.~D.} \bibnamefont{Heil}},
  \bibinfo{author}{\bibfnamefont{P.}~\bibnamefont{von Brentano}},
  \bibinfo{author}{\bibfnamefont{P.~A.} \bibnamefont{Butler}},
  \bibinfo{author}{\bibfnamefont{S.~D.} \bibnamefont{Hoblit}},
  \bibinfo{author}{\bibfnamefont{U.}~\bibnamefont{Kneissl}},
  \bibinfo{author}{\bibfnamefont{S.}~\bibnamefont{Lindenstruth}},
  \bibinfo{author}{\bibfnamefont{G.}~\bibnamefont{M\"uller}},
  \bibinfo{author}{\bibfnamefont{H.~H.} \bibnamefont{Pitz}},
  \bibinfo{author}{\bibfnamefont{K.~W.} \bibnamefont{Rose}},
  \bibinfo{author}{\bibfnamefont{W.}~\bibnamefont{Scharfe}},
  \bibinfo{author}{\bibfnamefont{M.}~\bibnamefont{Schumacher}}, \emph{et~al.},
  \bibinfo{year}{1989}, \bibinfo{journal}{Phys. Rev. Lett.}
  \textbf{\bibinfo{volume}{63}}, \bibinfo{pages}{609}.

\bibitem[{\citenamefont{Klein} \emph{et~al.}(2002)\citenamefont{Klein,
  Lisetskiy, Pietralla, Fransen, Gade, and von Brentano}}]{Klein:2002}
\bibinfo{author}{\bibnamefont{Klein}, \bibfnamefont{H.}},
  \bibinfo{author}{\bibfnamefont{A.~F.} \bibnamefont{Lisetskiy}},
  \bibinfo{author}{\bibfnamefont{N.}~\bibnamefont{Pietralla}},
  \bibinfo{author}{\bibfnamefont{C.}~\bibnamefont{Fransen}},
  \bibinfo{author}{\bibfnamefont{A.}~\bibnamefont{Gade}}, and
  \bibinfo{author}{\bibfnamefont{P.}~\bibnamefont{von Brentano}},
  \bibinfo{year}{2002}, \bibinfo{journal}{Phys. Rev. C}
  \textbf{\bibinfo{volume}{65}}, \bibinfo{pages}{044315}.

\bibitem[{\citenamefont{Kleinig} \emph{et~al.}(1998)\citenamefont{Kleinig,
  Nesterenkov, Reinhard, and Serra}}]{Kleinig:1998}
\bibinfo{author}{\bibnamefont{Kleinig}, \bibfnamefont{W.}},
  \bibinfo{author}{\bibfnamefont{V.~O.} \bibnamefont{Nesterenkov}},
  \bibinfo{author}{\bibfnamefont{P.-G.} \bibnamefont{Reinhard}}, and
  \bibinfo{author}{\bibfnamefont{L.}~\bibnamefont{Serra}},
  \bibinfo{year}{1998}, \bibinfo{journal}{Eur. Phys. J. D}
  \textbf{\bibinfo{volume}{4}}, \bibinfo{pages}{343}.

\bibitem[{\citenamefont{Kneissl} \emph{et~al.}(2006)\citenamefont{Kneissl,
  Pietralla, and Zilges}}]{Kneissl:2006}
\bibinfo{author}{\bibnamefont{Kneissl}, \bibfnamefont{U.}},
  \bibinfo{author}{\bibfnamefont{N.}~\bibnamefont{Pietralla}}, and
  \bibinfo{author}{\bibfnamefont{A.}~\bibnamefont{Zilges}},
  \bibinfo{year}{2006}, \bibinfo{journal}{J. Phys. G}
  \textbf{\bibinfo{volume}{32}}, \bibinfo{pages}{R217}.

\bibitem[{\citenamefont{Kneissl} \emph{et~al.}(1996)\citenamefont{Kneissl,
  Pitz, and Zilges}}]{Kneissl:1996}
\bibinfo{author}{\bibnamefont{Kneissl}, \bibfnamefont{U.}},
  \bibinfo{author}{\bibfnamefont{H.~H.} \bibnamefont{Pitz}}, and
  \bibinfo{author}{\bibfnamefont{A.}~\bibnamefont{Zilges}},
  \bibinfo{year}{1996}, \bibinfo{journal}{Prog. Part. Nucl. Phys.}
  \textbf{\bibinfo{volume}{37}}, \bibinfo{pages}{349}.

\bibitem[{\citenamefont{Knight} \emph{et~al.}(1984)\citenamefont{Knight,
  Clemenger, de~Heer, Saunders, Chou, and Cohen}}]{Knight:1984}
\bibinfo{author}{\bibnamefont{Knight}, \bibfnamefont{W.~D.}},
  \bibinfo{author}{\bibfnamefont{K.}~\bibnamefont{Clemenger}},
  \bibinfo{author}{\bibfnamefont{W.~A.} \bibnamefont{de~Heer}},
  \bibinfo{author}{\bibfnamefont{W.~A.} \bibnamefont{Saunders}},
  \bibinfo{author}{\bibfnamefont{M.~Y.} \bibnamefont{Chou}}, and
  \bibinfo{author}{\bibfnamefont{M.~L.} \bibnamefont{Cohen}},
  \bibinfo{year}{1984}, \bibinfo{journal}{Phys. Rev. Lett.}
  \textbf{\bibinfo{volume}{52}}, \bibinfo{pages}{2141}.

\bibitem[{\citenamefont{Kohstall} \emph{et~al.}(2005)\citenamefont{Kohstall,
  Belic, von Brentano, Fransen, Gade, Herzberg, Jolie, Kneissl, Linnemann,
  Nord, Pietralla, Pitz} \emph{et~al.}}]{Kohstall:2005}
\bibinfo{author}{\bibnamefont{Kohstall}, \bibfnamefont{C.}},
  \bibinfo{author}{\bibfnamefont{D.}~\bibnamefont{Belic}},
  \bibinfo{author}{\bibfnamefont{P.}~\bibnamefont{von Brentano}},
  \bibinfo{author}{\bibfnamefont{C.}~\bibnamefont{Fransen}},
  \bibinfo{author}{\bibfnamefont{A.}~\bibnamefont{Gade}},
  \bibinfo{author}{\bibfnamefont{R.-D.} \bibnamefont{Herzberg}},
  \bibinfo{author}{\bibfnamefont{J.}~\bibnamefont{Jolie}},
  \bibinfo{author}{\bibfnamefont{U.}~\bibnamefont{Kneissl}},
  \bibinfo{author}{\bibfnamefont{A.}~\bibnamefont{Linnemann}},
  \bibinfo{author}{\bibfnamefont{A.}~\bibnamefont{Nord}},
  \bibinfo{author}{\bibfnamefont{N.}~\bibnamefont{Pietralla}},
  \bibinfo{author}{\bibfnamefont{H.~H.} \bibnamefont{Pitz}}, \emph{et~al.},
  \bibinfo{year}{2005}, \bibinfo{journal}{Phys. Rev. C}
  \textbf{\bibinfo{volume}{72}}, \bibinfo{eid}{034302}.

\bibitem[{\citenamefont{Koskinen} \emph{et~al.}(1997)\citenamefont{Koskinen,
  Manninen, and Reimann}}]{Koskinen:1997}
\bibinfo{author}{\bibnamefont{Koskinen}, \bibfnamefont{M.}},
  \bibinfo{author}{\bibfnamefont{M.}~\bibnamefont{Manninen}}, and
  \bibinfo{author}{\bibfnamefont{S.~M.} \bibnamefont{Reimann}},
  \bibinfo{year}{1997}, \bibinfo{journal}{Phys. Rev. Lett.}
  \textbf{\bibinfo{volume}{79}}, \bibinfo{pages}{1389}.

\bibitem[{\citenamefont{Kurasawa and Suzuki}(1984)}]{Kurasawa:1984}
\bibinfo{author}{\bibnamefont{Kurasawa}, \bibfnamefont{H.}}, and
  \bibinfo{author}{\bibfnamefont{T.}~\bibnamefont{Suzuki}},
  \bibinfo{year}{1984}, \bibinfo{journal}{Phys. Lett. B}
  \textbf{\bibinfo{volume}{144}}, \bibinfo{pages}{151}.

\bibitem[{\citenamefont{Kuyucak and Stuchbery}(1995)}]{Kuyucak:1995}
\bibinfo{author}{\bibnamefont{Kuyucak}, \bibfnamefont{S.}}, and
  \bibinfo{author}{\bibfnamefont{A.~E.} \bibnamefont{Stuchbery}},
  \bibinfo{year}{1995}, \bibinfo{journal}{Phys. Lett. B}
  \textbf{\bibinfo{volume}{348}}, \bibinfo{pages}{315}.

\bibitem[{\citenamefont{van~der Laan} \emph{et~al.}(2008)\citenamefont{van~der
  Laan, Arenholz, Schmehl, and Schlom}}]{vanderLaan:2008}
\bibinfo{author}{\bibnamefont{van~der Laan}, \bibfnamefont{G.}},
  \bibinfo{author}{\bibfnamefont{E.}~\bibnamefont{Arenholz}},
  \bibinfo{author}{\bibfnamefont{A.}~\bibnamefont{Schmehl}}, and
  \bibinfo{author}{\bibfnamefont{D.~G.} \bibnamefont{Schlom}},
  \bibinfo{year}{2008}, \bibinfo{journal}{Phys. Rev. Lett.}
  \textbf{\bibinfo{volume}{100}}, \bibinfo{pages}{067403}.

\bibitem[{\citenamefont{Langanke and Mart\'inez-Pinedo}(2003)}]{Langanke:2003}
\bibinfo{author}{\bibnamefont{Langanke}, \bibfnamefont{K.}}, and
  \bibinfo{author}{\bibfnamefont{G.}~\bibnamefont{Mart\'inez-Pinedo}},
  \bibinfo{year}{2003}, \bibinfo{journal}{Rev. Mod. Phys.}
  \textbf{\bibinfo{volume}{75}}, \bibinfo{pages}{819}.

\bibitem[{\citenamefont{Langanke} \emph{et~al.}(2008)\citenamefont{Langanke,
  Mart\'{\i}nez-Pinedo, M\"{u}ller, Janka, Marek, Hix, Juodagalvis, and
  Sampaio}}]{Langanke:2008}
\bibinfo{author}{\bibnamefont{Langanke}, \bibfnamefont{K.}},
  \bibinfo{author}{\bibfnamefont{G.}~\bibnamefont{Mart\'{\i}nez-Pinedo}},
  \bibinfo{author}{\bibfnamefont{B.}~\bibnamefont{M\"{u}ller}},
  \bibinfo{author}{\bibfnamefont{H.-T.} \bibnamefont{Janka}},
  \bibinfo{author}{\bibfnamefont{A.}~\bibnamefont{Marek}},
  \bibinfo{author}{\bibfnamefont{W.~R.} \bibnamefont{Hix}},
  \bibinfo{author}{\bibfnamefont{A.}~\bibnamefont{Juodagalvis}}, and
  \bibinfo{author}{\bibfnamefont{J.~M.} \bibnamefont{Sampaio}},
  \bibinfo{year}{2008}, \bibinfo{journal}{Phys. Rev. Lett.}
  \textbf{\bibinfo{volume}{100}}, \bibinfo{eid}{011101}.

\bibitem[{\citenamefont{Langanke} \emph{et~al.}(2004)\citenamefont{Langanke,
  Mart\'{\i}nez-Pinedo, von Neumann-Cosel, and Richter}}]{Langanke:2004}
\bibinfo{author}{\bibnamefont{Langanke}, \bibfnamefont{K.}},
  \bibinfo{author}{\bibfnamefont{G.}~\bibnamefont{Mart\'{\i}nez-Pinedo}},
  \bibinfo{author}{\bibfnamefont{P.}~\bibnamefont{von Neumann-Cosel}}, and
  \bibinfo{author}{\bibfnamefont{A.}~\bibnamefont{Richter}},
  \bibinfo{year}{2004}, \bibinfo{journal}{Phys. Rev. Lett.}
  \textbf{\bibinfo{volume}{93}}, \bibinfo{eid}{202501}.

\bibitem[{\citenamefont{Laszewski} \emph{et~al.}(1988)\citenamefont{Laszewski,
  Alarcon, Dale, and Hoblit}}]{Laszewski:1988}
\bibinfo{author}{\bibnamefont{Laszewski}, \bibfnamefont{R.~M.}},
  \bibinfo{author}{\bibfnamefont{R.}~\bibnamefont{Alarcon}},
  \bibinfo{author}{\bibfnamefont{D.~S.} \bibnamefont{Dale}}, and
  \bibinfo{author}{\bibfnamefont{S.~D.} \bibnamefont{Hoblit}},
  \bibinfo{year}{1988}, \bibinfo{journal}{Phys. Rev. Lett.}
  \textbf{\bibinfo{volume}{61}}, \bibinfo{pages}{1710}.

\bibitem[{\citenamefont{Lehmann} \emph{et~al.}(1999)\citenamefont{Lehmann,
  Nord, de~Almeida~Pinto, Beck, Besserer, von Brentano, Drissi, Eckert,
  Herzberg, J\"ager, Jolie, Kneissl} \emph{et~al.}}]{Lehmann:1999}
\bibinfo{author}{\bibnamefont{Lehmann}, \bibfnamefont{H.}},
  \bibinfo{author}{\bibfnamefont{A.}~\bibnamefont{Nord}},
  \bibinfo{author}{\bibfnamefont{A.~E.} \bibnamefont{de~Almeida~Pinto}},
  \bibinfo{author}{\bibfnamefont{O.}~\bibnamefont{Beck}},
  \bibinfo{author}{\bibfnamefont{J.}~\bibnamefont{Besserer}},
  \bibinfo{author}{\bibfnamefont{P.}~\bibnamefont{von Brentano}},
  \bibinfo{author}{\bibfnamefont{S.}~\bibnamefont{Drissi}},
  \bibinfo{author}{\bibfnamefont{T.}~\bibnamefont{Eckert}},
  \bibinfo{author}{\bibfnamefont{R.-D.} \bibnamefont{Herzberg}},
  \bibinfo{author}{\bibfnamefont{D.}~\bibnamefont{J\"ager}},
  \bibinfo{author}{\bibfnamefont{J.}~\bibnamefont{Jolie}},
  \bibinfo{author}{\bibfnamefont{U.}~\bibnamefont{Kneissl}}, \emph{et~al.},
  \bibinfo{year}{1999}, \bibinfo{journal}{Phys. Rev. C}
  \textbf{\bibinfo{volume}{60}}, \bibinfo{pages}{024308}.

\bibitem[{\citenamefont{Lesher} \emph{et~al.}(2007)\citenamefont{Lesher, McKay,
  Mynk, Bandyopadhyay, Boukharouba, Fransen, Orce, McEllistrem, and
  Yates}}]{Lesher:2007}
\bibinfo{author}{\bibnamefont{Lesher}, \bibfnamefont{S.~R.}},
  \bibinfo{author}{\bibfnamefont{C.~J.} \bibnamefont{McKay}},
  \bibinfo{author}{\bibfnamefont{M.}~\bibnamefont{Mynk}},
  \bibinfo{author}{\bibfnamefont{D.}~\bibnamefont{Bandyopadhyay}},
  \bibinfo{author}{\bibfnamefont{N.}~\bibnamefont{Boukharouba}},
  \bibinfo{author}{\bibfnamefont{C.}~\bibnamefont{Fransen}},
  \bibinfo{author}{\bibfnamefont{J.~N.} \bibnamefont{Orce}},
  \bibinfo{author}{\bibfnamefont{M.~T.} \bibnamefont{McEllistrem}}, and
  \bibinfo{author}{\bibfnamefont{S.~W.} \bibnamefont{Yates}},
  \bibinfo{year}{2007}, \bibinfo{journal}{Phys. Rev. C}
  \textbf{\bibinfo{volume}{75}}, \bibinfo{pages}{034318}.

\bibitem[{\citenamefont{Li} \emph{et~al.}(2005)\citenamefont{Li, Pietralla,
  Fransen, von Garrel, Kneissl, Kohstall, Linnemann, Pitz, Rainovski, Richter,
  Scheck, Stedile} \emph{et~al.}}]{Li:2005}
\bibinfo{author}{\bibnamefont{Li}, \bibfnamefont{T.~C.}},
  \bibinfo{author}{\bibfnamefont{N.}~\bibnamefont{Pietralla}},
  \bibinfo{author}{\bibfnamefont{C.}~\bibnamefont{Fransen}},
  \bibinfo{author}{\bibfnamefont{H.}~\bibnamefont{von Garrel}},
  \bibinfo{author}{\bibfnamefont{U.}~\bibnamefont{Kneissl}},
  \bibinfo{author}{\bibfnamefont{C.}~\bibnamefont{Kohstall}},
  \bibinfo{author}{\bibfnamefont{A.}~\bibnamefont{Linnemann}},
  \bibinfo{author}{\bibfnamefont{H.~H.} \bibnamefont{Pitz}},
  \bibinfo{author}{\bibfnamefont{G.}~\bibnamefont{Rainovski}},
  \bibinfo{author}{\bibfnamefont{A.}~\bibnamefont{Richter}},
  \bibinfo{author}{\bibfnamefont{M.}~\bibnamefont{Scheck}},
  \bibinfo{author}{\bibfnamefont{F.}~\bibnamefont{Stedile}}, \emph{et~al.},
  \bibinfo{year}{2005}, \bibinfo{journal}{Phys. Rev. C}
  \textbf{\bibinfo{volume}{71}}, \bibinfo{eid}{044318}.

\bibitem[{\citenamefont{Li} \emph{et~al.}(2006)\citenamefont{Li, Pietralla,
  Tonchev, Ahmed, Ahn, Angell, Blackston, Costin, Keeter, Li, Lisetskiy,
  Mikhailov} \emph{et~al.}}]{Li:2006}
\bibinfo{author}{\bibnamefont{Li}, \bibfnamefont{T.~C.}},
  \bibinfo{author}{\bibfnamefont{N.}~\bibnamefont{Pietralla}},
  \bibinfo{author}{\bibfnamefont{A.~P.} \bibnamefont{Tonchev}},
  \bibinfo{author}{\bibfnamefont{M.~W.} \bibnamefont{Ahmed}},
  \bibinfo{author}{\bibfnamefont{T.}~\bibnamefont{Ahn}},
  \bibinfo{author}{\bibfnamefont{C.}~\bibnamefont{Angell}},
  \bibinfo{author}{\bibfnamefont{M.~A.} \bibnamefont{Blackston}},
  \bibinfo{author}{\bibfnamefont{A.}~\bibnamefont{Costin}},
  \bibinfo{author}{\bibfnamefont{K.~J.} \bibnamefont{Keeter}},
  \bibinfo{author}{\bibfnamefont{J.}~\bibnamefont{Li}},
  \bibinfo{author}{\bibfnamefont{A.}~\bibnamefont{Lisetskiy}},
  \bibinfo{author}{\bibfnamefont{S.}~\bibnamefont{Mikhailov}}, \emph{et~al.},
  \bibinfo{year}{2006}, \bibinfo{journal}{Phys. Rev. C}
  \textbf{\bibinfo{volume}{73}}, \bibinfo{eid}{054306}.

\bibitem[{\citenamefont{Lieb} \emph{et~al.}(1988)\citenamefont{Lieb, B\"orner,
  Dewey, Jolie, Robinson, Ulbig, and Winter}}]{Lieb:1988}
\bibinfo{author}{\bibnamefont{Lieb}, \bibfnamefont{K.~P.}},
  \bibinfo{author}{\bibfnamefont{H.~G.} \bibnamefont{B\"orner}},
  \bibinfo{author}{\bibfnamefont{M.~S.} \bibnamefont{Dewey}},
  \bibinfo{author}{\bibfnamefont{J.}~\bibnamefont{Jolie}},
  \bibinfo{author}{\bibfnamefont{S.~J.} \bibnamefont{Robinson}},
  \bibinfo{author}{\bibfnamefont{S.}~\bibnamefont{Ulbig}}, and
  \bibinfo{author}{\bibfnamefont{C.}~\bibnamefont{Winter}},
  \bibinfo{year}{1988}, \bibinfo{journal}{Phys. Lett. B}
  \textbf{\bibinfo{volume}{215}}, \bibinfo{pages}{50}.

\bibitem[{\citenamefont{Linnemann} \emph{et~al.}(2003)\citenamefont{Linnemann,
  von Brentano, Eberth, Enders, Fitzler, Fransen, Guliyev, Herzberg,
  K\"{a}ubler, Kuliev, von Neumann-Cosel, Pietralla}
  \emph{et~al.}}]{Linnemann:2003}
\bibinfo{author}{\bibnamefont{Linnemann}, \bibfnamefont{A.}},
  \bibinfo{author}{\bibfnamefont{P.}~\bibnamefont{von Brentano}},
  \bibinfo{author}{\bibfnamefont{J.}~\bibnamefont{Eberth}},
  \bibinfo{author}{\bibfnamefont{J.}~\bibnamefont{Enders}},
  \bibinfo{author}{\bibfnamefont{A.}~\bibnamefont{Fitzler}},
  \bibinfo{author}{\bibfnamefont{C.}~\bibnamefont{Fransen}},
  \bibinfo{author}{\bibfnamefont{E.}~\bibnamefont{Guliyev}},
  \bibinfo{author}{\bibfnamefont{R.-D.} \bibnamefont{Herzberg}},
  \bibinfo{author}{\bibfnamefont{L.}~\bibnamefont{K\"{a}ubler}},
  \bibinfo{author}{\bibfnamefont{A.~A.} \bibnamefont{Kuliev}},
  \bibinfo{author}{\bibfnamefont{P.}~\bibnamefont{von Neumann-Cosel}},
  \bibinfo{author}{\bibfnamefont{N.}~\bibnamefont{Pietralla}}, \emph{et~al.},
  \bibinfo{year}{2003}, \bibinfo{journal}{Phys. Lett. B}
  \textbf{\bibinfo{volume}{554}}, \bibinfo{pages}{15}.

\bibitem[{\citenamefont{Linnemann} \emph{et~al.}(2005)\citenamefont{Linnemann,
  Fransen, Gorska, Jolie, Kneissl, Knoch, M\"{u}cher, Pitz, Scheck, Scholl, and
  von Brentano}}]{Linnemann:2005}
\bibinfo{author}{\bibnamefont{Linnemann}, \bibfnamefont{A.}},
  \bibinfo{author}{\bibfnamefont{C.}~\bibnamefont{Fransen}},
  \bibinfo{author}{\bibfnamefont{M.}~\bibnamefont{Gorska}},
  \bibinfo{author}{\bibfnamefont{J.}~\bibnamefont{Jolie}},
  \bibinfo{author}{\bibfnamefont{U.}~\bibnamefont{Kneissl}},
  \bibinfo{author}{\bibfnamefont{P.}~\bibnamefont{Knoch}},
  \bibinfo{author}{\bibfnamefont{D.}~\bibnamefont{M\"{u}cher}},
  \bibinfo{author}{\bibfnamefont{H.~H.} \bibnamefont{Pitz}},
  \bibinfo{author}{\bibfnamefont{M.}~\bibnamefont{Scheck}},
  \bibinfo{author}{\bibfnamefont{C.}~\bibnamefont{Scholl}}, and
  \bibinfo{author}{\bibfnamefont{P.}~\bibnamefont{von Brentano}},
  \bibinfo{year}{2005}, \bibinfo{journal}{Phys. Rev. C}
  \textbf{\bibinfo{volume}{72}}, \bibinfo{eid}{064323}.

\bibitem[{\citenamefont{Linnemann} \emph{et~al.}(2007)\citenamefont{Linnemann,
  Fransen, Jolie, Kneissl, Knoch, Kohstall, M\"{u}cher, Pitz, Scheck, Scholl,
  Stedile, von Brentano} \emph{et~al.}}]{Linnemann:2007}
\bibinfo{author}{\bibnamefont{Linnemann}, \bibfnamefont{A.}},
  \bibinfo{author}{\bibfnamefont{C.}~\bibnamefont{Fransen}},
  \bibinfo{author}{\bibfnamefont{J.}~\bibnamefont{Jolie}},
  \bibinfo{author}{\bibfnamefont{U.}~\bibnamefont{Kneissl}},
  \bibinfo{author}{\bibfnamefont{P.}~\bibnamefont{Knoch}},
  \bibinfo{author}{\bibfnamefont{C.}~\bibnamefont{Kohstall}},
  \bibinfo{author}{\bibfnamefont{D.}~\bibnamefont{M\"{u}cher}},
  \bibinfo{author}{\bibfnamefont{H.~H.} \bibnamefont{Pitz}},
  \bibinfo{author}{\bibfnamefont{M.}~\bibnamefont{Scheck}},
  \bibinfo{author}{\bibfnamefont{C.}~\bibnamefont{Scholl}},
  \bibinfo{author}{\bibfnamefont{F.}~\bibnamefont{Stedile}},
  \bibinfo{author}{\bibfnamefont{P.}~\bibnamefont{von Brentano}},
  \emph{et~al.}, \bibinfo{year}{2007}, \bibinfo{journal}{Phys. Rev. C}
  \textbf{\bibinfo{volume}{75}}, \bibinfo{eid}{024310}.

\bibitem[{\citenamefont{Lipparini}(2003)}]{Lipparini:2003}
\bibinfo{author}{\bibnamefont{Lipparini}, \bibfnamefont{E.}},
  \bibinfo{year}{2003}, \emph{\bibinfo{title}{Modern Many-Particle Physics:
  Atomic Gases, Quantum Dots and Quantum Fluids}} (\bibinfo{publisher}{World
  Scientific}, \bibinfo{address}{Singapore}).

\bibitem[{\citenamefont{Lipparini and Richter}(1984)}]{Lipparini:1984}
\bibinfo{author}{\bibnamefont{Lipparini}, \bibfnamefont{E.}}, and
  \bibinfo{author}{\bibfnamefont{A.}~\bibnamefont{Richter}},
  \bibinfo{year}{1984}, \bibinfo{journal}{Phys. Lett. B}
  \textbf{\bibinfo{volume}{144}}, \bibinfo{pages}{13}.

\bibitem[{\citenamefont{Lipparini and Stringari}(1983)}]{Lipparini:1983}
\bibinfo{author}{\bibnamefont{Lipparini}, \bibfnamefont{E.}}, and
  \bibinfo{author}{\bibfnamefont{S.}~\bibnamefont{Stringari}},
  \bibinfo{year}{1983}, \bibinfo{journal}{Phys. Lett. B}
  \textbf{\bibinfo{volume}{130}}, \bibinfo{pages}{139}.

\bibitem[{\citenamefont{Lipparini and
  Stringari}(1989{\natexlab{a}})}]{Lipparini:1989b}
\bibinfo{author}{\bibnamefont{Lipparini}, \bibfnamefont{E.}}, and
  \bibinfo{author}{\bibfnamefont{S.}~\bibnamefont{Stringari}},
  \bibinfo{year}{1989}{\natexlab{a}}, \bibinfo{journal}{Phys. Rep.}
  \textbf{\bibinfo{volume}{175}}, \bibinfo{pages}{103}.

\bibitem[{\citenamefont{Lipparini and
  Stringari}(1989{\natexlab{b}})}]{Lipparini:1989a}
\bibinfo{author}{\bibnamefont{Lipparini}, \bibfnamefont{E.}}, and
  \bibinfo{author}{\bibfnamefont{S.}~\bibnamefont{Stringari}},
  \bibinfo{year}{1989}{\natexlab{b}}, \bibinfo{journal}{Phys. Rev. Lett.}
  \textbf{\bibinfo{volume}{63}}, \bibinfo{pages}{570}.

\bibitem[{\citenamefont{Lisantti} \emph{et~al.}(1984)\citenamefont{Lisantti,
  Tinsley, Drake, Bergqvist, Swenson, McDaniels, Bertrand, Gross, Horen, and
  Sjoreen}}]{Lisantti:1984}
\bibinfo{author}{\bibnamefont{Lisantti}, \bibfnamefont{J.}},
  \bibinfo{author}{\bibfnamefont{J.~R.} \bibnamefont{Tinsley}},
  \bibinfo{author}{\bibfnamefont{D.~M.} \bibnamefont{Drake}},
  \bibinfo{author}{\bibfnamefont{I.}~\bibnamefont{Bergqvist}},
  \bibinfo{author}{\bibfnamefont{L.~W.} \bibnamefont{Swenson}},
  \bibinfo{author}{\bibfnamefont{D.~K.} \bibnamefont{McDaniels}},
  \bibinfo{author}{\bibfnamefont{F.~E.} \bibnamefont{Bertrand}},
  \bibinfo{author}{\bibfnamefont{E.~E.} \bibnamefont{Gross}},
  \bibinfo{author}{\bibfnamefont{D.~J.} \bibnamefont{Horen}}, and
  \bibinfo{author}{\bibfnamefont{T.~P.} \bibnamefont{Sjoreen}},
  \bibinfo{year}{1984}, \bibinfo{journal}{Phys. Lett. B}
  \textbf{\bibinfo{volume}{147}}, \bibinfo{pages}{23}.

\bibitem[{\citenamefont{Lisetskiy} \emph{et~al.}(2007)\citenamefont{Lisetskiy,
  Caurier, Langanke, Mart\'{\i}nez-Pinedo, von Neumann-Cosel, Nowacki, and
  Richter}}]{Lisetskiy:2007}
\bibinfo{author}{\bibnamefont{Lisetskiy}, \bibfnamefont{A.~F.}},
  \bibinfo{author}{\bibfnamefont{E.}~\bibnamefont{Caurier}},
  \bibinfo{author}{\bibfnamefont{K.}~\bibnamefont{Langanke}},
  \bibinfo{author}{\bibfnamefont{G.}~\bibnamefont{Mart\'{\i}nez-Pinedo}},
  \bibinfo{author}{\bibfnamefont{P.}~\bibnamefont{von Neumann-Cosel}},
  \bibinfo{author}{\bibfnamefont{F.}~\bibnamefont{Nowacki}}, and
  \bibinfo{author}{\bibfnamefont{A.}~\bibnamefont{Richter}},
  \bibinfo{year}{2007}, \bibinfo{journal}{Nucl. Phys. A}
  \textbf{\bibinfo{volume}{789}}, \bibinfo{pages}{114}.

\bibitem[{\citenamefont{Lisetskiy} \emph{et~al.}(2000)\citenamefont{Lisetskiy,
  Pietralla, Fransen, Jolos, and von Brentano}}]{Lisetskiy:2000}
\bibinfo{author}{\bibnamefont{Lisetskiy}, \bibfnamefont{A.~F.}},
  \bibinfo{author}{\bibfnamefont{N.}~\bibnamefont{Pietralla}},
  \bibinfo{author}{\bibfnamefont{C.}~\bibnamefont{Fransen}},
  \bibinfo{author}{\bibfnamefont{R.~V.} \bibnamefont{Jolos}}, and
  \bibinfo{author}{\bibfnamefont{P.}~\bibnamefont{von Brentano}},
  \bibinfo{year}{2000}, \bibinfo{journal}{Nucl. Phys. A}
  \textbf{\bibinfo{volume}{677}}, \bibinfo{pages}{100}.

\bibitem[{\citenamefont{Liu and Zamick}(1987{\natexlab{a}})}]{Liu:1987a}
\bibinfo{author}{\bibnamefont{Liu}, \bibfnamefont{H.}}, and
  \bibinfo{author}{\bibfnamefont{L.}~\bibnamefont{Zamick}},
  \bibinfo{year}{1987}{\natexlab{a}}, \bibinfo{journal}{Nucl. Phys. A}
  \textbf{\bibinfo{volume}{467}}, \bibinfo{pages}{29}.

\bibitem[{\citenamefont{Liu and Zamick}(1987{\natexlab{b}})}]{Liu:1987c}
\bibinfo{author}{\bibnamefont{Liu}, \bibfnamefont{H.}}, and
  \bibinfo{author}{\bibfnamefont{L.}~\bibnamefont{Zamick}},
  \bibinfo{year}{1987}{\natexlab{b}}, \bibinfo{journal}{Phys. Rev. C}
  \textbf{\bibinfo{volume}{36}}, \bibinfo{pages}{2064}.

\bibitem[{\citenamefont{Liu and Zamick}(1987{\natexlab{c}})}]{Liu:1987b}
\bibinfo{author}{\bibnamefont{Liu}, \bibfnamefont{H.}}, and
  \bibinfo{author}{\bibfnamefont{L.}~\bibnamefont{Zamick}},
  \bibinfo{year}{1987}{\natexlab{c}}, \bibinfo{journal}{Phys. Rev. C}
  \textbf{\bibinfo{volume}{36}}, \bibinfo{pages}{2057}.

\bibitem[{\citenamefont{Lo~Iudice}(1988)}]{LoIudice:1988}
\bibinfo{author}{\bibnamefont{Lo~Iudice}, \bibfnamefont{N.}},
  \bibinfo{year}{1988}, \bibinfo{journal}{Phys. Rev. C}
  \textbf{\bibinfo{volume}{38}}, \bibinfo{pages}{2895}.

\bibitem[{\citenamefont{Lo~Iudice}(1996{\natexlab{a}})}]{LoIudice:1996a}
\bibinfo{author}{\bibnamefont{Lo~Iudice}, \bibfnamefont{N.}},
  \bibinfo{year}{1996}{\natexlab{a}}, \bibinfo{journal}{Nucl. Phys. A}
  \textbf{\bibinfo{volume}{605}}, \bibinfo{pages}{61}.

\bibitem[{\citenamefont{Lo~Iudice}(1996{\natexlab{b}})}]{LoIudice:1996b}
\bibinfo{author}{\bibnamefont{Lo~Iudice}, \bibfnamefont{N.}},
  \bibinfo{year}{1996}{\natexlab{b}}, \bibinfo{journal}{Phys. Rev. C}
  \textbf{\bibinfo{volume}{53}}, \bibinfo{pages}{2171}.

\bibitem[{\citenamefont{Lo~Iudice}(1997)}]{LoIudice:1997}
\bibinfo{author}{\bibnamefont{Lo~Iudice}, \bibfnamefont{N.}},
  \bibinfo{year}{1997}, \bibinfo{journal}{Phys. Part. Nucl.}
  \textbf{\bibinfo{volume}{28}}, \bibinfo{pages}{556}.

\bibitem[{\citenamefont{Lo~Iudice}(1998)}]{LoIudice:1998}
\bibinfo{author}{\bibnamefont{Lo~Iudice}, \bibfnamefont{N.}},
  \bibinfo{year}{1998}, \bibinfo{journal}{Phys. Rev. C}
  \textbf{\bibinfo{volume}{57}}, \bibinfo{pages}{1246}.

\bibitem[{\citenamefont{Lo~Iudice}(2000)}]{LoIudice:2000a}
\bibinfo{author}{\bibnamefont{Lo~Iudice}, \bibfnamefont{N.}},
  \bibinfo{year}{2000}, \bibinfo{journal}{Rivista Nuovo Cimento}
  \textbf{\bibinfo{volume}{23}}, \bibinfo{pages}{1}.

\bibitem[{\citenamefont{Lo~Iudice and Palumbo}(1978)}]{LoIudice:1978}
\bibinfo{author}{\bibnamefont{Lo~Iudice}, \bibfnamefont{N.}}, and
  \bibinfo{author}{\bibfnamefont{F.}~\bibnamefont{Palumbo}},
  \bibinfo{year}{1978}, \bibinfo{journal}{Phys. Rev. Lett.}
  \textbf{\bibinfo{volume}{41}}, \bibinfo{pages}{1532}.

\bibitem[{\citenamefont{Lo~Iudice and Palumbo}(1979)}]{LoIudice:1979}
\bibinfo{author}{\bibnamefont{Lo~Iudice}, \bibfnamefont{N.}}, and
  \bibinfo{author}{\bibfnamefont{F.}~\bibnamefont{Palumbo}},
  \bibinfo{year}{1979}, \bibinfo{journal}{Nucl. Phys. A}
  \textbf{\bibinfo{volume}{326}}, \bibinfo{pages}{193}.

\bibitem[{\citenamefont{Lo~Iudice and Richter}(1989)}]{LoIudice:1989}
\bibinfo{author}{\bibnamefont{Lo~Iudice}, \bibfnamefont{N.}}, and
  \bibinfo{author}{\bibfnamefont{A.}~\bibnamefont{Richter}},
  \bibinfo{year}{1989}, \bibinfo{journal}{Phys. Lett. B}
  \textbf{\bibinfo{volume}{228}}, \bibinfo{pages}{91}.

\bibitem[{\citenamefont{Lo~Iudice and Richter}(1993)}]{LoIudice:1993b}
\bibinfo{author}{\bibnamefont{Lo~Iudice}, \bibfnamefont{N.}}, and
  \bibinfo{author}{\bibfnamefont{A.}~\bibnamefont{Richter}},
  \bibinfo{year}{1993}, \bibinfo{journal}{Phys. Lett. B}
  \textbf{\bibinfo{volume}{304}}, \bibinfo{pages}{193}.

\bibitem[{\citenamefont{Lo~Iudice and Stoyanov}(2000)}]{LoIudice:2000b}
\bibinfo{author}{\bibnamefont{Lo~Iudice}, \bibfnamefont{N.}}, and
  \bibinfo{author}{\bibfnamefont{C.}~\bibnamefont{Stoyanov}},
  \bibinfo{year}{2000}, \bibinfo{journal}{Phys. Rev. C}
  \textbf{\bibinfo{volume}{62}}, \bibinfo{pages}{047302}.

\bibitem[{\citenamefont{Lo~Iudice and Stoyanov}(2002)}]{LoIudice:2002}
\bibinfo{author}{\bibnamefont{Lo~Iudice}, \bibfnamefont{N.}}, and
  \bibinfo{author}{\bibfnamefont{C.}~\bibnamefont{Stoyanov}},
  \bibinfo{year}{2002}, \bibinfo{journal}{Phys. Rev. C}
  \textbf{\bibinfo{volume}{65}}, \bibinfo{pages}{064304}.

\bibitem[{\citenamefont{Lo~Iudice and Stoyanov}(2004)}]{LoIudice:2004}
\bibinfo{author}{\bibnamefont{Lo~Iudice}, \bibfnamefont{N.}}, and
  \bibinfo{author}{\bibfnamefont{C.}~\bibnamefont{Stoyanov}},
  \bibinfo{year}{2004}, \bibinfo{journal}{Phys. Rev. C}
  \textbf{\bibinfo{volume}{69}}, \bibinfo{pages}{044312}.

\bibitem[{\citenamefont{Lo~Iudice and Stoyanov}(2006)}]{LoIudice:2006}
\bibinfo{author}{\bibnamefont{Lo~Iudice}, \bibfnamefont{N.}}, and
  \bibinfo{author}{\bibfnamefont{C.}~\bibnamefont{Stoyanov}},
  \bibinfo{year}{2006}, \bibinfo{journal}{Phys. Rev. C}
  \textbf{\bibinfo{volume}{73}}, \bibinfo{eid}{037305}.

\bibitem[{\citenamefont{Lo~Iudice} \emph{et~al.}(2009)\citenamefont{Lo~Iudice,
  Stoyanov, and Pietralla}}]{LoIudice:2009}
\bibinfo{author}{\bibnamefont{Lo~Iudice}, \bibfnamefont{N.}},
  \bibinfo{author}{\bibfnamefont{C.}~\bibnamefont{Stoyanov}}, and
  \bibinfo{author}{\bibfnamefont{N.}~\bibnamefont{Pietralla}},
  \bibinfo{year}{2009}, \bibinfo{journal}{Phys. Rev. C}
  \textbf{\bibinfo{volume}{80}}, \bibinfo{pages}{024311}.

\bibitem[{\citenamefont{Lo~Iudice} \emph{et~al.}(2008)\citenamefont{Lo~Iudice,
  Stoyanov, and Tarpanov}}]{LoIudice:2008}
\bibinfo{author}{\bibnamefont{Lo~Iudice}, \bibfnamefont{N.}},
  \bibinfo{author}{\bibfnamefont{C.}~\bibnamefont{Stoyanov}}, and
  \bibinfo{author}{\bibfnamefont{D.}~\bibnamefont{Tarpanov}},
  \bibinfo{year}{2008}, \bibinfo{journal}{Phys. Rev. C}
  \textbf{\bibinfo{volume}{77}}, \bibinfo{eid}{044310}.

\bibitem[{\citenamefont{L\"obner} \emph{et~al.}(1970)\citenamefont{L\"obner,
  Vetter, and H\"onig}}]{Loebner:1970}
\bibinfo{author}{\bibnamefont{L\"obner}, \bibfnamefont{K.~E.~G.}},
  \bibinfo{author}{\bibfnamefont{M.}~\bibnamefont{Vetter}}, and
  \bibinfo{author}{\bibfnamefont{V.}~\bibnamefont{H\"onig}},
  \bibinfo{year}{1970}, \bibinfo{journal}{Nucl. Data Tables A}
  \textbf{\bibinfo{volume}{7}}, \bibinfo{pages}{495}.

\bibitem[{\citenamefont{L\"uttge} \emph{et~al.}(1995)\citenamefont{L\"uttge,
  Hofmann, Horn, Neumeyer, Richter, Schrieder, Spamer, Stiller, Sober,
  Matthews, and Fagg}}]{Luettge:1995}
\bibinfo{author}{\bibnamefont{L\"uttge}, \bibfnamefont{C.}},
  \bibinfo{author}{\bibfnamefont{C.}~\bibnamefont{Hofmann}},
  \bibinfo{author}{\bibfnamefont{J.}~\bibnamefont{Horn}},
  \bibinfo{author}{\bibfnamefont{F.}~\bibnamefont{Neumeyer}},
  \bibinfo{author}{\bibfnamefont{A.}~\bibnamefont{Richter}},
  \bibinfo{author}{\bibfnamefont{G.}~\bibnamefont{Schrieder}},
  \bibinfo{author}{\bibfnamefont{E.}~\bibnamefont{Spamer}},
  \bibinfo{author}{\bibfnamefont{A.}~\bibnamefont{Stiller}},
  \bibinfo{author}{\bibfnamefont{D.~I.} \bibnamefont{Sober}},
  \bibinfo{author}{\bibfnamefont{S.~K.} \bibnamefont{Matthews}}, and
  \bibinfo{author}{\bibfnamefont{L.~W.} \bibnamefont{Fagg}},
  \bibinfo{year}{1995}, \bibinfo{journal}{Nucl. Instrum. Methods Phys. Res. A}
  \textbf{\bibinfo{volume}{366}}, \bibinfo{pages}{325}.

\bibitem[{\citenamefont{L\"uttge}
  \emph{et~al.}(1996{\natexlab{a}})\citenamefont{L\"uttge, von Neumann-Cosel,
  Neumeyer, Rangacharyulu, Richter, Schrieder, Spamer, Sober, Matthews, and
  Brown}}]{Luettge:1996b}
\bibinfo{author}{\bibnamefont{L\"uttge}, \bibfnamefont{C.}},
  \bibinfo{author}{\bibfnamefont{P.}~\bibnamefont{von Neumann-Cosel}},
  \bibinfo{author}{\bibfnamefont{F.}~\bibnamefont{Neumeyer}},
  \bibinfo{author}{\bibfnamefont{C.}~\bibnamefont{Rangacharyulu}},
  \bibinfo{author}{\bibfnamefont{A.}~\bibnamefont{Richter}},
  \bibinfo{author}{\bibfnamefont{G.}~\bibnamefont{Schrieder}},
  \bibinfo{author}{\bibfnamefont{E.}~\bibnamefont{Spamer}},
  \bibinfo{author}{\bibfnamefont{D.~I.} \bibnamefont{Sober}},
  \bibinfo{author}{\bibfnamefont{S.~K.} \bibnamefont{Matthews}}, and
  \bibinfo{author}{\bibfnamefont{B.~A.} \bibnamefont{Brown}},
  \bibinfo{year}{1996}{\natexlab{a}}, \bibinfo{journal}{Phys. Rev. C}
  \textbf{\bibinfo{volume}{53}}, \bibinfo{pages}{127}.

\bibitem[{\citenamefont{L\"uttge}
  \emph{et~al.}(1996{\natexlab{b}})\citenamefont{L\"uttge, von Neumann-Cosel,
  Neumeyer, and Richter}}]{Luettge:1996a}
\bibinfo{author}{\bibnamefont{L\"uttge}, \bibfnamefont{C.}},
  \bibinfo{author}{\bibfnamefont{P.}~\bibnamefont{von Neumann-Cosel}},
  \bibinfo{author}{\bibfnamefont{F.}~\bibnamefont{Neumeyer}}, and
  \bibinfo{author}{\bibfnamefont{A.}~\bibnamefont{Richter}},
  \bibinfo{year}{1996}{\natexlab{b}}, \bibinfo{journal}{Nucl. Phys. A}
  \textbf{\bibinfo{volume}{606}}, \bibinfo{pages}{183}.

\bibitem[{\citenamefont{Marag\`o} \emph{et~al.}(2001)\citenamefont{Marag\`o,
  Hechenblaikner, Hodby, and Foot}}]{Marago:2001}
\bibinfo{author}{\bibnamefont{Marag\`o}, \bibfnamefont{O.}},
  \bibinfo{author}{\bibfnamefont{G.}~\bibnamefont{Hechenblaikner}},
  \bibinfo{author}{\bibfnamefont{E.}~\bibnamefont{Hodby}}, and
  \bibinfo{author}{\bibfnamefont{C.}~\bibnamefont{Foot}}, \bibinfo{year}{2001},
  \bibinfo{journal}{Phys. Rev. Lett.} \textbf{\bibinfo{volume}{86}},
  \bibinfo{pages}{3938}.

\bibitem[{\citenamefont{Marag\`o} \emph{et~al.}(2000)\citenamefont{Marag\`o,
  Hopkins, Arlt, Hodby, Hechenblaikner, and Foot}}]{Marago:2000}
\bibinfo{author}{\bibnamefont{Marag\`o}, \bibfnamefont{O.~M.}},
  \bibinfo{author}{\bibfnamefont{S.~A.} \bibnamefont{Hopkins}},
  \bibinfo{author}{\bibfnamefont{J.}~\bibnamefont{Arlt}},
  \bibinfo{author}{\bibfnamefont{E.}~\bibnamefont{Hodby}},
  \bibinfo{author}{\bibfnamefont{G.}~\bibnamefont{Hechenblaikner}}, and
  \bibinfo{author}{\bibfnamefont{C.~J.} \bibnamefont{Foot}},
  \bibinfo{year}{2000}, \bibinfo{journal}{Phys. Rev. Lett.}
  \textbf{\bibinfo{volume}{84}}, \bibinfo{pages}{2056}.

\bibitem[{\citenamefont{Margraf} \emph{et~al.}(1990)\citenamefont{Margraf,
  Degener, Friedrichs, Heil, Jung, Kneissl, Lindenstruth, Pitz, Schacht,
  Seemann, Stock, Wesselborg} \emph{et~al.}}]{Margraf:1990}
\bibinfo{author}{\bibnamefont{Margraf}, \bibfnamefont{J.}},
  \bibinfo{author}{\bibfnamefont{A.}~\bibnamefont{Degener}},
  \bibinfo{author}{\bibfnamefont{H.}~\bibnamefont{Friedrichs}},
  \bibinfo{author}{\bibfnamefont{R.~D.} \bibnamefont{Heil}},
  \bibinfo{author}{\bibfnamefont{A.}~\bibnamefont{Jung}},
  \bibinfo{author}{\bibfnamefont{U.}~\bibnamefont{Kneissl}},
  \bibinfo{author}{\bibfnamefont{S.}~\bibnamefont{Lindenstruth}},
  \bibinfo{author}{\bibfnamefont{H.~H.} \bibnamefont{Pitz}},
  \bibinfo{author}{\bibfnamefont{H.}~\bibnamefont{Schacht}},
  \bibinfo{author}{\bibfnamefont{U.}~\bibnamefont{Seemann}},
  \bibinfo{author}{\bibfnamefont{R.}~\bibnamefont{Stock}},
  \bibinfo{author}{\bibfnamefont{C.}~\bibnamefont{Wesselborg}}, \emph{et~al.},
  \bibinfo{year}{1990}, \bibinfo{journal}{Phys. Rev. C}
  \textbf{\bibinfo{volume}{42}}, \bibinfo{pages}{771}.

\bibitem[{\citenamefont{Margraf} \emph{et~al.}(1995)\citenamefont{Margraf,
  Eckert, Rittner, Bauske, Beck, Kneissl, Maser, Pitz, Schiller, von Brentano,
  Fischer, Herzberg} \emph{et~al.}}]{Margraf:1995}
\bibinfo{author}{\bibnamefont{Margraf}, \bibfnamefont{J.}},
  \bibinfo{author}{\bibfnamefont{T.}~\bibnamefont{Eckert}},
  \bibinfo{author}{\bibfnamefont{M.}~\bibnamefont{Rittner}},
  \bibinfo{author}{\bibfnamefont{I.}~\bibnamefont{Bauske}},
  \bibinfo{author}{\bibfnamefont{O.}~\bibnamefont{Beck}},
  \bibinfo{author}{\bibfnamefont{U.}~\bibnamefont{Kneissl}},
  \bibinfo{author}{\bibfnamefont{H.}~\bibnamefont{Maser}},
  \bibinfo{author}{\bibfnamefont{H.~H.} \bibnamefont{Pitz}},
  \bibinfo{author}{\bibfnamefont{A.}~\bibnamefont{Schiller}},
  \bibinfo{author}{\bibfnamefont{P.}~\bibnamefont{von Brentano}},
  \bibinfo{author}{\bibfnamefont{R.}~\bibnamefont{Fischer}},
  \bibinfo{author}{\bibfnamefont{R.-D.} \bibnamefont{Herzberg}}, \emph{et~al.},
  \bibinfo{year}{1995}, \bibinfo{journal}{Phys. Rev. C}
  \textbf{\bibinfo{volume}{52}}, \bibinfo{pages}{2429}.

\bibitem[{\citenamefont{Margraf} \emph{et~al.}()\citenamefont{Margraf, Heil,
  Kneissl, Maier, Pitz, Friedrichs, Lindenstruth, Schlitt, Wesselborg, von
  Brentano, Herzberg, and Zilges}}]{Margraf:1993}
\bibinfo{author}{\bibnamefont{Margraf}, \bibfnamefont{J.}},
  \bibinfo{author}{\bibfnamefont{R.~D.} \bibnamefont{Heil}},
  \bibinfo{author}{\bibfnamefont{U.}~\bibnamefont{Kneissl}},
  \bibinfo{author}{\bibfnamefont{U.}~\bibnamefont{Maier}},
  \bibinfo{author}{\bibfnamefont{H.~H.} \bibnamefont{Pitz}},
  \bibinfo{author}{\bibfnamefont{H.}~\bibnamefont{Friedrichs}},
  \bibinfo{author}{\bibfnamefont{S.}~\bibnamefont{Lindenstruth}},
  \bibinfo{author}{\bibfnamefont{B.}~\bibnamefont{Schlitt}},
  \bibinfo{author}{\bibfnamefont{C.}~\bibnamefont{Wesselborg}},
  \bibinfo{author}{\bibfnamefont{P.}~\bibnamefont{von Brentano}},
  \bibinfo{author}{\bibfnamefont{R.-D.} \bibnamefont{Herzberg}}, and
  \bibinfo{author}{\bibfnamefont{j.~. P. v. . . p. . . n. . . y. . . d. . P. p.
  .~A.} \bibnamefont{Zilges}, \bibfnamefont{A.}}, ????

\bibitem[{\citenamefont{Mart\'inez-Pinedo}
  \emph{et~al.}(1996)\citenamefont{Mart\'inez-Pinedo, Poves, Caurier, and
  Zuker}}]{Martinez-Pinedo:1996}
\bibinfo{author}{\bibnamefont{Mart\'inez-Pinedo}, \bibfnamefont{G.}},
  \bibinfo{author}{\bibfnamefont{A.}~\bibnamefont{Poves}},
  \bibinfo{author}{\bibfnamefont{E.}~\bibnamefont{Caurier}}, and
  \bibinfo{author}{\bibfnamefont{A.~P.} \bibnamefont{Zuker}},
  \bibinfo{year}{1996}, \bibinfo{journal}{Phys. Rev. C}
  \textbf{\bibinfo{volume}{53}}, \bibinfo{pages}{R2602}.

\bibitem[{\citenamefont{McCullen} \emph{et~al.}(1964)\citenamefont{McCullen,
  Bayman, and Zamick}}]{McCullen:1964}
\bibinfo{author}{\bibnamefont{McCullen}, \bibfnamefont{J.~D.}},
  \bibinfo{author}{\bibfnamefont{B.~F.} \bibnamefont{Bayman}}, and
  \bibinfo{author}{\bibfnamefont{L.}~\bibnamefont{Zamick}},
  \bibinfo{year}{1964}, \bibinfo{journal}{Phys. Rev.}
  \textbf{\bibinfo{volume}{134}}, \bibinfo{pages}{B515}.

\bibitem[{\citenamefont{Minguzzi and Tosi}(2001)}]{Minguzzi:2001}
\bibinfo{author}{\bibnamefont{Minguzzi}, \bibfnamefont{A.}}, and
  \bibinfo{author}{\bibfnamefont{M.~P.} \bibnamefont{Tosi}},
  \bibinfo{year}{2001}, \bibinfo{journal}{Phys. Rev. A}
  \textbf{\bibinfo{volume}{63}}, \bibinfo{pages}{023609}.

\bibitem[{\citenamefont{Mizusaki} \emph{et~al.}(1996)\citenamefont{Mizusaki,
  Honma, and Otsuka}}]{Mizusaki:1996}
\bibinfo{author}{\bibnamefont{Mizusaki}, \bibfnamefont{T.}},
  \bibinfo{author}{\bibfnamefont{M.}~\bibnamefont{Honma}}, and
  \bibinfo{author}{\bibfnamefont{T.}~\bibnamefont{Otsuka}},
  \bibinfo{year}{1996}, \bibinfo{journal}{Phys. Rev. C}
  \textbf{\bibinfo{volume}{53}}, \bibinfo{pages}{2786}.

\bibitem[{\citenamefont{Mizusaki} \emph{et~al.}(1991)\citenamefont{Mizusaki,
  Otsuka, and Sugita}}]{Mizusaki:1991}
\bibinfo{author}{\bibnamefont{Mizusaki}, \bibfnamefont{T.}},
  \bibinfo{author}{\bibfnamefont{T.}~\bibnamefont{Otsuka}}, and
  \bibinfo{author}{\bibfnamefont{M.}~\bibnamefont{Sugita}},
  \bibinfo{year}{1991}, \bibinfo{journal}{Phys. Rev. C}
  \textbf{\bibinfo{volume}{44}}, \bibinfo{pages}{R1277}.

\bibitem[{\citenamefont{Mizusaki} \emph{et~al.}(1999)\citenamefont{Mizusaki,
  Otsuka, Utsuno, Honma, and Sebe}}]{Mizusaki:1999}
\bibinfo{author}{\bibnamefont{Mizusaki}, \bibfnamefont{T.}},
  \bibinfo{author}{\bibfnamefont{T.}~\bibnamefont{Otsuka}},
  \bibinfo{author}{\bibfnamefont{Y.}~\bibnamefont{Utsuno}},
  \bibinfo{author}{\bibfnamefont{M.}~\bibnamefont{Honma}}, and
  \bibinfo{author}{\bibfnamefont{T.}~\bibnamefont{Sebe}}, \bibinfo{year}{1999},
  \bibinfo{journal}{Phys. Rev. C} \textbf{\bibinfo{volume}{59}},
  \bibinfo{pages}{R1846}.

\bibitem[{\citenamefont{Modugno} \emph{et~al.}(2002)\citenamefont{Modugno,
  Modugno, Riboli, Roati, and Inguscio}}]{Modugno:2002}
\bibinfo{author}{\bibnamefont{Modugno}, \bibfnamefont{G.}},
  \bibinfo{author}{\bibfnamefont{M.}~\bibnamefont{Modugno}},
  \bibinfo{author}{\bibfnamefont{F.}~\bibnamefont{Riboli}},
  \bibinfo{author}{\bibfnamefont{G.}~\bibnamefont{Roati}}, and
  \bibinfo{author}{\bibfnamefont{M.}~\bibnamefont{Inguscio}},
  \bibinfo{year}{2002}, \bibinfo{journal}{Phys. Rev. Lett.}
  \textbf{\bibinfo{volume}{89}}, \bibinfo{pages}{190404}.

\bibitem[{\citenamefont{Mukhopadhyay}
  \emph{et~al.}(2008)\citenamefont{Mukhopadhyay, Scheck, Crider, Choudry,
  Elhami, Peters, McEllistrem, Orce, and Yates}}]{Mukhopadhyay:2008}
\bibinfo{author}{\bibnamefont{Mukhopadhyay}, \bibfnamefont{S.}},
  \bibinfo{author}{\bibfnamefont{M.}~\bibnamefont{Scheck}},
  \bibinfo{author}{\bibfnamefont{B.}~\bibnamefont{Crider}},
  \bibinfo{author}{\bibfnamefont{S.~N.} \bibnamefont{Choudry}},
  \bibinfo{author}{\bibfnamefont{E.}~\bibnamefont{Elhami}},
  \bibinfo{author}{\bibfnamefont{E.}~\bibnamefont{Peters}},
  \bibinfo{author}{\bibfnamefont{M.~T.} \bibnamefont{McEllistrem}},
  \bibinfo{author}{\bibfnamefont{J.~N.} \bibnamefont{Orce}}, and
  \bibinfo{author}{\bibfnamefont{S.~W.} \bibnamefont{Yates}},
  \bibinfo{year}{2008}, \bibinfo{journal}{Phys. Rev. C}
  \textbf{\bibinfo{volume}{78}}, \bibinfo{eid}{034317}.

\bibitem[{\citenamefont{Nakatsukasa}
  \emph{et~al.}(1994)\citenamefont{Nakatsukasa, Matsuyanagi, Hamamoto, and
  Nazarewicz}}]{Nakatsukasa:1994}
\bibinfo{author}{\bibnamefont{Nakatsukasa}, \bibfnamefont{T.}},
  \bibinfo{author}{\bibfnamefont{K.}~\bibnamefont{Matsuyanagi}},
  \bibinfo{author}{\bibfnamefont{I.}~\bibnamefont{Hamamoto}}, and
  \bibinfo{author}{\bibfnamefont{W.}~\bibnamefont{Nazarewicz}},
  \bibinfo{year}{1994}, \bibinfo{journal}{Nucl. Phys. A}
  \textbf{\bibinfo{volume}{573}}, \bibinfo{pages}{333}.

\bibitem[{\citenamefont{Nesterenko}
  \emph{et~al.}(1997)\citenamefont{Nesterenko, Kleinig, Gudkov, Lo~Iudice, and
  Kvasil}}]{Nesterenko:1997}
\bibinfo{author}{\bibnamefont{Nesterenko}, \bibfnamefont{V.~O.}},
  \bibinfo{author}{\bibfnamefont{W.}~\bibnamefont{Kleinig}},
  \bibinfo{author}{\bibfnamefont{V.~V.} \bibnamefont{Gudkov}},
  \bibinfo{author}{\bibfnamefont{N.}~\bibnamefont{Lo~Iudice}}, and
  \bibinfo{author}{\bibfnamefont{J.}~\bibnamefont{Kvasil}},
  \bibinfo{year}{1997}, \bibinfo{journal}{Phys. Rev. A}
  \textbf{\bibinfo{volume}{56}}, \bibinfo{pages}{607}.

\bibitem[{\citenamefont{Nesterenko}
  \emph{et~al.}(1999)\citenamefont{Nesterenko, Kleinig, de~Souza~Cruz, and
  Lo~Iudice}}]{Nesterenko:1999}
\bibinfo{author}{\bibnamefont{Nesterenko}, \bibfnamefont{V.~O.}},
  \bibinfo{author}{\bibfnamefont{W.}~\bibnamefont{Kleinig}},
  \bibinfo{author}{\bibfnamefont{F.~F.} \bibnamefont{de~Souza~Cruz}}, and
  \bibinfo{author}{\bibfnamefont{N.}~\bibnamefont{Lo~Iudice}},
  \bibinfo{year}{1999}, \bibinfo{journal}{Phys. Rev. Lett.}
  \textbf{\bibinfo{volume}{83}}, \bibinfo{pages}{57}.

\bibitem[{\citenamefont{Nesterenko}
  \emph{et~al.}(2000)\citenamefont{Nesterenko, Marinelli, de~Souza~Cruz,
  Kleinig, and Reinhard}}]{Nesterenko:2000}
\bibinfo{author}{\bibnamefont{Nesterenko}, \bibfnamefont{V.~O.}},
  \bibinfo{author}{\bibfnamefont{J.~R.} \bibnamefont{Marinelli}},
  \bibinfo{author}{\bibfnamefont{F.~F.} \bibnamefont{de~Souza~Cruz}},
  \bibinfo{author}{\bibfnamefont{W.}~\bibnamefont{Kleinig}}, and
  \bibinfo{author}{\bibfnamefont{P.-G.} \bibnamefont{Reinhard}},
  \bibinfo{year}{2000}, \bibinfo{journal}{Phys. Rev. Lett.}
  \textbf{\bibinfo{volume}{85}}, \bibinfo{pages}{3141}.

\bibitem[{\citenamefont{von Neumann-Cosel}(1997)}]{vonNeumann-Cosel:1997a}
\bibinfo{author}{\bibnamefont{von Neumann-Cosel}, \bibfnamefont{P.}},
  \bibinfo{year}{1997}, \bibinfo{journal}{Prog. Part. Nucl. Phys.}
  \textbf{\bibinfo{volume}{38}}, \bibinfo{pages}{213}.

\bibitem[{\citenamefont{von Neumann-Cosel} \emph{et~al.}(2009)\citenamefont{von
  Neumann-Cosel, Adachi, Bertulani, Carter, Dozono, Fujita, Fujita, Fujita,
  Hashimoto, Hatanaka, Itoh, Kalmykov} \emph{et~al.}}]{vonNeumann-Cosel:2009}
\bibinfo{author}{\bibnamefont{von Neumann-Cosel}, \bibfnamefont{P.}},
  \bibinfo{author}{\bibfnamefont{T.}~\bibnamefont{Adachi}},
  \bibinfo{author}{\bibfnamefont{C.~A.} \bibnamefont{Bertulani}},
  \bibinfo{author}{\bibfnamefont{J.}~\bibnamefont{Carter}},
  \bibinfo{author}{\bibfnamefont{M.}~\bibnamefont{Dozono}},
  \bibinfo{author}{\bibfnamefont{H.}~\bibnamefont{Fujita}},
  \bibinfo{author}{\bibfnamefont{K.}~\bibnamefont{Fujita}},
  \bibinfo{author}{\bibfnamefont{Y.}~\bibnamefont{Fujita}},
  \bibinfo{author}{\bibfnamefont{H.}~\bibnamefont{Hashimoto}},
  \bibinfo{author}{\bibfnamefont{K.}~\bibnamefont{Hatanaka}},
  \bibinfo{author}{\bibfnamefont{M.}~\bibnamefont{Itoh}},
  \bibinfo{author}{\bibfnamefont{Y.}~\bibnamefont{Kalmykov}}, \emph{et~al.},
  \bibinfo{year}{2009}, in \emph{\bibinfo{booktitle}{Proceedings of the 13th
  International Symposium on Capture Gamma-Ray Spectroscopy and Related
  Topics}}, edited by \bibinfo{editor}{\bibfnamefont{J.}~\bibnamefont{Jolie}},
  \bibinfo{editor}{\bibfnamefont{A.}~\bibnamefont{Zilges}},
  \bibinfo{editor}{\bibfnamefont{N.}~\bibnamefont{Warr}}, and
  \bibinfo{editor}{\bibfnamefont{A.}~\bibnamefont{Blazhev}}
  (\bibinfo{publisher}{AIP, Vol.~1090}), p. \bibinfo{pages}{404}.

\bibitem[{\citenamefont{von Neumann-Cosel and
  Ginocchio}(2000)}]{vonNeumann-Cosel:2000a}
\bibinfo{author}{\bibnamefont{von Neumann-Cosel}, \bibfnamefont{P.}}, and
  \bibinfo{author}{\bibfnamefont{J.~N.} \bibnamefont{Ginocchio}},
  \bibinfo{year}{2000}, \bibinfo{journal}{Phys. Rev. C}
  \textbf{\bibinfo{volume}{62}}, \bibinfo{pages}{014308}.

\bibitem[{\citenamefont{von Neumann-Cosel} \emph{et~al.}(1995)\citenamefont{von
  Neumann-Cosel, Ginocchio, Bauer, and Richter}}]{vonNeumann-Cosel:1995}
\bibinfo{author}{\bibnamefont{von Neumann-Cosel}, \bibfnamefont{P.}},
  \bibinfo{author}{\bibfnamefont{J.~N.} \bibnamefont{Ginocchio}},
  \bibinfo{author}{\bibfnamefont{H.}~\bibnamefont{Bauer}}, and
  \bibinfo{author}{\bibfnamefont{A.}~\bibnamefont{Richter}},
  \bibinfo{year}{1995}, \bibinfo{journal}{Phys. Rev. Lett.}
  \textbf{\bibinfo{volume}{75}}, \bibinfo{pages}{4178}.

\bibitem[{\citenamefont{von Neumann-Cosel} \emph{et~al.}(2000)\citenamefont{von
  Neumann-Cosel, Gr\"{a}f, Kr\"{a}mer, Richter, and
  Spamer}}]{vonNeumann-Cosel:2000b}
\bibinfo{author}{\bibnamefont{von Neumann-Cosel}, \bibfnamefont{P.}},
  \bibinfo{author}{\bibfnamefont{H.-D.} \bibnamefont{Gr\"{a}f}},
  \bibinfo{author}{\bibfnamefont{U.}~\bibnamefont{Kr\"{a}mer}},
  \bibinfo{author}{\bibfnamefont{A.}~\bibnamefont{Richter}}, and
  \bibinfo{author}{\bibfnamefont{E.}~\bibnamefont{Spamer}},
  \bibinfo{year}{2000}, \bibinfo{journal}{Nucl. Phys. A}
  \textbf{\bibinfo{volume}{669}}, \bibinfo{pages}{3}.

\bibitem[{\citenamefont{von Neumann-Cosel} \emph{et~al.}(1999)\citenamefont{von
  Neumann-Cosel, Neumeyer, Nishizaki, Ponomarev, Rangacharyulu, Reitz, Richter,
  Schrieder, Sober, Waindzoch, and Wambach}}]{vonNeumann-Cosel:1999}
\bibinfo{author}{\bibnamefont{von Neumann-Cosel}, \bibfnamefont{P.}},
  \bibinfo{author}{\bibfnamefont{F.}~\bibnamefont{Neumeyer}},
  \bibinfo{author}{\bibfnamefont{S.}~\bibnamefont{Nishizaki}},
  \bibinfo{author}{\bibfnamefont{V.~Y.} \bibnamefont{Ponomarev}},
  \bibinfo{author}{\bibfnamefont{C.}~\bibnamefont{Rangacharyulu}},
  \bibinfo{author}{\bibfnamefont{B.}~\bibnamefont{Reitz}},
  \bibinfo{author}{\bibfnamefont{A.}~\bibnamefont{Richter}},
  \bibinfo{author}{\bibfnamefont{G.}~\bibnamefont{Schrieder}},
  \bibinfo{author}{\bibfnamefont{D.~I.} \bibnamefont{Sober}},
  \bibinfo{author}{\bibfnamefont{T.}~\bibnamefont{Waindzoch}}, and
  \bibinfo{author}{\bibfnamefont{J.}~\bibnamefont{Wambach}},
  \bibinfo{year}{1999}, \bibinfo{journal}{Phys. Rev. Lett.}
  \textbf{\bibinfo{volume}{82}}, \bibinfo{pages}{1105}.

\bibitem[{\citenamefont{von Neumann-Cosel} \emph{et~al.}(1998)\citenamefont{von
  Neumann-Cosel, Poves, Retamosa, and Richter}}]{vonNeumann-Cosel:1998}
\bibinfo{author}{\bibnamefont{von Neumann-Cosel}, \bibfnamefont{P.}},
  \bibinfo{author}{\bibfnamefont{A.}~\bibnamefont{Poves}},
  \bibinfo{author}{\bibfnamefont{J.}~\bibnamefont{Retamosa}}, and
  \bibinfo{author}{\bibfnamefont{A.}~\bibnamefont{Richter}},
  \bibinfo{year}{1998}, \bibinfo{journal}{Phys. Lett. B}
  \textbf{\bibinfo{volume}{443}}, \bibinfo{pages}{1}.

\bibitem[{\citenamefont{von Neumann-Cosel} \emph{et~al.}(1997)\citenamefont{von
  Neumann-Cosel, Richter, Fujita, and Anderson}}]{vonNeumann-Cosel:1997b}
\bibinfo{author}{\bibnamefont{von Neumann-Cosel}, \bibfnamefont{P.}},
  \bibinfo{author}{\bibfnamefont{A.}~\bibnamefont{Richter}},
  \bibinfo{author}{\bibfnamefont{Y.}~\bibnamefont{Fujita}}, and
  \bibinfo{author}{\bibfnamefont{B.~D.} \bibnamefont{Anderson}},
  \bibinfo{year}{1997}, \bibinfo{journal}{Phys. Rev. C}
  \textbf{\bibinfo{volume}{55}}, \bibinfo{pages}{532}.

\bibitem[{\citenamefont{Nojarov} \emph{et~al.}(1986)\citenamefont{Nojarov,
  Bochnacki, and Faessler}}]{Nojarov:1986}
\bibinfo{author}{\bibnamefont{Nojarov}, \bibfnamefont{R.}},
  \bibinfo{author}{\bibfnamefont{Z.}~\bibnamefont{Bochnacki}}, and
  \bibinfo{author}{\bibfnamefont{A.}~\bibnamefont{Faessler}},
  \bibinfo{year}{1986}, \bibinfo{journal}{Z. Phys. A}
  \textbf{\bibinfo{volume}{324}}, \bibinfo{pages}{289}.

\bibitem[{\citenamefont{Nojarov and Faessler}(1990)}]{Nojarov:1990}
\bibinfo{author}{\bibnamefont{Nojarov}, \bibfnamefont{R.}}, and
  \bibinfo{author}{\bibfnamefont{A.}~\bibnamefont{Faessler}},
  \bibinfo{year}{1990}, \bibinfo{journal}{Z. Phys. A}
  \textbf{\bibinfo{volume}{336}}, \bibinfo{pages}{151}.

\bibitem[{\citenamefont{Nojarov} \emph{et~al.}(1987)\citenamefont{Nojarov,
  Faessler, and Civitarese}}]{Nojarov:1987}
\bibinfo{author}{\bibnamefont{Nojarov}, \bibfnamefont{R.}},
  \bibinfo{author}{\bibfnamefont{A.}~\bibnamefont{Faessler}}, and
  \bibinfo{author}{\bibfnamefont{O.}~\bibnamefont{Civitarese}},
  \bibinfo{year}{1987}, \bibinfo{journal}{Phys. Lett. B}
  \textbf{\bibinfo{volume}{183}}, \bibinfo{pages}{122}.

\bibitem[{\citenamefont{Nojarov} \emph{et~al.}(1995)\citenamefont{Nojarov,
  Faessler, and Dingfelder}}]{Nojarov:1995b}
\bibinfo{author}{\bibnamefont{Nojarov}, \bibfnamefont{R.}},
  \bibinfo{author}{\bibfnamefont{A.}~\bibnamefont{Faessler}}, and
  \bibinfo{author}{\bibfnamefont{M.}~\bibnamefont{Dingfelder}},
  \bibinfo{year}{1995}, \bibinfo{journal}{Phys. Rev. C}
  \textbf{\bibinfo{volume}{51}}, \bibinfo{pages}{2449}.

\bibitem[{\citenamefont{Nojarov} \emph{et~al.}(1991)\citenamefont{Nojarov,
  Faessler, and Lipas}}]{Nojarov:1991}
\bibinfo{author}{\bibnamefont{Nojarov}, \bibfnamefont{R.}},
  \bibinfo{author}{\bibfnamefont{A.}~\bibnamefont{Faessler}}, and
  \bibinfo{author}{\bibfnamefont{P.~O.} \bibnamefont{Lipas}},
  \bibinfo{year}{1991}, \bibinfo{journal}{Nucl. Phys. A}
  \textbf{\bibinfo{volume}{533}}, \bibinfo{pages}{381}.

\bibitem[{\citenamefont{Nojarov} \emph{et~al.}(1992)\citenamefont{Nojarov,
  Faessler, and Lipas}}]{Nojarov:1992}
\bibinfo{author}{\bibnamefont{Nojarov}, \bibfnamefont{R.}},
  \bibinfo{author}{\bibfnamefont{A.}~\bibnamefont{Faessler}}, and
  \bibinfo{author}{\bibfnamefont{P.~O.} \bibnamefont{Lipas}},
  \bibinfo{year}{1992}, \bibinfo{journal}{Nucl. Phys. A}
  \textbf{\bibinfo{volume}{537}}, \bibinfo{pages}{707 (erratum)}.

\bibitem[{\citenamefont{Nord} \emph{et~al.}(2003)\citenamefont{Nord, Enders,
  de~Almeida~Pinto, Belic, von Brentano, Fransen, Kneissl, Kohstall, Linneman,
  von Neumann-Cosel, Pietralla, Pitz} \emph{et~al.}}]{Nord:2003}
\bibinfo{author}{\bibnamefont{Nord}, \bibfnamefont{A.}},
  \bibinfo{author}{\bibfnamefont{J.}~\bibnamefont{Enders}},
  \bibinfo{author}{\bibfnamefont{A.~E.} \bibnamefont{de~Almeida~Pinto}},
  \bibinfo{author}{\bibfnamefont{D.}~\bibnamefont{Belic}},
  \bibinfo{author}{\bibfnamefont{P.}~\bibnamefont{von Brentano}},
  \bibinfo{author}{\bibfnamefont{C.}~\bibnamefont{Fransen}},
  \bibinfo{author}{\bibfnamefont{U.}~\bibnamefont{Kneissl}},
  \bibinfo{author}{\bibfnamefont{C.}~\bibnamefont{Kohstall}},
  \bibinfo{author}{\bibfnamefont{A.}~\bibnamefont{Linneman}},
  \bibinfo{author}{\bibfnamefont{P.}~\bibnamefont{von Neumann-Cosel}},
  \bibinfo{author}{\bibfnamefont{N.}~\bibnamefont{Pietralla}},
  \bibinfo{author}{\bibfnamefont{H.~H.} \bibnamefont{Pitz}}, \emph{et~al.},
  \bibinfo{year}{2003}, \bibinfo{journal}{Phys. Rev. C}
  \textbf{\bibinfo{volume}{67}}, \bibinfo{pages}{034307}.

\bibitem[{\citenamefont{Nord} \emph{et~al.}(1996)\citenamefont{Nord, Schiller,
  Eckert, Beck, Besserer, von Brentano, Fischer, Herzberg, J\"ager, Kneissl,
  Margraf, Maser} \emph{et~al.}}]{Nord:1996}
\bibinfo{author}{\bibnamefont{Nord}, \bibfnamefont{A.}},
  \bibinfo{author}{\bibfnamefont{A.}~\bibnamefont{Schiller}},
  \bibinfo{author}{\bibfnamefont{T.}~\bibnamefont{Eckert}},
  \bibinfo{author}{\bibfnamefont{O.}~\bibnamefont{Beck}},
  \bibinfo{author}{\bibfnamefont{J.}~\bibnamefont{Besserer}},
  \bibinfo{author}{\bibfnamefont{P.}~\bibnamefont{von Brentano}},
  \bibinfo{author}{\bibfnamefont{R.}~\bibnamefont{Fischer}},
  \bibinfo{author}{\bibfnamefont{R.-D.} \bibnamefont{Herzberg}},
  \bibinfo{author}{\bibfnamefont{D.}~\bibnamefont{J\"ager}},
  \bibinfo{author}{\bibfnamefont{U.}~\bibnamefont{Kneissl}},
  \bibinfo{author}{\bibfnamefont{J.}~\bibnamefont{Margraf}},
  \bibinfo{author}{\bibfnamefont{H.}~\bibnamefont{Maser}}, \emph{et~al.},
  \bibinfo{year}{1996}, \bibinfo{journal}{Phys. Rev. C}
  \textbf{\bibinfo{volume}{54}}, \bibinfo{pages}{2287}.

\bibitem[{\citenamefont{Oda and Muto}(1987)}]{Oda:1987}
\bibinfo{author}{\bibnamefont{Oda}, \bibfnamefont{M.~H., T.}}, and
  \bibinfo{author}{\bibfnamefont{K.}~\bibnamefont{Muto}}, \bibinfo{year}{1987},
  \bibinfo{journal}{Phys. Lett. B} \textbf{\bibinfo{volume}{190}},
  \bibinfo{pages}{14}.

\bibitem[{\citenamefont{Otsuka and Mizusaki}(1999)}]{Otsuka:1999}
\bibinfo{author}{\bibnamefont{Otsuka}, \bibfnamefont{M.~H., T.}}, and
  \bibinfo{author}{\bibfnamefont{T.}~\bibnamefont{Mizusaki}},
  \bibinfo{year}{1999}, \bibinfo{journal}{J. Phys. G}
  \textbf{\bibinfo{volume}{25}}, \bibinfo{pages}{699}.

\bibitem[{\citenamefont{Otsuka}(1990)}]{Otsuka:1990b}
\bibinfo{author}{\bibnamefont{Otsuka}, \bibfnamefont{T.}},
  \bibinfo{year}{1990}, \bibinfo{journal}{Nucl. Phys. A}
  \textbf{\bibinfo{volume}{507}}, \bibinfo{pages}{129c}.

\bibitem[{\citenamefont{Otsuka}(1992)}]{Otsuka:1992a}
\bibinfo{author}{\bibnamefont{Otsuka}, \bibfnamefont{T.}},
  \bibinfo{year}{1992}, \bibinfo{journal}{Hyperfine Int.}
  \textbf{\bibinfo{volume}{74}}, \bibinfo{pages}{93}.

\bibitem[{\citenamefont{Otsuka and Ginocchio}(1985)}]{Otsuka:1985}
\bibinfo{author}{\bibnamefont{Otsuka}, \bibfnamefont{T.}}, and
  \bibinfo{author}{\bibfnamefont{J.~N.} \bibnamefont{Ginocchio}},
  \bibinfo{year}{1985}, \bibinfo{journal}{Phys. Rev. Lett.}
  \textbf{\bibinfo{volume}{54}}, \bibinfo{pages}{777}.

\bibitem[{\citenamefont{Otsuka} \emph{et~al.}(1998)\citenamefont{Otsuka, Honma,
  and Mizusaki}}]{Otsuka:1998}
\bibinfo{author}{\bibnamefont{Otsuka}, \bibfnamefont{T.}},
  \bibinfo{author}{\bibfnamefont{M.}~\bibnamefont{Honma}}, and
  \bibinfo{author}{\bibfnamefont{T.}~\bibnamefont{Mizusaki}},
  \bibinfo{year}{1998}, \bibinfo{journal}{Phys. Rev. Lett.}
  \textbf{\bibinfo{volume}{81}}, \bibinfo{pages}{1588}.

\bibitem[{\citenamefont{Otsuka and Kim}(1994)}]{Otsuka:1994b}
\bibinfo{author}{\bibnamefont{Otsuka}, \bibfnamefont{T.}}, and
  \bibinfo{author}{\bibfnamefont{K.-H.} \bibnamefont{Kim}},
  \bibinfo{year}{1994}, \bibinfo{journal}{Phys. Rev. C}
  \textbf{\bibinfo{volume}{50}}, \bibinfo{pages}{R1768}.

\bibitem[{\citenamefont{Otsuka} \emph{et~al.}(1990)\citenamefont{Otsuka, Pan,
  and Arima}}]{Otsuka:1990a}
\bibinfo{author}{\bibnamefont{Otsuka}, \bibfnamefont{T.}},
  \bibinfo{author}{\bibfnamefont{X.~W.} \bibnamefont{Pan}}, and
  \bibinfo{author}{\bibfnamefont{A.}~\bibnamefont{Arima}},
  \bibinfo{year}{1990}, \bibinfo{journal}{Phys. Lett. B}
  \textbf{\bibinfo{volume}{247}}, \bibinfo{pages}{191}.

\bibitem[{\citenamefont{Otten}(1989)}]{Otten:1989}
\bibinfo{author}{\bibnamefont{Otten}, \bibfnamefont{E.}}, \bibinfo{year}{1989},
  \emph{\bibinfo{title}{Treatise on Heavy-Ion Science, Vol.~8}}
  (\bibinfo{publisher}{Plenum}, \bibinfo{address}{New York}).

\bibitem[{\citenamefont{Petermann} \emph{et~al.}(2010)\citenamefont{Petermann,
  Langanke, Mart\'{\i}nez-Pinedo, von Neumann-Cosel, Nowacki, and
  Richter}}]{Petermann:2008}
\bibinfo{author}{\bibnamefont{Petermann}, \bibfnamefont{I.}},
  \bibinfo{author}{\bibfnamefont{K.}~\bibnamefont{Langanke}},
  \bibinfo{author}{\bibfnamefont{G.}~\bibnamefont{Mart\'{\i}nez-Pinedo}},
  \bibinfo{author}{\bibfnamefont{P.}~\bibnamefont{von Neumann-Cosel}},
  \bibinfo{author}{\bibfnamefont{F.}~\bibnamefont{Nowacki}}, and
  \bibinfo{author}{\bibfnamefont{A.}~\bibnamefont{Richter}},
  \bibinfo{year}{2010}, \bibinfo{journal}{Phys. Rev. C}
  \textbf{\bibinfo{volume}{81}}, \bibinfo{pages}{014308}.

\bibitem[{\citenamefont{Pietralla}
  \emph{et~al.}(2001{\natexlab{a}})\citenamefont{Pietralla, Barton~III,
  Kr\"{u}cken, Beausang, Caprio, Casten, Cooper, Hecht, Newman, Novak, and
  Zamfir}}]{Pietralla:2001}
\bibinfo{author}{\bibnamefont{Pietralla}, \bibfnamefont{N.}},
  \bibinfo{author}{\bibfnamefont{C.~J.} \bibnamefont{Barton~III}},
  \bibinfo{author}{\bibfnamefont{R.}~\bibnamefont{Kr\"{u}cken}},
  \bibinfo{author}{\bibfnamefont{C.~W.} \bibnamefont{Beausang}},
  \bibinfo{author}{\bibfnamefont{M.~A.} \bibnamefont{Caprio}},
  \bibinfo{author}{\bibfnamefont{R.~F.} \bibnamefont{Casten}},
  \bibinfo{author}{\bibfnamefont{J.~R.} \bibnamefont{Cooper}},
  \bibinfo{author}{\bibfnamefont{A.~A.} \bibnamefont{Hecht}},
  \bibinfo{author}{\bibfnamefont{H.}~\bibnamefont{Newman}},
  \bibinfo{author}{\bibfnamefont{J.~R.} \bibnamefont{Novak}}, and
  \bibinfo{author}{\bibfnamefont{N.~V.} \bibnamefont{Zamfir}},
  \bibinfo{year}{2001}{\natexlab{a}}, \bibinfo{journal}{Phys. Rev. C}
  \textbf{\bibinfo{volume}{64}}, \bibinfo{pages}{031301(R)}.

\bibitem[{\citenamefont{Pietralla}
  \emph{et~al.}(2001{\natexlab{b}})\citenamefont{Pietralla, Berant, Litvinenko,
  Hartman, Mikhailov, Pinayev, Swift, Ahmed, Kelley, Nelson, Prior, Sabourov}
  \emph{et~al.}}]{Pietralla:2002}
\bibinfo{author}{\bibnamefont{Pietralla}, \bibfnamefont{N.}},
  \bibinfo{author}{\bibfnamefont{Z.}~\bibnamefont{Berant}},
  \bibinfo{author}{\bibfnamefont{V.~N.} \bibnamefont{Litvinenko}},
  \bibinfo{author}{\bibfnamefont{S.}~\bibnamefont{Hartman}},
  \bibinfo{author}{\bibfnamefont{F.~F.} \bibnamefont{Mikhailov}},
  \bibinfo{author}{\bibfnamefont{I.~V.} \bibnamefont{Pinayev}},
  \bibinfo{author}{\bibfnamefont{G.}~\bibnamefont{Swift}},
  \bibinfo{author}{\bibfnamefont{M.~W.} \bibnamefont{Ahmed}},
  \bibinfo{author}{\bibfnamefont{J.~H.} \bibnamefont{Kelley}},
  \bibinfo{author}{\bibfnamefont{S.~O.} \bibnamefont{Nelson}},
  \bibinfo{author}{\bibfnamefont{R.}~\bibnamefont{Prior}},
  \bibinfo{author}{\bibfnamefont{K.}~\bibnamefont{Sabourov}}, \emph{et~al.},
  \bibinfo{year}{2001}{\natexlab{b}}, \bibinfo{journal}{Phys. Rev. Lett.}
  \textbf{\bibinfo{volume}{88}}, \bibinfo{pages}{012502}.

\bibitem[{\citenamefont{Pietralla}
  \emph{et~al.}(1998{\natexlab{a}})\citenamefont{Pietralla, von Brentano,
  Gelberg, Otsuka, Richter, Smirnova, and Wiedenh\"over}}]{Pietralla:1998a}
\bibinfo{author}{\bibnamefont{Pietralla}, \bibfnamefont{N.}},
  \bibinfo{author}{\bibfnamefont{P.}~\bibnamefont{von Brentano}},
  \bibinfo{author}{\bibfnamefont{A.}~\bibnamefont{Gelberg}},
  \bibinfo{author}{\bibfnamefont{T.}~\bibnamefont{Otsuka}},
  \bibinfo{author}{\bibfnamefont{A.}~\bibnamefont{Richter}},
  \bibinfo{author}{\bibfnamefont{N.}~\bibnamefont{Smirnova}}, and
  \bibinfo{author}{\bibfnamefont{I.}~\bibnamefont{Wiedenh\"over}},
  \bibinfo{year}{1998}{\natexlab{a}}, \bibinfo{journal}{Phys. Rev. C}
  \textbf{\bibinfo{volume}{58}}, \bibinfo{pages}{191}.

\bibitem[{\citenamefont{Pietralla}
  \emph{et~al.}(1998{\natexlab{b}})\citenamefont{Pietralla, von Brentano,
  Herzberg, Kneissl, Lo~Iudice, Maser, Pitz, and Zilges}}]{Pietralla:1998b}
\bibinfo{author}{\bibnamefont{Pietralla}, \bibfnamefont{N.}},
  \bibinfo{author}{\bibfnamefont{P.}~\bibnamefont{von Brentano}},
  \bibinfo{author}{\bibfnamefont{R.-D.} \bibnamefont{Herzberg}},
  \bibinfo{author}{\bibfnamefont{U.}~\bibnamefont{Kneissl}},
  \bibinfo{author}{\bibfnamefont{N.}~\bibnamefont{Lo~Iudice}},
  \bibinfo{author}{\bibfnamefont{H.}~\bibnamefont{Maser}},
  \bibinfo{author}{\bibfnamefont{H.~H.} \bibnamefont{Pitz}}, and
  \bibinfo{author}{\bibfnamefont{A.}~\bibnamefont{Zilges}},
  \bibinfo{year}{1998}{\natexlab{b}}, \bibinfo{journal}{Phys. Rev. C}
  \textbf{\bibinfo{volume}{58}}, \bibinfo{pages}{184}.

\bibitem[{\citenamefont{Pietralla} \emph{et~al.}(2008)\citenamefont{Pietralla,
  von Brentano, and Lisetsky}}]{Pietralla:2008}
\bibinfo{author}{\bibnamefont{Pietralla}, \bibfnamefont{N.}},
  \bibinfo{author}{\bibfnamefont{P.}~\bibnamefont{von Brentano}}, and
  \bibinfo{author}{\bibfnamefont{A.~F.} \bibnamefont{Lisetsky}},
  \bibinfo{year}{2008}, \bibinfo{journal}{Prog. Part. Nucl. Phys.}
  \textbf{\bibinfo{volume}{60}}, \bibinfo{pages}{225}.

\bibitem[{\citenamefont{Pietralla} \emph{et~al.}(1999)\citenamefont{Pietralla,
  Fransen, Belic, von Brentano, Frie\ss{}ner, Kneissl, Linnemann, Nord, Pitz,
  Otsuka, Schneider, Werner} \emph{et~al.}}]{Pietralla:1999}
\bibinfo{author}{\bibnamefont{Pietralla}, \bibfnamefont{N.}},
  \bibinfo{author}{\bibfnamefont{C.}~\bibnamefont{Fransen}},
  \bibinfo{author}{\bibfnamefont{D.}~\bibnamefont{Belic}},
  \bibinfo{author}{\bibfnamefont{P.}~\bibnamefont{von Brentano}},
  \bibinfo{author}{\bibfnamefont{C.}~\bibnamefont{Frie\ss{}ner}},
  \bibinfo{author}{\bibfnamefont{U.}~\bibnamefont{Kneissl}},
  \bibinfo{author}{\bibfnamefont{A.}~\bibnamefont{Linnemann}},
  \bibinfo{author}{\bibfnamefont{A.}~\bibnamefont{Nord}},
  \bibinfo{author}{\bibfnamefont{H.~H.} \bibnamefont{Pitz}},
  \bibinfo{author}{\bibfnamefont{T.}~\bibnamefont{Otsuka}},
  \bibinfo{author}{\bibfnamefont{I.}~\bibnamefont{Schneider}},
  \bibinfo{author}{\bibfnamefont{V.}~\bibnamefont{Werner}}, \emph{et~al.},
  \bibinfo{year}{1999}, \bibinfo{journal}{Phys. Rev. Lett.}
  \textbf{\bibinfo{volume}{83}}, \bibinfo{pages}{1303}.

\bibitem[{\citenamefont{Pietralla} \emph{et~al.}(2000)\citenamefont{Pietralla,
  Fransen, von Brentano, Dewald, Fitzler, Frie\ss{}ner, and
  Gableske}}]{Pietralla:2000}
\bibinfo{author}{\bibnamefont{Pietralla}, \bibfnamefont{N.}},
  \bibinfo{author}{\bibfnamefont{C.}~\bibnamefont{Fransen}},
  \bibinfo{author}{\bibfnamefont{P.}~\bibnamefont{von Brentano}},
  \bibinfo{author}{\bibfnamefont{A.}~\bibnamefont{Dewald}},
  \bibinfo{author}{\bibfnamefont{A.}~\bibnamefont{Fitzler}},
  \bibinfo{author}{\bibfnamefont{C.}~\bibnamefont{Frie\ss{}ner}}, and
  \bibinfo{author}{\bibfnamefont{J.}~\bibnamefont{Gableske}},
  \bibinfo{year}{2000}, \bibinfo{journal}{Phys. Rev. Lett.}
  \textbf{\bibinfo{volume}{84}}, \bibinfo{pages}{3775}.

\bibitem[{\citenamefont{Pitaevskii and Stringari}(2003)}]{Pitaevskii:2003}
\bibinfo{author}{\bibnamefont{Pitaevskii}, \bibfnamefont{L.}}, and
  \bibinfo{author}{\bibfnamefont{S.}~\bibnamefont{Stringari}},
  \bibinfo{year}{2003}, \emph{\bibinfo{title}{Bose-Einstein Condensation}},
  International Series of Monographs on Physics, Vol.~116
  (\bibinfo{publisher}{Oxford Science}, \bibinfo{address}{Clarendon}).

\bibitem[{\citenamefont{Poves} \emph{et~al.}(1989)\citenamefont{Poves,
  Retamosa, and Moya~de Guerra}}]{Poves:1989}
\bibinfo{author}{\bibnamefont{Poves}, \bibfnamefont{A.}},
  \bibinfo{author}{\bibfnamefont{J.}~\bibnamefont{Retamosa}}, and
  \bibinfo{author}{\bibfnamefont{E.}~\bibnamefont{Moya~de Guerra}},
  \bibinfo{year}{1989}, \bibinfo{journal}{Phys. Rev. C}
  \textbf{\bibinfo{volume}{39}}, \bibinfo{pages}{1639}.

\bibitem[{\citenamefont{Poves} \emph{et~al.}(2001)\citenamefont{Poves,
  S\'{a}nchez-Solano, Caurier, and Nowacki}}]{Poves:2001}
\bibinfo{author}{\bibnamefont{Poves}, \bibfnamefont{A.}},
  \bibinfo{author}{\bibfnamefont{J.}~\bibnamefont{S\'{a}nchez-Solano}},
  \bibinfo{author}{\bibfnamefont{E.}~\bibnamefont{Caurier}}, and
  \bibinfo{author}{\bibfnamefont{F.}~\bibnamefont{Nowacki}},
  \bibinfo{year}{2001}, \bibinfo{journal}{Nucl. Phys. A}
  \textbf{\bibinfo{volume}{694}}, \bibinfo{pages}{157}.

\bibitem[{\citenamefont{Puente and Serra}(1999)}]{Puente:1999}
\bibinfo{author}{\bibnamefont{Puente}, \bibfnamefont{A.}}, and
  \bibinfo{author}{\bibfnamefont{L.}~\bibnamefont{Serra}},
  \bibinfo{year}{1999}, \bibinfo{journal}{Phys. Rev. Lett.}
  \textbf{\bibinfo{volume}{83}}, \bibinfo{pages}{3266}.

\bibitem[{\citenamefont{Raduta and Delion}(1990)}]{Raduta:1990a}
\bibinfo{author}{\bibnamefont{Raduta}, \bibfnamefont{A.~A.}}, and
  \bibinfo{author}{\bibfnamefont{D.~S.} \bibnamefont{Delion}},
  \bibinfo{year}{1990}, \bibinfo{journal}{Nucl. Phys. A}
  \textbf{\bibinfo{volume}{513}}, \bibinfo{pages}{11}.

\bibitem[{\citenamefont{Raduta and Lo~Iudice}(1989)}]{Raduta:1989c}
\bibinfo{author}{\bibnamefont{Raduta}, \bibfnamefont{A.~A.}}, and
  \bibinfo{author}{\bibfnamefont{N.}~\bibnamefont{Lo~Iudice}},
  \bibinfo{year}{1989}, \bibinfo{journal}{Z. Phys. A}
  \textbf{\bibinfo{volume}{334}}, \bibinfo{pages}{403}.

\bibitem[{\citenamefont{Rainovski} \emph{et~al.}(2006)\citenamefont{Rainovski,
  Pietralla, Ahn, Lister, Janssens, Carpenter, Zhu, and
  Barton~III}}]{Rainovski:2006}
\bibinfo{author}{\bibnamefont{Rainovski}, \bibfnamefont{G.}},
  \bibinfo{author}{\bibfnamefont{N.}~\bibnamefont{Pietralla}},
  \bibinfo{author}{\bibfnamefont{T.}~\bibnamefont{Ahn}},
  \bibinfo{author}{\bibfnamefont{C.~J.} \bibnamefont{Lister}},
  \bibinfo{author}{\bibfnamefont{R.~V.~F.} \bibnamefont{Janssens}},
  \bibinfo{author}{\bibfnamefont{M.~P.} \bibnamefont{Carpenter}},
  \bibinfo{author}{\bibfnamefont{S.}~\bibnamefont{Zhu}}, and
  \bibinfo{author}{\bibfnamefont{C.~J.} \bibnamefont{Barton~III}},
  \bibinfo{year}{2006}, \bibinfo{journal}{Phys. Rev. Lett.}
  \textbf{\bibinfo{volume}{96}}, \bibinfo{eid}{122501}.

\bibitem[{\citenamefont{Raman} \emph{et~al.}(1991)\citenamefont{Raman, Fagg,
  and Hicks}}]{Raman:1991}
\bibinfo{author}{\bibnamefont{Raman}, \bibfnamefont{S.}},
  \bibinfo{author}{\bibfnamefont{L.~W.} \bibnamefont{Fagg}}, and
  \bibinfo{author}{\bibfnamefont{R.~S.} \bibnamefont{Hicks}},
  \bibinfo{year}{1991} (\bibinfo{publisher}{World Scientific},
  \bibinfo{address}{Singapore}), International Review of Nuclear Physics,
  Vol.~7, p. \bibinfo{pages}{355}.

\bibitem[{\citenamefont{Rangacharyulu}
  \emph{et~al.}(1991)\citenamefont{Rangacharyulu, Richter, W\"ortche, Ziegler,
  and Casten}}]{Rangacharyulu:1991}
\bibinfo{author}{\bibnamefont{Rangacharyulu}, \bibfnamefont{C.}},
  \bibinfo{author}{\bibfnamefont{A.}~\bibnamefont{Richter}},
  \bibinfo{author}{\bibfnamefont{H.~J.} \bibnamefont{W\"ortche}},
  \bibinfo{author}{\bibfnamefont{W.}~\bibnamefont{Ziegler}}, and
  \bibinfo{author}{\bibfnamefont{R.~F.} \bibnamefont{Casten}},
  \bibinfo{year}{1991}, \bibinfo{journal}{Phys. Rev. C}
  \textbf{\bibinfo{volume}{43}}, \bibinfo{pages}{R949}.

\bibitem[{\citenamefont{Reitz} \emph{et~al.}(2002)\citenamefont{Reitz, van~den
  Berg, Frekers, Hofmann, de~Huu, Kalmykov, Lenske, von Neumann-Cosel,
  Ponomarev, Rakers, Richter, Schrieder} \emph{et~al.}}]{Reitz:2002}
\bibinfo{author}{\bibnamefont{Reitz}, \bibfnamefont{B.}},
  \bibinfo{author}{\bibfnamefont{A.~M.} \bibnamefont{van~den Berg}},
  \bibinfo{author}{\bibfnamefont{D.}~\bibnamefont{Frekers}},
  \bibinfo{author}{\bibfnamefont{F.}~\bibnamefont{Hofmann}},
  \bibinfo{author}{\bibfnamefont{M.}~\bibnamefont{de~Huu}},
  \bibinfo{author}{\bibfnamefont{Y.}~\bibnamefont{Kalmykov}},
  \bibinfo{author}{\bibfnamefont{H.}~\bibnamefont{Lenske}},
  \bibinfo{author}{\bibfnamefont{P.}~\bibnamefont{von Neumann-Cosel}},
  \bibinfo{author}{\bibfnamefont{V.~Y.} \bibnamefont{Ponomarev}},
  \bibinfo{author}{\bibfnamefont{S.}~\bibnamefont{Rakers}},
  \bibinfo{author}{\bibfnamefont{A.}~\bibnamefont{Richter}},
  \bibinfo{author}{\bibfnamefont{G.}~\bibnamefont{Schrieder}}, \emph{et~al.},
  \bibinfo{year}{2002}, \bibinfo{journal}{Phys. Lett. B}
  \textbf{\bibinfo{volume}{532}}, \bibinfo{pages}{179}.

\bibitem[{\citenamefont{Reitz} \emph{et~al.}(1999)\citenamefont{Reitz, Hofmann,
  von Neumann-Cosel, Neumeyer, Rangacharyulu, Richter, Schrieder, Sober, and
  Brown}}]{Reitz:1999}
\bibinfo{author}{\bibnamefont{Reitz}, \bibfnamefont{B.}},
  \bibinfo{author}{\bibfnamefont{F.}~\bibnamefont{Hofmann}},
  \bibinfo{author}{\bibfnamefont{P.}~\bibnamefont{von Neumann-Cosel}},
  \bibinfo{author}{\bibfnamefont{F.}~\bibnamefont{Neumeyer}},
  \bibinfo{author}{\bibfnamefont{C.}~\bibnamefont{Rangacharyulu}},
  \bibinfo{author}{\bibfnamefont{A.}~\bibnamefont{Richter}},
  \bibinfo{author}{\bibfnamefont{G.}~\bibnamefont{Schrieder}},
  \bibinfo{author}{\bibfnamefont{D.~I.} \bibnamefont{Sober}}, and
  \bibinfo{author}{\bibfnamefont{B.~A.} \bibnamefont{Brown}},
  \bibinfo{year}{1999}, \bibinfo{journal}{Phys. Rev. Lett.}
  \textbf{\bibinfo{volume}{82}}, \bibinfo{pages}{291}.

\bibitem[{\citenamefont{Retamosa} \emph{et~al.}(1990)\citenamefont{Retamosa,
  Udias, Poves, and Moya~de Guerra}}]{Retamosa:1990}
\bibinfo{author}{\bibnamefont{Retamosa}, \bibfnamefont{J.}},
  \bibinfo{author}{\bibfnamefont{J.~M.} \bibnamefont{Udias}},
  \bibinfo{author}{\bibfnamefont{A.}~\bibnamefont{Poves}}, and
  \bibinfo{author}{\bibfnamefont{E.}~\bibnamefont{Moya~de Guerra}},
  \bibinfo{year}{1990}, \bibinfo{journal}{Nucl. Phys. A}
  \textbf{\bibinfo{volume}{511}}, \bibinfo{pages}{221}.

\bibitem[{\citenamefont{RIA}(2003)}]{RIA:2003}
\bibinfo{author}{\bibnamefont{RIA}}, \bibinfo{year}{2003},
  \emph{\bibinfo{title}{The Intellectual Challenges of RIA: A White Paper from
  the RIA Users Community}}
  (\bibinfo{publisher}{http://www.orau.org/ria/pdf/intell.pdf}).

\bibitem[{\citenamefont{RIA}(2006)}]{RIA:2006}
\bibinfo{author}{\bibnamefont{RIA}}, \bibinfo{year}{2006},
  \emph{\bibinfo{title}{Isotope Science Facility at Michigan State University}}
  (\bibinfo{publisher}{http://www.nscl.msu.edu/future/nsclwhitepaper2006}).

\bibitem[{\citenamefont{Richter}(1983)}]{Richter:1983}
\bibinfo{author}{\bibnamefont{Richter}, \bibfnamefont{A.}},
  \bibinfo{year}{1983}, in \emph{\bibinfo{booktitle}{Proceedings of the
  International Conference on Nuclear Physics, Vol. 2}}, edited by
  \bibinfo{editor}{\bibfnamefont{P.}~\bibnamefont{Blasi}} and
  \bibinfo{editor}{\bibfnamefont{R.~A.} \bibnamefont{Ricci}}
  (\bibinfo{publisher}{Tipografica Compositori}, \bibinfo{address}{Bologna}),
  p. \bibinfo{pages}{189}.

\bibitem[{\citenamefont{Richter}(1985)}]{Richter:1985}
\bibinfo{author}{\bibnamefont{Richter}, \bibfnamefont{A.}},
  \bibinfo{year}{1985}, \bibinfo{journal}{Prog. Part. Nucl. Phys.}
  \textbf{\bibinfo{volume}{13}}, \bibinfo{pages}{1}.

\bibitem[{\citenamefont{Richter}(1990)}]{Richter:1990a}
\bibinfo{author}{\bibnamefont{Richter}, \bibfnamefont{A.}},
  \bibinfo{year}{1990}, \bibinfo{journal}{Nucl. Phys. A}
  \textbf{\bibinfo{volume}{507}}, \bibinfo{pages}{99c}.

\bibitem[{\citenamefont{Richter}(1991)}]{Richter:1991}
\bibinfo{author}{\bibnamefont{Richter}, \bibfnamefont{A.}},
  \bibinfo{year}{1991}, \bibinfo{journal}{Nucl. Phys. A}
  \textbf{\bibinfo{volume}{522}}, \bibinfo{pages}{139c}.

\bibitem[{\citenamefont{Richter}(1993)}]{Richter:1993a}
\bibinfo{author}{\bibnamefont{Richter}, \bibfnamefont{A.}},
  \bibinfo{year}{1993}, \bibinfo{journal}{Nucl. Phys. A}
  \textbf{\bibinfo{volume}{553}}, \bibinfo{pages}{417c}.

\bibitem[{\citenamefont{Richter}(1994)}]{Richter:1993b}
\bibinfo{author}{\bibnamefont{Richter}, \bibfnamefont{A.}},
  \bibinfo{year}{1994}, in \emph{\bibinfo{booktitle}{Proceedings of the 4th
  Internationbal Spring Seminar on Nuclear Physics 'The Building Blocks of
  Nuclear Structure'}}, edited by
  \bibinfo{editor}{\bibfnamefont{A.}~\bibnamefont{Covello}}
  (\bibinfo{publisher}{World Scientific}, \bibinfo{address}{Singapore}), p.
  \bibinfo{pages}{335}.

\bibitem[{\citenamefont{Richter}(1995)}]{Richter:1995}
\bibinfo{author}{\bibnamefont{Richter}, \bibfnamefont{A.}},
  \bibinfo{year}{1995}, \bibinfo{journal}{Prog. Part. Nucl. Phys.}
  \textbf{\bibinfo{volume}{34}}, \bibinfo{pages}{261}.

\bibitem[{\citenamefont{Richter and Kn\"{u}pfer}(1980)}]{Richter:1980}
\bibinfo{author}{\bibnamefont{Richter}, \bibfnamefont{A.}}, and
  \bibinfo{author}{\bibfnamefont{W.}~\bibnamefont{Kn\"{u}pfer}},
  \bibinfo{year}{1980}, in \emph{\bibinfo{booktitle}{Electron and Pion
  Interactions with Nuclei at Intermediate Energies}}, edited by
  \bibinfo{editor}{\bibfnamefont{W.}~\bibnamefont{Bertozzi}},
  \bibinfo{editor}{\bibfnamefont{S.}~\bibnamefont{Costa}}, and
  \bibinfo{editor}{\bibfnamefont{C.}~\bibnamefont{Schaerf}}
  (\bibinfo{publisher}{Harwood Academic}, \bibinfo{address}{Newark, New
  Jersey}), p. \bibinfo{pages}{241}.

\bibitem[{\citenamefont{Richter} \emph{et~al.}(1990)\citenamefont{Richter,
  Weiss, H\"ausser, and Brown}}]{Richter:1990b}
\bibinfo{author}{\bibnamefont{Richter}, \bibfnamefont{A.}},
  \bibinfo{author}{\bibfnamefont{A.}~\bibnamefont{Weiss}},
  \bibinfo{author}{\bibfnamefont{O.}~\bibnamefont{H\"ausser}}, and
  \bibinfo{author}{\bibfnamefont{B.~A.} \bibnamefont{Brown}},
  \bibinfo{year}{1990}, \bibinfo{journal}{Phys. Rev. Lett.}
  \textbf{\bibinfo{volume}{65}}, \bibinfo{pages}{2519}.

\bibitem[{\citenamefont{Ring and Schuck}(1980)}]{Ring:1980}
\bibinfo{author}{\bibnamefont{Ring}, \bibfnamefont{P.}}, and
  \bibinfo{author}{\bibfnamefont{P.}~\bibnamefont{Schuck}},
  \bibinfo{year}{1980}, \emph{\bibinfo{title}{The Nuclear Many-Body Problem}}
  (\bibinfo{publisher}{Springer}, \bibinfo{address}{Heidelberg}).

\bibitem[{\citenamefont{Rohozinski and Greiner}(1985)}]{Rohozinski:1985}
\bibinfo{author}{\bibnamefont{Rohozinski}, \bibfnamefont{S.~G.}}, and
  \bibinfo{author}{\bibfnamefont{W.}~\bibnamefont{Greiner}},
  \bibinfo{year}{1985}, \bibinfo{journal}{Z. Phys. A}
  \textbf{\bibinfo{volume}{322}}, \bibinfo{pages}{271}.

\bibitem[{\citenamefont{Rompf} \emph{et~al.}(1998)\citenamefont{Rompf,
  Beuschel, Draayer, Scheid, and Hirsch}}]{Rompf:1998}
\bibinfo{author}{\bibnamefont{Rompf}, \bibfnamefont{D.}},
  \bibinfo{author}{\bibfnamefont{T.}~\bibnamefont{Beuschel}},
  \bibinfo{author}{\bibfnamefont{J.~P.} \bibnamefont{Draayer}},
  \bibinfo{author}{\bibfnamefont{W.}~\bibnamefont{Scheid}}, and
  \bibinfo{author}{\bibfnamefont{J.~G.} \bibnamefont{Hirsch}},
  \bibinfo{year}{1998}, \bibinfo{journal}{Phys. Rev. C}
  \textbf{\bibinfo{volume}{57}}, \bibinfo{pages}{1703}.

\bibitem[{\citenamefont{Rowe}(1970)}]{Rowe:1970}
\bibinfo{author}{\bibnamefont{Rowe}, \bibfnamefont{D.}}, \bibinfo{year}{1970},
  \emph{\bibinfo{title}{Nuclear Collective Motion}}
  (\bibinfo{publisher}{Methuen and Co.}, \bibinfo{address}{London}).

\bibitem[{\citenamefont{Sambataro and Dieperink}(1981)}]{Sambataro:1981}
\bibinfo{author}{\bibnamefont{Sambataro}, \bibfnamefont{M.}}, and
  \bibinfo{author}{\bibfnamefont{A.}~\bibnamefont{Dieperink}},
  \bibinfo{year}{1981}, \bibinfo{journal}{Phys. Lett. B}
  \textbf{\bibinfo{volume}{107}}, \bibinfo{pages}{249}.

\bibitem[{\citenamefont{Sambataro} \emph{et~al.}(1984)\citenamefont{Sambataro,
  Scholten, Dieperink, and Picciotto}}]{Sambataro:1984}
\bibinfo{author}{\bibnamefont{Sambataro}, \bibfnamefont{M.}},
  \bibinfo{author}{\bibfnamefont{O.}~\bibnamefont{Scholten}},
  \bibinfo{author}{\bibfnamefont{A.~E.~L.} \bibnamefont{Dieperink}}, and
  \bibinfo{author}{\bibfnamefont{G.}~\bibnamefont{Picciotto}},
  \bibinfo{year}{1984}, \bibinfo{journal}{Nucl. Phys. A}
  \textbf{\bibinfo{volume}{423}}, \bibinfo{pages}{333}.

\bibitem[{\citenamefont{Sarriguren}
  \emph{et~al.}(1996)\citenamefont{Sarriguren, Moya~de Guerra, and
  Nojarov}}]{Sarriguren:1996}
\bibinfo{author}{\bibnamefont{Sarriguren}, \bibfnamefont{P.}},
  \bibinfo{author}{\bibfnamefont{E.}~\bibnamefont{Moya~de Guerra}}, and
  \bibinfo{author}{\bibfnamefont{R.}~\bibnamefont{Nojarov}},
  \bibinfo{year}{1996}, \bibinfo{journal}{Phys. Rev. C}
  \textbf{\bibinfo{volume}{54}}, \bibinfo{pages}{690}.

\bibitem[{\citenamefont{Sarriguren}
  \emph{et~al.}(1993)\citenamefont{Sarriguren, de~Guerra, Nojarov, and
  Faessler}}]{Sarriguren:1993}
\bibinfo{author}{\bibnamefont{Sarriguren}, \bibfnamefont{P.}},
  \bibinfo{author}{\bibfnamefont{E.~M.} \bibnamefont{de~Guerra}},
  \bibinfo{author}{\bibfnamefont{R.}~\bibnamefont{Nojarov}}, and
  \bibinfo{author}{\bibfnamefont{A.}~\bibnamefont{Faessler}},
  \bibinfo{year}{1993}, \bibinfo{journal}{J. Phys. G}
  \textbf{\bibinfo{volume}{19}}, \bibinfo{pages}{291}.

\bibitem[{\citenamefont{Sarriguren}
  \emph{et~al.}(1994)\citenamefont{Sarriguren, de~Guerra, Nojarov, and
  Faessler}}]{Sarriguren:1994}
\bibinfo{author}{\bibnamefont{Sarriguren}, \bibfnamefont{P.}},
  \bibinfo{author}{\bibfnamefont{E.~M.} \bibnamefont{de~Guerra}},
  \bibinfo{author}{\bibfnamefont{R.}~\bibnamefont{Nojarov}}, and
  \bibinfo{author}{\bibfnamefont{A.}~\bibnamefont{Faessler}},
  \bibinfo{year}{1994}, \bibinfo{journal}{J. Phys. G}
  \textbf{\bibinfo{volume}{20}}, \bibinfo{pages}{315}.

\bibitem[{\citenamefont{Savran} \emph{et~al.}(2005)\citenamefont{Savran,
  M\"uller, Zilges, Babilon, Ahmed, Kelley, Tonchev, Tornow, Weller, Pietralla,
  Li, Pinayev} \emph{et~al.}}]{Savran:2005}
\bibinfo{author}{\bibnamefont{Savran}, \bibfnamefont{D.}},
  \bibinfo{author}{\bibfnamefont{S.}~\bibnamefont{M\"uller}},
  \bibinfo{author}{\bibfnamefont{A.}~\bibnamefont{Zilges}},
  \bibinfo{author}{\bibfnamefont{M.}~\bibnamefont{Babilon}},
  \bibinfo{author}{\bibfnamefont{M.~W.} \bibnamefont{Ahmed}},
  \bibinfo{author}{\bibfnamefont{J.~H.} \bibnamefont{Kelley}},
  \bibinfo{author}{\bibfnamefont{A.}~\bibnamefont{Tonchev}},
  \bibinfo{author}{\bibfnamefont{W.}~\bibnamefont{Tornow}},
  \bibinfo{author}{\bibfnamefont{H.~R.} \bibnamefont{Weller}},
  \bibinfo{author}{\bibfnamefont{N.}~\bibnamefont{Pietralla}},
  \bibinfo{author}{\bibfnamefont{J.}~\bibnamefont{Li}},
  \bibinfo{author}{\bibfnamefont{I.~V.} \bibnamefont{Pinayev}}, \emph{et~al.},
  \bibinfo{year}{2005}, \bibinfo{journal}{Phys. Rev. C}
  \textbf{\bibinfo{volume}{71}}, \bibinfo{pages}{034304}.

\bibitem[{\citenamefont{Scheck} \emph{et~al.}(2008)\citenamefont{Scheck,
  Choudry, Elhami, McEllistrem, Mukhopadhyay, Orce, and Yates}}]{Scheck:2008}
\bibinfo{author}{\bibnamefont{Scheck}, \bibfnamefont{M.}},
  \bibinfo{author}{\bibfnamefont{S.~N.} \bibnamefont{Choudry}},
  \bibinfo{author}{\bibfnamefont{E.}~\bibnamefont{Elhami}},
  \bibinfo{author}{\bibfnamefont{M.~T.} \bibnamefont{McEllistrem}},
  \bibinfo{author}{\bibfnamefont{S.}~\bibnamefont{Mukhopadhyay}},
  \bibinfo{author}{\bibfnamefont{J.~N.} \bibnamefont{Orce}}, and
  \bibinfo{author}{\bibfnamefont{S.~W.} \bibnamefont{Yates}},
  \bibinfo{year}{2008}, \bibinfo{journal}{Phys. Rev. C}
  \textbf{\bibinfo{volume}{78}}, \bibinfo{pages}{034302}.

\bibitem[{\citenamefont{Scheck} \emph{et~al.}(2004)\citenamefont{Scheck, von
  Garrel, Tsoneva, Belic, von Brentano, Fransen, Gade, Jolie, Kneissl,
  Kohstall, Linnemann, Nord} \emph{et~al.}}]{Scheck:2004}
\bibinfo{author}{\bibnamefont{Scheck}, \bibfnamefont{M.}},
  \bibinfo{author}{\bibfnamefont{H.}~\bibnamefont{von Garrel}},
  \bibinfo{author}{\bibfnamefont{N.}~\bibnamefont{Tsoneva}},
  \bibinfo{author}{\bibfnamefont{D.}~\bibnamefont{Belic}},
  \bibinfo{author}{\bibfnamefont{P.}~\bibnamefont{von Brentano}},
  \bibinfo{author}{\bibfnamefont{C.}~\bibnamefont{Fransen}},
  \bibinfo{author}{\bibfnamefont{A.}~\bibnamefont{Gade}},
  \bibinfo{author}{\bibfnamefont{J.}~\bibnamefont{Jolie}},
  \bibinfo{author}{\bibfnamefont{U.}~\bibnamefont{Kneissl}},
  \bibinfo{author}{\bibfnamefont{C.}~\bibnamefont{Kohstall}},
  \bibinfo{author}{\bibfnamefont{A.}~\bibnamefont{Linnemann}},
  \bibinfo{author}{\bibfnamefont{A.}~\bibnamefont{Nord}}, \emph{et~al.},
  \bibinfo{year}{2004}, \bibinfo{journal}{Phys. Rev. C}
  \textbf{\bibinfo{volume}{70}}, \bibinfo{pages}{044319}.

\bibitem[{\citenamefont{Schlegel} \emph{et~al.}(1996)\citenamefont{Schlegel,
  von Neumann-Cosel, Richter, and Van~Isacker}}]{Schlegel:1996}
\bibinfo{author}{\bibnamefont{Schlegel}, \bibfnamefont{C.}},
  \bibinfo{author}{\bibfnamefont{P.}~\bibnamefont{von Neumann-Cosel}},
  \bibinfo{author}{\bibfnamefont{A.}~\bibnamefont{Richter}}, and
  \bibinfo{author}{\bibfnamefont{P.}~\bibnamefont{Van~Isacker}},
  \bibinfo{year}{1996}, \bibinfo{journal}{Phys. Lett. B}
  \textbf{\bibinfo{volume}{375}}, \bibinfo{pages}{21}.

\bibitem[{\citenamefont{Scholten} \emph{et~al.}(1984)\citenamefont{Scholten,
  Dieperink, Heyde, and Van~Isacker}}]{Scholten:1984}
\bibinfo{author}{\bibnamefont{Scholten}, \bibfnamefont{O.}},
  \bibinfo{author}{\bibfnamefont{A.~E.~L.} \bibnamefont{Dieperink}},
  \bibinfo{author}{\bibfnamefont{K.}~\bibnamefont{Heyde}}, and
  \bibinfo{author}{\bibfnamefont{P.}~\bibnamefont{Van~Isacker}},
  \bibinfo{year}{1984}, \bibinfo{journal}{Phys. Lett. B}
  \textbf{\bibinfo{volume}{149}}, \bibinfo{pages}{279}.

\bibitem[{\citenamefont{Scholten}
  \emph{et~al.}(1985{\natexlab{a}})\citenamefont{Scholten, Heyde, and
  Van~Isacker}}]{Scholten:1985a}
\bibinfo{author}{\bibnamefont{Scholten}, \bibfnamefont{O.}},
  \bibinfo{author}{\bibfnamefont{K.}~\bibnamefont{Heyde}}, and
  \bibinfo{author}{\bibfnamefont{P.}~\bibnamefont{Van~Isacker}},
  \bibinfo{year}{1985}{\natexlab{a}}, \bibinfo{journal}{Phys. Rev. Lett.}
  \textbf{\bibinfo{volume}{55}}, \bibinfo{pages}{1866}.

\bibitem[{\citenamefont{Scholten}
  \emph{et~al.}(1985{\natexlab{b}})\citenamefont{Scholten, Heyde, Van~Isacker,
  Jolie, Moreau, Waroquier, and Sau}}]{Scholten:1985b}
\bibinfo{author}{\bibnamefont{Scholten}, \bibfnamefont{O.}},
  \bibinfo{author}{\bibfnamefont{K.}~\bibnamefont{Heyde}},
  \bibinfo{author}{\bibfnamefont{P.}~\bibnamefont{Van~Isacker}},
  \bibinfo{author}{\bibfnamefont{J.}~\bibnamefont{Jolie}},
  \bibinfo{author}{\bibfnamefont{J.}~\bibnamefont{Moreau}},
  \bibinfo{author}{\bibfnamefont{M.}~\bibnamefont{Waroquier}}, and
  \bibinfo{author}{\bibfnamefont{J.}~\bibnamefont{Sau}},
  \bibinfo{year}{1985}{\natexlab{b}}, \bibinfo{journal}{Nucl. Phys. A}
  \textbf{\bibinfo{volume}{438}}, \bibinfo{pages}{41}.

\bibitem[{\citenamefont{Scholtz} \emph{et~al.}(1989)\citenamefont{Scholtz,
  Nojarov, and Faessler}}]{Scholtz:1989b}
\bibinfo{author}{\bibnamefont{Scholtz}, \bibfnamefont{F.~G.}},
  \bibinfo{author}{\bibfnamefont{R.}~\bibnamefont{Nojarov}}, and
  \bibinfo{author}{\bibfnamefont{A.}~\bibnamefont{Faessler}},
  \bibinfo{year}{1989}, \bibinfo{journal}{Phys. Rev. Lett.}
  \textbf{\bibinfo{volume}{63}}, \bibinfo{pages}{1356}.

\bibitem[{\citenamefont{Sch\"uller}
  \emph{et~al.}(1996)\citenamefont{Sch\"uller, Biese, Keller, Steinebach,
  Heitmann, Grambow, and Eberl}}]{Schueller:1996}
\bibinfo{author}{\bibnamefont{Sch\"uller}, \bibfnamefont{C.}},
  \bibinfo{author}{\bibfnamefont{G.}~\bibnamefont{Biese}},
  \bibinfo{author}{\bibfnamefont{K.}~\bibnamefont{Keller}},
  \bibinfo{author}{\bibfnamefont{C.}~\bibnamefont{Steinebach}},
  \bibinfo{author}{\bibfnamefont{D.}~\bibnamefont{Heitmann}},
  \bibinfo{author}{\bibfnamefont{P.}~\bibnamefont{Grambow}}, and
  \bibinfo{author}{\bibfnamefont{K.}~\bibnamefont{Eberl}},
  \bibinfo{year}{1996}, \bibinfo{journal}{Phys. Rev. B}
  \textbf{\bibinfo{volume}{54}}, \bibinfo{pages}{R17304}.

\bibitem[{\citenamefont{Schwengner}
  \emph{et~al.}(2007)\citenamefont{Schwengner, Rusev, Benouaret, Beyer, Erhard,
  Grosse, Junghans, Klug, Kosev, Kostov, Nair, Nankov}
  \emph{et~al.}}]{Schwengner:2007}
\bibinfo{author}{\bibnamefont{Schwengner}, \bibfnamefont{R.}},
  \bibinfo{author}{\bibfnamefont{G.}~\bibnamefont{Rusev}},
  \bibinfo{author}{\bibfnamefont{N.}~\bibnamefont{Benouaret}},
  \bibinfo{author}{\bibfnamefont{R.}~\bibnamefont{Beyer}},
  \bibinfo{author}{\bibfnamefont{M.}~\bibnamefont{Erhard}},
  \bibinfo{author}{\bibfnamefont{E.}~\bibnamefont{Grosse}},
  \bibinfo{author}{\bibfnamefont{A.~R.} \bibnamefont{Junghans}},
  \bibinfo{author}{\bibfnamefont{J.}~\bibnamefont{Klug}},
  \bibinfo{author}{\bibfnamefont{K.}~\bibnamefont{Kosev}},
  \bibinfo{author}{\bibfnamefont{L.}~\bibnamefont{Kostov}},
  \bibinfo{author}{\bibfnamefont{C.}~\bibnamefont{Nair}},
  \bibinfo{author}{\bibfnamefont{N.}~\bibnamefont{Nankov}}, \emph{et~al.},
  \bibinfo{year}{2007}, \bibinfo{journal}{Phys. Rev. C}
  \textbf{\bibinfo{volume}{76}}, \bibinfo{pages}{034321}.

\bibitem[{\citenamefont{Schwengner}
  \emph{et~al.}(1997{\natexlab{a}})\citenamefont{Schwengner, Winter, Schauer,
  Grinberg, Becker, von Brentano, Eberth, Enders, von Egidy, Herzberg, Huxel,
  K\"{a}ubler} \emph{et~al.}}]{Schwengner:1997a}
\bibinfo{author}{\bibnamefont{Schwengner}, \bibfnamefont{R.}},
  \bibinfo{author}{\bibfnamefont{G.}~\bibnamefont{Winter}},
  \bibinfo{author}{\bibfnamefont{W.}~\bibnamefont{Schauer}},
  \bibinfo{author}{\bibfnamefont{M.}~\bibnamefont{Grinberg}},
  \bibinfo{author}{\bibfnamefont{F.}~\bibnamefont{Becker}},
  \bibinfo{author}{\bibfnamefont{P.}~\bibnamefont{von Brentano}},
  \bibinfo{author}{\bibfnamefont{J.}~\bibnamefont{Eberth}},
  \bibinfo{author}{\bibfnamefont{J.}~\bibnamefont{Enders}},
  \bibinfo{author}{\bibfnamefont{T.}~\bibnamefont{von Egidy}},
  \bibinfo{author}{\bibfnamefont{R.-D.} \bibnamefont{Herzberg}},
  \bibinfo{author}{\bibfnamefont{N.}~\bibnamefont{Huxel}},
  \bibinfo{author}{\bibfnamefont{L.}~\bibnamefont{K\"{a}ubler}}, \emph{et~al.},
  \bibinfo{year}{1997}{\natexlab{a}}, \bibinfo{journal}{Nucl. Phys. A}
  \textbf{\bibinfo{volume}{620}}, \bibinfo{pages}{277}.

\bibitem[{\citenamefont{Schwengner}
  \emph{et~al.}(1997{\natexlab{b}})\citenamefont{Schwengner, Winter, Schauer,
  Grinberg, Becker, von Brentano, Eberth, Enders, von Egidy, Herzberg, Huxel,
  K\"{a}ubler} \emph{et~al.}}]{Schwengner:1997b}
\bibinfo{author}{\bibnamefont{Schwengner}, \bibfnamefont{R.}},
  \bibinfo{author}{\bibfnamefont{G.}~\bibnamefont{Winter}},
  \bibinfo{author}{\bibfnamefont{W.}~\bibnamefont{Schauer}},
  \bibinfo{author}{\bibfnamefont{M.}~\bibnamefont{Grinberg}},
  \bibinfo{author}{\bibfnamefont{F.}~\bibnamefont{Becker}},
  \bibinfo{author}{\bibfnamefont{P.}~\bibnamefont{von Brentano}},
  \bibinfo{author}{\bibfnamefont{J.}~\bibnamefont{Eberth}},
  \bibinfo{author}{\bibfnamefont{J.}~\bibnamefont{Enders}},
  \bibinfo{author}{\bibfnamefont{T.}~\bibnamefont{von Egidy}},
  \bibinfo{author}{\bibfnamefont{R.-D.} \bibnamefont{Herzberg}},
  \bibinfo{author}{\bibfnamefont{N.}~\bibnamefont{Huxel}},
  \bibinfo{author}{\bibfnamefont{L.}~\bibnamefont{K\"{a}ubler}}, \emph{et~al.},
  \bibinfo{year}{1997}{\natexlab{b}}, \bibinfo{journal}{Nucl. Phys. A}
  \textbf{\bibinfo{volume}{624}}, \bibinfo{pages}{776(E)}.

\bibitem[{\citenamefont{Serra} \emph{et~al.}(1999)\citenamefont{Serra, Puente,
  and Lipparini}}]{Serra:1999}
\bibinfo{author}{\bibnamefont{Serra}, \bibfnamefont{L.}},
  \bibinfo{author}{\bibfnamefont{A.}~\bibnamefont{Puente}}, and
  \bibinfo{author}{\bibfnamefont{E.}~\bibnamefont{Lipparini}},
  \bibinfo{year}{1999}, \bibinfo{journal}{Phys. Rev. B}
  \textbf{\bibinfo{volume}{60}}, \bibinfo{pages}{R13966}.

\bibitem[{\citenamefont{Shevchenko}
  \emph{et~al.}(2009)\citenamefont{Shevchenko, Burda, Carter, Cooper, Fearick,
  F\"{o}rtsch, Fujita, Fujita, Kalmykov, Lacroix, Lawrie, von Neumann-Cosel}
  \emph{et~al.}}]{Shevchenko:2009}
\bibinfo{author}{\bibnamefont{Shevchenko}, \bibfnamefont{A.}},
  \bibinfo{author}{\bibfnamefont{O.}~\bibnamefont{Burda}},
  \bibinfo{author}{\bibfnamefont{J.}~\bibnamefont{Carter}},
  \bibinfo{author}{\bibfnamefont{G.~R.~J.} \bibnamefont{Cooper}},
  \bibinfo{author}{\bibfnamefont{R.~W.} \bibnamefont{Fearick}},
  \bibinfo{author}{\bibfnamefont{S.~V.} \bibnamefont{F\"{o}rtsch}},
  \bibinfo{author}{\bibfnamefont{H.}~\bibnamefont{Fujita}},
  \bibinfo{author}{\bibfnamefont{Y.}~\bibnamefont{Fujita}},
  \bibinfo{author}{\bibfnamefont{Y.}~\bibnamefont{Kalmykov}},
  \bibinfo{author}{\bibfnamefont{D.}~\bibnamefont{Lacroix}},
  \bibinfo{author}{\bibfnamefont{J.~J.} \bibnamefont{Lawrie}},
  \bibinfo{author}{\bibfnamefont{P.}~\bibnamefont{von Neumann-Cosel}},
  \emph{et~al.}, \bibinfo{year}{2009}, \bibinfo{journal}{Phys. Rev. C}
  \textbf{\bibinfo{volume}{79}}, \bibinfo{eid}{044305}.

\bibitem[{\citenamefont{Shevchenko}
  \emph{et~al.}(2008)\citenamefont{Shevchenko, Carter, Cooper, Fearick,
  Kalmykov, von Neumann-Cosel, Ponomarev, Richter, Usman, and
  Wambach}}]{Shevchenko:2008}
\bibinfo{author}{\bibnamefont{Shevchenko}, \bibfnamefont{A.}},
  \bibinfo{author}{\bibfnamefont{J.}~\bibnamefont{Carter}},
  \bibinfo{author}{\bibfnamefont{G.~R.~J.} \bibnamefont{Cooper}},
  \bibinfo{author}{\bibfnamefont{R.~W.} \bibnamefont{Fearick}},
  \bibinfo{author}{\bibfnamefont{Y.}~\bibnamefont{Kalmykov}},
  \bibinfo{author}{\bibfnamefont{P.}~\bibnamefont{von Neumann-Cosel}},
  \bibinfo{author}{\bibfnamefont{V.~Y.} \bibnamefont{Ponomarev}},
  \bibinfo{author}{\bibfnamefont{A.}~\bibnamefont{Richter}},
  \bibinfo{author}{\bibfnamefont{I.}~\bibnamefont{Usman}}, and
  \bibinfo{author}{\bibfnamefont{J.}~\bibnamefont{Wambach}},
  \bibinfo{year}{2008}, \bibinfo{journal}{Phys. Rev. C}
  \textbf{\bibinfo{volume}{77}}, \bibinfo{eid}{024302}.

\bibitem[{\citenamefont{Shevchenko}
  \emph{et~al.}(2004)\citenamefont{Shevchenko, Carter, Fearick, F\"{o}rtsch,
  Fujita, Fujita, Kalmykov, Lacroix, Lawrie, von Neumann-Cosel, Neveling,
  Ponomarev} \emph{et~al.}}]{Shevchenko:2004}
\bibinfo{author}{\bibnamefont{Shevchenko}, \bibfnamefont{A.}},
  \bibinfo{author}{\bibfnamefont{J.}~\bibnamefont{Carter}},
  \bibinfo{author}{\bibfnamefont{R.~W.} \bibnamefont{Fearick}},
  \bibinfo{author}{\bibfnamefont{S.~V.} \bibnamefont{F\"{o}rtsch}},
  \bibinfo{author}{\bibfnamefont{H.}~\bibnamefont{Fujita}},
  \bibinfo{author}{\bibfnamefont{Y.}~\bibnamefont{Fujita}},
  \bibinfo{author}{\bibfnamefont{Y.}~\bibnamefont{Kalmykov}},
  \bibinfo{author}{\bibfnamefont{D.}~\bibnamefont{Lacroix}},
  \bibinfo{author}{\bibfnamefont{J.~J.} \bibnamefont{Lawrie}},
  \bibinfo{author}{\bibfnamefont{P.}~\bibnamefont{von Neumann-Cosel}},
  \bibinfo{author}{\bibfnamefont{R.}~\bibnamefont{Neveling}},
  \bibinfo{author}{\bibfnamefont{V.~Y.} \bibnamefont{Ponomarev}},
  \emph{et~al.}, \bibinfo{year}{2004}, \bibinfo{journal}{Phys. Rev. Lett.}
  \textbf{\bibinfo{volume}{93}}, \bibinfo{eid}{122501}.

\bibitem[{\citenamefont{Shimizu} \emph{et~al.}(2001)\citenamefont{Shimizu,
  Otsuka, Mizusaki, and Honma}}]{Shimizu:2001}
\bibinfo{author}{\bibnamefont{Shimizu}, \bibfnamefont{N.}},
  \bibinfo{author}{\bibfnamefont{T.}~\bibnamefont{Otsuka}},
  \bibinfo{author}{\bibfnamefont{T.}~\bibnamefont{Mizusaki}}, and
  \bibinfo{author}{\bibfnamefont{M.}~\bibnamefont{Honma}},
  \bibinfo{year}{2001}, \bibinfo{journal}{Phys. Rev. Lett.}
  \textbf{\bibinfo{volume}{86}}, \bibinfo{pages}{1171}.

\bibitem[{\citenamefont{Shriner~Jr.}
  \emph{et~al.}(1991)\citenamefont{Shriner~Jr., Mitchell, and von
  Egidy}}]{Shriner:1991}
\bibinfo{author}{\bibnamefont{Shriner~Jr.}, \bibfnamefont{J.~F.}},
  \bibinfo{author}{\bibfnamefont{G.~E.} \bibnamefont{Mitchell}}, and
  \bibinfo{author}{\bibfnamefont{T.}~\bibnamefont{von Egidy}},
  \bibinfo{year}{1991}, \bibinfo{journal}{Z. Phys. A}
  \textbf{\bibinfo{volume}{338}}, \bibinfo{pages}{309}.

\bibitem[{\citenamefont{Sieja} \emph{et~al.}(2009)\citenamefont{Sieja,
  Mart\'\i{}nez-Pinedo, Coquard, and Pietralla}}]{Sieja:2009}
\bibinfo{author}{\bibnamefont{Sieja}, \bibfnamefont{K.}},
  \bibinfo{author}{\bibfnamefont{G.}~\bibnamefont{Mart\'\i{}nez-Pinedo}},
  \bibinfo{author}{\bibfnamefont{L.}~\bibnamefont{Coquard}}, and
  \bibinfo{author}{\bibfnamefont{N.}~\bibnamefont{Pietralla}},
  \bibinfo{year}{2009}, \bibinfo{journal}{Phys. Rev. C}
  \textbf{\bibinfo{volume}{80}}, \bibinfo{pages}{054311}.

\bibitem[{\citenamefont{Sikorski and Merkt}(1989)}]{Sikorski:1989}
\bibinfo{author}{\bibnamefont{Sikorski}, \bibfnamefont{C.}}, and
  \bibinfo{author}{\bibfnamefont{U.}~\bibnamefont{Merkt}},
  \bibinfo{year}{1989}, \bibinfo{journal}{Phys. Rev. Lett.}
  \textbf{\bibinfo{volume}{62}}, \bibinfo{pages}{2164}.

\bibitem[{\citenamefont{Smith} \emph{et~al.}(1995)\citenamefont{Smith, Pan,
  Feng, and Guidry}}]{Smith:1995}
\bibinfo{author}{\bibnamefont{Smith}, \bibfnamefont{B.~H.}},
  \bibinfo{author}{\bibfnamefont{X.-W.} \bibnamefont{Pan}},
  \bibinfo{author}{\bibfnamefont{D.~H.} \bibnamefont{Feng}}, and
  \bibinfo{author}{\bibfnamefont{M.}~\bibnamefont{Guidry}},
  \bibinfo{year}{1995}, \bibinfo{journal}{Phys. Rev. Lett.}
  \textbf{\bibinfo{volume}{75}}, \bibinfo{pages}{3086}.

\bibitem[{\citenamefont{Sober} \emph{et~al.}(1985)\citenamefont{Sober, Metsch,
  Kn\"upfer, Eulenberg, K\"uchler, Richter, Spamer, and Steffen}}]{Sober:1985}
\bibinfo{author}{\bibnamefont{Sober}, \bibfnamefont{D.~I.}},
  \bibinfo{author}{\bibfnamefont{B.~C.} \bibnamefont{Metsch}},
  \bibinfo{author}{\bibfnamefont{W.}~\bibnamefont{Kn\"upfer}},
  \bibinfo{author}{\bibfnamefont{G.}~\bibnamefont{Eulenberg}},
  \bibinfo{author}{\bibfnamefont{G.}~\bibnamefont{K\"uchler}},
  \bibinfo{author}{\bibfnamefont{A.}~\bibnamefont{Richter}},
  \bibinfo{author}{\bibfnamefont{E.}~\bibnamefont{Spamer}}, and
  \bibinfo{author}{\bibfnamefont{W.}~\bibnamefont{Steffen}},
  \bibinfo{year}{1985}, \bibinfo{journal}{Phys. Rev. C}
  \textbf{\bibinfo{volume}{31}}, \bibinfo{pages}{2054}.

\bibitem[{\citenamefont{Soloviev}(1992)}]{Soloviev:1992}
\bibinfo{author}{\bibnamefont{Soloviev}, \bibfnamefont{V.~G.}},
  \bibinfo{year}{1992}, \emph{\bibinfo{title}{Theory of the Atomic Nucleus:
  Quasiparticles and Phonons}} (\bibinfo{publisher}{IOP},
  \bibinfo{address}{Bristol}).

\bibitem[{\citenamefont{Soloviev} \emph{et~al.}(1996)\citenamefont{Soloviev,
  Sushkov, and Shirikova}}]{Soloviev:1996b}
\bibinfo{author}{\bibnamefont{Soloviev}, \bibfnamefont{V.~G.}},
  \bibinfo{author}{\bibfnamefont{A.~V.} \bibnamefont{Sushkov}}, and
  \bibinfo{author}{\bibfnamefont{N.~Y.} \bibnamefont{Shirikova}},
  \bibinfo{year}{1996}, \bibinfo{journal}{Phys. Rev. C}
  \textbf{\bibinfo{volume}{53}}, \bibinfo{pages}{1022}.

\bibitem[{\citenamefont{Soloviev}
  \emph{et~al.}(1997{\natexlab{a}})\citenamefont{Soloviev, Sushkov, and
  Shirikova}}]{Soloviev:1997a}
\bibinfo{author}{\bibnamefont{Soloviev}, \bibfnamefont{V.~G.}},
  \bibinfo{author}{\bibfnamefont{A.~V.} \bibnamefont{Sushkov}}, and
  \bibinfo{author}{\bibfnamefont{N.~Y.} \bibnamefont{Shirikova}},
  \bibinfo{year}{1997}{\natexlab{a}}, \bibinfo{journal}{Prog. Part. Nucl.
  Phys.} \textbf{\bibinfo{volume}{38}}, \bibinfo{pages}{53}.

\bibitem[{\citenamefont{Soloviev}
  \emph{et~al.}(1997{\natexlab{b}})\citenamefont{Soloviev, Sushkov, and
  Shirikova}}]{Soloviev:1997b}
\bibinfo{author}{\bibnamefont{Soloviev}, \bibfnamefont{V.~G.}},
  \bibinfo{author}{\bibfnamefont{A.~V.} \bibnamefont{Sushkov}}, and
  \bibinfo{author}{\bibfnamefont{N.~Y.} \bibnamefont{Shirikova}},
  \bibinfo{year}{1997}{\natexlab{b}}, \bibinfo{journal}{Phys. Rev. C}
  \textbf{\bibinfo{volume}{56}}, \bibinfo{pages}{2528}.

\bibitem[{\citenamefont{Soloviev}
  \emph{et~al.}(1997{\natexlab{c}})\citenamefont{Soloviev, Sushkov, Shirikova,
  and Lo~Iudice}}]{Soloviev:1997c}
\bibinfo{author}{\bibnamefont{Soloviev}, \bibfnamefont{V.~G.}},
  \bibinfo{author}{\bibfnamefont{A.~V.} \bibnamefont{Sushkov}},
  \bibinfo{author}{\bibfnamefont{N.~Y.} \bibnamefont{Shirikova}}, and
  \bibinfo{author}{\bibfnamefont{N.}~\bibnamefont{Lo~Iudice}},
  \bibinfo{year}{1997}{\natexlab{c}}, \bibinfo{journal}{Nucl. Phys. A}
  \textbf{\bibinfo{volume}{613}}, \bibinfo{pages}{45}.

\bibitem[{\citenamefont{Steffen} \emph{et~al.}(1980)\citenamefont{Steffen,
  Gr\"af, Gross, Meuer, Richter, Spamer, Titze, and
  Kn\"{u}pfer}}]{Steffen:1980}
\bibinfo{author}{\bibnamefont{Steffen}, \bibfnamefont{W.}},
  \bibinfo{author}{\bibfnamefont{H.-D.} \bibnamefont{Gr\"af}},
  \bibinfo{author}{\bibfnamefont{W.}~\bibnamefont{Gross}},
  \bibinfo{author}{\bibfnamefont{D.}~\bibnamefont{Meuer}},
  \bibinfo{author}{\bibfnamefont{A.}~\bibnamefont{Richter}},
  \bibinfo{author}{\bibfnamefont{E.}~\bibnamefont{Spamer}},
  \bibinfo{author}{\bibfnamefont{O.}~\bibnamefont{Titze}}, and
  \bibinfo{author}{\bibfnamefont{W.}~\bibnamefont{Kn\"{u}pfer}},
  \bibinfo{year}{1980}, \bibinfo{journal}{Phys. Lett. B}
  \textbf{\bibinfo{volume}{95}}, \bibinfo{pages}{23}.

\bibitem[{\citenamefont{Strenz} \emph{et~al.}(1994)\citenamefont{Strenz,
  Bockelmann, Hirler, Abstreiter, B\"ohm, and Weimann}}]{Strenz:1994}
\bibinfo{author}{\bibnamefont{Strenz}, \bibfnamefont{R.}},
  \bibinfo{author}{\bibfnamefont{U.}~\bibnamefont{Bockelmann}},
  \bibinfo{author}{\bibfnamefont{F.}~\bibnamefont{Hirler}},
  \bibinfo{author}{\bibfnamefont{G.}~\bibnamefont{Abstreiter}},
  \bibinfo{author}{\bibfnamefont{G.}~\bibnamefont{B\"ohm}}, and
  \bibinfo{author}{\bibfnamefont{G.}~\bibnamefont{Weimann}},
  \bibinfo{year}{1994}, \bibinfo{journal}{Phys. Rev. Lett.}
  \textbf{\bibinfo{volume}{73}}, \bibinfo{pages}{3022}.

\bibitem[{\citenamefont{Sugawara-Tanabe and
  Arima}(1989)}]{Sugawara-Tanabe:1989}
\bibinfo{author}{\bibnamefont{Sugawara-Tanabe}, \bibfnamefont{K.}}, and
  \bibinfo{author}{\bibfnamefont{A.}~\bibnamefont{Arima}},
  \bibinfo{year}{1989}, \bibinfo{journal}{Phys. Lett. B}
  \textbf{\bibinfo{volume}{229}}, \bibinfo{pages}{327}.

\bibitem[{\citenamefont{Suzuki and Rowe}(1977)}]{Suzuki:1977}
\bibinfo{author}{\bibnamefont{Suzuki}, \bibfnamefont{T.}}, and
  \bibinfo{author}{\bibfnamefont{D.}~\bibnamefont{Rowe}}, \bibinfo{year}{1977},
  \bibinfo{journal}{Nucl. Phys. A} \textbf{\bibinfo{volume}{289}},
  \bibinfo{pages}{461}.

\bibitem[{\citenamefont{Takayanagi}
  \emph{et~al.}(1988)\citenamefont{Takayanagi, Shimizu, and
  Arima}}]{Takayanagi:1988}
\bibinfo{author}{\bibnamefont{Takayanagi}, \bibfnamefont{K.}},
  \bibinfo{author}{\bibfnamefont{K.}~\bibnamefont{Shimizu}}, and
  \bibinfo{author}{\bibfnamefont{A.}~\bibnamefont{Arima}},
  \bibinfo{year}{1988}, \bibinfo{journal}{Nucl. Phys. A}
  \textbf{\bibinfo{volume}{481}}, \bibinfo{pages}{313}.

\bibitem[{\citenamefont{Tamii} \emph{et~al.}(2009)\citenamefont{Tamii, Fujita,
  Matsubara, Adachi, Carter, Dozono, Fujita, Fujita, Hashimoto, Hatanaka,
  Itahashi, Itoh} \emph{et~al.}}]{Tamii:2009}
\bibinfo{author}{\bibnamefont{Tamii}, \bibfnamefont{A.}},
  \bibinfo{author}{\bibfnamefont{Y.}~\bibnamefont{Fujita}},
  \bibinfo{author}{\bibfnamefont{H.}~\bibnamefont{Matsubara}},
  \bibinfo{author}{\bibfnamefont{T.}~\bibnamefont{Adachi}},
  \bibinfo{author}{\bibfnamefont{J.}~\bibnamefont{Carter}},
  \bibinfo{author}{\bibfnamefont{M.}~\bibnamefont{Dozono}},
  \bibinfo{author}{\bibfnamefont{H.}~\bibnamefont{Fujita}},
  \bibinfo{author}{\bibfnamefont{K.}~\bibnamefont{Fujita}},
  \bibinfo{author}{\bibfnamefont{H.}~\bibnamefont{Hashimoto}},
  \bibinfo{author}{\bibfnamefont{K.}~\bibnamefont{Hatanaka}},
  \bibinfo{author}{\bibfnamefont{T.}~\bibnamefont{Itahashi}},
  \bibinfo{author}{\bibfnamefont{M.}~\bibnamefont{Itoh}}, \emph{et~al.},
  \bibinfo{year}{2009}, \bibinfo{journal}{Nuclear Instrum. and Methods in Phys.
  Research A} \textbf{\bibinfo{volume}{605}}, \bibinfo{pages}{326 }.

\bibitem[{\citenamefont{Towner}(1987)}]{Towner:1987}
\bibinfo{author}{\bibnamefont{Towner}, \bibfnamefont{I.~S.}},
  \bibinfo{year}{1987}, \bibinfo{journal}{Phys. Rep.}
  \textbf{\bibinfo{volume}{155}}, \bibinfo{pages}{263}.

\bibitem[{\citenamefont{Utsuno} \emph{et~al.}(1999)\citenamefont{Utsuno,
  Otsuka, Mizusaki, and Honma}}]{Utsuno:1999}
\bibinfo{author}{\bibnamefont{Utsuno}, \bibfnamefont{Y.}},
  \bibinfo{author}{\bibfnamefont{T.}~\bibnamefont{Otsuka}},
  \bibinfo{author}{\bibfnamefont{T.}~\bibnamefont{Mizusaki}}, and
  \bibinfo{author}{\bibfnamefont{M.}~\bibnamefont{Honma}},
  \bibinfo{year}{1999}, \bibinfo{journal}{Phys. Rev. C}
  \textbf{\bibinfo{volume}{60}}, \bibinfo{pages}{054315}.

\bibitem[{\citenamefont{Van~Isacker and Frank}(1989)}]{VanIsacker:1989}
\bibinfo{author}{\bibnamefont{Van~Isacker}, \bibfnamefont{P.}}, and
  \bibinfo{author}{\bibfnamefont{A.}~\bibnamefont{Frank}},
  \bibinfo{year}{1989}, \bibinfo{journal}{Phys. Lett. B}
  \textbf{\bibinfo{volume}{225}}, \bibinfo{pages}{1}.

\bibitem[{\citenamefont{Van~Isacker}
  \emph{et~al.}(1986)\citenamefont{Van~Isacker, Heyde, Jolie, and
  Sevrin}}]{VanIsacker:1986}
\bibinfo{author}{\bibnamefont{Van~Isacker}, \bibfnamefont{P.}},
  \bibinfo{author}{\bibfnamefont{K.}~\bibnamefont{Heyde}},
  \bibinfo{author}{\bibfnamefont{J.}~\bibnamefont{Jolie}}, and
  \bibinfo{author}{\bibfnamefont{A.}~\bibnamefont{Sevrin}},
  \bibinfo{year}{1986}, \bibinfo{journal}{Ann. Phys. (N.Y.)}
  \textbf{\bibinfo{volume}{171}}, \bibinfo{pages}{253}.

\bibitem[{\citenamefont{Van~Isacker}
  \emph{et~al.}(1992)\citenamefont{Van~Isacker, Nagarajan, and
  Warner}}]{VanIsacker:1992}
\bibinfo{author}{\bibnamefont{Van~Isacker}, \bibfnamefont{P.}},
  \bibinfo{author}{\bibfnamefont{M.~A.} \bibnamefont{Nagarajan}}, and
  \bibinfo{author}{\bibfnamefont{D.~D.} \bibnamefont{Warner}},
  \bibinfo{year}{1992}, \bibinfo{journal}{Phys. Rev. C}
  \textbf{\bibinfo{volume}{45}}, \bibinfo{pages}{R13}.

\bibitem[{\citenamefont{Vanhoy} \emph{et~al.}(1995)\citenamefont{Vanhoy,
  Anthony, Haas, Benedict, Meehan, Hicks, Davoren, and
  Lundstedt}}]{Vanhoy:1995}
\bibinfo{author}{\bibnamefont{Vanhoy}, \bibfnamefont{J.~R.}},
  \bibinfo{author}{\bibfnamefont{J.~M.} \bibnamefont{Anthony}},
  \bibinfo{author}{\bibfnamefont{B.~M.} \bibnamefont{Haas}},
  \bibinfo{author}{\bibfnamefont{B.~H.} \bibnamefont{Benedict}},
  \bibinfo{author}{\bibfnamefont{B.~T.} \bibnamefont{Meehan}},
  \bibinfo{author}{\bibfnamefont{S.~F.} \bibnamefont{Hicks}},
  \bibinfo{author}{\bibfnamefont{C.~M.} \bibnamefont{Davoren}}, and
  \bibinfo{author}{\bibfnamefont{C.~L.} \bibnamefont{Lundstedt}},
  \bibinfo{year}{1995}, \bibinfo{journal}{Phys. Rev. C}
  \textbf{\bibinfo{volume}{52}}, \bibinfo{pages}{2387}.

\bibitem[{\citenamefont{Vi\~nas} \emph{et~al.}(2001)\citenamefont{Vi\~nas,
  Roth, Schuck, and Wambach}}]{Vinas:2001}
\bibinfo{author}{\bibnamefont{Vi\~nas}, \bibfnamefont{X.}},
  \bibinfo{author}{\bibfnamefont{R.}~\bibnamefont{Roth}},
  \bibinfo{author}{\bibfnamefont{P.}~\bibnamefont{Schuck}}, and
  \bibinfo{author}{\bibfnamefont{J.}~\bibnamefont{Wambach}},
  \bibinfo{year}{2001}, \bibinfo{journal}{Phys. Rev. A}
  \textbf{\bibinfo{volume}{64}}, \bibinfo{pages}{055601}.

\bibitem[{\citenamefont{Warner and Van~Isacker}(1997)}]{Warner:1997}
\bibinfo{author}{\bibnamefont{Warner}, \bibfnamefont{D.~D.}}, and
  \bibinfo{author}{\bibfnamefont{P.}~\bibnamefont{Van~Isacker}},
  \bibinfo{year}{1997}, \bibinfo{journal}{Phys. Lett. B}
  \textbf{\bibinfo{volume}{395}}, \bibinfo{pages}{145}.

\bibitem[{\citenamefont{Weidenm\"uller and
  Mitchell}(2009)}]{Weidenmueller:2009}
\bibinfo{author}{\bibnamefont{Weidenm\"uller}, \bibfnamefont{H.~A.}}, and
  \bibinfo{author}{\bibfnamefont{G.~E.} \bibnamefont{Mitchell}},
  \bibinfo{year}{2009}, \bibinfo{journal}{Rev. Mod. Phys.}
  \textbf{\bibinfo{volume}{81}}, \bibinfo{pages}{539}.

\bibitem[{\citenamefont{Weller} \emph{et~al.}(2009)\citenamefont{Weller, Ahmed,
  Gao, Tornow, Wu, Gai, and Miskimen}}]{Weller:2009}
\bibinfo{author}{\bibnamefont{Weller}, \bibfnamefont{H.~R.}},
  \bibinfo{author}{\bibfnamefont{M.~W.} \bibnamefont{Ahmed}},
  \bibinfo{author}{\bibfnamefont{H.}~\bibnamefont{Gao}},
  \bibinfo{author}{\bibfnamefont{W.}~\bibnamefont{Tornow}},
  \bibinfo{author}{\bibfnamefont{Y.~K.} \bibnamefont{Wu}},
  \bibinfo{author}{\bibfnamefont{M.}~\bibnamefont{Gai}}, and
  \bibinfo{author}{\bibfnamefont{R.}~\bibnamefont{Miskimen}},
  \bibinfo{year}{2009}, \bibinfo{journal}{Prog. Part. Nucl. Phys.}
  \textbf{\bibinfo{volume}{62}}, \bibinfo{pages}{257}.

\bibitem[{\citenamefont{Werner} \emph{et~al.}(2002)\citenamefont{Werner, Belic,
  von Brentano, Fransen, Gade, von Garrel, Jolie, Kneissl, Kohstall, Linneman,
  Lisetsky, Pietralla} \emph{et~al.}}]{Werner:2002}
\bibinfo{author}{\bibnamefont{Werner}, \bibfnamefont{V.}},
  \bibinfo{author}{\bibfnamefont{D.}~\bibnamefont{Belic}},
  \bibinfo{author}{\bibfnamefont{P.}~\bibnamefont{von Brentano}},
  \bibinfo{author}{\bibfnamefont{C.}~\bibnamefont{Fransen}},
  \bibinfo{author}{\bibfnamefont{A.}~\bibnamefont{Gade}},
  \bibinfo{author}{\bibfnamefont{H.}~\bibnamefont{von Garrel}},
  \bibinfo{author}{\bibfnamefont{J.}~\bibnamefont{Jolie}},
  \bibinfo{author}{\bibfnamefont{U.}~\bibnamefont{Kneissl}},
  \bibinfo{author}{\bibfnamefont{C.}~\bibnamefont{Kohstall}},
  \bibinfo{author}{\bibfnamefont{A.}~\bibnamefont{Linneman}},
  \bibinfo{author}{\bibfnamefont{A.~F.} \bibnamefont{Lisetsky}},
  \bibinfo{author}{\bibfnamefont{N.}~\bibnamefont{Pietralla}}, \emph{et~al.},
  \bibinfo{year}{2002}, \bibinfo{journal}{Phys. Lett. B}
  \textbf{\bibinfo{volume}{550}}, \bibinfo{pages}{140}.

\bibitem[{\citenamefont{Werner} \emph{et~al.}(2008)\citenamefont{Werner,
  Benczer-Koller, Kumbartzki, Holt, Boutachkov, Stefanova, Perry, Pietralla,
  Ai, Aleksandrova, Anderson, Cakirli} \emph{et~al.}}]{Werner:2008}
\bibinfo{author}{\bibnamefont{Werner}, \bibfnamefont{V.}},
  \bibinfo{author}{\bibfnamefont{N.}~\bibnamefont{Benczer-Koller}},
  \bibinfo{author}{\bibfnamefont{G.}~\bibnamefont{Kumbartzki}},
  \bibinfo{author}{\bibfnamefont{J.~D.} \bibnamefont{Holt}},
  \bibinfo{author}{\bibfnamefont{P.}~\bibnamefont{Boutachkov}},
  \bibinfo{author}{\bibfnamefont{E.}~\bibnamefont{Stefanova}},
  \bibinfo{author}{\bibfnamefont{M.}~\bibnamefont{Perry}},
  \bibinfo{author}{\bibfnamefont{N.}~\bibnamefont{Pietralla}},
  \bibinfo{author}{\bibfnamefont{H.}~\bibnamefont{Ai}},
  \bibinfo{author}{\bibfnamefont{K.}~\bibnamefont{Aleksandrova}},
  \bibinfo{author}{\bibfnamefont{G.}~\bibnamefont{Anderson}},
  \bibinfo{author}{\bibfnamefont{R.~B.} \bibnamefont{Cakirli}}, \emph{et~al.},
  \bibinfo{year}{2008}, \bibinfo{journal}{Phys. Rev. C}
  \textbf{\bibinfo{volume}{78}}, \bibinfo{eid}{031301(R)}.

\bibitem[{\citenamefont{Wesselborg}
  \emph{et~al.}(1988)\citenamefont{Wesselborg, von Brentano, Zell, Heil, Pitz,
  Berg, Kneissl, Lindenstruth, Seemann, and Stock}}]{Wesselborg:1988}
\bibinfo{author}{\bibnamefont{Wesselborg}, \bibfnamefont{C.}},
  \bibinfo{author}{\bibfnamefont{P.}~\bibnamefont{von Brentano}},
  \bibinfo{author}{\bibfnamefont{K.~O.} \bibnamefont{Zell}},
  \bibinfo{author}{\bibfnamefont{R.~D.} \bibnamefont{Heil}},
  \bibinfo{author}{\bibfnamefont{H.~H.} \bibnamefont{Pitz}},
  \bibinfo{author}{\bibfnamefont{U.~E.~P.} \bibnamefont{Berg}},
  \bibinfo{author}{\bibfnamefont{U.}~\bibnamefont{Kneissl}},
  \bibinfo{author}{\bibfnamefont{S.}~\bibnamefont{Lindenstruth}},
  \bibinfo{author}{\bibfnamefont{U.}~\bibnamefont{Seemann}}, and
  \bibinfo{author}{\bibfnamefont{R.}~\bibnamefont{Stock}},
  \bibinfo{year}{1988}, \bibinfo{journal}{Phys. Lett. B}
  \textbf{\bibinfo{volume}{207}}, \bibinfo{pages}{22}.

\bibitem[{\citenamefont{Wesselborg}
  \emph{et~al.}(1986)\citenamefont{Wesselborg, Schiffer, Zell, von Brentano,
  Bohle, Richter, Berg, Brinkm\"{o}ller, R\"{o}mer, Osterfeld, and
  Yabe}}]{Wesselborg:1986}
\bibinfo{author}{\bibnamefont{Wesselborg}, \bibfnamefont{C.}},
  \bibinfo{author}{\bibfnamefont{K.}~\bibnamefont{Schiffer}},
  \bibinfo{author}{\bibfnamefont{K.~O.} \bibnamefont{Zell}},
  \bibinfo{author}{\bibfnamefont{P.}~\bibnamefont{von Brentano}},
  \bibinfo{author}{\bibfnamefont{D.}~\bibnamefont{Bohle}},
  \bibinfo{author}{\bibfnamefont{A.}~\bibnamefont{Richter}},
  \bibinfo{author}{\bibfnamefont{G.~P.~A.} \bibnamefont{Berg}},
  \bibinfo{author}{\bibfnamefont{B.}~\bibnamefont{Brinkm\"{o}ller}},
  \bibinfo{author}{\bibfnamefont{J.~G.~M.} \bibnamefont{R\"{o}mer}},
  \bibinfo{author}{\bibfnamefont{F.}~\bibnamefont{Osterfeld}}, and
  \bibinfo{author}{\bibfnamefont{M.}~\bibnamefont{Yabe}}, \bibinfo{year}{1986},
  \bibinfo{journal}{Z. Phys. A} \textbf{\bibinfo{volume}{323}},
  \bibinfo{pages}{485}.

\bibitem[{\citenamefont{Wiedenh\"over}
  \emph{et~al.}(1997)\citenamefont{Wiedenh\"over, Gelberg, Otsuka, Pietralla,
  Gableske, Dewald, and von Brentano}}]{Wiedenhoever:1997}
\bibinfo{author}{\bibnamefont{Wiedenh\"over}, \bibfnamefont{I.}},
  \bibinfo{author}{\bibfnamefont{A.}~\bibnamefont{Gelberg}},
  \bibinfo{author}{\bibfnamefont{T.}~\bibnamefont{Otsuka}},
  \bibinfo{author}{\bibfnamefont{N.}~\bibnamefont{Pietralla}},
  \bibinfo{author}{\bibfnamefont{J.}~\bibnamefont{Gableske}},
  \bibinfo{author}{\bibfnamefont{A.}~\bibnamefont{Dewald}}, and
  \bibinfo{author}{\bibfnamefont{P.}~\bibnamefont{von Brentano}},
  \bibinfo{year}{1997}, \bibinfo{journal}{Phys. Rev. C}
  \textbf{\bibinfo{volume}{56}}, \bibinfo{pages}{R2354}.

\bibitem[{\citenamefont{Williams} \emph{et~al.}(2009)\citenamefont{Williams,
  Casperson, Werner, Ai, Boutachkov, Chamberlain, G\"urdal, Heinz, McCutchan,
  Qian, and Winkler}}]{Williams:2009}
\bibinfo{author}{\bibnamefont{Williams}, \bibfnamefont{E.}},
  \bibinfo{author}{\bibfnamefont{R.~J.} \bibnamefont{Casperson}},
  \bibinfo{author}{\bibfnamefont{V.}~\bibnamefont{Werner}},
  \bibinfo{author}{\bibfnamefont{H.}~\bibnamefont{Ai}},
  \bibinfo{author}{\bibfnamefont{P.}~\bibnamefont{Boutachkov}},
  \bibinfo{author}{\bibfnamefont{M.}~\bibnamefont{Chamberlain}},
  \bibinfo{author}{\bibfnamefont{G.}~\bibnamefont{G\"urdal}},
  \bibinfo{author}{\bibfnamefont{A.}~\bibnamefont{Heinz}},
  \bibinfo{author}{\bibfnamefont{E.~A.} \bibnamefont{McCutchan}},
  \bibinfo{author}{\bibfnamefont{J.}~\bibnamefont{Qian}}, and
  \bibinfo{author}{\bibfnamefont{R.}~\bibnamefont{Winkler}},
  \bibinfo{year}{2009}, \bibinfo{journal}{Phys. Rev. C}
  \textbf{\bibinfo{volume}{80}}, \bibinfo{pages}{054309}.

\bibitem[{\citenamefont{Willis} \emph{et~al.}(1989)\citenamefont{Willis,
  Morlet, Marty, Djalali, Bohle, Diesener, Richter, and Stein}}]{Willis:1989}
\bibinfo{author}{\bibnamefont{Willis}, \bibfnamefont{A.}},
  \bibinfo{author}{\bibfnamefont{M.}~\bibnamefont{Morlet}},
  \bibinfo{author}{\bibfnamefont{N.}~\bibnamefont{Marty}},
  \bibinfo{author}{\bibfnamefont{C.}~\bibnamefont{Djalali}},
  \bibinfo{author}{\bibfnamefont{D.}~\bibnamefont{Bohle}},
  \bibinfo{author}{\bibfnamefont{H.}~\bibnamefont{Diesener}},
  \bibinfo{author}{\bibfnamefont{A.}~\bibnamefont{Richter}}, and
  \bibinfo{author}{\bibfnamefont{H.}~\bibnamefont{Stein}},
  \bibinfo{year}{1989}, \bibinfo{journal}{Nuclear Physics A}
  \textbf{\bibinfo{volume}{499}}, \bibinfo{pages}{367}.

\bibitem[{\citenamefont{Wolf} \emph{et~al.}(1987)\citenamefont{Wolf, Casten,
  and Warner}}]{Wolf:1987}
\bibinfo{author}{\bibnamefont{Wolf}, \bibfnamefont{A.}},
  \bibinfo{author}{\bibfnamefont{R.~F.} \bibnamefont{Casten}}, and
  \bibinfo{author}{\bibfnamefont{D.~D.} \bibnamefont{Warner}},
  \bibinfo{year}{1987}, \bibinfo{journal}{Phys. Lett. B}
  \textbf{\bibinfo{volume}{190}}, \bibinfo{pages}{19}.

\bibitem[{\citenamefont{W\"ortche}(1994)}]{Woertche:1994}
\bibinfo{author}{\bibnamefont{W\"ortche}, \bibfnamefont{H.~J.}},
  \bibinfo{year}{1994}, \bibinfo{type}{Doctoral thesis},
  \bibinfo{school}{Technische Universit\"at Darmstadt}.

\bibitem[{\citenamefont{Yevetska} \emph{et~al.}(2010)\citenamefont{Yevetska,
  Enders, Fritzsche, von Neumann-Cosel, Oberstedt, Richter, Savran, and
  Sonnabend}}]{Yevetska:2009}
\bibinfo{author}{\bibnamefont{Yevetska}, \bibfnamefont{O.}},
  \bibinfo{author}{\bibfnamefont{J.}~\bibnamefont{Enders}},
  \bibinfo{author}{\bibfnamefont{M.}~\bibnamefont{Fritzsche}},
  \bibinfo{author}{\bibfnamefont{P.}~\bibnamefont{von Neumann-Cosel}},
  \bibinfo{author}{\bibfnamefont{S.}~\bibnamefont{Oberstedt}},
  \bibinfo{author}{\bibfnamefont{A.}~\bibnamefont{Richter}},
  \bibinfo{author}{\bibfnamefont{D.}~\bibnamefont{Savran}}, and
  \bibinfo{author}{\bibfnamefont{K.}~\bibnamefont{Sonnabend}},
  \bibinfo{year}{2010}, \bibinfo{journal}{Phys. Rev. C} , \bibinfo{pages}{(in
  press)}.

\bibitem[{\citenamefont{Zamick}(1985)}]{Zamick:1985}
\bibinfo{author}{\bibnamefont{Zamick}, \bibfnamefont{L.}},
  \bibinfo{year}{1985}, \bibinfo{journal}{Phys. Rev. C}
  \textbf{\bibinfo{volume}{31}}, \bibinfo{pages}{1955}.

\bibitem[{\citenamefont{Zamick}(1986{\natexlab{a}})}]{Zamick:1986b}
\bibinfo{author}{\bibnamefont{Zamick}, \bibfnamefont{L.}},
  \bibinfo{year}{1986}{\natexlab{a}}, \bibinfo{journal}{Phys. Lett. B}
  \textbf{\bibinfo{volume}{167}}, \bibinfo{pages}{1}.

\bibitem[{\citenamefont{Zamick}(1986{\natexlab{b}})}]{Zamick:1986a}
\bibinfo{author}{\bibnamefont{Zamick}, \bibfnamefont{L.}},
  \bibinfo{year}{1986}{\natexlab{b}}, \bibinfo{journal}{Phys. Rev. C}
  \textbf{\bibinfo{volume}{33}}, \bibinfo{pages}{691}.

\bibitem[{\citenamefont{Zamick and Zheng}(1992)}]{Zamick:1992}
\bibinfo{author}{\bibnamefont{Zamick}, \bibfnamefont{L.}}, and
  \bibinfo{author}{\bibfnamefont{D.~C.} \bibnamefont{Zheng}},
  \bibinfo{year}{1992}, \bibinfo{journal}{Phys. Rev. C}
  \textbf{\bibinfo{volume}{46}}, \bibinfo{pages}{2106}.

\bibitem[{\citenamefont{Zawischa}(1998)}]{Zawischa:1998}
\bibinfo{author}{\bibnamefont{Zawischa}, \bibfnamefont{D.}},
  \bibinfo{year}{1998}, \bibinfo{journal}{J. Phys. G}
  \textbf{\bibinfo{volume}{24}}, \bibinfo{pages}{683}.

\bibitem[{\citenamefont{Zawischa} \emph{et~al.}(1990)\citenamefont{Zawischa,
  Macfarlane, and Speth}}]{Zawischa:1990a}
\bibinfo{author}{\bibnamefont{Zawischa}, \bibfnamefont{D.}},
  \bibinfo{author}{\bibfnamefont{M.}~\bibnamefont{Macfarlane}}, and
  \bibinfo{author}{\bibfnamefont{J.}~\bibnamefont{Speth}},
  \bibinfo{year}{1990}, \bibinfo{journal}{Phys. Rev. C}
  \textbf{\bibinfo{volume}{42}}, \bibinfo{pages}{1461}.

\bibitem[{\citenamefont{Zawischa and Speth}(1990)}]{Zawischa:1990b}
\bibinfo{author}{\bibnamefont{Zawischa}, \bibfnamefont{D.}}, and
  \bibinfo{author}{\bibfnamefont{J.}~\bibnamefont{Speth}},
  \bibinfo{year}{1990}, \bibinfo{journal}{Phys. Lett. B}
  \textbf{\bibinfo{volume}{252}}, \bibinfo{pages}{4}.

\bibitem[{\citenamefont{Zawischa and Speth}(1994)}]{Zawischa:1994a}
\bibinfo{author}{\bibnamefont{Zawischa}, \bibfnamefont{D.}}, and
  \bibinfo{author}{\bibfnamefont{J.}~\bibnamefont{Speth}},
  \bibinfo{year}{1994}, \bibinfo{journal}{Nucl. Phys. A}
  \textbf{\bibinfo{volume}{569}}, \bibinfo{pages}{343}.

\bibitem[{\citenamefont{Ziegler} \emph{et~al.}(1993)\citenamefont{Ziegler,
  Huxel, von Neumann-Cosel, Rangacharyulu, Richter, Spieler, De~Coster, and
  Heyde}}]{Ziegler:1993}
\bibinfo{author}{\bibnamefont{Ziegler}, \bibfnamefont{W.}},
  \bibinfo{author}{\bibfnamefont{N.}~\bibnamefont{Huxel}},
  \bibinfo{author}{\bibfnamefont{P.}~\bibnamefont{von Neumann-Cosel}},
  \bibinfo{author}{\bibfnamefont{C.}~\bibnamefont{Rangacharyulu}},
  \bibinfo{author}{\bibfnamefont{A.}~\bibnamefont{Richter}},
  \bibinfo{author}{\bibfnamefont{C.}~\bibnamefont{Spieler}},
  \bibinfo{author}{\bibfnamefont{C.}~\bibnamefont{De~Coster}}, and
  \bibinfo{author}{\bibfnamefont{K.}~\bibnamefont{Heyde}},
  \bibinfo{year}{1993}, \bibinfo{journal}{Nucl. Phys.}
  \textbf{\bibinfo{volume}{A564}}, \bibinfo{pages}{366}.

\bibitem[{\citenamefont{Ziegler} \emph{et~al.}(1990)\citenamefont{Ziegler,
  Rangacharyulu, Richter, and Spieler}}]{Ziegler:1990}
\bibinfo{author}{\bibnamefont{Ziegler}, \bibfnamefont{W.}},
  \bibinfo{author}{\bibfnamefont{C.}~\bibnamefont{Rangacharyulu}},
  \bibinfo{author}{\bibfnamefont{A.}~\bibnamefont{Richter}}, and
  \bibinfo{author}{\bibfnamefont{C.}~\bibnamefont{Spieler}},
  \bibinfo{year}{1990}, \bibinfo{journal}{Phys. Rev. Lett.}
  \textbf{\bibinfo{volume}{65}}, \bibinfo{pages}{2515}.

\bibitem[{\citenamefont{Zilges} \emph{et~al.}(1996)\citenamefont{Zilges, von
  Brentano, Herzberg, Kneissl, Margraf, and Pitz}}]{Zilges:1996}
\bibinfo{author}{\bibnamefont{Zilges}, \bibfnamefont{A.}},
  \bibinfo{author}{\bibfnamefont{P.}~\bibnamefont{von Brentano}},
  \bibinfo{author}{\bibfnamefont{R.-D.} \bibnamefont{Herzberg}},
  \bibinfo{author}{\bibfnamefont{U.}~\bibnamefont{Kneissl}},
  \bibinfo{author}{\bibfnamefont{J.}~\bibnamefont{Margraf}}, and
  \bibinfo{author}{\bibfnamefont{H.~H.} \bibnamefont{Pitz}},
  \bibinfo{year}{1996}, \bibinfo{journal}{Nucl. Phys. A}
  \textbf{\bibinfo{volume}{599}}, \bibinfo{pages}{147c}.

\bibitem[{\citenamefont{Zilges} \emph{et~al.}(1990)\citenamefont{Zilges, von
  Brentano, Richter, Heil, Kneissl, Pitz, and Wesselborg}}]{Zilges:1990b}
\bibinfo{author}{\bibnamefont{Zilges}, \bibfnamefont{A.}},
  \bibinfo{author}{\bibfnamefont{P.}~\bibnamefont{von Brentano}},
  \bibinfo{author}{\bibfnamefont{A.}~\bibnamefont{Richter}},
  \bibinfo{author}{\bibfnamefont{R.~D.} \bibnamefont{Heil}},
  \bibinfo{author}{\bibfnamefont{U.}~\bibnamefont{Kneissl}},
  \bibinfo{author}{\bibfnamefont{H.~H.} \bibnamefont{Pitz}}, and
  \bibinfo{author}{\bibfnamefont{C.}~\bibnamefont{Wesselborg}},
  \bibinfo{year}{1990}, \bibinfo{journal}{Phys. Rev. C}
  \textbf{\bibinfo{volume}{42}}, \bibinfo{pages}{1945}.

\end{thebibliography}

\end{document}